\documentclass[12pt,a4wide]{book}
\usepackage{graphicx}
\usepackage{hyperref}
\usepackage{amsmath}
\usepackage{makecell}
\usepackage{parskip}
\usepackage{mathtools}
\usepackage{float}
\usepackage{algorithm}
\usepackage{algpseudocode}
\usepackage{enumerate}
\usepackage{adjustbox}
\usepackage{longtable}
\usepackage{enumitem}
\usepackage{booktabs}
\usepackage{amssymb}
\usepackage{pdfpages}
\usepackage{adjustbox}
\usepackage{rotating}
\usepackage{amssymb}
\usepackage{mathrsfs}
\usepackage{multirow}
\usepackage[font=small,labelfont=bf]{caption}
\usepackage[font=footnotesize,caption=false]{subfig}
\usepackage{ lscape}
\usepackage{etoolbox}
\usepackage{array, booktabs, caption}
\usepackage[font=small,labelfont=bf]{caption}
\usepackage[font=footnotesize,caption=false]{subfig}
\usepackage{makecell}

\usepackage[T1]{fontenc}
\usepackage{algpseudocode}
\usepackage[symbol]{footmisc}

\usepackage{makecell}
\usepackage{mathtools}
\usepackage[english]{babel}
\usepackage[T1]{fontenc}
\usepackage{enumerate}
\usepackage{longtable}
\usepackage{enumitem}
\usepackage{booktabs}
\usepackage{amssymb}
\usepackage{makecell}

\usepackage[T1]{fontenc}
\usepackage{color}
\usepackage{xcolor}

\usepackage{graphicx}
\usepackage{mathtools}
\usepackage{amsfonts}
\usepackage{amssymb}
\usepackage{graphics}
\usepackage{color}
\usepackage{pgfplots}
\usepackage{hyperref}
\usepackage{listings}

\usepackage{booktabs}
\usepackage{pgfplots}
\usepackage{amssymb}
\usepackage{mathrsfs}
\usepackage{amsmath}
\usepackage{hyperref}
\usepackage{longtable}
\usepackage{multirow}
\usetikzlibrary{arrows,automata}

\usepackage{float}
\usepackage[caption = false]{subfig}

\usepackage{listings}

\usepackage{url}
\usepackage{multirow}
\usepackage{multicol}
\usepackage{amssymb}
\usepackage{mathrsfs}
\usepackage{enumerate}
\usepackage{enumitem}
\newsavebox{\measurebox}

\usepackage{cite}
\usepackage{epsf} 
\usepackage{latexsym} 
\usepackage{graphicx}
\usepackage{amssymb}
\usepackage{amsmath,amssymb}
\usepackage{amsfonts}
\usepackage{fancyhdr}
\usepackage{enumitem}
\usepackage{chngcntr}
\usepackage{floatpag}
\usepackage{latexsym}
\usepackage{tabularx, booktabs} 
\usepackage[title,titletoc,toc]{appendix}
\usepackage{subfig}
\usepackage{amsmath}
\usepackage{mathrsfs}
\usepackage{afterpage}
\newcommand\blankpage{%
    \null
    \thispagestyle{empty}%
    \addtocounter{page}{-1}%
    \newpage}
    
\makeatletter
\newcommand\fs@norules{\def\@fs@cfont{\bfseries}\let\@fs@capt\floatc@ruled
  \def\@fs@pre{}%
  \def\@fs@post{}%
  \def\@fs@mid{\kern3pt}%
  \let\@fs@iftopcapt\iftrue}      
 
\usepackage{longtable}  
  
\usepackage{pgfplots}  
\usepackage{multirow}
  
\usepackage{colortbl}

\definecolor{webgreen}{rgb}{0, 0.5, 0} 
\definecolor{webblue}{rgb}{0, 0, 0.5} 
\definecolor{webred}{rgb}{0.0, 0, 0} 
\definecolor{webred}{rgb}{0.5, 0, 0} 
\definecolor{webblack}{rgb}{0, 0, 0} 

\usepackage{booktabs}
\usepackage{makecell}

\usepackage{longtable}

\fancyhfoffset[L]{0.5cm}

\makeatother

\usepackage[toc,nonumberlist]{glossaries}
\makeglossaries
\glsdisablehyper

\newcolumntype{Y}{>{\centering\arraybackslash}X}
\newtheorem{theoremm}{Theorem}[chapter]
\newtheorem{eqed}{Example}[chapter]
\newtheorem {lemmaa}{Lemma}[chapter]

\newtheorem{defnn}{Definition}[chapter]
\newtheorem {corollaryy}{Corollary}[chapter]
\newtheorem {axiomm}{Axiom}[chapter]
\newtheorem {inferencee}{Inference}[chapter]

\newtheorem {hypothesiss}{Hypothesis}[chapter]
\newtheorem {conjecturee}{Conjecture}[chapter]

\newtheorem {prp}{Property}[chapter]
\newenvironment{example}{\begin{eqed} \rm}{\hfill$\Box$ \end{eqed}}

\newenvironment{infprf}{\noindent {\bf Informal Proof :\ } }{$\Box$ }

\newenvironment{Informal Proof}{\begin{prp} \sl}{\end{prp}}

\newenvironment{proof of correctness}{\noindent {\bf Proof of Correctness :\ } }{\hfill$\Box$ }

\newenvironment{sketch of proof}{\noindent {\bf Sketch of proof :\ } }{\hfill$\Box$ }

\newtheorem {definition}{Definition}

\newtheorem {procd}{Procedure}

\newcommand{\eat}[1]{}

\definecolor{mygreen}{rgb}{0,0.6,0}
\definecolor{mygray}{rgb}{0.5,0.5,0.5}
\definecolor{mymauve}{rgb}{0.58,0,0.82}
\begin{document}
\sloppy

\pagestyle{empty}
\pagestyle{empty} 
	\begin{center} 
	\Large{\textbf{\textsf{{Cellular Automata: Temporal Stochasticity and Computability}}}} \\ 
	\end{center}
	\begin{center} 
	\vspace{4.0cm}
	 \textbf{\normalsize{SUBRATA PAUL}} \\
	\vspace{4.8cm}
        \begin{center}
 		 \includegraphics[height=4cm, width=4cm]{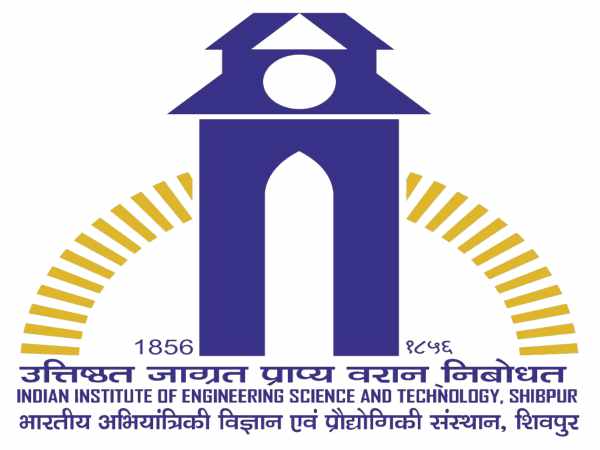}
        \end{center} 
        \vspace{1.2cm}     
      \scriptsize{\textbf{DEPARTMENT OF INFORMATION TECHNOLOGY}}\\ 
      \scriptsize\textbf{{INDIAN INSTITUTE OF ENGINEERING SCIENCE AND TECHNOLOGY, SHIBPUR}} \\
     \scriptsize{\textbf{HOWRAH, WEST BENGAL, INDIA-711103}}\\
               \vspace{0.5cm} 
   \large{\textbf{2022}} \\  
  	\end{center} 

\newpage


\pagestyle{empty} 
	\begin{center} 
	\Huge{\textbf{\textsf{{Cellular Automata: Temporal Stochasticity and Computability}}}} \\
        \vspace{1.5cm} 
        \Large{\textbf{Subrata Paul}}\\   
        	 \vspace{0.3cm} 
     \large{{Registration No. 2020ITM011}} \\
	     \vspace{1.2cm}
        \small{\em{A report submitted in partial fulfillment for the degree of}}\\     
        \vspace{0.12in} 
        \small{\textbf{Masters of Technology}}\\ 
        \small{\textbf{in}}\\ 
        \small{\textbf{Information Technology}}\\ 
        
     \vspace{1.0cm} 
	\large{\em{Under the supervision of}}\\ 
	\vspace{0.10cm} 
	\large{\textbf{Dr. Sukanta Das}}\\ 
	\small{Associate Professor}\\
	\small{Department of Information Technology}\\
	{Indian Institute of Engineering Science and Technology, Shibpur}\\       
        
        \vspace{0.45in} 
        \begin{center}
		\includegraphics[height=4cm, width=4cm]{logo}
        \end{center}
		\vspace{0.3cm}

	\normalsize{\textbf{Department of Information Technology}}\\ 
\normalsize{\textbf{Indian Institute of Engineering Science and Technology, Shibpur}}\\
	\normalsize{\textbf{Howrah, West Bengal, India -- 711103}}\\ 
    \normalsize{\textbf{2022}}\\ 
	\end{center} 

\newpage

\pagestyle{empty} 
\begin{center} 
\begin{figure}[h]
\vspace{-1.5cm} 
\centering
	\includegraphics[height=2.5cm, width=2.5cm]{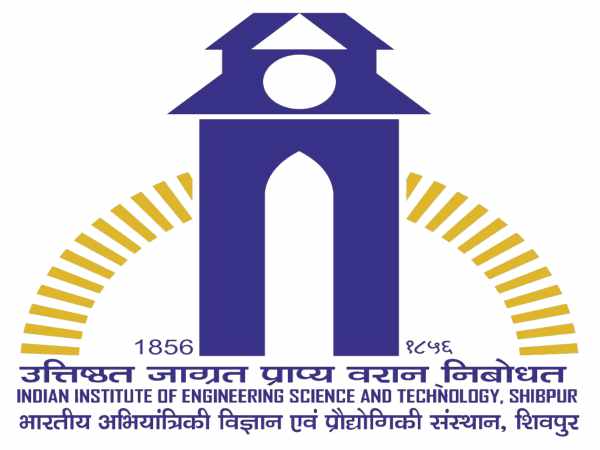}
\end{figure}
\small{\textbf{Department of Information Technology}}\\ 
	\small{\textbf{Indian Institute of Engineering Science and Technology, Shibpur}}\\
	\small{\textbf{Howrah, West Bengal, India -- 711103}}\\  
\vspace{0.40in} 

{\Large \bf CERTIFICATE OF APPROVAL}\\ 
\end{center} 
\vspace{0.22in} 
\normalsize{It is certified that, the thesis entitled \emph{\bf ``Cellular Automata: Temporal Stochasticity and Computability''}, is a record of bonafide work carried out under my guidance and supervision by \textbf{Subrata Paul} in the Department of Information Technology of Indian Institute of Engineering Science and Technology, Shibpur. 
 
In my opinion, the thesis has fulfilled the requirements for the degree of M.Tech in Information Technology of Indian Institute of Engineering Science and Technology, Shibpur. The work has reached the standard necessary for submission and, to the best of my knowledge, the results embodied in this thesis have not been submitted for the award of any other degree or diploma.

\vspace{0.2in} 
\begin{flushright}
	\includegraphics[height=2\baselineskip]{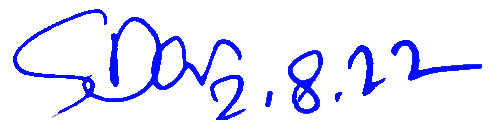} \par
			\textbf{(Dr. Sukanta Das)}  \\
		\small{Associate~ Professor} \\
	\small{Dept. of Information Technology} \\
		\small{Indian Institute of Engineering Science and Technology,} \\
		\small{Shibpur, Howrah, West Bengal, India --711103} \\
\end{flushright}
\vspace{0.1in}
\hspace{-0.2in}{Counter signed by:}
\begin{flushright}
	\includegraphics[height=4\baselineskip]{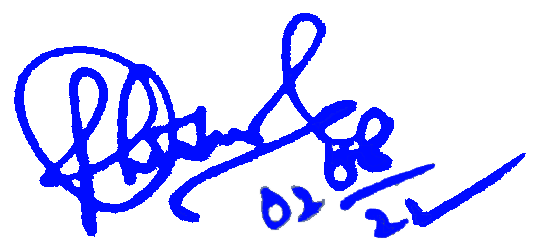} \par
	\textbf{(Dr. Prasun Ghosal)}  \\
	\small{Associate~ Professor \& Head} \\
	\small{Dept. of Information Technology} \\
	\small{Indian Institute of Engineering Science and Technology,} \\
	\small{Shibpur, Howrah, West Bengal, India --711103} \\
\end{flushright}

	


\vspace{1mm}
} 
\newpage 

\newpage

\vspace*{\fill}  

\begin{quote}

\centering

\begin{large}


\textbf{\Large{\textit{Dedicated}}}\\

\vspace{0.25cm}

\textbf{\Large{\textit{to}}}\\

\vspace{0.25cm}
\textbf{\Large{\textit{My Mother}}}\\
\vspace{0.25cm}
\textbf{\Large{\textit{and}}}\\
\vspace{0.25cm}
\textbf{\Large{\textit{to all the struggling people over the world}}}\\
\vspace{0.25cm}

\vspace{0.25cm}


\vspace{0.25cm}


\end{large}

\end{quote}

\vspace*{\fill} 

\clearpage

\pagestyle{plain}
\pagenumbering{roman}
\afterpage{\blankpage}

\cleardoublepage
\phantomsection
\addcontentsline{toc}{chapter}{Acknowledgement}
\begin{center}
\vspace{-5.5cm}
 \textbf{\Large Acknowledgement}
\end{center}

Foremost, I would like to express my sincere gratitude to my advisor Dr. Sukanta Das, Associate Professor, Department of Information Technology, Indian Institute of Engineering Science and Technology (IIEST), Shibpur, for his continuous support and help in all the stages of preparing this dissertation. I am also thankful for his patience, motivation, enthusiasm, and immense knowledge. His guidance helped me all the time for writing this thesis. I could not have imagined having a better advisor and mentor for my thesis. During this journey, I have learned lot of things from him, especially how to be disciplined in research and in life.

I'd also like to take this opportunity to express my deep respect and appreciation to Dr. Kamalika Bhattacharjee, Assistant Professor, Department of Computer Science, NIT Tiruchirapally, Tiruchirapally, for her valuable suggestions and advice, which have helped me to be more analytical and rigorous in using scientific methodology. At the same time, I want to express my unending gratitude to Prof. Biplab K. Sikdar, Professor, Department of Computer Science and Technology, IIEST, Shibpur, and Dr. Souvik Roy, C3iHub, Indian Institute of Technology, Kanpur, whose intellectual interactions have benefited me tremendously. 

All the reported works were accomplished through joint efforts. In the research ``Affinity Classification Problem by Stochastic Cellular Automata'', where I worked with Dr. Kamalika Bhattacharjee. Dr. Souvik Roy developed the Temporally Stochastic theory. I've worked with him to investigate the dynamics and one use for it. 

I am grateful for the financial support the Indian Institute of Engineering Science and Technology, Shibpur provided for my research during the tenure of my M.Tech.

I am grateful to the current head of department, Associate Prof. Prasun Ghosal, of the Department of Information Technology at IIEST, Shibpur, as well as all the other respected professors for being so kind as to provide their support at various phases. In addition to my advisor, I would like to express my deepest appreciation to each and every member of the M.Tech committee for their insights and technical suggestions. I want to express my gratitude to the department's technical and non-technical employees (Malay-sir, Suman-da, and Dinu-da) for their support and dedication.

I also want to express my sincere thanks to all of my friends for their continuous encouragement. In this list, I'd want to specifically thank my friend Bishwayan, whose encouragement helped me immensely, especially during a difficult period. He never left my side, pushing and inspiring me through both the good and bad times. A special mention should be made to my friends Ranit, Pritam, Sourav, Shilpa, and Prashant for their friendship, love, and emotional guidance. I also thank my labmates Rakesh da, Partha, Vaibhav, Reema, Raju, Subham, and Heamant for their support all throughout the last two years. I sincerely appreciate the love and support from my friends Sohinee, Subhajit, Sarani, Puja, Sarbajaya, Krishnendu, and Anindita.

Most importantly, I would like to express my heartfelt respect to my parents (Late Bimal Chandra Paul and Ms. Jyotsna Rani Pal) for their support, sacrifice, and inspiration from the very beginning of my academic career. I would also like to thank one of my teachers (late Arunima Mukharjee ) for her encouragement, support, and sacrifice during my whole academic career. She was truly my mentor, philosopher, and guide in every aspect. I want to acknowledge the guidance of the teachers and mentors of primary and secondary education, with whom I stepped into the world of education.

\vspace{35mm}

\noindent{\bf Dated:$~$}  \includegraphics[height=1.2\baselineskip]{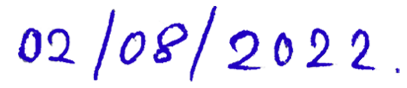}\hspace{1.9 in} \includegraphics[height=1.9\baselineskip]{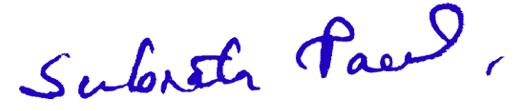} \\
{\bf Indian Institute of Engineering Science} \hspace*{0.4in} \dotfill \\
{\bf and Technology, Shibpur} \hspace*{2.0in} {\bf(Subrata Paul)} \\
{\bf Howrah, West Bengal, India} \hspace*{3.3cm} {\textbf{Reg. No.: 2020ITM011}}


\newpage
\cleardoublepage
\phantomsection
\addcontentsline{toc}{chapter}{Abstract}
\afterpage{\blankpage}
\chapter*{Abstract}
\label{abstr}

\textbf{T}his work introduces temporally stochasticity in cellular automata and the behavior of such cellular automata. The work also explores the computational ability of such cellular automaton that illustrates  the computability of solving the affinity classification problem. In addition to that, a cellular automaton, defined over Cayley tree, is shown as the classical searching problem solver.

\textbf{T}he proposed temporally stochastic cellular automata deals with two elementary cellular automata rules, say $f$ and $g$. The $f$ is the default rule, however, $g$ is temporally applied to the overall system with some probability $\tau$ which acts as a noise in the system. As the mathematical analysis of such a system is a difficult task, we use qualitative and quantitative simulations to get insights into the dynamics of this temporally stochastic cellular automata. We are interested in exploring the questions $-$ is it possible that two periodic (resp. chaotic) rules together depict chaotic (resp. periodic) dynamics. To answer these questions, we fully classify temporally stochastic cellular automata. Here, we are also particularly interested in studying phase transition and various types of class transition dynamics.

\textbf{A}fter exploring the dynamics of temporally stochastic cellular automata (TSCAs), we study the dynamical behavior of these temporally stochastic cellular automata (TSCAs) to identify the TSCAs that converge to a fixed point from any seed. We apply each of the convergent TSCAs to some standard datasets and observe the effectiveness of each TSCA as a pattern classifier. It is observed that the proposed TSCA-based classifier shows competitive performance in comparison with existing classifier algorithms.

\textbf{T}he work introduces a new problem in the field of cellular automata, named as, \emph{affinity classification problem} which is a generalization of the \emph{density classification problem}. To solve this problem, we use temporally stochastic cellular automata where two rules are stochastically applied in each step to all cells of the automata. Our model is defined on a two-dimensional grid and has affection capability. We show that this model can be used in several applications, like modeling self-healing systems.

Finally, the work introduces a new model of computing unit developed around cellular automata to reduce the workload of the Central Processing Unit (CPU) of a machine to compute. Each cell of the computing unit acts as a tiny processing element with attached memory. Such a CA is implemented on the Cayley Tree to realize efficient solutions for diverse computational problems. To prove the effectiveness of this model, this work targets solutions for the Searching problems. However, such a cellular structure can be employed to solve various computational problems, Searching problem is one of them.


\cleardoublepage
\tableofcontents

\cleardoublepage
\phantomsection
\addcontentsline{toc}{chapter}{\listfigurename}
\listoffigures

\cleardoublepage
\phantomsection
\addcontentsline{toc}{chapter}{\listtablename}
\listoftables


%
\clearpage

\pagestyle{fancy}
\pagenumbering{arabic} 
\chapter{Introduction}
\label{chap1}
The growth of science has always been captivated by the wonders of nature. The way that nature operates is incredibly unpredictable, with each living thing playing a specific role in determining overall behavior. However, the Turing Machine~\cite{turing1937computable, Turing}, whose computation was controlled by a centralized control tape head, formed the foundation of the mathematical model used from the beginning of the modern computer era.  Even von Neumann's proposed computer design is managed by the CPU, which is also a centrally controlled mechanism.

From the very first computer to modern smart-phones, all operated in a centralized manner. We may see patterns in the natural world, such as those found in snowflakes, ant motion, sea shells, etc., where we observe that centralized control may be emerging. For example, it appears that a leader may be in charge of the entire colony.  However, each ant makes its own decisions and performs its own duty. 

In the early $20^{th}$ century, a new field of research known as Network Science was introduced in this area to examine individualities and parallelism. Various models have been presented throughout this period, many of which are bio-inspired and provide distributed decentralized computing. One of the most significant advancements in this area was the invention of cellular automata. Decentralization has been a widely used notion in computation since the first widely utilized distributed systems, like Ethernet~\cite{Clark21, Maarten7}, were introduced. Since the distributed system \emph{Internet} caused a ``paradigm shift'', the idea of decentralization has cropped up often in almost all areas of human endeavor.

An array of networked, yet independent, computer components make up a distributed system. These components only communicate with one another to coordinate their functions. From the standpoint of a process, a distributed system may be seen as a collection of geographically scattered processes that only communicate via message exchange. As a result, the processes in the system can only speak to one another while doing a computational task. In a distributed computing architecture, the supervision and control of the computation are not exercised by a single entity. The components and processes of a distributed computation may be recognized by their unique identifiers. A central organization is required for a system with detectable individual identities, or a ``non-anonymous system'', in order to give the processes their distinctive individuality. The fundamental tenets of distributed control are violated by this. As a result, a distributed system must be anonymous by definition~\cite{Maarten7, tel73}. A number of formal models have already been published for distributed systems~\cite{consys1,  Milner, Reisig}, providing useful insights. Conversely, because of their innate parallelism, cellular automata (CAs) can always be a natural choice for distributed computing frameworks.

In the early 1950's Jon von Neumann first introduced a self-reproducing Automata \cite{Neuma66}; which was later named as ``Cellular Automata''. He introduced constructive universality in cellular automata to study the implementability of self-reproducing machines and the concept of computational universality. A computing machine is said to be computationally universal if it is capable of simulating any other computing machine; for example, John von Neumann’s universal constructor is a machine capable of emulating other machines that can be embedded in its cellular automaton. Computational universality and constructive universality are conceptually related properties, but a machine does not need to possess a universal computer to be a universal constructor. Jon von Neumann, in fact, demonstrated that it is possible to implement a Turing machine in his cellular automaton, but pointed out that a Turing machine is not a necessary component of the universal constructor.

The study of biological phenomena has become a focus for all researchers in the domains of science, philosophy, technology, and other related fields. In this direction, Christopher Langton introduced a research area titled ``Artificial Life'' that uses cellular automata (CA) as a natural basis for the artificial life model. Some of the CAs have inherited some characteristics of biological systems, such as \emph{self-replication, self-organization, self-healing}, etc. Conway's ``Game of Life''~\cite{Gardner70} is an important example of such type of CA. The focus will now shift to another attribute termed \emph{affection}. As is well known, every biological system will readily show fondness towards anything. 

Over the globe, scientists are struggling in their own way to make \emph{intelligent machines}. In general, a living system is known as an \emph{intelligent machine}. The Turing Test (TT), a behavioral test developed by Alan Turing in 1950~\cite{TT1950}, is one of the fundamental tools to analyze if a machine is intelligent. A machine's performance in TT is evaluated based on how it responds to a series of questions, and passing TT implies that the machine is acting intelligently. According to Turing, such a machine should be recognized as a thinking machine. However, this test is based on a functional approach, where the machine is labeled \emph{intelligent} if it successfully completes specific tasks, or acts intelligently. However, this method does not care if the system has intrinsic intelligence like a living being. The Chinese Room Argument~\cite{Searle1980}, which distinguishes between what is intelligent and what behaves intelligently, is well known. In fact, this problem inspired the development of the strong AI paradigm ~\cite{Kenneth1988}. By going beyond this functional perspective, we want to further the notion of intelligent machines in our study. If a machine has similar properties to a living system, then the machine is called an \emph{intelligent machine}. It is crucial to define the term \emph{intelligent} before providing a response to the question, ``How intelligent has a machine become?''. Living systems are frequently referred to as intelligent systems. If a machine has similar properties to a living system, then the machine is called \emph{intelligent machine}. By looking at how the terms \emph{machine} and \emph{intelligent} are typically used, it is able to structure the definitions to represent that use as closely as possible. It is very challenging to escape the conclusion that the meaning and the answer to the question, \emph{``How intelligent has a machine become?''}.

\section{Motivation and Objective of the thesis}
Exploring the processing power of decentralized computational models, namely distributed computing on cellular automata, is the objective of this thesis. Each cell in a cellular automaton (CA) is made up of a finite automaton that communicates with its neighbors in order to move to its next state~\cite{Neuma66}. The CA is distributed across a regular grid. The appeal of a CA is that it generates complicated (global) behavior from simple local interactions. Some people even assert that nature functions as a quantum information processing system~\cite{PhysRevLett.88.237901}, with CA serving as the mechanism for this processing~\cite{Zuse1982,wolfram2002new}. In fact, using CAs as models for concurrency and distributed systems was one of the pioneering efforts in CAs. According to published research, a few distributed systems issues have also been solved computationally using CAs~\cite{Cor,Smith71}. In this dissertation, we investigate how cellular automata (CAs) can be used to resolve the following issues:
\begin{itemize}
	\item Watch how a system behaves when noise has an impact on it.
	\item Such a system is capable of classifying patterns.
	\item Affinity classification problem,
	\item Searching problem.
\end{itemize}
When a system evolves in real life and noise enters the system, the behavior of the system may alter. In this direction, researchers have proposed stochastic CA in the literature~\cite{ Heinz9, fates13, fates:LIPIcs}, where the system employs, several rules for different cells, rather than a single local rule. This dissertation examines these kinds of CAs from noisy real-time systems. A variant of the CA model, temporary stochastic cellular automata, a new form of stochastic CA, has been proposed for studying these kinds of changes. The vast majority of scientific research is based on classical CAs. However, this dissertation shows a willingness to explore a variant of CA, which we refer to as temporally stochastic cellular automata. In order to do that, analysis of the CA's behavior and categorization into different classes is our objective. Using this classification, we choose the convergent TSCAs and utilize them to construct a pattern classifier. For the same purpose, certain algorithms have been developed in the past. But compared to the existing techniques, our proposed model produces outcomes that are competitive.

All researchers had the vision of making an intelligent system from the beginning. Living systems are often used to describe intelligent systems. When a machine can imitate a living system, it is referred to be an intelligent system. One can find that the strength of the living system is \emph{affection}. In our study, we introduced a cellular automata model with affection capabilities, and we believe that this feature lends our model some intelligence.
The next state of a cell in a standard cellular automaton (CA) depends only on the nearest neighborhood configuration from the previous time step, making them ahistoric (memoryless). However, by taking automata with memory capacity into consideration~\cite{alonso2009memory,CPLX_CPLX21495,doi:10.,Alonso-Sanz2013,Seck12,Das2022}, the conventional CA framework has been expanded. The update criteria of the CA are not changed, but each cell now contains a history of all previous iterations according to a weighted mean of all previous states. The process of performing computations entirely within memory is known as in-memory computation (or in-memory computing). This phrase often refers to extensive, sophisticated computations that must be performed on a cluster of machines using specialist systems software. As a cluster, the machines pool their memory so that the computation is basically performed over several machines and makes use of the combined memory of all the machines. In \cite{Das2022}, a variant of CA is introduced to implement those types of notions utilizing cellular automata, where a memory unit is attached to each CA cell and an additional processing unit is also added to the memory to aid in calculation and lessen the workload on the CPU. Using this notion, we develop a model to solve well-known \emph{searching problem}.
\section{Contribution of the thesis}
In light of the aforementioned goal, we have conducted this study. The following is a summary of the major findings from the research activities:
\begin{itemize}
	\item We have explored the behaviors of temporary stochastic cellular automata, where we deal with elementary cellular automata (ECAs). 
	\item We have found that whereas certain stochastic CAs are insensitive to the temporal noise rate, others are impacted by temporal noise. But even these CAs have produced a wide range of outcomes. 
	
	\item It is noteworthy that stochastic CAs with (at least one) chaotic rule have often shown less resistance during phase transition (i.e. the critical value of the noise rate is low). But during phase transition, the stochastic CAs devoid of any chaotic rule has shown greater resilience (i.e., the critical value of the noise rate is high). This is another intriguing finding from this research.
	\item We have found the convergent temporally stochastic cellular automata (TSCAs) that were utilized in the development of a two-class pattern classifier after evaluating their dynamics. In this situation, the proposed design of the TSCA-based two-class pattern classifier gives a competitive performance in contrast to existing common approaches.
	\item We have introduced a new problem known as the affinity classification problem. We have developed a devoted machine that is integrated into a two-dimensional cellular automaton with periodic boundary conditions and Moore neighborhood dependence. Our model may be described by four parameters: $K; \phi(x);\psi(x);$ and $p$, and has affection capabilities to a converging point, all-1 or all-0.
	\item By adjusting its parameters alone, the dedicated machine may partially solve the density classification problem.
	\item Using this concept, we may develop a self-healing system. We are aware that self-healing allows any species to survive and develop.
	\item To reduce the workload on the CPU, a new kind of cellular automata-based model has been developed in which each cell of the CA has a memory and an additional processing unit attached to it.
	\item In this type of CA, each cell has some additional processing power that helps with the circumstances at hand. We have shown how the model can solve the search problem. The model is able to decide the search's outcome by perceiving just one node.
	
	\item In the CA over Cayley tree, the arrangement of elements in the array does not affect the flow of the scheme. No matter if the elements in the array are already sorted, reverse sorted, or randomly placed, the scheme works the same for all these cases, and thus the time complexity for all such cases is the same, $O(k + log(n))$ where $k$ is the key element and $n$ is a total number of elements presents in the set of elements.

\end{itemize}

\section{Organization of the thesis}
In this section, we provide an organization of the thesis along with a summary of each chapter. The contribution of the thesis to the aforementioned topic is divided. Before going to discuss the different chapters, the introduction of the CAs is provided in Chapter~\ref{chap2}. 
\begin{itemize}
	\item \textbf{Chapter~\ref{chap2}.} This chapter describes the principle of CAs, different forms of CAs, and a brief history of CAs. After that, we concentrate on some non-classical CAs, such as stochastic CAs, automata networks, and non-uniform CAs. Finally, we briefly discuss the history of CAs as well as various computing tasks and social applications.		
	\item \textbf{Chapter~\ref{chap3}.} This chapter addresses a non-conventional CA variant where noise has a temporal impact on the entire system. In reality, one of the two rules, let's say $f$ and $g$, can be used to update a cell at a time step. You may think of the $f$ as the CA's default rule. In contrast, $g$ is applied temporally to the entire system with a certain probability and behaves as noise in the system. We named the variant as \emph{Temporary Stochastic CA}.
	\item \textbf{Chapter~\ref{chap4}.} This is an extension of the previous chapter, where an application of \emph{Temporary Stochastic CA} is discussed. The convergence property of TSCA is discussed in this chapter. With the help of this convergent TSCA, we build a pattern classifier. On a few standard datasets, we deploy each convergent TSCA and analyze the performance as a pattern classifier. It has been noted that when compared to previous classifier algorithms, the proposed TSCA-based classifier shows competitive performance.
	\item \textbf{Chapter~\ref{chap5}.} This chapter introduces \emph{affinity classification problem} as a generalization of the \emph{density classification problem}. Formally, the problem can be stated as, for given an initial configuration, and a binary cellular automaton that converges to $all-1$ configuration if the density of $1$s is more than $\rho$. Otherwise, it converges to $all-0$ configuration. A variant of the CA model is proposed to solve this kind of problem.
	
	\item \textbf{Chapter~\ref{chap6}.} This chapter introduces a new model of computing unit developed around cellular automata, where each cell of the computing unit acts as a tiny processing element with attached memory. The Cayley Tree is used to implement such a CA in order to achieve effective solutions for various computing problems. This study focuses on finding solutions to the searching problems to demonstrate the usefulness of this paradigm. Such cellular structures may, however, be used to address a variety of computing problems, the searching problem being one of them.
	\item \textbf{Chapter~\ref{chap7}.} This chapter wraps up the thesis with a few unresolved problems that might be addressed later.
	
\end{itemize}

\chapter{Survey on Cellular Automata}
\label{chap2}
Cellular Automaton (CA) is a discrete, abstract computational system that consists of a regular network of finite state automata, formally known as cells. A cell changes its state depending on the state of its neighbors using a local update rule and all the cells change their states simultaneously using the same update rule.

For the last fifty years, cellular automata have been rigorously studied in various fields because of their simple structures. Some theoretical aspects of cellular automata include chaos in CAs~\cite{Devaney,langton90,DCTMitchell93,Cattaneocht,Supreeti_2018_chaos,KAMILYA2019116}, reversibility~\cite{Siap11,Kari90,Kari94,marti2011reversibility,DIGREGORIO75,Bhattacharjee16b,Sarkar12,SethiD14,tome1994necessary,0305-4470-37-22-006,kamalikaThesis,CayLeyid02,uguz2013reversibility,BhattacharjeeNR16,Ángel13,cinkir2011reversibility,Fats2018,Sarkar11}, reachability problems~\cite{naskar12}, primitive polynomials~\cite{ADAK202179,Bardell1992} etc. Also, CAs have been applied to model physical systems~\cite{Suzudo2004185,ulam1962some,uguz14,White51175,liang2001s,BAUDAINS2013211,10.2307/2094159,GUNJI1990317,Margolus198481}, VLSI design and test~\cite{Pries86,vlsi00d,Horte89c,Tsali91,biplab,SukantaTH,santanu00,Chowd92d,Tsali90,vlsi02a,ubist,Misra92b,vlsi00b,vlsi00a,Chandrama,aspdac04,Bardell,dubrova1999multiple,MitraDCN96}, cryptography~\cite{Sethi2016,Seredynski2004753,wolfram1985cryptography,das2013car30}, pattern classification~\cite{CPLX:CPLX21749,maji2003theory,DasMNS09,Sethi6641432,Sethi2015}, random number generator~\cite{Horte89a,Tomassini96,MARSAGLIA19931,L'Ecuyer:2007:TCL:1268776.1268777,opre.48.2.308.12385,S1064827598349033,Rukhin10statisticaltest,Matsumoto:1998:MTE:272991.272995,soto1999statistical,Mitra2014,BhattacharjeeIJMPC2018,Panneton:2005:XRN:1113316.1113319,Doi1004061,Park:1988:RNG:63039.63042,JAMES1990329,LEcuyer:1997,L'Ecuyer:2005:FRN:1162708.1162732,HELLEKALEK1998485,rukhin2001statistical,Rotenberg:1960:NPN:321008.321019,fishman1990multiplicative,doi:10.1137/0907002,Eichenauer1986,L'Ecuyer:1988:EPC:62959.62969,10.2307.2347988,doi:10.1287.opre.44.5.816,doi:10.1287.opre.47.1.159,nance1978some,marsaglia1985current,RNG,BHATTACHARJEE2022100471,saito2011TinyMT,Tsali91,comer2012random}, compression~\cite{Bhatt95,Lafe,lafe2002method,ShawSM04,ShawDS06,Chandrama,ShawMSSRC04}, image processing~\cite{Rosin06,Rosin2010790,Jin2012538,Soto2008,Mariot2017,Okba11,Wongthanavasu03,Sadeghi12,ShawDS06,ye2008novel,Sato2009}, number-conserving~\cite{Morita98,Morita:99,boccara98,das2011characterization,Boccara02,Durand03,Moreira03,Formenti2003269,Kohyama01011989-28,GOLES20113616}, computability~\cite{Kutrib2011,short12,Mark1,Chopard2012,Darabos7,Soto2008,langton90,Smith71,Durand-Lose98,lindgren90,wolfram84,toffoli77,DCTMitchell93,Culik90,Margolus198481,Hem82,naka,tome1994necessary}, natural computing~\cite{Sethi2015,Sato2009,Alberto12}, biological systems~\cite{RUXTON98,Mark1,gaylord1995computer,ermentrout1993cellular}, medical science~\cite{AMTuring,ermentrout1993cellular,RUXTON98,Dabbaghian2012,dhande1984ternary}, social networking~\cite{6108515,Beltran11,Sakoda89791,Epstein96,liebrand1996frontiers,Andreas98,moreno1957first,david2004social,Hegselmann1996,Nowak1996,DABBAGHIAN20101752,Benito1012898,Das7477,LANG201412,10.1007/978-3-642-40495-5_3,iet.cp.2013.2283} etc.

\section{Cellular Automata}
A cellular automaton (CA) is made up of a number of cells arranged in a regular network. Each of a CA's cells is a finite automaton that uses a finite state-set, called $S$. At particular times and locations, the CAs alter. Throughout evolution, a CA cell changes based on the state of its neighbors. In other words, a cell updates its state using a next-state function, also referred to as a local rule. The neighbors of the cell's current state serve as the function's inputs. The aggregation of all cell states throughout time is referred to as the CA's configuration. A CA alternates between configurations as a result as it evolves.
\begin{definition}
	A cellular automaton is a quadruple ($\mathcal{L}$, $\mathcal{S}$, $\mathcal{M}$, $\mathcal{R}$) where,
	\begin{itemize}
		\item $\mathcal{L} \subseteq \mathbb{Z}^\mathcal{D}$ is the $\mathcal{D}-$dimensional cellular space. A set of cells are placed in the locations of $\mathcal{L}$ to form a regular network.
		\item  $\mathcal{S}$ is the finite set of states; e.g. $\mathcal{S} = \{0, 1,\cdots, d-1\}$.
		\item $\mathcal{M}=(\vec{v_1},\vec{v_2},\cdots,\vec{v_m})$ is the neighborhood vector of $m$ distinct elements of $\mathcal{L}$ which associates one cell to it's neighbors.
		\item The local rule of the CA is $\mathcal{R}:\mathcal{S}^m\rightarrow\mathcal{S}$. A cell's subsequent state is determined by the expression $f(s_1,s_2,s_3,\cdots,s_m)$ where $s_1,s_2,s_3,\cdots,s_m$ denotes the states of it's $m$ neighbors.
		\end{itemize}   
\end{definition}
Cellular automata can be of many types. One of a cellular automaton's most fundamental properties is the kind of grid that it evolves on. The most fundamental grids of this kind are one-dimensional arrays (1D CA). In case of two dimensional CA, grids of square, triangular, and hexagonal shapes can be used. Additionally, cellular automata may be built on Cartesian grids in any numbers of dimensions, with the integer lattice in dimensions being the most popular option. For example, The Wolfram's cellular automaton are implemented on a one-dimensional integer lattice.

The three fundamental characteristics of a classical CA are locality, synchronicity, and uniformity. 
\begin{itemize}
	\item According to the concept of locality, the computation of CAs using a rule is a local computation. A cell's state varies in response to local interactions only.
	\item Synchronicity refers to the simultaneous updating of all local computation-performing cells.
	\item The employment of the same local rule by all cells is referred to as uniformity.
\end{itemize}

Therefore, the local rule, which is distinct throughout the lattice, is used to carry out the actual calculation. The radius of a CA represents the quantity of succeeding cells in a direction that a cell depends on. A CA cell, for instance, depends on its two successive left cells and its two subsequent right cells if the radius of the CA is, let's say, $2$. The CA is a $(2 + 2 + 1) = 5-$neighborhood CA in the scenario.
\begin{figure}[hbt!]\centering 
	\includegraphics[width=0.50\textwidth]{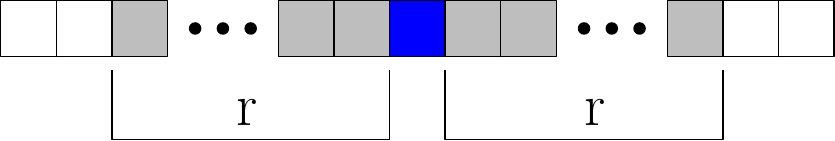} 
	\caption{neighborhood dependence in a one-dimensional cellular automaton with radious $r$.}
	\label{r}
\end{figure}

The neighborhood dependence of a one-dimensional CA is shown in Fig.~\ref{r}. A cell uses $r$ left neighbors and $r$ right neighbors in order to move to the next state.

\begin{figure}[hbt!]
	\subfloat[]{
		\begin{minipage}[c][1\width]{
				0.30\textwidth}
			\label{von}
			\centering
			\includegraphics[width=1\textwidth]{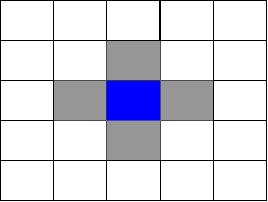}
	\end{minipage}}
	\hfill 	
	\subfloat[]{
		\begin{minipage}[c][1\width]{
				0.30\textwidth}
			\label{moor}
			\centering
			\includegraphics[width=1\textwidth]{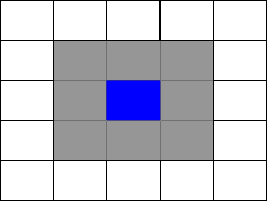}
	\end{minipage}}
	\caption{The neighborhood dependencies for two-dimensional cellular automata; (a) Von numann neighborhood; (b) Moore neighborhood.}
\label{fig:neighbors}
\end{figure}

However, (John) von Neumann neighborhood dependency and Moore neighborhood dependency are typically used to identify the neighborhood of a two-dimensional CA. John von Neumann's suggested CA is a two-dimensional CA with square grids. A cell represents each square box. Each cell has one of $29$ possible states. Each cell's subsequent state is determined by the current states of its four orthogonal neighbors and itself (see Fig.~\ref{von}). Later, the CAs are simplified with fewer states even if they have maintained their capacity to reproduce itself and computational efforts; see~\cite{Arbib66,Codd68,Banks71,thatcher1964,langton1984self,banks1970universality,Smith71,iirgen1987simple,Morita89,Culik90,Goucher2010a,martin1994universal,DUBACQ0202,ollinger2002quest,ollinger2003intrinsic,cook2004universality} for illustration.

Four non-orthogonal cells are also taken into consideration as neighbors in the Moore neighborhood~\cite{moore1962machine} of two dimensional CAs, which uses a $9$-neighborhood dependence.  Moore neighborhood dependence of a CA is depicted in Fig.~\ref{moor}. The renowned Game of Life, a CA that Martin Gardner popularised and John Conway presented, was developed using this type of neighborhood structure~\cite{Gardner70}.

\begin{figure*}[hbt!]
	\begin{center}
		\begin{tabular}{cc}
			\includegraphics[width=0.4\textwidth]{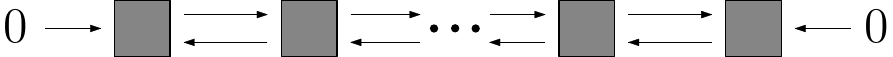}  &   \includegraphics[width=0.4\textwidth]{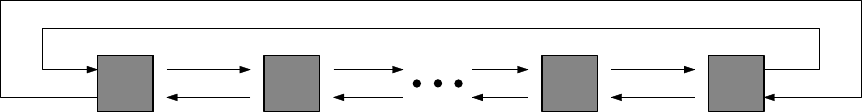}  \\ 
			
			(a) Null Boundary   & (b) Periodic Boundary \\
			\includegraphics[width=0.4\textwidth]{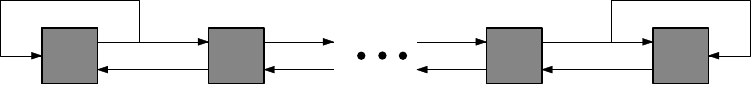}&
			
			\includegraphics[width=0.4\textwidth]{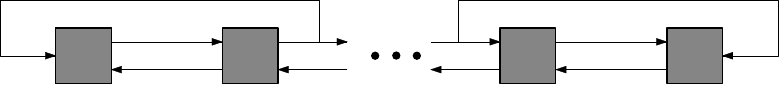} \\
			
			(c) Adiabatic Boundary & (d) Reflexive Boundary \\
			 \includegraphics[width=0.4\textwidth]{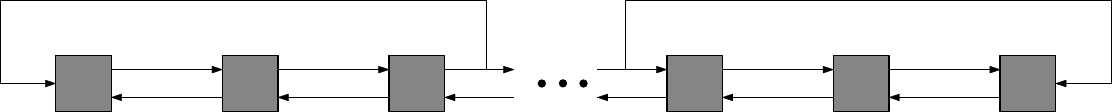} & \\   
			 
			  (e) Intermediate Boundary &\\
		\end{tabular}
		\caption{Boundary conditions of one-dimensional finite CAs. (a) Null Boundary ; (b) Periodic Boundary; 	(c) Adiabatic Boundary; (d) Reflexive Boundary;  (e) Intermediate Boundary. }
		\label{neigh}
	\end{center}
\end{figure*}

The existence of a barrier and boundary conditions is irrelevant because the cellular space is often infinite. The assumption that cellular space is finite, which undoubtedly has bounds, is made in several publications. If the automata are to be put into practise, finite CAs are necessary. A finite CA is often investigated with two boundary conditions, that is, open and periodic. In this study, we primarily take into account finite CAs with periodic boundary conditions. In a few instances, nevertheless, we also research dynamics with an open boundary condition. The missing neighbors of extreme cells (leftmost and rightmost cell in one-dimensional CAs) are typically given some fixed states in case of open boundary CAs. The most common open boundary condition is null boundary, which ensures that any missing neighbors of terminal cells are always in state $0$ (Null). The works~\cite{Wolfr83, Horte89a,Tsali91, ppc1,das2010scalable} shows some use of CAs with null boundary consition. On the other hand, with a periodic boundary condition, the boundary cells are immediate neighbors of certain other boundary cells. For one-dimensional+ CAs, the neighboring cells on the right and left are shown in Fig.~\ref{neigh} as an example, where the left neighbor of the leftmost cell (resp. the right neighbor of the rightmost cell) is the rightmost cell (resp. leftmost cell). Some researchers have also looked at periodic boundary conditions for higher dimensional CAs~\cite{Palas1,Jin2012538,uguz2013reversibility}.

Initially conceived using a two-dimensional lattice of cells~\cite{Toffo87,cattell1996synthesis,Siap11,imai2000computation,morita2016universality,margenstern1999polynomial,Margenstern200199,Packa85b,uguz14}, cellular automata have also been proposed in greater dimensional space~\cite{gandin19973d,Palas1,miller2005two,mo20143,dennunzio2014multidimensional,tomassini29,Darabos7,Marr20}. Even yet, many researchers from all around the world have committed their careers to studying one-dimensional CAs~\cite{Amoroso72,Pries86,sutner1991bruijn,sikdar2002design}. Elementary Cellular Automata (ECAs) are a subclass of CAs that Wolfram has studied~\cite{Wolfr83,wolfram2002new}. ECAs are one-dimensional, two-state, and dependent on three nearest neighbors. ECAs and their modifications have received a significant amount of attention in cellular automata research during the past three decades.

\subsection{Elementary Cellular Automata}
One of the simplest structures with complicated behavior is a one-dimensional, two-state, three-neighborhood CA. Because of this, there is a significant demand for CAs among academics that study natural phenomena. These CAs are referred to as Wolfram's CAs or Elementary CAs (ECAs). Wolfram conducted significant research on the dynamics of these CAs and categorized the CAs based on their dynamical tendencies. ECAs are further categorised and described by other scholars for theoretical advancement and application standpoint in part because of this study.

In elementary CAs, each cell evolves its state based on the states of its left neighbor, self, and right neighbor. This results in a one-dimensional infinite lattice of cells. That is, the radius $(r)$ of a cell is $1$ and $S = \{0, 1\}$. A cell is subjected to a function $f$ or local rule, depending on which the cell changes its state from one state to the next. Tabular representations are often possible for the local rule $f$. Table~\ref{Table:ECA} displays two basic principles with $S = 0$ and $1$, and $r = 1$. Usually, the rules are given as $d-ary$ (binary for ECAs) or their corresponding decimal integers.

\begin{table}[ht]
	\centering
	\begin{adjustbox}{width=0.8\columnwidth,center}
		\begin{tabular}{cccccccccc} \hline \hline
		RMT&\makecell{111\\(7)}&\makecell{110\\(6)}&\makecell{101\\(5)}&\makecell{100\\(4)}&\makecell{011\\(3)}&\makecell{010\\(2)}&\makecell{001\\(1)}&\makecell{000\\(0)}&Rule\\\hline
		&0&0&0&1&1&1&1&0&30\\
		f&0&1&0&1&1&0&1&0&90\\
		&1&0&0&1&0&1&1&0&150\\\hline\hline
		\end{tabular}
	\end{adjustbox}
	\caption{Elementary CA rules $30,90$ and $150$.}
	\label{Table:ECA}
\end{table}

There are $2^{2^3}$ possible functions for CAs, with two states and three neighborhoods. There are $2^8 = 256$ elementary CA rules, according to this. The phrase \emph{Rule Min Term}, which is explained in the next para, is referred to by the acronym \emph{RMT} in Table~\ref{Table:ECA}. \emph{RMTs} of ECA rule $30, 90$ and $150$ are shown in the Table~\ref{Table:ECA}.

\begin{definition}
	(Rule Min Term (RMT))~\cite{Bhattacharjee16b}: Let $f : S^{2r+1} \rightarrow S$ be the local rule of a one-dimensional CA. An input $(x_{-r},\cdots, x_0,\cdots, x_r) \in s^{2r+1}$ to $f$ is	called a Rule Min Term (RMT). An RMT is commonly represented by a decimal	number $r = x_{-r}.d^{2r} + x_{-r+1}.d^{2r-1} + \cdots + x_r$ and $f(x_{-r}, \cdots, x_0, \cdots, x_r)$ is represented by $f(r)$. RMT $r$ is called passive if $f(x_{-r}, \cdots, x_0, \cdots, x_r)=x_0$. Otherwise the RMT is called active RMT.
\end{definition} 

The first row of Table~\ref{Table:ECA} shows the \emph{RMTs}. For a $d$-state CA, total number of RMTs is $d^{2r+1}$. Hence, number of  \emph{RMTs}, in case of an ECA, is $2^3 = 8$ and number of rules is $2^8=256$. We often omit commas within an \emph{RMT}, if it does
not lead to any confusion. That is, an RMT $(x_1, x_2, x_3)$ is generally presented
as $x_1x_2x_3$. In ECA rule $30$, the RMTs $0 (000), 2 (010), 3 (011)$ and $5 (110)$ are passive (see Table~\ref{Table:ECA}), whereas the rest \emph{RMTs} are active.

\subsection{Wolfram's Classification of Cellular Automata}
A one-dimensional, two-state, three-neighborhood CA has one of the most basic structures yet is capable of displaying sophisticated behavior. As a result, researchers have a strong demand for such CAs in order to study naturally occurring phenomena. The term `Wolfram's CAs'' or ``Elementary CAs'' refers to such CAs. According to their dynamical tendencies, Wolfram categorized the CAs after a thorough study of their dynamics. Different scholars have further classified and described ECAs in light of this work, both from a theoretical development and practical standpoint~\cite{ppc1,FSSPCA3,mitch98a,Sarkar00Survey,binder1993phase,ninagawa2014classifying,KARI20053,BhattacharjeeNR16}.

The fundamental unit of an elementary cellular automaton is a finite automaton defined over a one-dimensional array. Two states are used by the automaton, and they are updated synchronously based on their own state and the states of their two nearest neighbors. In this paper~\cite{wolfram84b}, according to the outcomes of the system's evolution from a random initial state, Stephen Wolfram proposed classifying cellular automaton rules into four types:
\begin{enumerate}
	\item[] \textbf{Class I.} A homogeneous state of CA is developing,
	\item[] \textbf{Class II.} CA is periodically developing,
	\item[] \textbf{Class III.} CA is chaotically developing,
	\item[] \textbf{Class IV.} comprises all preceding instances; a class of complex rules.
\end{enumerate}	
This classification was initially only for one-dimensional cellular automata, but in~\cite{Packa85b}, he extended it to two-dimensional cellular automata.

According to this classification, it makes intuitive sense to assume that \emph{class IV} cellular automata will only exhibit gliders, Turing-universality, and comparable complex behaviour. Wolfram states that ``These comparisons lead to the assumption that \emph{class IV} automata are characterized by the possibility for universal computing'' after analyzing the existence of glider-like structures in a \emph{class IV} automaton and their resemblance to structures in Conway's Life.

A cellular automaton's capacity for colors (or unique states) must also be stated. Usually an integer, the easiest option for this number is (binary). Colors 0 and 1 in a binary automaton are typically referred to as ``white'' and ``black'', respectively. However, continuous range cellular automata may also be taken into consideration.
\begin{figure}[hbt!]\centering 
	\includegraphics[width=0.50\textwidth]{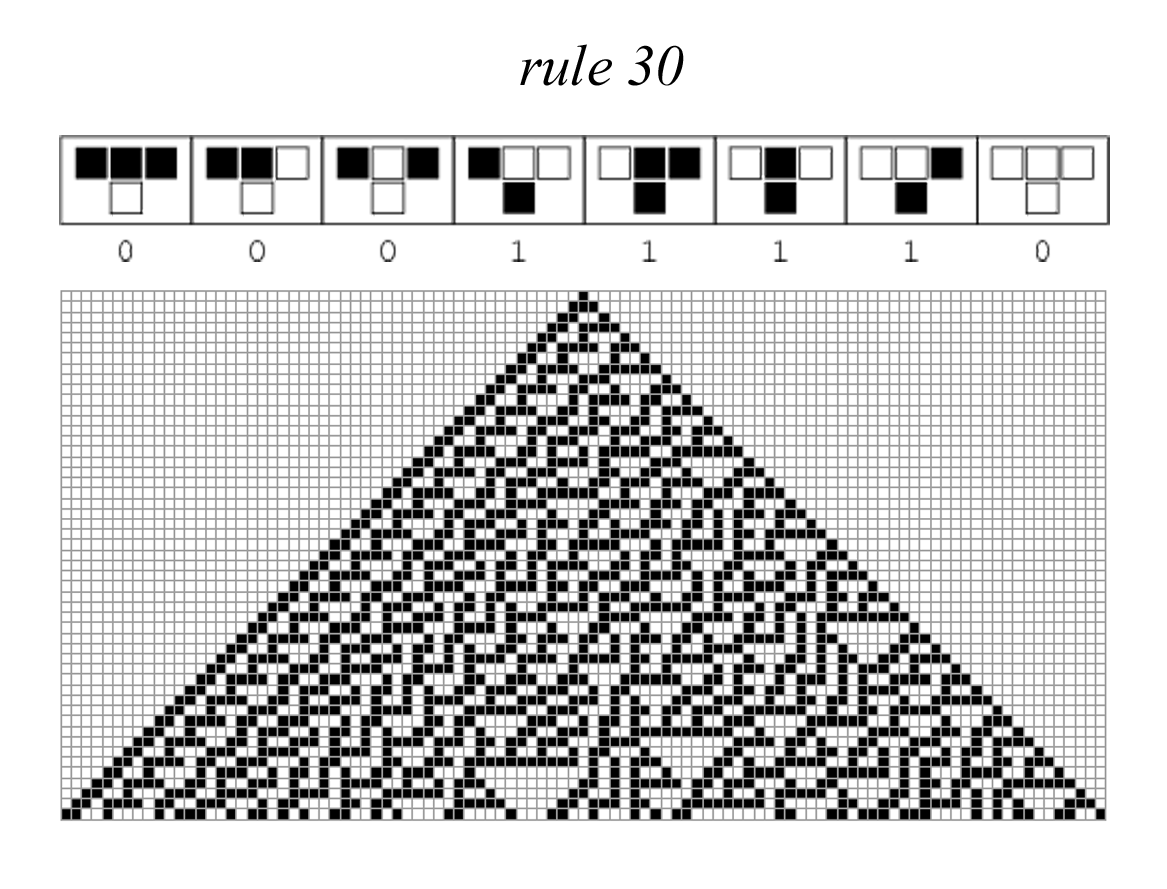} 
	\caption{Space-time diagram of a $1-$D Cellular Automaton of (Wolfram rule $30$).}
	\label{rule30}
\end{figure}

A space-time diagram of an elementary CA, ECA $30$ is shown in Fig.~\ref{rule30}, where the development of the CA starts with a single $1$. White cells denote state $0$, whereas black cells denote state $1$.

\subsection{Li-Packard Classification}

The first initiative to categorize each rule's behavior based on observation was made by Wolfram. It provides academics with a fresh perspective on how to better understand the dynamics of CAs. The classification, however, does not totally separate the rules of one class from another. There are several rules where two kinds of behavior are superimposed. For example, a CA may behave chaotically in some areas, or two chaotic sections may be divided by a wall. Local chaos is one type of such activity. Wolfram's classification was somewhat changed by Li and Packard in 1990, and they divided the ECA rules into five categories: null, fixed point, periodic, locally chaotic, and chaotic. Rule space analysis served as the foundation for this categorization. A rule's likelihood of being linked to another rule is evaluated. The classes is reported in Table~\ref{Table:Li_Packard}.

\begin{table}[ht]
	\centering
	\begin{adjustbox}{width=0.8\columnwidth,center}
		\begin{tabular}{c|c} \hline 
			Classes & ECA Rules \\ \hline \hline
			Null & 0, 8, 32, 40, 128, 136, 160, 168\\\hline
			Fixed Point & \makecell{2, 4, 10, 12, 13, 24, 34, 36, 42, 44, 46, 56, 57, 58,\\
				72, 76, 77, 78, 104, 130, 132, 138, 140, 152, 162, 164,\\
				170, 172, 184, 200, 204, 232} \\\hline
			Periodic & \makecell{1, 3, 5, 6, 7, 9, 11, 14, 15, 19, 23, 25, 27, 28, 29, 33,\\35, 37, 38, 41, 43, 50, 51, 74, 108, 131, 133, 134,\\ 142, 156, 178} \\\hline
			Localy Chaotic & 26, 73, 154\\\hline
			Chaotic & 18, 22, 30, 45, 54, 60, 90, 105, 106, 129, 137, 146, 150, 161\\\hline
			
		\end{tabular}
	\end{adjustbox}
	\caption{Elementary CA rules classification according to Li $\&$ Packard~\cite{Li90thestructure}.}
	\label{Table:Li_Packard}
\end{table}

One new class is introduced among the five classes of ECA rules, i.e., locally chaotic. It displays an intriguing case of CA behavior. Chaos is typically caused by infinitely long CAs. However, this class represents chaos inside a small space, in contrast to other classes that are similar to Wolfram's classes. A cell's neighbors get information that contributes to it, but information cannot cross a wall. This obstruction, sometimes referred to as the blocking word, is where the information is impeded. Despite the fact that a behaviour inside a defined region is always predictable, the authors have characterized it as chaotic by looking at how information spreads among the cells there. Rule 154 is an example of a locally chaotic ECA rule. 

\subsection{von Neumann's Universal constructor}
Robbert Oppenheimer, Enrico Fermi, Niels Bohr, Hans Bethe, Richard Feynman, Eugene Wigner, John Von Neumann, and several more scientists collaborated on the Manhattan project at Los Alamos during the Second World War. Because the team included of the top scientists, it was amazing. The scientific objective for such project was also quite clear: ``Building bomb in a race against the Nazis''. However, some scientists started to consider the difficulty of computers about the same period and computer simulation also served as a bridge between theory and experimentation. Jon von Neumann was also interested in self-replicating machine and cellular automata around that time.

The mathematician Stanislas Ulam, who conducted the initial experiments on one of the earliest stored programme computers at Los Alamos, is credited as being the genuine inventor of artificial growth and evolution research. Ulam was fascinated by the development patterns of geometrical forms in two and three dimensions that were produced by incredibly straightforward recursive procedures. John von Neumann was motivated by Ulam's inventions, which helped him develop the first cellular automaton model. From there, the initial definition of artificial life emerges, i.e., one of the fundamental concepts in artificial life is the notion of complexity emerging through the combination of basic principles.

John von Neumann developed the idea of cellular automata at the start of the 1950s~\cite{Neuma66}. von Neumann was interested in the possibility of reproduction and explored for the logical prerequisites necessary for a non-trivial self-reproduction. He had originally developed a kinematic model of a robot floating in a lake with all the parts required to construct further robots. He imagined the robot gathering parts and putting them together to create a duplicate of itself. In essence, von Neumann was successful in demonstrating how the floating robot might replicate itself, but sadly, a large portion of his research was hampered by the issue of motion in the lake. Ulam's method was therefore adopted by von Neumann, who furthered the abstraction.

Instead of simulating self-reproduction at the genetic level, his approach was to abstract it in its coherent manner. If self-reproduction can be described as a logical series of processes, then there is a universal Turing Machine that is capable of self-reproduction. A two-dimensional cellular automaton with $29$ possible states was defined by John Von Neumann where, a transition rule is applied to the current cell and its four orthogonal neighbors to determine the state of each cell.


von Neumann demonstrated that, rather than relying on some unexplained quality of matter, one of life's primary characteristics could be described using logical concepts. Sadly, he passed away in 1957 before completing his proof. His contributions were finished and revised by Arthur Burks, who collaborated with him on the logical design of the EDVAC (one of the earliest computers). John von Neumann is today regarded as the pioneer of the Artificial Life concept for deriving from the natural self-reproduction its logical (computational) form.

\subsection{Conway’s Game of Life}
In von Neumann’s approach~\cite{Neuma66} there were $29$ states of each cell so the complexity of computing of each cell state is much higher. Many researchers were trying to reduce the computational complexity, actually wanted to reduce the number of states without compromising the property (Self-replication) of that machine. In the year 1970 John Horton Conway, a young mathematician at Gonville and Caius College of the Cambridge University, proposed Game of Life is certainly the best example of the idea that complex worlds could emerge from simple rules. Conway adapted Ulam and von Neumann’s approach based on cellular automata. In Conway’s Game of Life there were only $2$ states of each cell which are capable to perform Self-reproduction.  The state of each cell (\emph{alive/dead}) is the result of two rules applied on the cell and its eight neighbors.

Life’s rules are marvelously simple:
\begin{itemize} 
	\item If the number of \emph{alive} cells is exactly three, the current cell will be \emph{alive} in the next generation.
	\item If the number of \emph{alive}  cells is zero, one, four, five, six, seven or eight, the cell will be \emph{dead}  in the next generation.
\end{itemize}
\begin{figure}[hbt!]\centering 
	\includegraphics[width=0.8\textwidth]{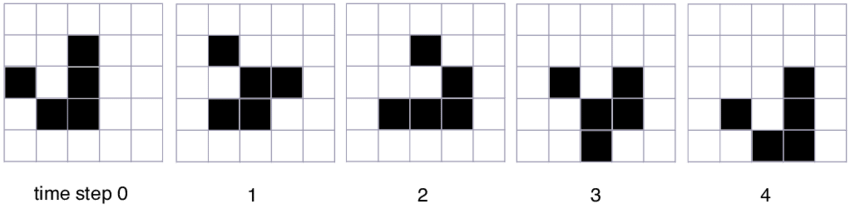} 
	\caption{Subsequent stages of the glider pattern on Game of Life}
	\label{gol}
\end{figure}
Life has been experimented with extensively. Many of the configurations which emerge seem to have a \emph{life} of their own. One of the most remarkable example of life’s structures is the \emph{glider}, a configuration of period four which displaces itself diagonally. This CA can also perform self-replication by following the above rules.

\subsection{Codd’s Cellular Automaton}
Cellular Automata, introduced in 1968 by British computer scientist Edgar F. Codd~\cite{Codd68}, uses $8$ rather than $29$ states to replicate the computation and building universality of von Neumann's CA. Similar to von Neumann's \emph{universal constructor}, Codd demonstrated that it was possible to construct a self-replicating machine in his CA, but he never provided a full implementation.

\begin{figure}[hbt!]\centering 
	\includegraphics[width=0.8\textwidth]{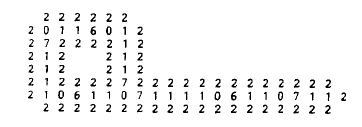} 
	\caption{Codd's path extension machine }
	\label{codd}
\end{figure}

Fig.~\ref{codd} depicts the loop instruction using $8$ states CA introduced by Edgar F. Codd. This loop is a self replicating loop and the instructions for building a loop might not fit within a loop that size. Codd's signal sequence, which is needed to build another loop (Self-replication), does not appear to allow for this as he needs the sequence $7-0-1-1-6-0$ merely to lengthen the route by one cell. How might a loop store enough data to develop a structure that is the same size as itself if it takes six cells to store the information needed to generate a single cell of the new machine? We only need to store the instructions necessary to generate one side and one corner of the storage loop if we make it a perfect square, is the solution.The procedure of making a side and a corner will be repeated four times as these instructions go around the loop four times.

The instructions to construct one side and one corner are too lengthy to fit in a loop even when using Codd's signal sequences.
\subsection{Langton’s Self-Reproducing Automata}
The writings of von Neumann, Ulam, and Conway had a big influence on Christopher Langton. In late 1971, he coined the phrase \emph{Artificial Life}, Christopher Langton came at Los Alamos in August $1986$ to begin a postdoctoral position at the Center for Nonlinear Studies. He planned the inaugural ALife workshop at the Los Alamos National Laboratory in September 1987. The Santa Fe Institute, Apple Computer Inc., and the Center for Nonlinear Studies all provided financial support for the program. It gathered $160$ experts from many fields, including anthropology, computer science, biology, and physics who were all interested in simulating and synthesizing biological systems.

Another of Burks' students, Christopher Langton, developed a self-replicating pattern in 1984 based on a very basic Codd's automaton configuration known as the periodic emitter, which was in turn derived from the periodic pulser organ in von Neumann's 29-state automaton. Christopher Langton proved that the ability to build anything at all was not a need for self-reproduction. Some of Codd's signals had their meaning changed by Langton, giving them additional potency on their own. Instead of modifying array configurations as indicated in Codd's section to change the meaning of signals, transition rules that govern how the array's configurations behave must be changed. By using this method, he shortens the entire sequence required to build one side and one corner to the point that it can now fit inside the size of the loop that it produces~\cite{langton1984self}.
\begin{figure}[hbt!]\centering 
	\includegraphics[width=0.7\textwidth]{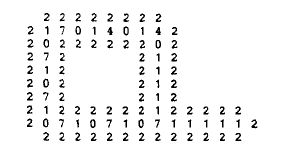} 
	\caption{Self reproducing loop with Initial sequence~\cite{langton1984self}: $7\;0 - 7 \;0 - 7 \;0 - 7 \;0 - 7 \;0 - 7\; 0 - 4 \;0 - 4 \;0$ }
	\label{langton1}
\end{figure}

As a result, Codd's requirement that the sequence $7\; 0 - 6\; 0$ be used to expand the data route by one cell has been replaced with the $7\; 0$ signal, which can trigger this extension on its own. Another distinction is that Langton's signal allowed signals to be just three cells apart, but Codd's signal required the machine to be spaced four cells apart. Allow two consecutive  $4\; 0$ signals to complete a left path extension in place of Codd's necessary  sequence of $4\; 0 - 4\; 0 - 5\; 0- 6\; 0$.

The automaton is based on eight-state cells which are used -
\begin{enumerate}
	\item as information to replicate in the cellular environment, resulting in the birth of a child, and
	 \item as instructions to be carried out in accordance with the transition rule.
\end{enumerate}

\begin{figure}[hbt!]\centering 
	\includegraphics[width=0.8\textwidth]{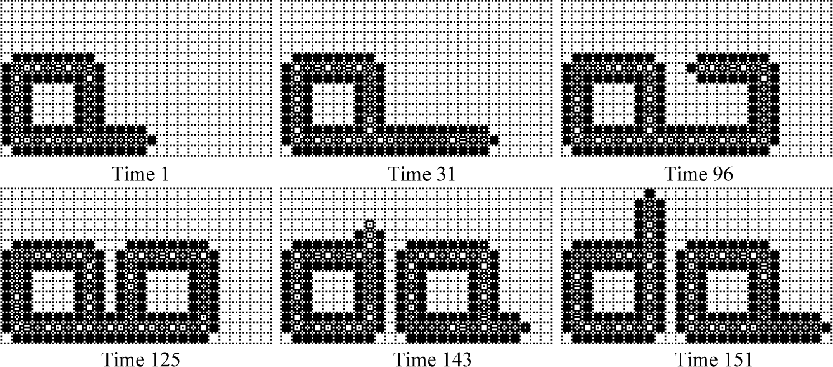} 
	\caption{Langton’s self-reproducing automaton~\cite{langton1984self}.}
	\label{langton2}
\end{figure}

The basic structure has successfully replicated itself after $151$ time steps~\cite{langton1984self}. Then, each of these \emph{loops} goes on to replicate itself in a similar way, growing a colony of \emph{loops} (see Fig.~\ref{langton2}) in the process. The genotype codes for the elements of a dynamic process in the cell, and it is this dynamic process that is principally responsible for computing the expression of the genotype during development. This experiment illustrates the essence of what happens in natural development.

\section{Artificial life}

In 1986, American computer scientist Christopher Langton coined the phrase \emph{Artificial Life} while planning the inaugural ``Workshop on the Synthesis and Simulation of Living Systems''~\cite{langton86}. Since then, the concept of artificial life has permeated computer science, gaming, artificial intelligence research, and other fields. The Web and the way it enables networked computers to produce complex habitats in which artificial creatures may survive and evolve have played a significant role in its development.
The study of artificial systems that display behaviour like that of living things is known as artificial life. It is the endeavor to provide an explanation for life in all of its conceivable forms, without restriction to the specific instances that have developed on Earth. This comprises computer simulations, biological and chemical investigations, and entirely theoretical undertakings. Investigations focus on processes at the molecular, societal, and evolutionary levels ~\cite{copeland2004essential,Gotts09,Jeanson2008E,banda2,Bersini94,Sughimura14}. Extraction of biological systems' logical form is the ultimate objective.
\subsection{Langton's loop}
Christopher Langton developed a two-dimensional, self-replicating cellular automaton in 1984~\cite{langton1984self}. The replicating pattern consists of a loop that contains genomic information's. A series of cells that flow through the loop's arm and eventually merge to form another loop are the genetic information. Some components of the genetic code are designed to make the loop make three left turns before closing or dying and ceasing to reproduce. The colony of loops that is replicated in an infinite two-dimensional space has no size restriction. This system is regarded as a top illustration of synthetic self-reproduction.
In This directions, Langton explored some variations of CAs and its abilities~\cite{langton86, LangtonII,langton90}. In the nest section Langton's ant, a similar variation of CA, is discussed.

\begin{figure}\centering    
	\includegraphics[width=0.5\textwidth]{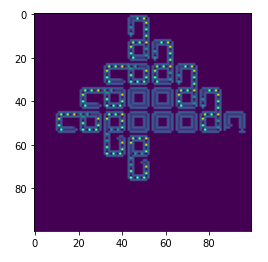}
	\captionof{figure}{Self reproduction of langton's loop}
	\label{langtonLoop}
\end{figure}
After multiple (more than $250$) iterations, we may observe certain dead loops in the cellular space during the process. When the genetic information runs in a loop around its arms, it is automatically generated. Six $7$s are found in the loop, followed by two $4$s. As a consequence, it expands straight forth before turning $90$ degrees counterclockwise in response to the $4$s command. The command is repeated until all four sides of the new loop have been formed since this pattern of $7$s and $4$s keeps revolving around the loop. The other states are accustomed to advancing to new loops. A $5$ signal is delivered to the parent loop after it has developed a completely formed child to instruct it to rotate $90$ degrees and begin developing a child on the opposite side. Similar to $6$, which instructs the youngster to continue and have children of their own, is $6$. A loop begins developing the next child once its genome has been utilized $6$ three times: once to construct an arm, once for each of its child's four sides, and once to transfer the genome to its child (its clone). The parents will pass away once all four children have been produced, leaving behind a spiral shell and an empty genome. The entire situation spirals outward around a point of no return.
\subsection{Langton's Ant}
The Langton's ant is a two-dimensional  4-state universal Turing Machine. It was invented by Chris Langton in 1986. It is basically an ant, sitting on a square lattice of cells, which are initially white. The ant moves on the plane and changes the color of cells, creating patterns on it. But the movement of the ant is not random. It follows the following set of rules-
\begin{itemize}
	\item If the ant is on a black square, it turns right 90 degrees and moves forward one unit.
	\item If the ant is on a white square, it turns left 90 degrees and moves forward one unit.
	\item When the ant leaves a square, it inverts the color.
\end{itemize}
As the ant starts its journey, it creates a black and white pattern while moving. Initially, the changes are not distinctive, but as we iterate them over and over again, a beautiful pattern emerges. But if we further increase the number of iterations (approximately $10000$), the ant starts repeating its path with a gradual shift instead of making new patterns. Thus, we obtain a \emph{highway} like pattern that is infinite. The ant keeps moving on that highway and gives the following pattern.

\begin{figure}\centering    
	\includegraphics[width=0.5\textwidth]{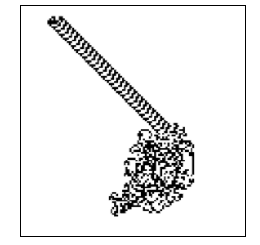}
	\captionof{figure}{Langton's Ant after $13000$ iterations.}
	\label{langtonAnt}
\end{figure}

\section{Summary}
This chapter has included a brief review of cellular automata, a brief survey of various models have developed using cellular automata. To understand the behavior, scientists have investigated dynamical systems in a variety of approaches. To distinguish between the dynamical behaviors, CAs have been classified in several ways. Many researchers have studied the topology of self-replicating cellular automata, which provides insight into the Artificial Life. The parametrization process may be used to forecast the behavior of CAs. One may partially describe the behavior of CAs by defining parameters. However, whether the CA is homogeneous, periodic, or chaotic, parameters may fail to appropriately identify it. Therefore, developing a more precise parameter is constantly needed in the research community. Studies have also been conducted for traditional CAs. There has not been any literature on understanding the behavior of Temporally Stochastic CAs (TSCAs). It offers a new field for researchers to find the dynamics of TSCAs and develop parameters for them.

\chapter{Temporally Stochastic Elementary Cellular Automata : Classes and Dynamics}
\label{chap3}

\section{Introduction}
During $1950$'s, Alan Turing in his article on morphogenesis~\cite{Turing} argued that the question of randomness is fundamental to understand the laws of life. To study such natural phenomenon, stochastic cellular automata (CAs) have been introduced by several CA researchers \cite{Fates17, arrighi2013stochastic, fates13}, where the update rules are chosen randomly from a set of rules. Though, traditionally cellular automata are deterministic and uniform~\cite{BhattacharjeeNR16}, i.e. all the CAs cells are updated by simultaneously identical local rule. 

In this direction, the present chapter explores another variation of non-conventional CA where the overall system temporally affected by a noise. In reality, at a time step, a cell can be updated using one of the two rules, say $f$ and $g$. The $f$ can be considered as the default rule for the CA. On the other hand, $g$ is temporally applied to the overall system with some probability which acts as a noise in the system. According to Turing (regarding morphogenesis, \cite{Turing}), the system chooses one of the symmetric direction of evolution which depicts one of the possible ways to evolve the system with randomness. The proposed framework of cellular automata is suitable for the study of such natural phenomenon. We refer to these automata as \emph{temporally stochastic cellular automata}. 

In this chapter, we deal with elementary cellular automata (ECAs) which are a one-dimensional array of finite automata. Each automaton takes two states and updates its state in discrete time depending on its own state and states of its two closest neighbors. All cells update their states synchronously. Wolfram \cite{wolfram2002new, Wolfram94} have introduced following general classification of the ECAs rules;
\begin{itemize}
	\item[] Class I: evolving to a homogeneous configuration;
	\item[] Class II: evolving periodically;
	\item[] Class III: evolving chaotically; and
	\item[] Class IV: class of complex rules.
\end{itemize}
Later, Li and Packard have identified that some periodic rules are locally chaotic \cite{Li90thestructure, Genaro13}. Obeying these classifications, we target to answer the following question -- When $f$ and $g$ belong to the same class, is it possible that during the evolution of temporally stochastic CA, it shows the behavior of a different class? That is, is it possible that two periodic (resp. chaotic) rules together depict a kind of closeness towards chaos (resp. simplicity)? Further, behavior of which class will dominate when $f$ and $g$ belong to different class? Another rich issue, is it possible to get any {\em phase transition} for a critical value of noise (i.e. probability of rule $g$). Throughout the journey, the classifications of Wolfram \cite{wolfram2002new, Wolfram94} and  Li-Packard \cite{Li90thestructure} are our point of reference. As we will see, this study is sufficiently rich to provide many of these kinds of worthy examples, with potential applications in the study of physical, chemical and biological systems. 

In the above background, this chapter is organized as follows. The temporally stochastic cellular automata is introduced in Section~\ref{S2}. The main result of the work in Section~\ref{S3}, followed by the detailed results under the temporal noise environment in Section~\ref{S4} are presented. These result sections answer all the questions that we have raised above. Finally, Section~\ref{S5} summarize this chapter and discuss the wide variety of interesting insight of the temporally stochastic system.

\section{Temporally Stochastic CAs}
\label{S2}

In this work, we consider one-dimensional three-neighborhood binary cellular automata with periodic boundary condition which is commonly known as elementary cellular automata (ECAs). The cells are arranged as a ring and the set of indices that represent the cells is denoted by $\mathcal{L}$ = $\mathbb{Z}/n\mathbb{Z}$, where $n$ is the number of cells. At each time step $t \in \mathbb{N}$, a cell is assigned a state and we denote by $Q$ = \{$0,1$\} the set of states. The collection of all states at given time is called a configuration. If $x$ is a configuration then $x =$($x_i$)$_{i \in \mathcal{L}}$ where $x_i$ is the state of cell $i \in \mathcal{L}$. The set of all configurations is denoted by $Q^{\mathcal{L}}$. 

Here, a cell changes its state depending on left neighbor, self and right neighbor. At each time step, the updates are made synchronously according to a local rule $f:$ $Q^3 \rightarrow Q$. Given a local function $f$ and a set of cells $\mathcal{L}$, one can define the global function $G: Q^{\mathcal{L}} \rightarrow Q^{\mathcal{L}}$ such that, the image $y = (y_i)_{i \in \mathcal{L}} = G(x)$ of a configuration $x = (x_i)_{i \in \mathcal{L}} \in Q^{\mathcal{L}}$ is given by,
\begin{equation}
	\nonumber
	\forall i \in \mathcal{L}, y_i = f(x_{i-1},x_i,x_{i+1})
\end{equation}

Each rule $f$ is associated with a `decimal code' $w$, where $w$ = $f$($0,0,0$) $\cdot$ $2^0$ + $f$($0,0,1$) $\cdot$ $2^1$ + $\cdots$ + $f$($1,1,1$) $\cdot$ $2^7$, for the naming purpose. There are $2^8$ = $256$ ECA rules in two-state three-neighborhood dependency. Through the use of left/right reflexion and $0/1$ complementarity, it is possible to narrow down the $256$ ECA rule space to $88$ classes, each represented by the rule of smallest number, i.e. the minimal representative ECA rule \cite{Li90thestructure}. In our work, we consider $88$ minimal representative ECA rules.

Let us now discuss temporally stochastic cellular automata where at a time step, a cell can be updated using one of the two rules $f$ and $g$. Here, $f$ is the default rule for the CA, whereas, $g$ is the noise and is applied with some probability. That is, rule $g$ is applied with probability $\tau$ $\in$ [$0,1$]  whereas the rule $f$ is applied with probability ($1 - \tau$). We call $\tau$ as the {\em temporal noise rate}. This way of looking at these rules makes both of them temporally stochastic. Therefore,  

\vspace{.3cm}
$ y = \begin{cases} G_g (x) \hspace{1cm} & \mbox{with probability}  \hspace{.2cm} \tau \\ G_f (x) & \mbox{with probability } 1 - \tau \end{cases}$
\vspace{.3cm}

where, $G_g$($x$)$\mid_i$ =  $g$($x_{i-1}$,$x_i$,$x_{i+1}$) and $G_f$($x$)$\mid_i$ =  $f$($x_{i-1}$,$x_i$,$x_{i+1}$). We write ($f,g$)[$\tau$] to denote the proposed system specification. 

In this study, we are particularly interested in the {\em qualitative} transformation (i.e. visible change in space-time diagram) that a cellular system may undergo when one progressively varies the noise rate. In this qualitative approach, we need to look at the evolution of the configurations, i.e. the space-time diagrams, by eye over a few time steps.  This traditional approach can provide a good visual comparison. However, we also study the ratio of cell with state one in the formal {\em quantitative} approach to understand phase transition kind of dynamics more properly. The density of a configuration $x \in Q^{\mathcal{L}}$ can be written as $d$($x$) = $\#_1 x$/$\mid$x$\mid$, where $\#_1 x$ is the number of $1$'s in the configuration $x$ and $\mid$x$\mid$ is the size of the configuration. Without loss of generality, we start with a configuration of initial density $0.5$ which is constructed using {\em Bernoulli} process. In both qualitative and quantitative study, we start with a configuration of CA size $500$. Here, we let the system evolve during $2000$ time steps and average the density parameter value for (last) $100$ time steps. It is also possible that the system may show different behavior for different run. Therefore, for each instance (i.e. for each ($f,g$)[$\tau$]), we study the cellular system's dynamics for $25$ times. Although the results reported here are based on CA size $500$, we repeat the experiment for various other sizes to cross-verify the result and observe that the dynamical behavior of an automaton remains almost same for the other sizes. Therefore, in general, we claim that the cellular system's almost always show similar dynamics for any CA size.

\section{Main result} 
\label{S3}
As this study uses Wolfram's \cite{wolfram2002new, Wolfram94} and  Li-Packard's \cite{Li90thestructure} classification, we first note down the Wolfram's classification for $88$ minimal representative ECAs in Table~\ref{Table1}. According to Li-Packard, Wolfram's class II CAs show fixed point, periodic, locally chaotic behavior which are respectively marked with black, underlined and bold in Table~\ref{Table1}. Here, $f$ and $g$ are taken from the $88$ minimal representative ECAs set. Hence, working with $^{88}C_2$ = $\frac{88 \times 87}{2}$ = $3828$ couples of ($f,g$) is sufficient after considering the exchange symmetry between $f$ and $g$.

\begin{table}
	\centering
	\begin{adjustbox}{width=\columnwidth,center}
		\begin{tabular}{ccccccccccccc} \hline  
			Class I & 0 & 8 & 32 & 40 & 128 & 136 & 160 & 168 &  &  &  & \\ 
			Class II & 2 & 4 & 10 & 12 & 13 & 24 & 34 & 36 & 42 & 44 & 46 & 56\\ 
			& 57 & 58 & 72 & 76 & 77 & 78 & 104 & 130 & 132 & 138 & 140 & 152\\
			& 162 & 164 & 170 & 172 & 184 & 200 & 204 & 232 & \underline{1} & \underline{3} & \underline{5}  &  \underline{6}\\
			& \underline{7} & \underline{9} & \underline{11} & \underline{14} & \underline{15} & \underline{19} & \underline{23} & \underline{25} & \underline{27} & \underline{28} & \underline{29}  &  \underline{33}\\
			& \underline{35} & \underline{37} & \underline{38} & \underline{43} & \underline{50} & \underline{51} & \underline{62} & \underline{74} & \underline{94} & \underline{108} & \underline{134}  &  \underline{142}\\
			& \underline{156} & \underline{178} & \textbf{26} & \textbf{73} & \textbf{154} &  &  &  &  &  &   &  \\
			Class III & 18 & 22 & 30 & 45 & 60 & 90 & 105 & 122 & 126 & 146 & 150 & \\ 
			Class IV & 41 & 54 & 106 & 110 &  & &  & &  &  &  & \\
			\hline
		\end{tabular}
	\end{adjustbox}
		\caption{Wolfram's classification for $88$ minimal representative ECAs.}			
		\label{Table1}
\end{table}

Therefore, this study captures the dynamics of $3828$ couples of ($f,g$) following the qualitative and quantitative (if necessary) approach where the dynamics of ($f,g$) targets to identify the similarities in  Wolfram's classes. However, during this mapping, to make this first experience easier, we merge Wolfram's class III and IV, i.e. chaotic and complex behavior, and locally chaotic behavior  (Li-Packard's class) together. Obviously, theoretically the dynamics of chaotic and locally chaotic rules are different. However, according to space-time diagram, locally chaotic rules show more closeness towards chaotic and complex dynamics in comparison with periodic dynamics. Hence, we map the dynamics of ($f,g$) into following three classes.
\begin{itemize}
	\item[-] Class A, which is similar to Wolfram's class I (or, Li-Packard's Uniform (U) behavior);
	\item[-] Class B, which is similar to Wolfram's class II, except three Locally Chaotic (LC) dynamics (that is, Li-Packard's Fixed Point (FP) and Periodic (P) behavior); and
	\item[-] Class C, which is similar to Wolfram's class III (Chaotic (C)) and class IV (Complex (CO)) along with those three locally chaotic behavior. 
\end{itemize}

For later part of the work, we shall use these class names (A,B,C) for individual ECA rules also. That is, rules of class I will be treated as rules of class A, and so on. However, we can not always be able to map the dynamics for many couples of ($f,g$) into above three classes. These couples of ($f,g$) show the well known {\em phase transition} \cite{ROY2019600} and {\em transition of class} (similar to \cite{Martinez12}) dynamics. 

\begin{figure}[hbt!]\centering 
	\includegraphics[width=5.8in]{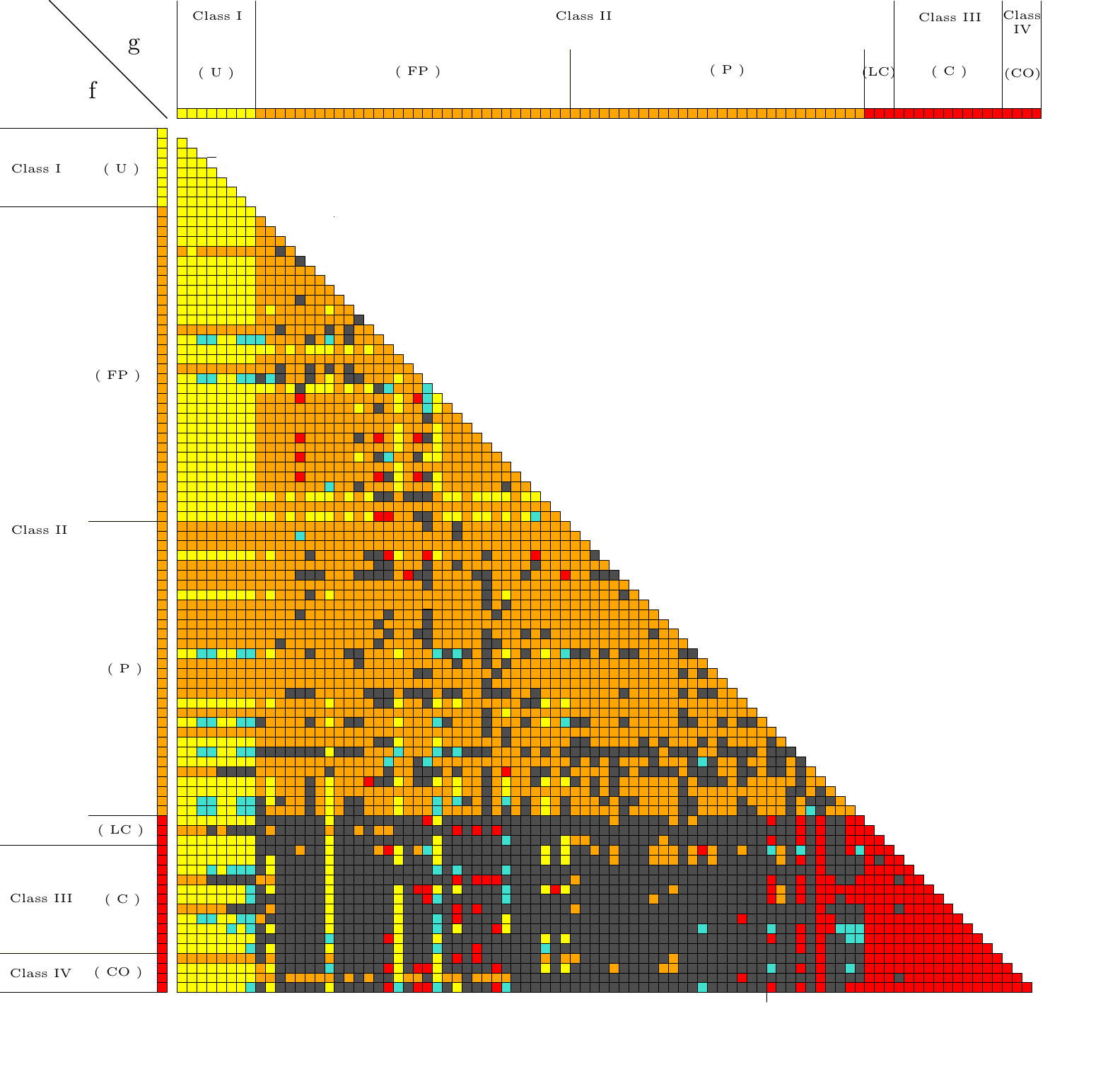} 
	\caption{Summarized behavior of temporally stochastic cellular automata.}
	\label{Fig1}
\end{figure}

Overall, we have performed a large number of experiments on the pairs of rules. The summary of the outcome is note in Fig.~\ref{Fig1} where class A,~B and C are marked by yellow, orange and red respectively. The $88$ rules are plotted horizontally and vertically where the vertical line shows the $f$ rules and $g$ is represented by the horizontal line. Each box on the line (vertical and horizontal) represents a rule. The rules are numbered as per the sequence of Table~\ref{Table1}. That is, rule $0$ is the first rule and rule $110$ is the $88$th rule. A cell ($i,j$) in the figure depicts the behavior of the stochastic CA ($f,g$) where $f$ and $g$ are the rules represented by the $i$th box in the vertical line and $j$th box in the horizontal line respectively. Here, we consider the exchange symmetry between $f$ and $g$. So while stochastic CA ($f,g$) is plotted, the place for the CA ($g,f$) remains blank (white) in Fig.~\ref{Fig1}.
Additionally, the place for the CA ($f,f$) is kept blank (white) in the Fig.~\ref{Fig1}. To illustrate the above discussion clearly, the following matrix depicts the partial representation of Fig.~\ref{Fig1}.

\begin{table}[!htbp] 
	\centering
	\scriptsize
	\begin{tabular}{c|cccccc} \hline 
		& 0 & 8 & 32 & 40 & 128 & $\cdots$ \\ \hline
		0 & & & & & & \\ 
		8 & (8,0) & & & & & \\ 
		32 & (32,0) & (32,8) & & & & \\ 
		40 & (40,0) & (40,8) & (40,32) & & & \\ 
		128 & (128,0) & (128,8) & (128,32) & (128,40) & & \\ 
		$\vdots $ & $\vdots $ & $\vdots $ & $\vdots $ & $\vdots $ & $\vdots $ & \\ 
	\end{tabular}
\end{table}

Now, there are two possibilities for a couple ($f,g$) -- $f$ and $g$ belong to the same class; and $f$ and $g$ are from different class. We denote the class of $f$ and $g$ as $\mathcal{C}$($f$) and $\mathcal{C}$($g$) respectively and class of ($f,g$) is denoted by $\mathcal{C}$(($f,g$)). According to Fig~\ref{Fig1}, following are the rich set of observations regarding the dynamics of those temporally stochastic rules.

\begin{itemize}
	
	\item[1.] If $\mathcal{C}$($f$) = $\mathcal{C}$($g$), then the first possibility is $\mathcal{C}$(($f,g$)) = $\mathcal{C}$($f$), dynamics of which observed in large number of stochastic CA ($f,g$). Under this case, the CA ($f,g$) where $\mathcal{C}$($f$) = $\mathcal{C}$($g$) = class A, evolves to a homogeneous configuration as ECA $f$ and ECA $g$ do. 
	
	\item[2.] When $\mathcal{C}$($f$) = $\mathcal{C}$($g$), then it may be possible that, $\mathcal{C}$(($f,g$)) $\neq$ $\mathcal{C}$($f$). That is, these stochastic CAs ($f,g$) are massively affected by the noise. Under this umbrella of dynamics, the most two interesting observations are:
	\begin{itemize}
		\item[-] Couple of two periodic rules (class B) show kind of closeness towards chaotic/complex/locally chaotic dynamics (class C). Observe that, in Fig.~\ref{Fig1}, many couple of orange CAs show red dynamics.
		\item[-] On the contrary, couple of two periodic rules (class B) show uniform (evolving to homogeneous configuration, i.e. class A) dynamics. See, many couples of orange CAs depict yellow dynamics in Fig.~\ref{Fig1}. 
	\end{itemize}
	
	\item[3.] Next a stochastic CA ($f,g$) where $\mathcal{C}$($f$) $\neq$ $\mathcal{C}$($g$), shows the dynamics where one of the rule's class dominates, i.e. $\mathcal{C}$(($f,g$)) = $\mathcal{C}$($f$) or $\mathcal{C}$(($f,g$)) = $\mathcal{C}$($g$). Under this case, the CA ($f,g$) where $\mathcal{C}$($f$) = class A and $\mathcal{C}$($g$) = class C, shows a behavior with $\mathcal{C}$(($f,g$)) = $\mathcal{C}$($f$). Similarly, the stochastic CA ($f,g$) with $\mathcal{C}$($f$) = class B and $\mathcal{C}$($g$) = class C shows the dynamics as $\mathcal{C}$(($f,g$)) = $\mathcal{C}$($g$), see Fig.~\ref{Fig1}.

	\item[4.] On the other hand, a stochastic CA ($f,g$) where $\mathcal{C}$($f$) $\neq$ $\mathcal{C}$($g$), shows the dynamics where none of the rule's class dominates, i.e. $\mathcal{C}$(($f,g$)) $\neq$ $\mathcal{C}$($f$) and $\mathcal{C}$(($f,g$)) $\neq$ $\mathcal{C}$($g$). Here, the CA ($f,g$) where $\mathcal{C}$($f$) = class A and $\mathcal{C}$($g$) = class C, shows a behavior with $\mathcal{C}$(($f,g$)) = class B. Similarly, the CA ($f,g$) with $\mathcal{C}$($f$) = class B and $\mathcal{C}$($g$) = class C shows the dynamics as $\mathcal{C}$(($f,g$)) = class A, see Fig.~\ref{Fig1}.
	
\end{itemize}

Till now we have not mentioned about the temporal noise rate ($\tau$). That is, for the above cases, if we progressively vary the temporal noise rate, the cellular system's dynamics remains unchanged.
To sum up, these stochastic CAs ($f,g$) are not temporal noise rate ($\tau$) sensitive \footnote{Note that, the cellular system may behaves differently for sufficiently small value of $\tau$ ($\tau \approx 0.01$).} However, there are cases (following) where the system is $\tau$ sensitive. 

\begin{itemize}
	\item[5.] Interestingly, some stochastic CAs ($f,g$) show a discontinuity after a critical value of temporal noise rate ($\tau$). This type of brutal change of behavior is well known as second-order {\em phase transition}. In this case, there exists a critical temporal noise rate, say $\tau_c$, which separates a behavior where the system converges to $0^{\mathcal{L}}$ (passive phase) and a behavior with a stationary non-zero density (active phase). In Fig.~\ref{Fig1}, the CA ($f,g$) with this phase transition behavior are marked by blue.
	
	\item[6.] Lastly, for a set of stochastic CAs ($f,g$), the class dynamics of the system changes after a critical value of $\tau$, say $\tau_t$. That is,  $\mathcal{C}$(($f,g$))[$\tau$] $\neq$ $\mathcal{C}$(($f,g$))[$\tau'$] where $\tau \in$ [0,$\tau_t$] and $\tau' \in$ [$\tau_t$,1]. Here, a CA ($f,g$) with $\mathcal{C}$($f$) = class B and $\mathcal{C}$($g$) = class C shows periodic behavior, but slowly transforms into chaotic dynamics after critical value of noise $\tau_t$. Fig~\ref{Fig1} notes these kind of behavior in black. In the rest of the chapter, we will denote this dynamics as {\em class transition}.
\end{itemize}

\begin{figure*}[hbt!]
	\begin{center}
		\scalebox{0.8}{
			\begin{tabular}{ccccc}
				ECA 22 & ECA 18 & (22,18)[0.1] & (22,18)[0.5] & (22,18)[0.9] \\[6pt]
				\includegraphics[width=31mm]{TEM_IMAGE/NR22-eps-converted-to.pdf} & \includegraphics[width=31mm]{TEM_IMAGE/NR18-eps-converted-to.pdf} &   \includegraphics[width=31mm]{TEM_IMAGE/NR22_18_1-eps-converted-to.pdf} &   \includegraphics[width=31mm]{TEM_IMAGE/NR22_18_5-eps-converted-to.pdf} &   \includegraphics[width=31mm]{TEM_IMAGE/NR22_18_9-eps-converted-to.pdf} \\
				ECA 150 & ECA 126 & (150,126)[0.1] & (150,126)[0.5] & (150,126)[0.9] \\
				\includegraphics[width=31mm]{TEM_IMAGE/NR150-eps-converted-to.pdf} & \includegraphics[width=31mm]{TEM_IMAGE/NR126-eps-converted-to.pdf} &   \includegraphics[width=31mm]{TEM_IMAGE/NR150_126_1-eps-converted-to.pdf} &   \includegraphics[width=31mm]{TEM_IMAGE/NR150_126_5-eps-converted-to.pdf}   &   \includegraphics[width=31mm]{TEM_IMAGE/NR150_126_9-eps-converted-to.pdf} \\
				ECA 43 & ECA 77 & (43,77)[0.1] & (43,77)[0.5] & (43,77)[0.9] \\
				\includegraphics[width=31mm]{TEM_IMAGE/NR43-eps-converted-to.pdf} & \includegraphics[width=31mm]{TEM_IMAGE/NR77-eps-converted-to.pdf} &   \includegraphics[width=31mm]{TEM_IMAGE/NR43_77_1-eps-converted-to.pdf} &   \includegraphics[width=31mm]{TEM_IMAGE/NR43_77_5-eps-converted-to.pdf}  &   \includegraphics[width=31mm]{TEM_IMAGE/NR43_77_9-eps-converted-to.pdf} \\
		\end{tabular}}
		\caption{Stochastic CAs ($f,g$) dynamics when $\mathcal{C}$(($f,g$)) = $\mathcal{C}$($f$) = $\mathcal{C}$($g$).}
		\label{Fig2}
	\end{center}
\end{figure*}

\begin{figure*}[hbt!]
	\begin{center}
		\scalebox{0.8}{
			\begin{tabular}{ccccc}
				ECA 33 & ECA 5 & (33,5)[0.1] & (33,5)[0.5] & (33,5)[0.9] \\[6pt]
				\includegraphics[width=31mm]{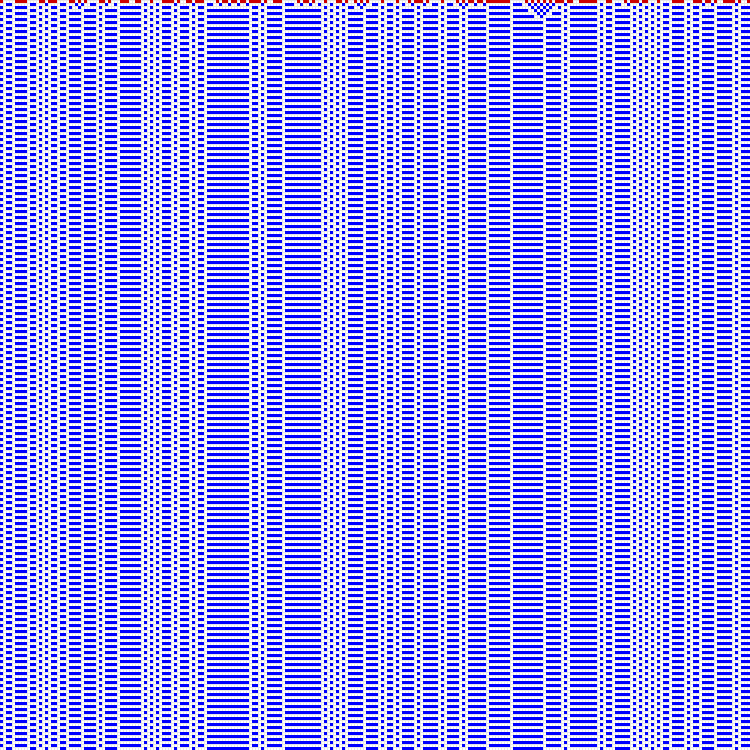} & \includegraphics[width=31mm]{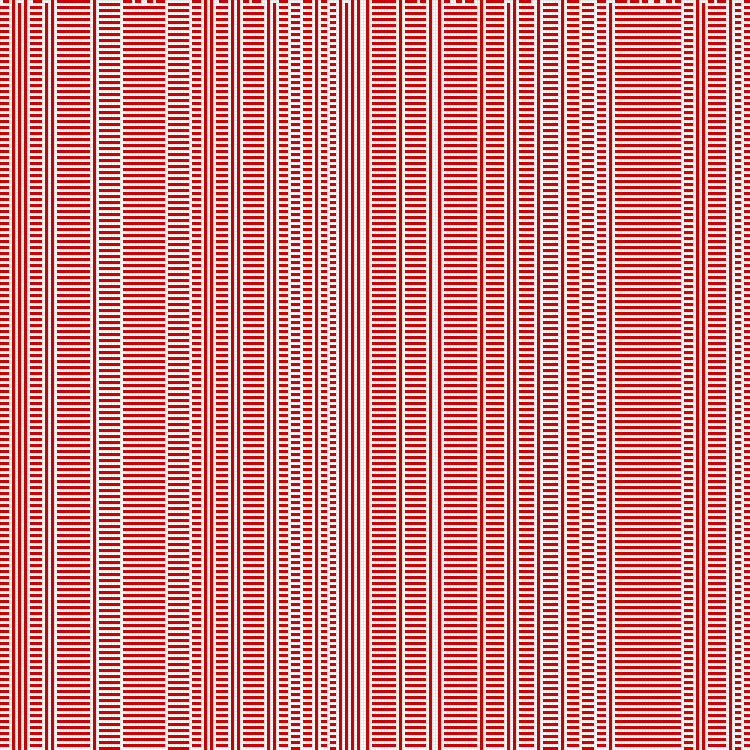} &   \includegraphics[width=31mm]{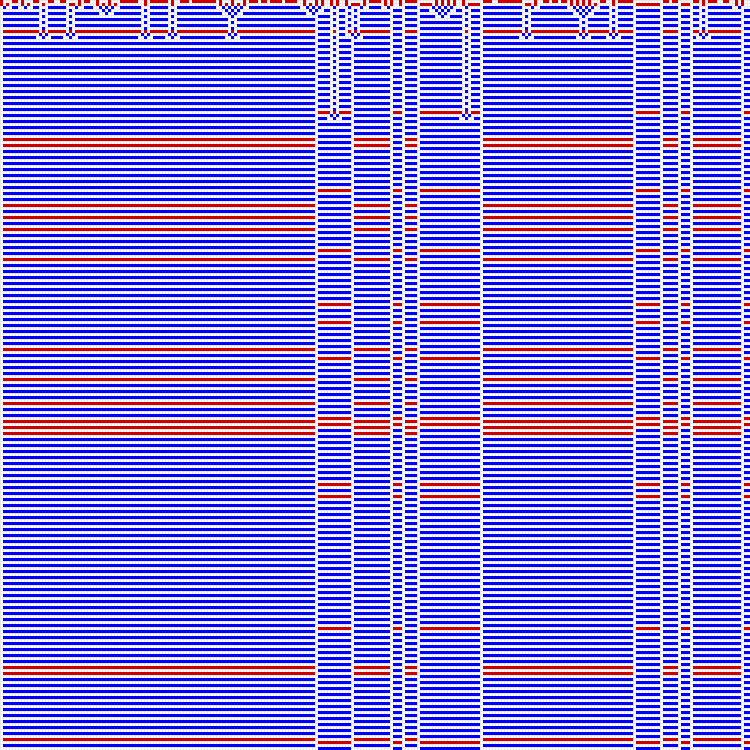} &   \includegraphics[width=31mm]{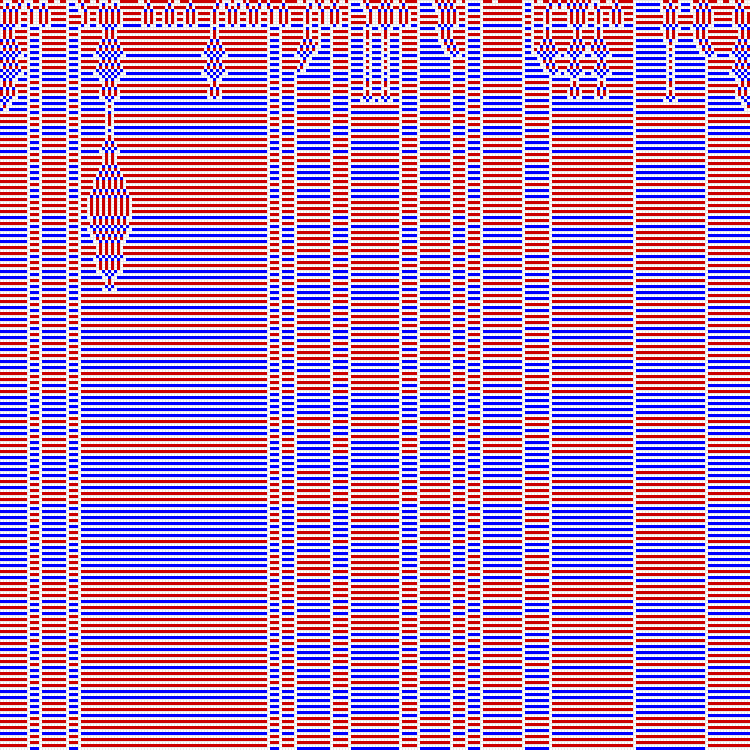} &   \includegraphics[width=31mm]{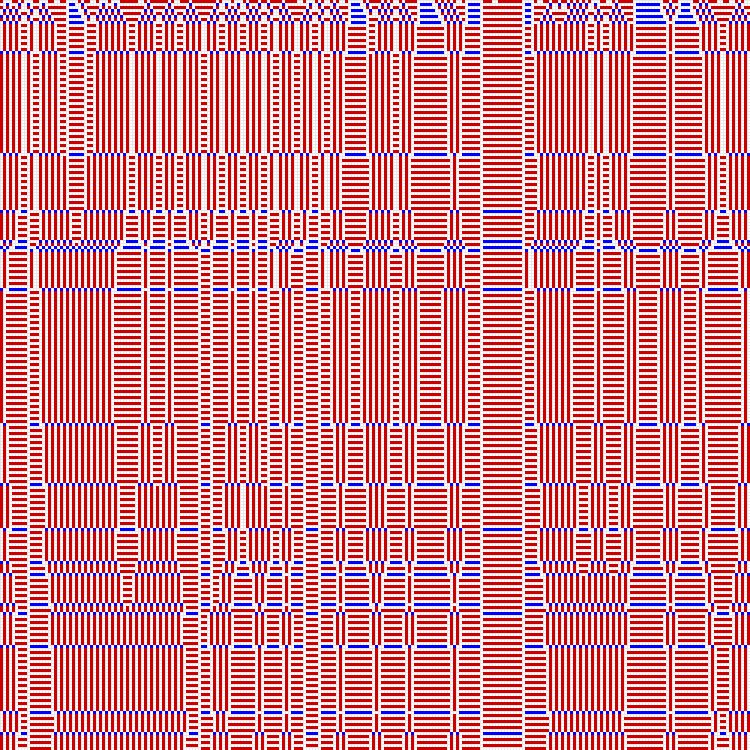} \\
				ECA 45 & ECA 30 & (45,30)[0.1] & (45,30)[0.5] & (45,30)[0.9] \\[6pt]
				\includegraphics[width=31mm]{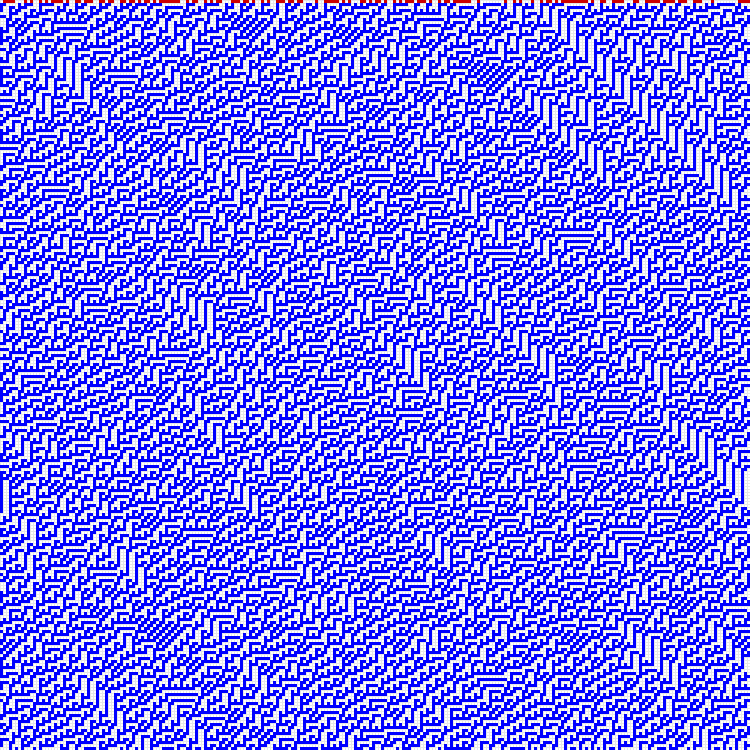} & \includegraphics[width=31mm]{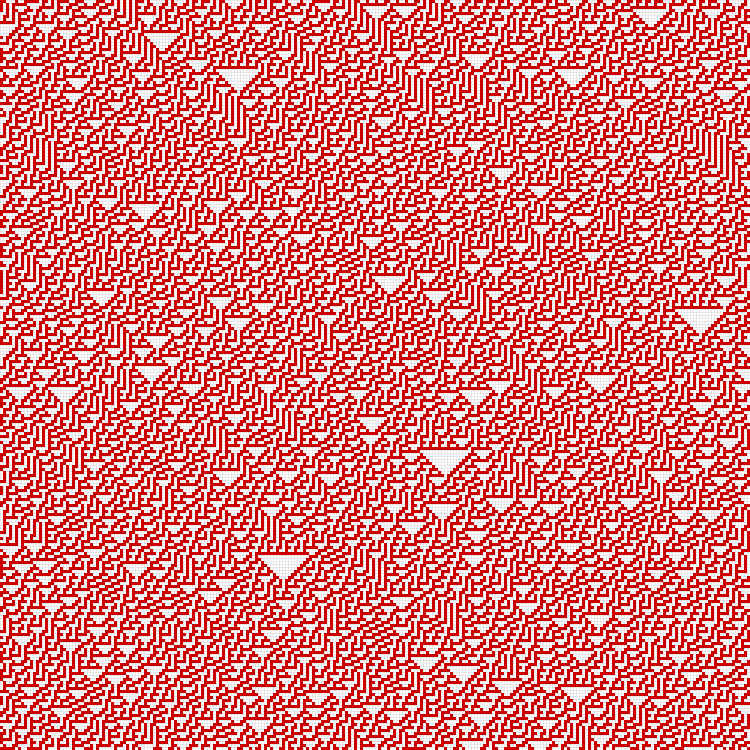} &   \includegraphics[width=31mm]{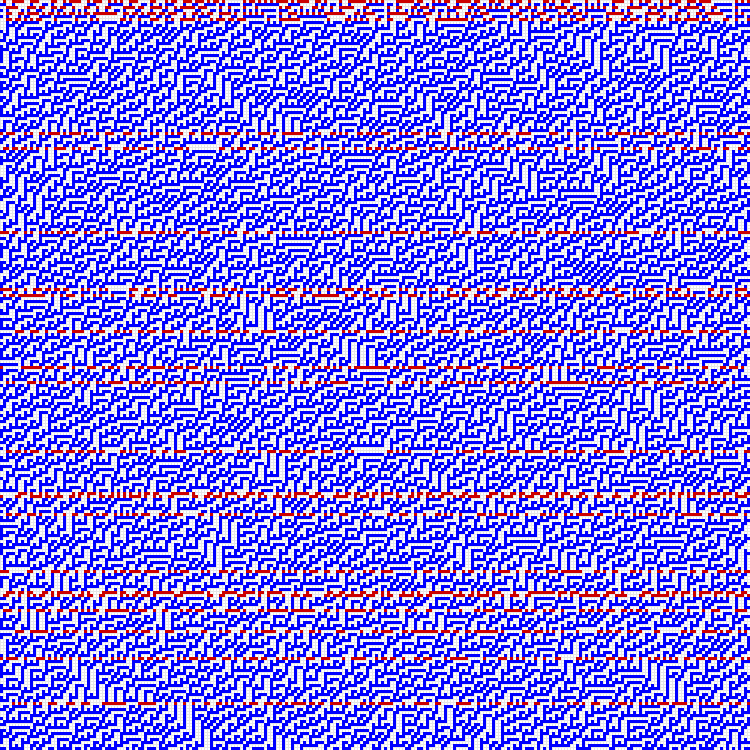} &   \includegraphics[width=31mm]{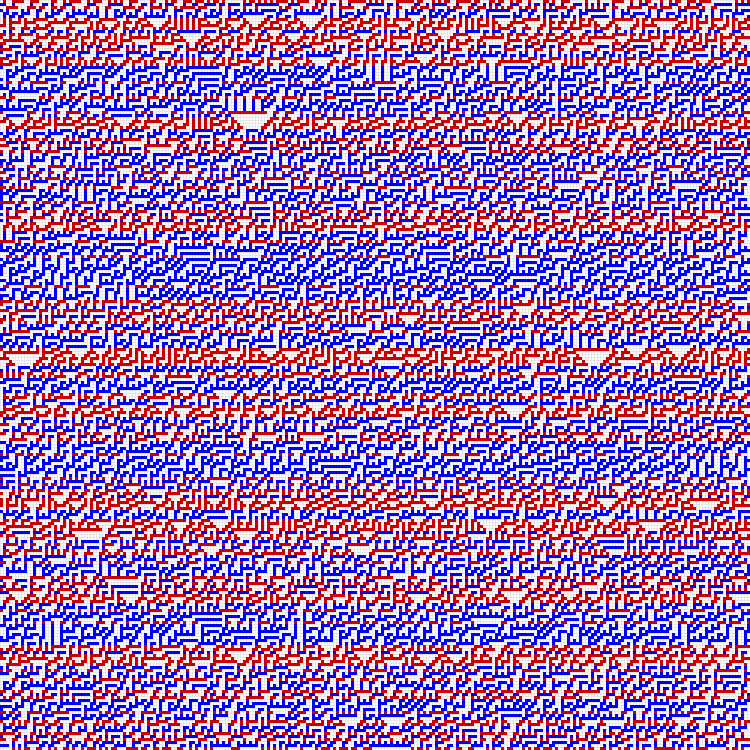} &   \includegraphics[width=31mm]{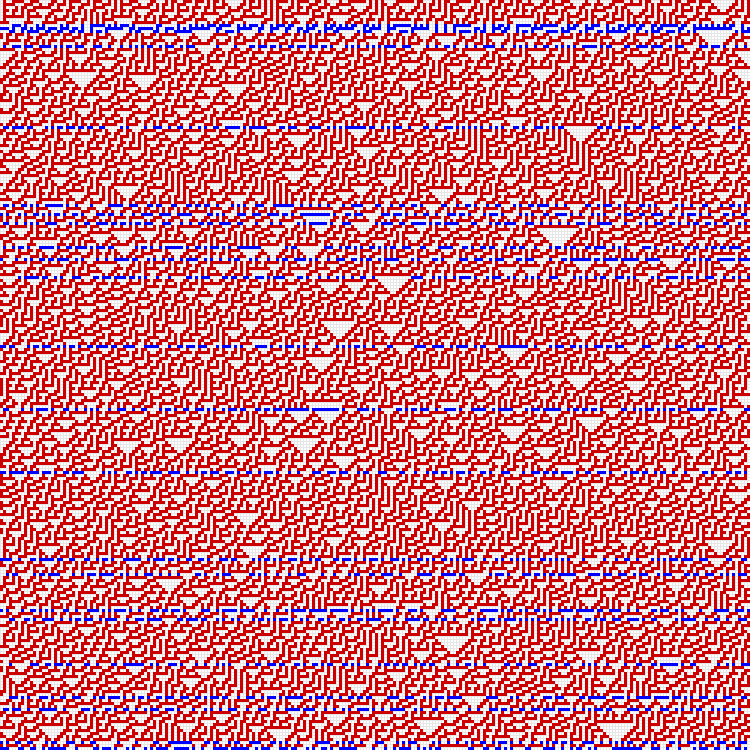} \\
				ECA 172 & ECA 140 & (172,140)[0.1] & (172,140)[0.5] & (172,140)[0.9] \\[6pt]
				\includegraphics[width=31mm]{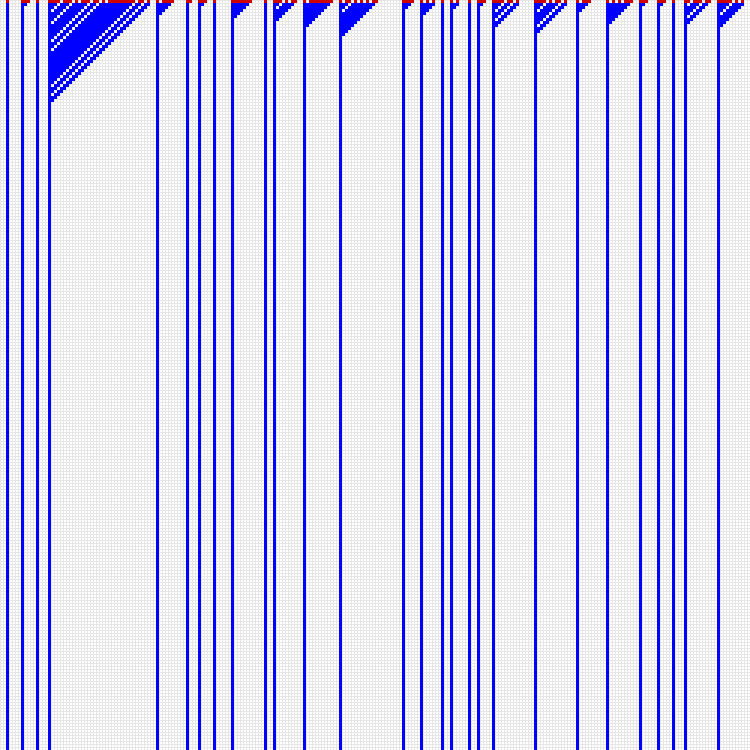} & \includegraphics[width=31mm]{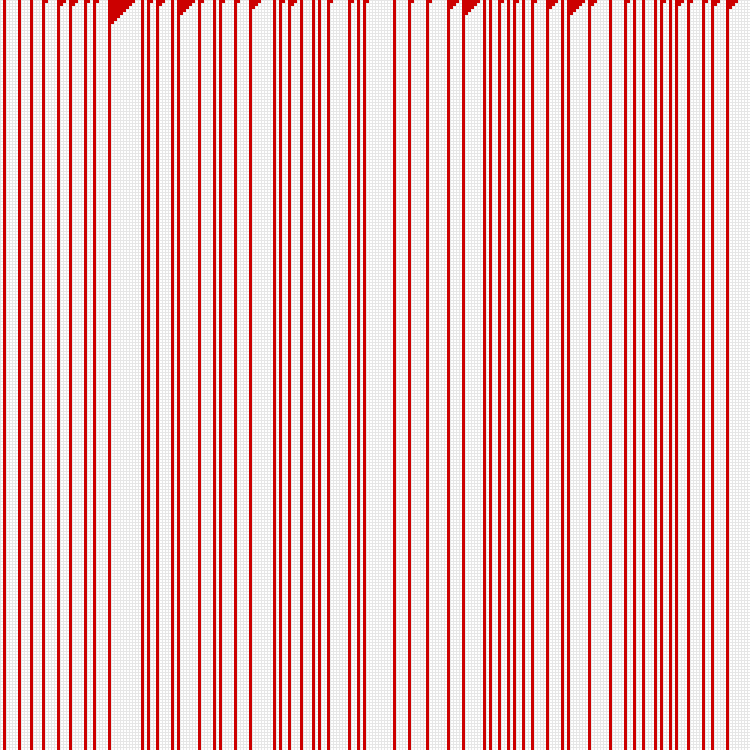} &   \includegraphics[width=31mm]{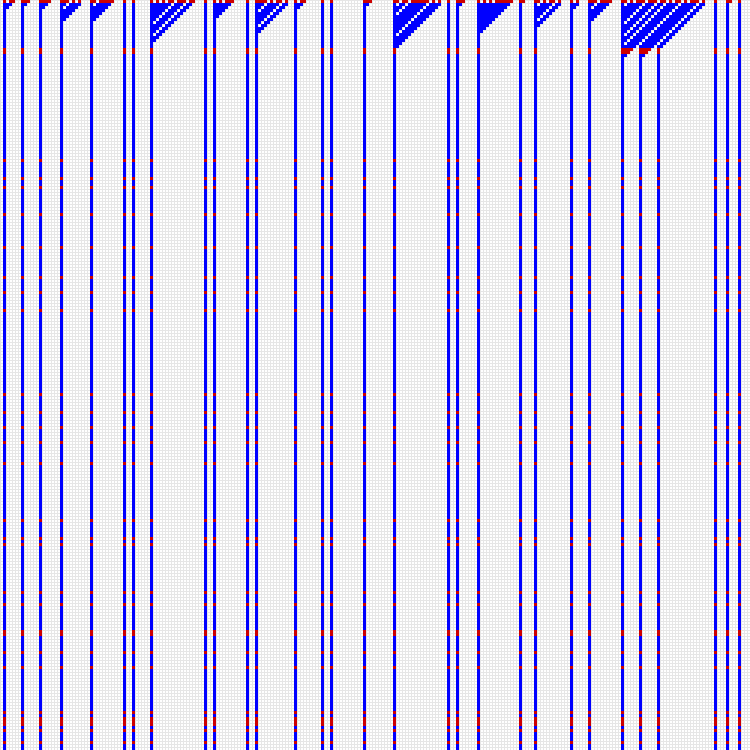} &   \includegraphics[width=31mm]{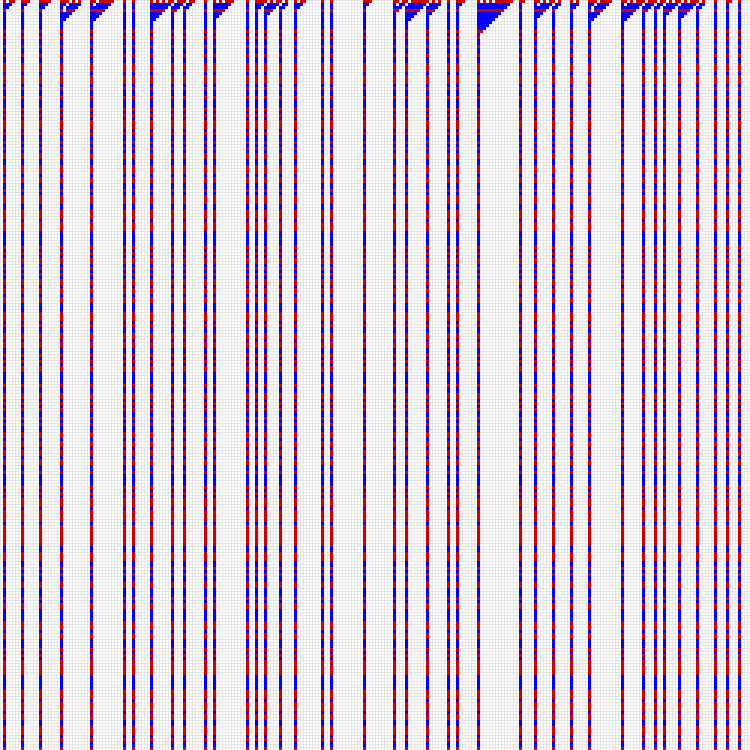} &   \includegraphics[width=31mm]{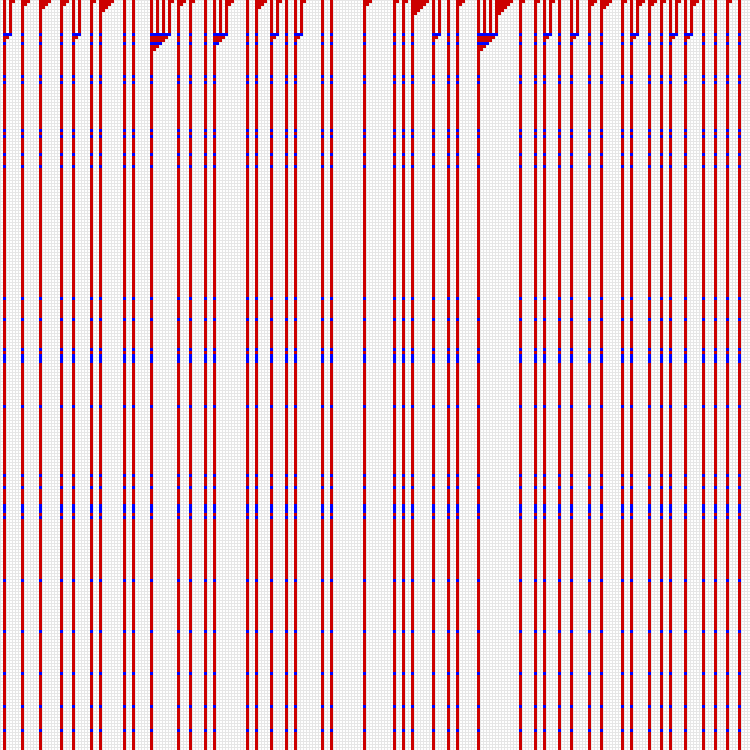} \\
		\end{tabular}}
		\caption{Stochastic CAs ($f,g$) dynamics when $\mathcal{C}$(($f,g$)) = $\mathcal{C}$($f$) = $\mathcal{C}$($g$).}
		\label{TSCA1}
	\end{center}
\end{figure*}
The above discussion indicates the rich set of possibility in dynamics for the temporal stochastic CAs. In the next section, we revisit these dynamics with examples, and in more detail. 

\section{Detailed results}
\label{S4}

Let us now detail out the results that we have presented in previous section. Here we pick up suitable examples for different cases to illustrate the behavior of the automaton.

\begin{figure*}[hbt!]
	\begin{center}
		\scalebox{0.8}{
			\begin{tabular}{ccccc}
				ECA 164 & ECA 131 & (164,131)[0.1] & (164,131)[0.2] & (164,131)[0.3] \\[6pt]
				\includegraphics[width=31mm]{TEM_IMAGE/NR164-eps-converted-to.pdf} & \includegraphics[width=31mm]{TEM_IMAGE/NR131-eps-converted-to.pdf} & \includegraphics[width=31mm]{TEM_IMAGE/NR164_131_1-eps-converted-to.pdf} & \includegraphics[width=31mm]{TEM_IMAGE/NR164_131_2-eps-converted-to.pdf} & \includegraphics[width=31mm]{TEM_IMAGE/NR164_131_3-eps-converted-to.pdf} \\
				
				ECA 164 & ECA 13 & (164,13)[0.1] & (164,13)[0.2] & (164,13)[0.3] \\
				
				\includegraphics[width=31mm]{TEM_IMAGE/NR164-eps-converted-to.pdf}  &  \includegraphics[width=31mm]{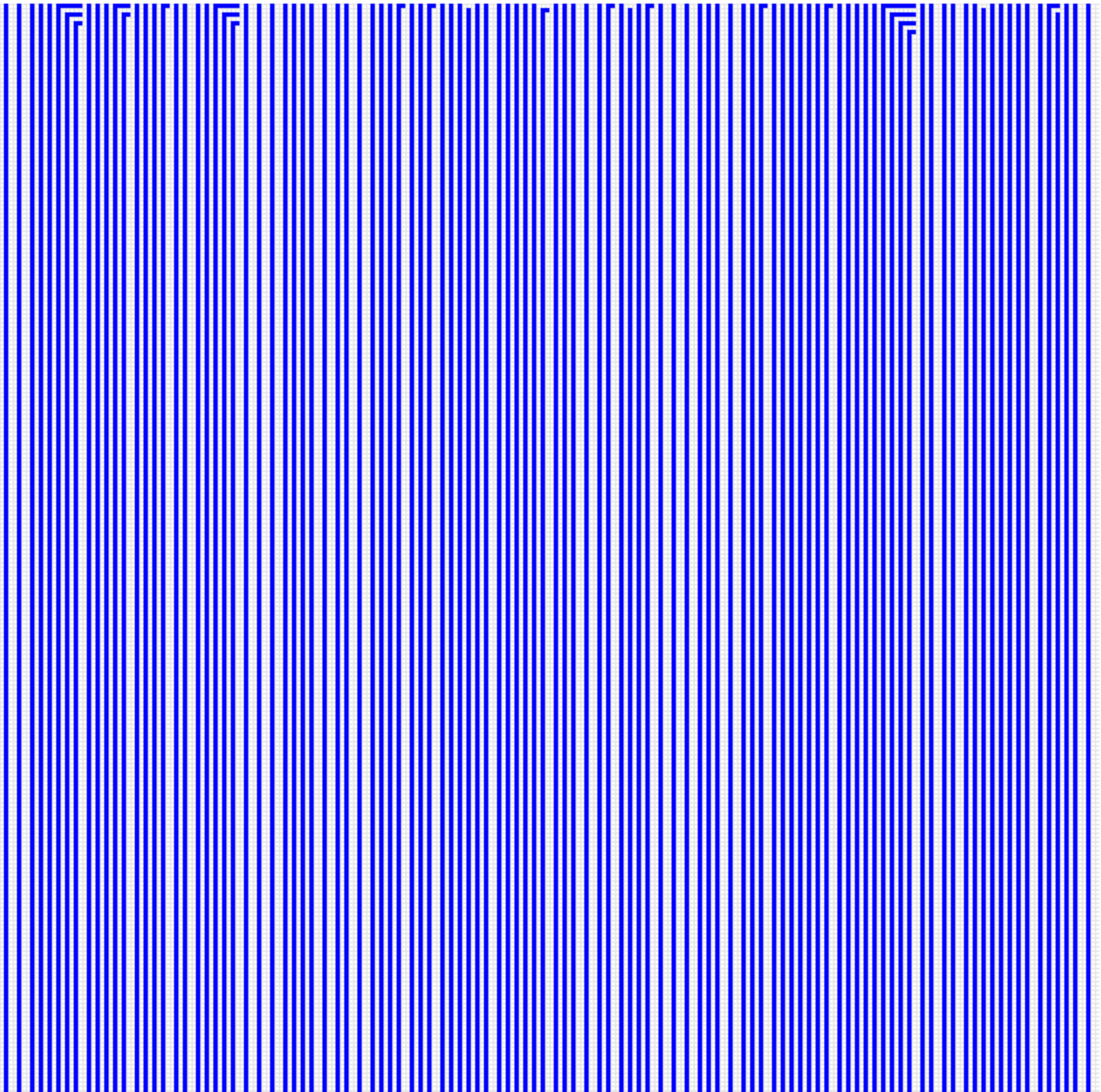}  &   \includegraphics[width=31mm]{TEM_IMAGE/NR164_13_1-eps-converted-to.pdf}  &  \includegraphics[width=31mm]{TEM_IMAGE/NR164_13_2-eps-converted-to.pdf}  & \includegraphics[width=31mm]{TEM_IMAGE/NR164_13_3-eps-converted-to.pdf} \\
				
				ECA 200 & ECA 130 & (200,130)[0.1] & (200,130)[0.2] & (200,130)[0.3] \\
				
				\includegraphics[width=31mm]{TEM_IMAGE/NR200-eps-converted-to.pdf} & \includegraphics[width=31mm]{TEM_IMAGE/NR130-eps-converted-to.pdf}  & \includegraphics[width=31mm]{TEM_IMAGE/NR200_130_1-eps-converted-to.pdf} & \includegraphics[width=31mm]{TEM_IMAGE/NR200_130_2-eps-converted-to.pdf}  & \includegraphics[width=31mm]{TEM_IMAGE/NR200_130_3-eps-converted-to.pdf} \\
		\end{tabular}}
		\caption{Stochastic CAs ($f,g$) dynamics when $\mathcal{C}$(($f,g$)) $\neq$ $\mathcal{C}$($f$) and $\mathcal{C}$($f$) = $\mathcal{C}$($g$).}
		\label{Fig3}
	\end{center}
\end{figure*}

\begin{figure*}[hbt!]
	\begin{center}
		\scalebox{0.8}{
			\begin{tabular}{ccccc}
				ECA 9 & ECA 77 & (9,77)[0.1] & (9,77)[0.5] & (9,77)[0.9] \\[6pt]
				\includegraphics[width=31mm]{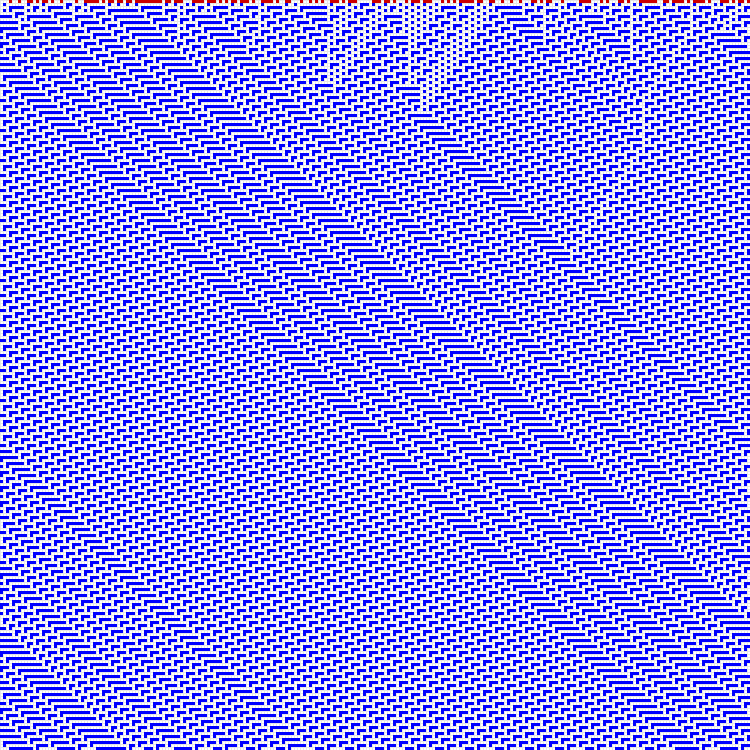} & \includegraphics[width=31mm]{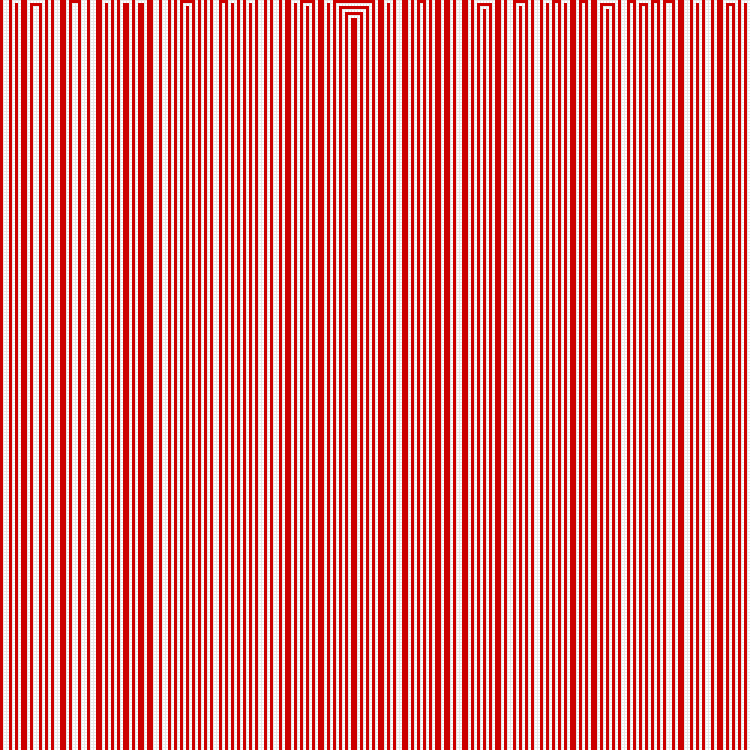} &   \includegraphics[width=31mm]{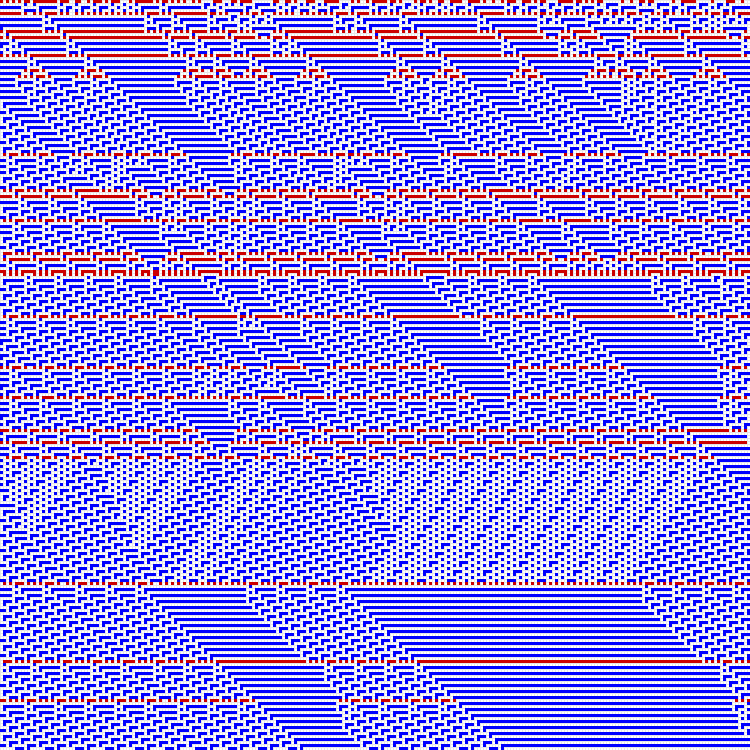} &   \includegraphics[width=31mm]{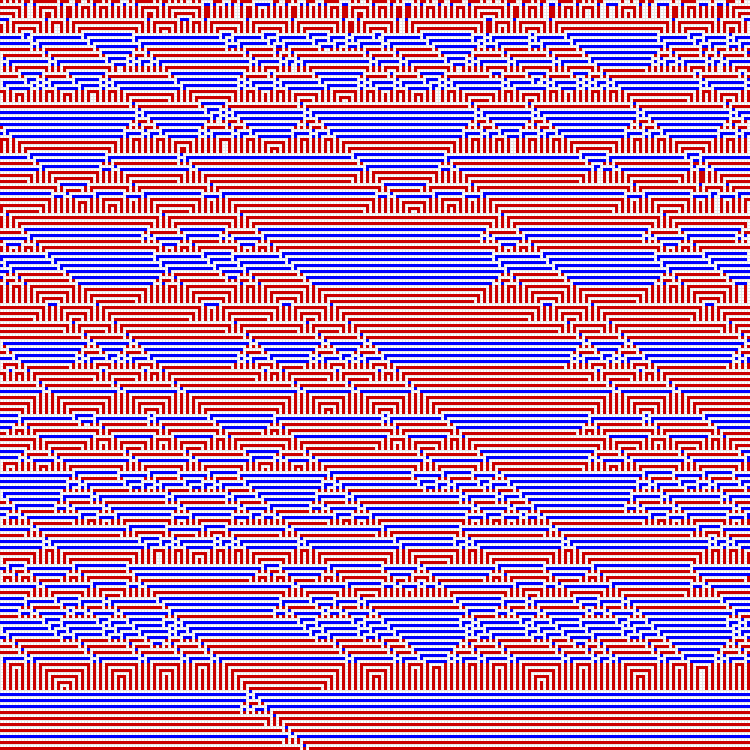} &   \includegraphics[width=31mm]{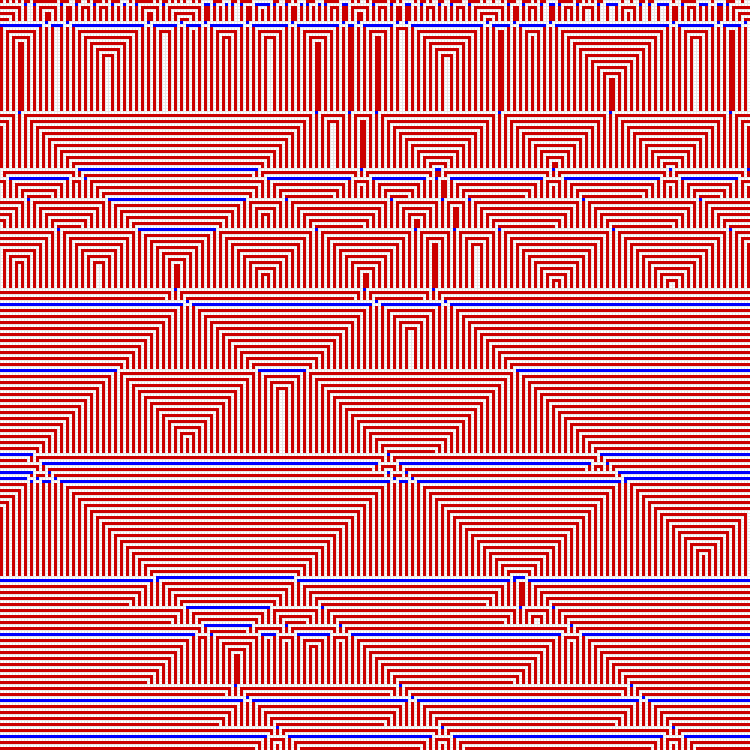} \\
				ECA 130 & ECA 13 & (130,13)[0.1] & (130,13)[0.5] & (130,13)[0.9] \\[6pt]
				\includegraphics[width=31mm]{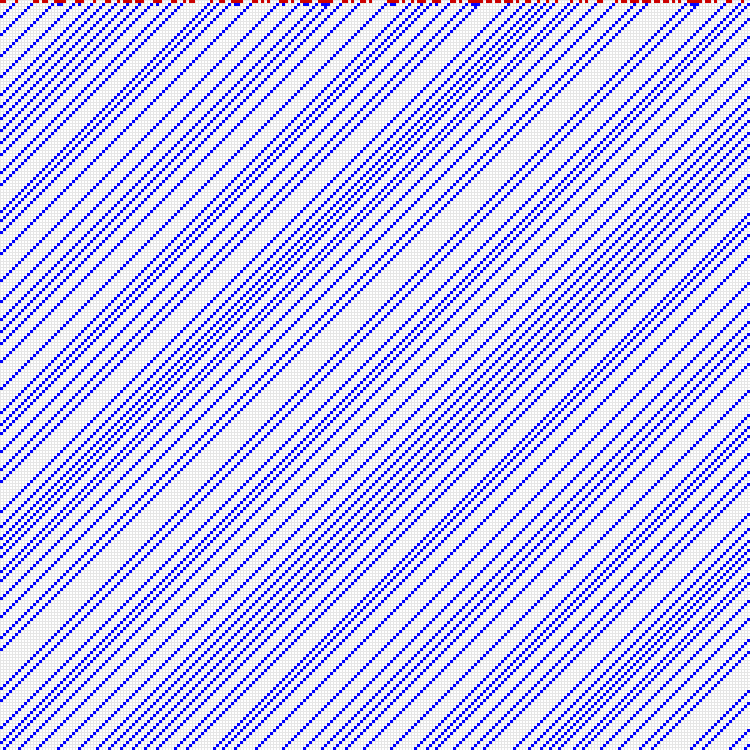} & \includegraphics[width=31mm]{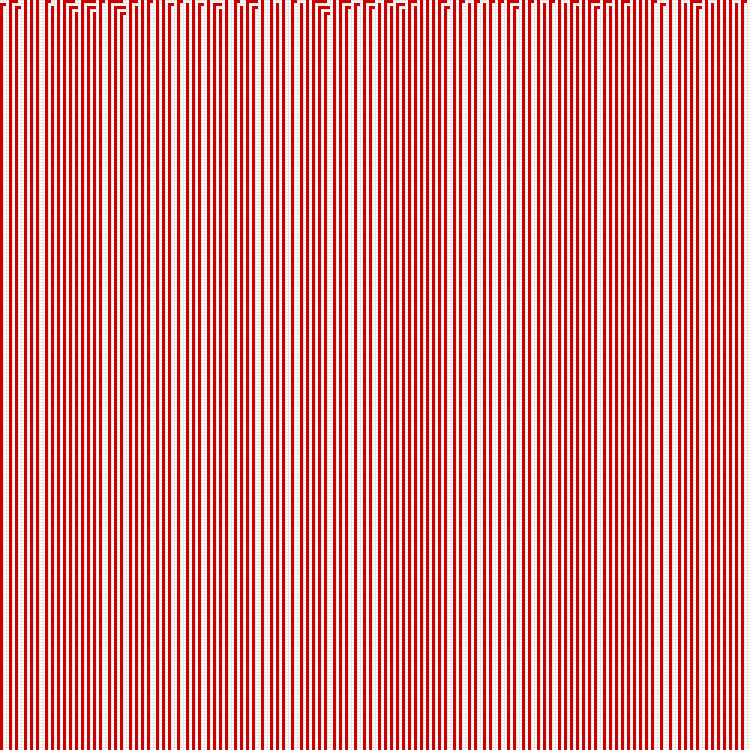} &   \includegraphics[width=31mm]{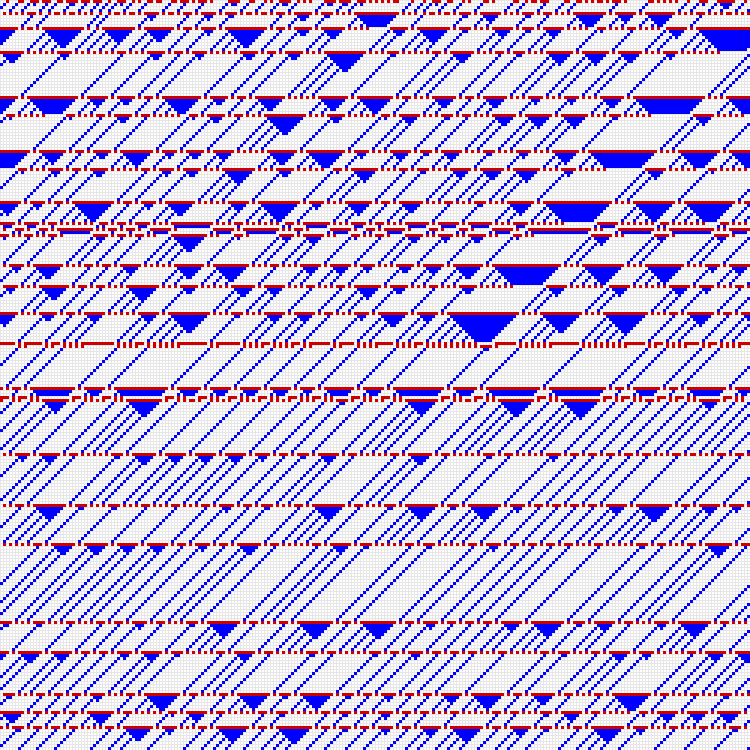} &   \includegraphics[width=31mm]{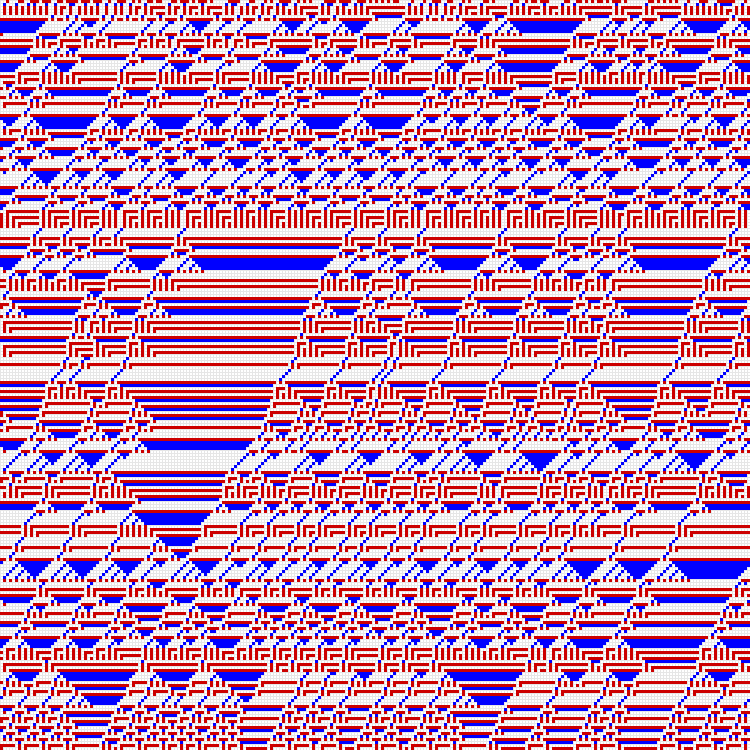} &   \includegraphics[width=31mm]{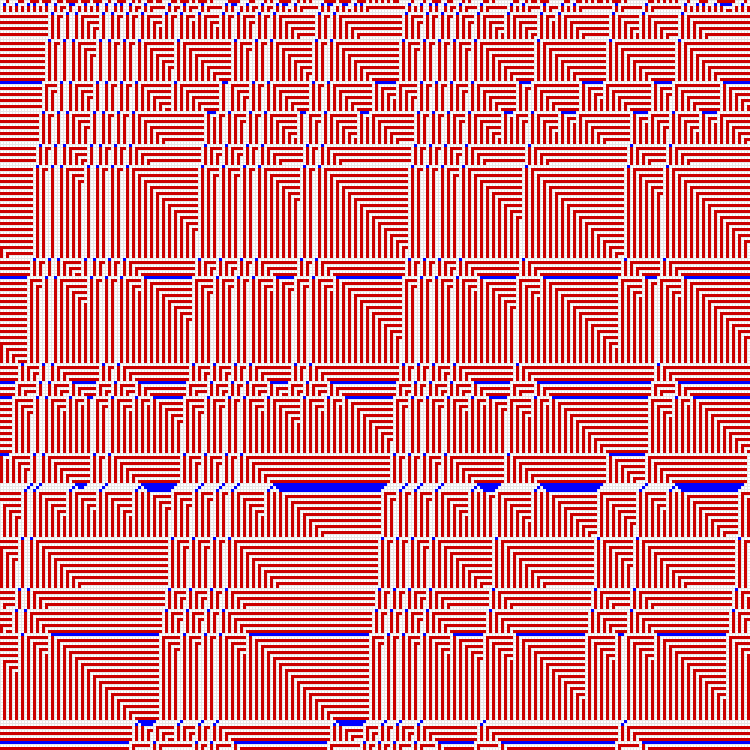} \\
				ECA 172 & ECA 77 & (172,77)[0.1] & (172,77)[0.5] & (172,77)[0.9] \\[6pt]
				\includegraphics[width=31mm]{TEM_IMAGE/Added/rule172.png} & \includegraphics[width=31mm]{TEM_IMAGE/Added/rule77v1.png} &   \includegraphics[width=31mm]{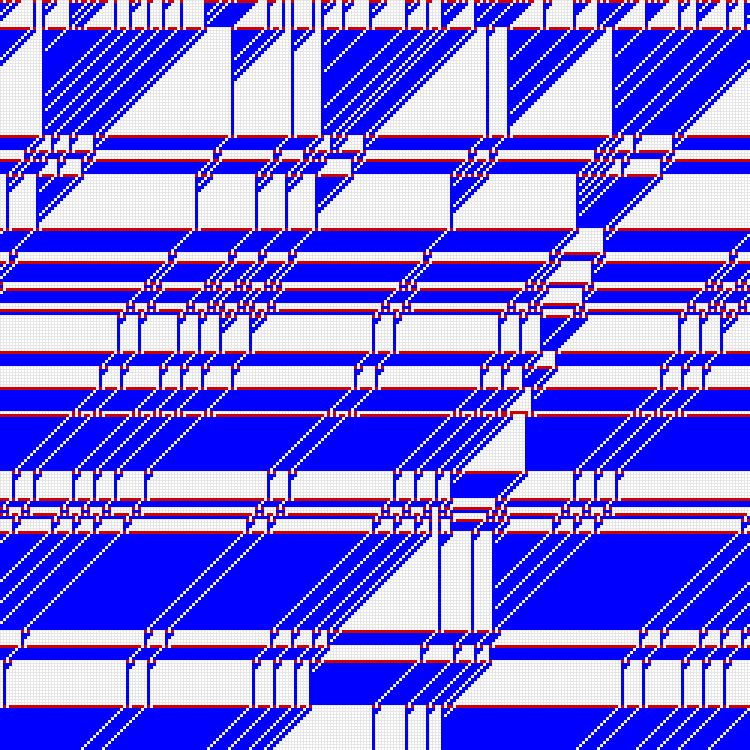} &   \includegraphics[width=31mm]{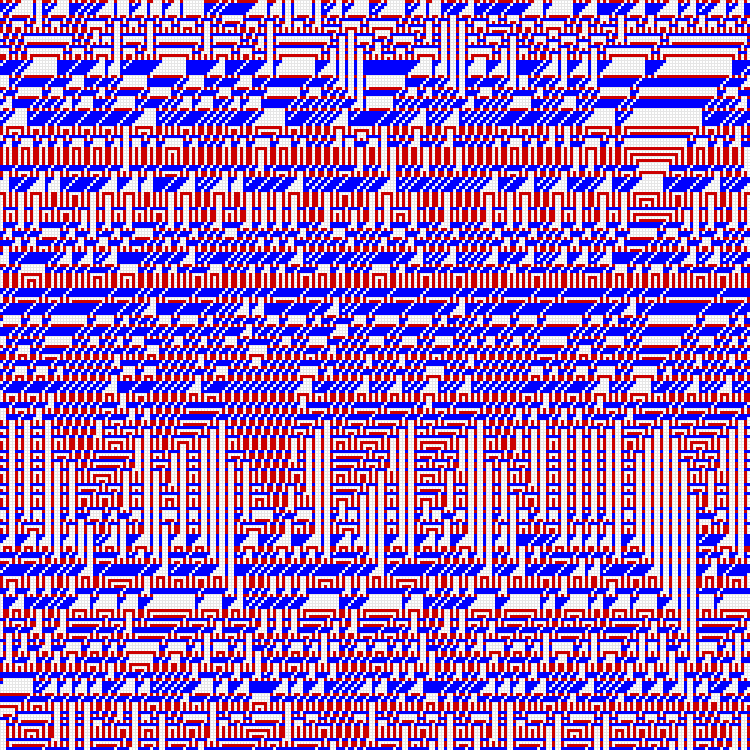} &   \includegraphics[width=31mm]{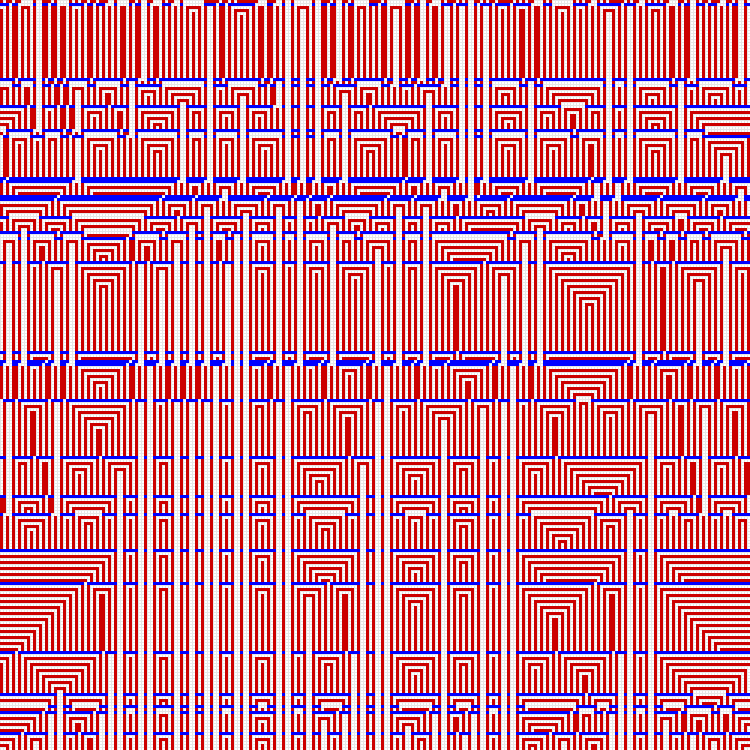} \\
		\end{tabular}}
		\caption{Stochastic CAs ($f,g$) dynamics when $\mathcal{C}$(($f,g$)) $\neq$ $\mathcal{C}$($f$) and $\mathcal{C}$($f$) = $\mathcal{C}$($g$).}
		\label{TSCA2}
	\end{center}
\end{figure*}

\begin{figure*}[hbt!]
	\begin{center}
		\scalebox{0.8}{
			\begin{tabular}{ccccc}
				ECA 77 & ECA 130 & (77,130)[0.1] & (77,130)[0.5] & (77,130)[0.9] \\[6pt]
				\includegraphics[width=31mm]{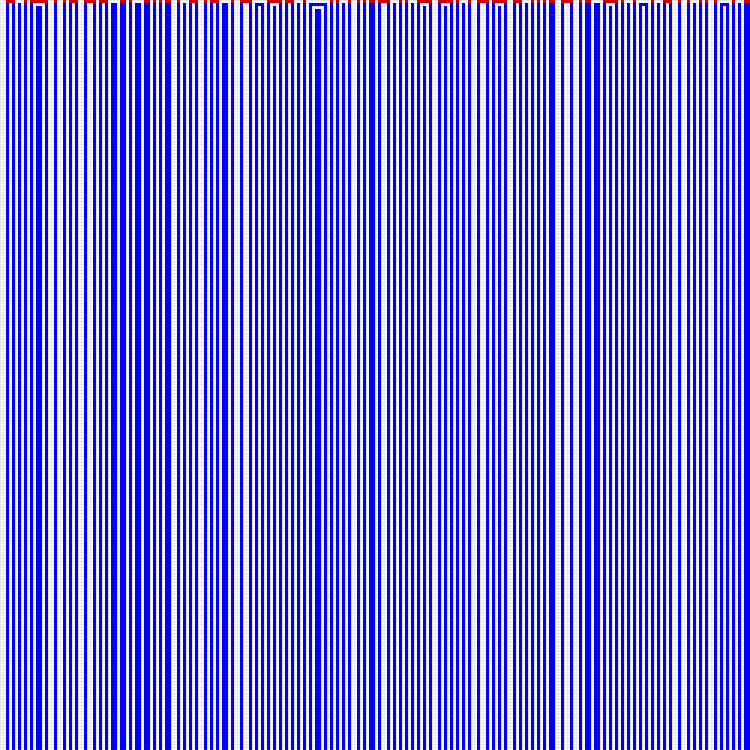} & \includegraphics[width=31mm]{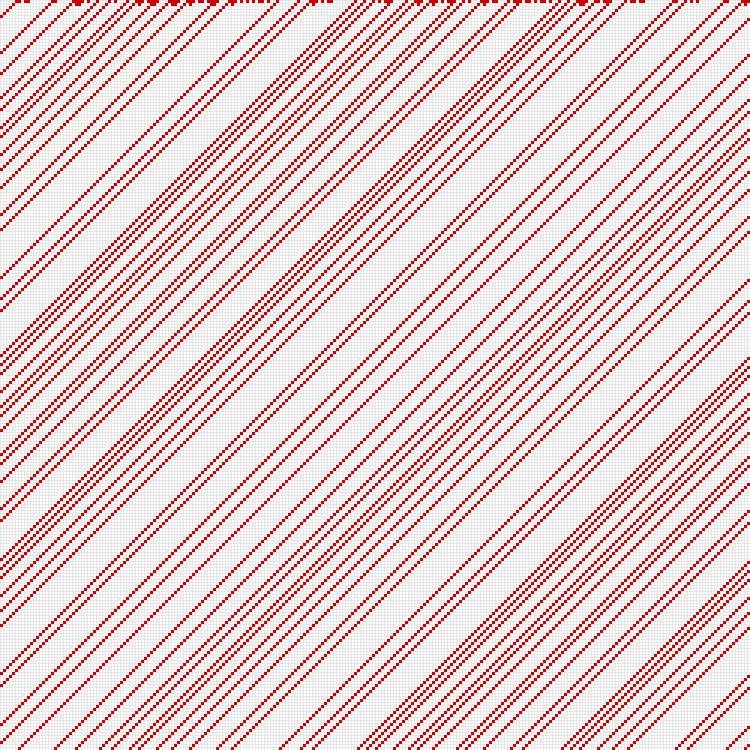} &   \includegraphics[width=31mm]{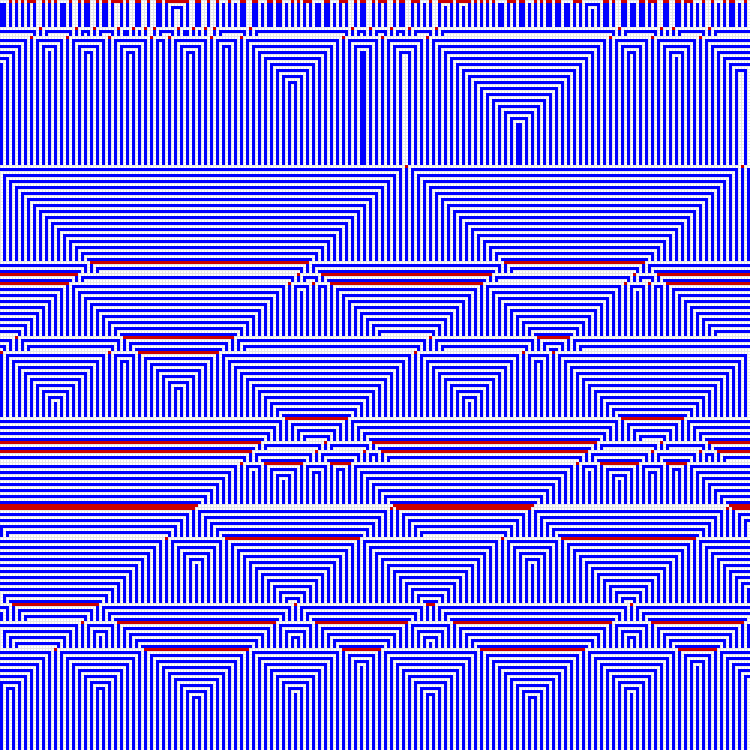} &   \includegraphics[width=31mm]{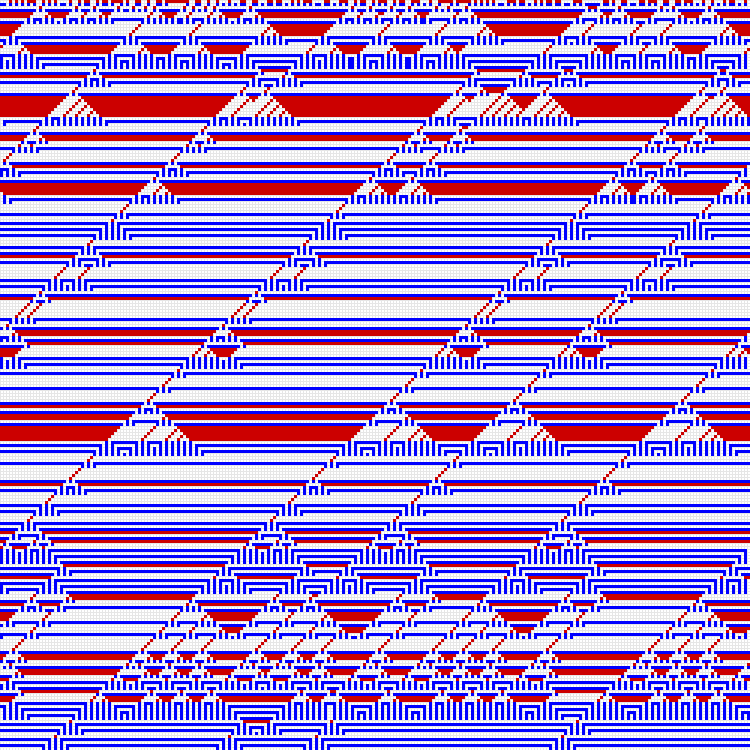} &   \includegraphics[width=31mm]{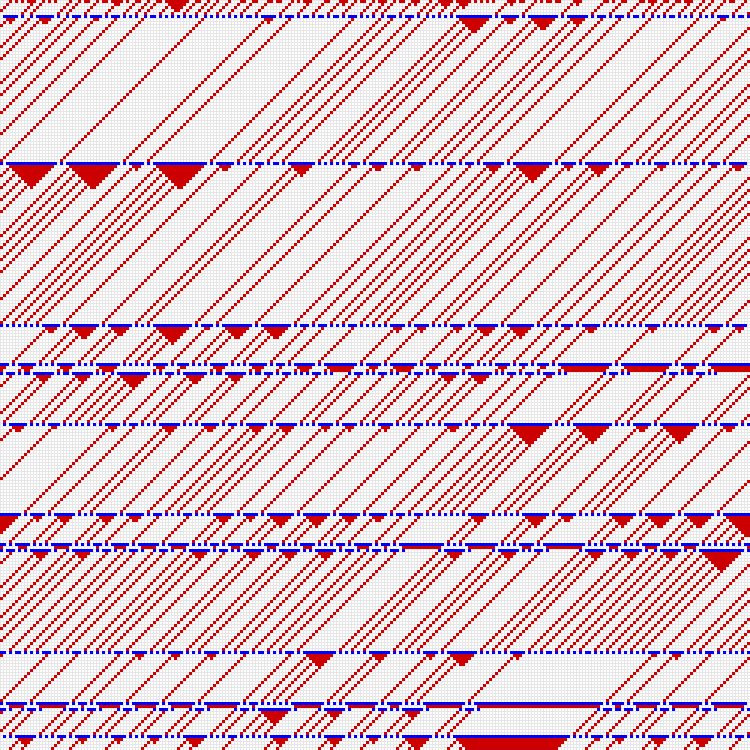} \\
				ECA 38 & ECA 232 & (38,232)[0.1] & (38,232)[0.5] & (38,232)[0.9] \\[6pt]
				\includegraphics[width=31mm]{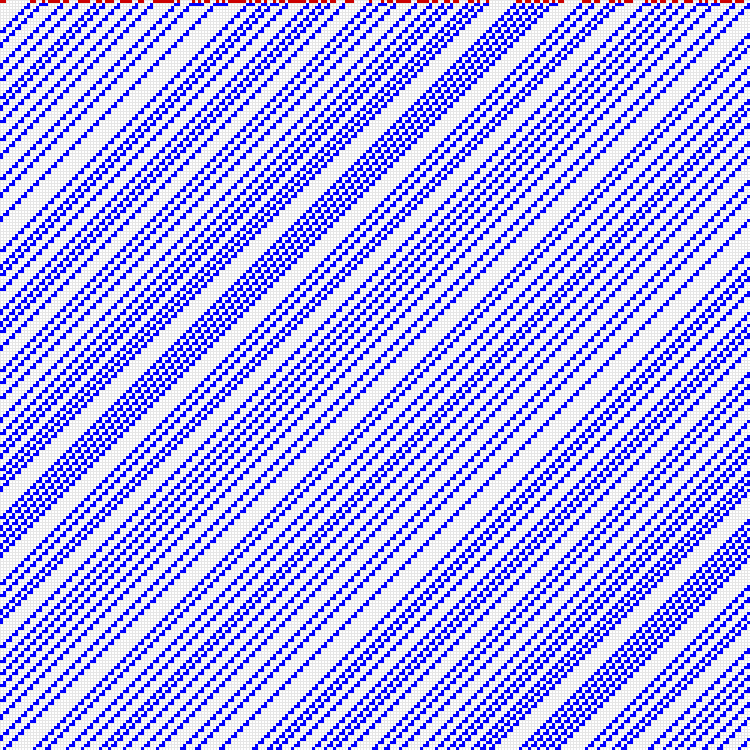} & \includegraphics[width=31mm]{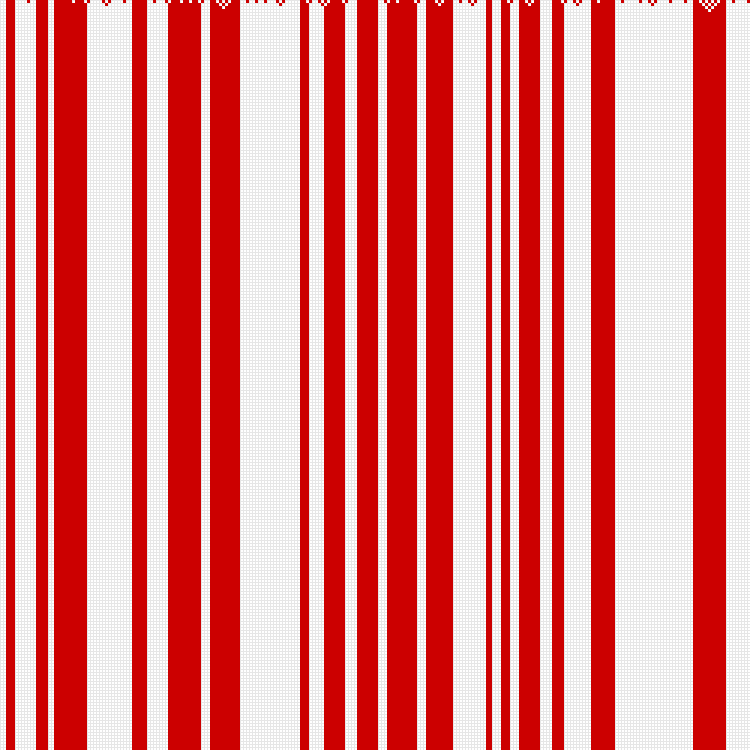} &   \includegraphics[width=31mm]{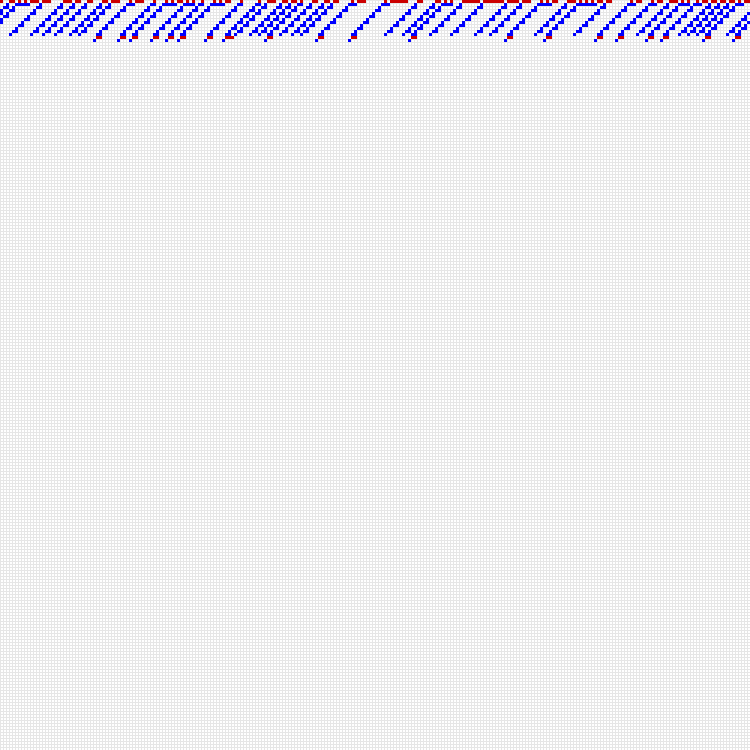} &   \includegraphics[width=31mm]{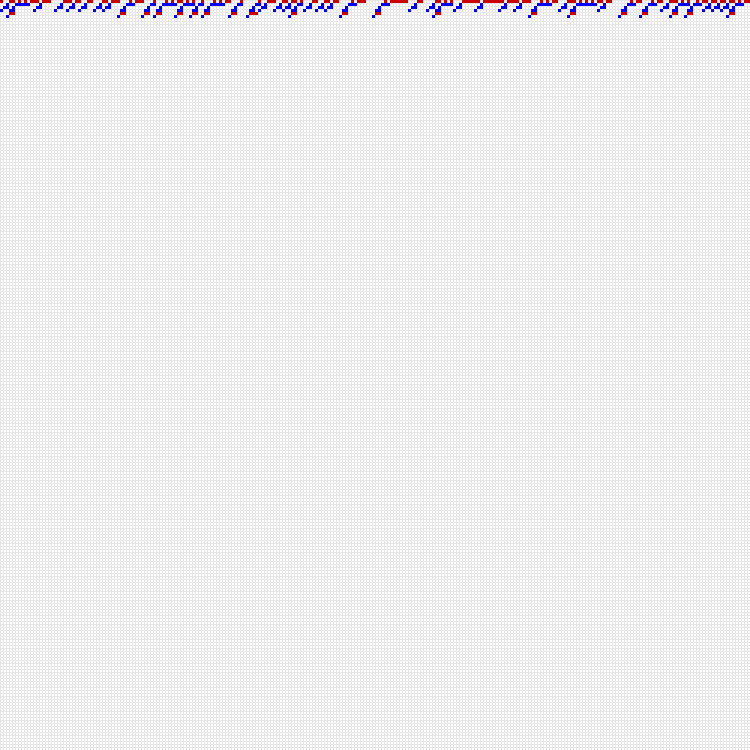} &   \includegraphics[width=31mm]{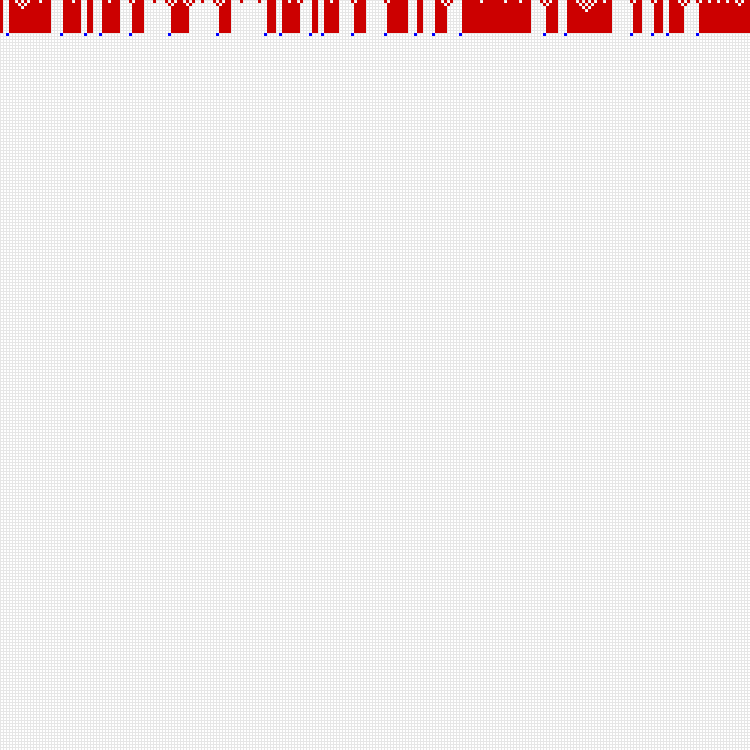} \\
				ECA 134 & ECA 164 & (134,164)[0.1] & (134,164)[0.5] & (134,164)[0.9] \\[6pt]
				\includegraphics[width=31mm]{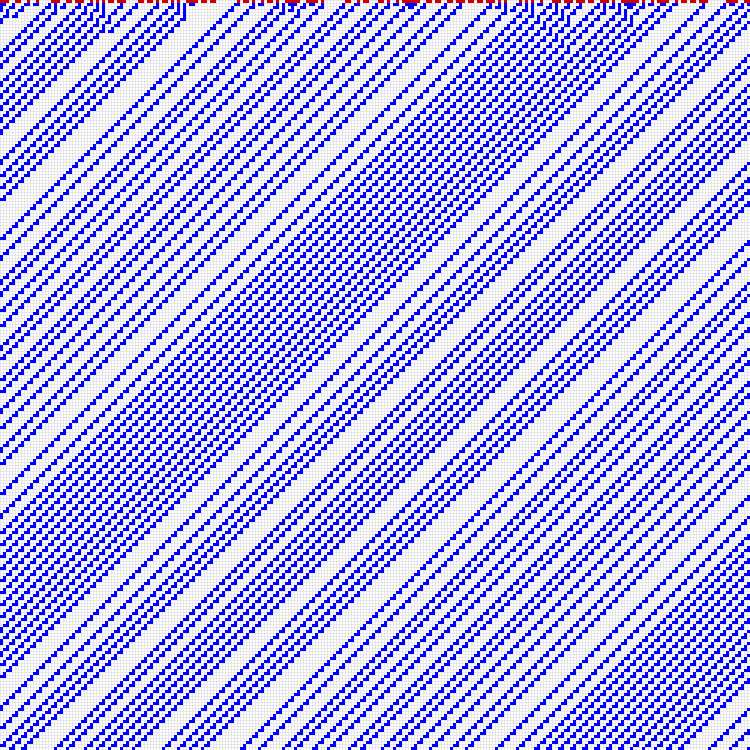} & \includegraphics[width=31mm]{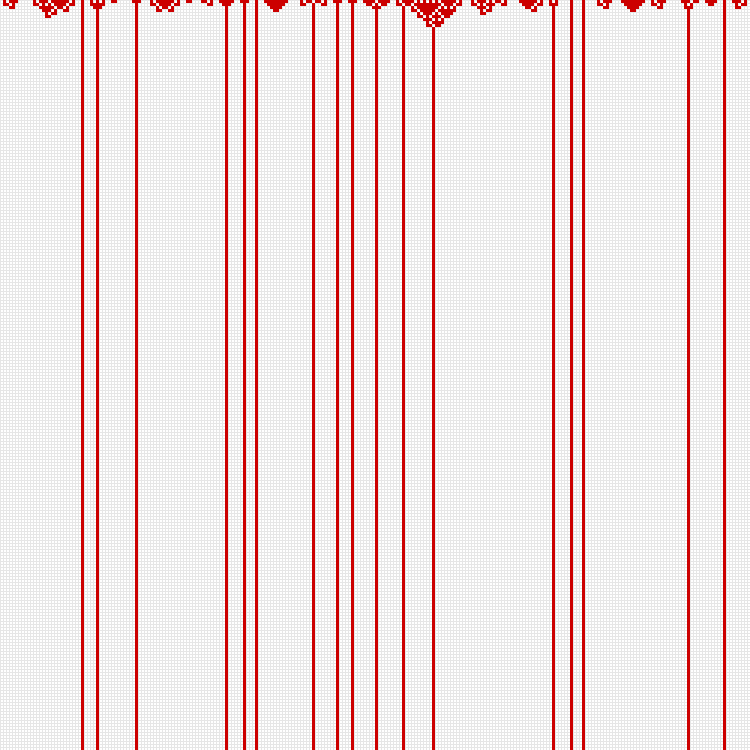} &   \includegraphics[width=31mm]{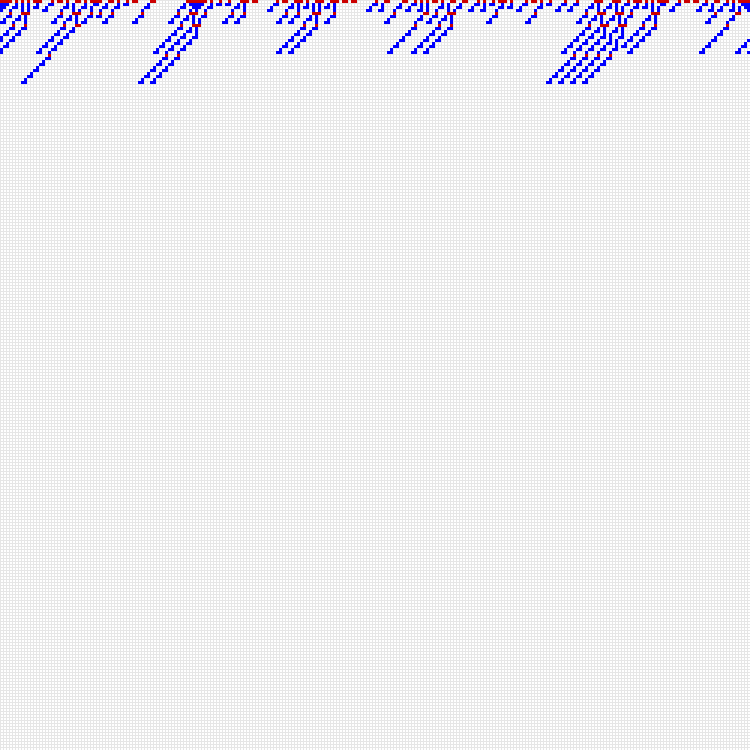} &   \includegraphics[width=31mm]{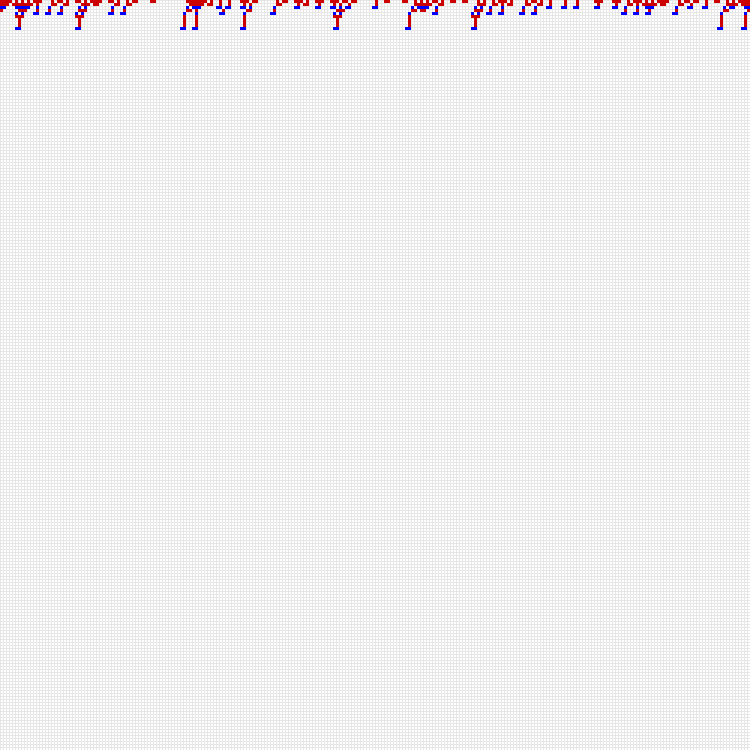} &   \includegraphics[width=31mm]{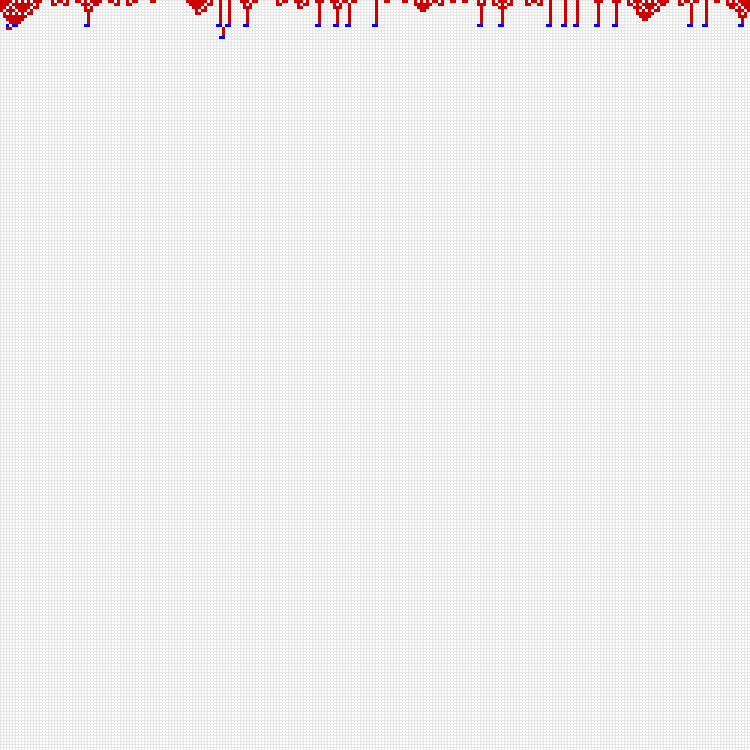} \\
		\end{tabular}}
		\caption{Stochastic CAs ($f,g$) dynamics when $\mathcal{C}$(($f,g$)) $\neq$ $\mathcal{C}$($f$) and $\mathcal{C}$($f$) = $\mathcal{C}$($g$).}
		\label{TSCA3}
	\end{center}
\end{figure*}

\begin{figure*}[hbt!]
	\begin{center}
	\begin{adjustbox}{width=\columnwidth,center}
		\begin{tabular}{ccccc}
			ECA 122 & ECA 37 & (122,37)[0.1] & (122,37)[0.5] & (122,37)[0.9] \\ [6pt]
			\includegraphics[width=31mm]{TEM_IMAGE/NR122-eps-converted-to.pdf} & \includegraphics[width=31mm]{TEM_IMAGE/NR37-eps-converted-to.pdf} &   \includegraphics[width=31mm]{TEM_IMAGE/NR122_37_1-eps-converted-to.pdf} &   \includegraphics[width=31mm]{TEM_IMAGE/NR122_37_5-eps-converted-to.pdf}  &   \includegraphics[width=31mm]{TEM_IMAGE/NR122_37_9-eps-converted-to.pdf} \\
			
			ECA 22 & ECA 7 & (22,7)[0.1] & (22,7)[0.5] & (22,7)[0.9] \\
			\includegraphics[width=31mm]{TEM_IMAGE/NR22X-eps-converted-to.pdf} & \includegraphics[width=31mm]{TEM_IMAGE/NR7-eps-converted-to.pdf}  &   \includegraphics[width=31mm]{TEM_IMAGE/NR22_7_2-eps-converted-to.pdf} &   \includegraphics[width=31mm]{TEM_IMAGE/NR22_7_5-eps-converted-to.pdf} &   \includegraphics[width=31mm]{TEM_IMAGE/NR22_7_9-eps-converted-to.pdf} \\
			
			ECA 22 & ECA 128 & (22,128)[0.1] & (22,128)[0.5] & (22,128)[0.9] \\
			\includegraphics[width=31mm]{TEM_IMAGE/DD22-eps-converted-to.pdf} & \includegraphics[width=31mm]{TEM_IMAGE/DD128-eps-converted-to.pdf}  &   \includegraphics[width=31mm]{TEM_IMAGE/DD22_128_1-eps-converted-to.pdf} &   \includegraphics[width=31mm]{TEM_IMAGE/DD22_128_5-eps-converted-to.pdf} &   \includegraphics[width=31mm]{TEM_IMAGE/DD22_128_9-eps-converted-to.pdf} \\
			
			ECA 11 & ECA 8 & (11,8)[0.1] & (11,8)[0.5] & (11,8)[0.9] \\
			\includegraphics[width=31mm]{TEM_IMAGE/DD11-eps-converted-to.pdf} & \includegraphics[width=31mm]{TEM_IMAGE/DD8-eps-converted-to.pdf}  &   \includegraphics[width=31mm]{TEM_IMAGE/DD11_8_1-eps-converted-to.pdf} &   \includegraphics[width=31mm]{TEM_IMAGE/DD11_8_5-eps-converted-to.pdf} &   \includegraphics[width=31mm]{TEM_IMAGE/DD11_8_9-eps-converted-to.pdf} \\
			
			ECA 44 & ECA 40 & (44,40)[0.1] & (44,40)[0.5] & (44,40)[0.9] \\
			\includegraphics[width=31mm]{TEM_IMAGE/DD44-eps-converted-to.pdf} & \includegraphics[width=31mm]{TEM_IMAGE/DD40-eps-converted-to.pdf}  &   \includegraphics[width=31mm]{TEM_IMAGE/DD44_40_1-eps-converted-to.pdf} &   \includegraphics[width=31mm]{TEM_IMAGE/DD44_40_5-eps-converted-to.pdf} &   \includegraphics[width=31mm]{TEM_IMAGE/DD44_40_9-eps-converted-to.pdf} \\
		\end{tabular}
	\end{adjustbox}
		\caption{Stochastic CAs ($f,g$) dynamics where either $\mathcal{C}$(($f,g$)) = $\mathcal{C}$($f$) or $\mathcal{C}$(($f,g$)) = $\mathcal{C}$($g$). Here, $\mathcal{C}$($f$) $\neq$ $\mathcal{C}$($g$).}
		\label{Fig4}
	\end{center}
\end{figure*}

\begin{figure*}[hbt!]
	\begin{center}
		\scalebox{0.8}{
			\begin{tabular}{ccccc}
				ECA 41 & ECA 23 & (41,23)[0.1] & (41,23)[0.5] & (41,23)[0.9] \\[6pt]
				\includegraphics[width=31mm]{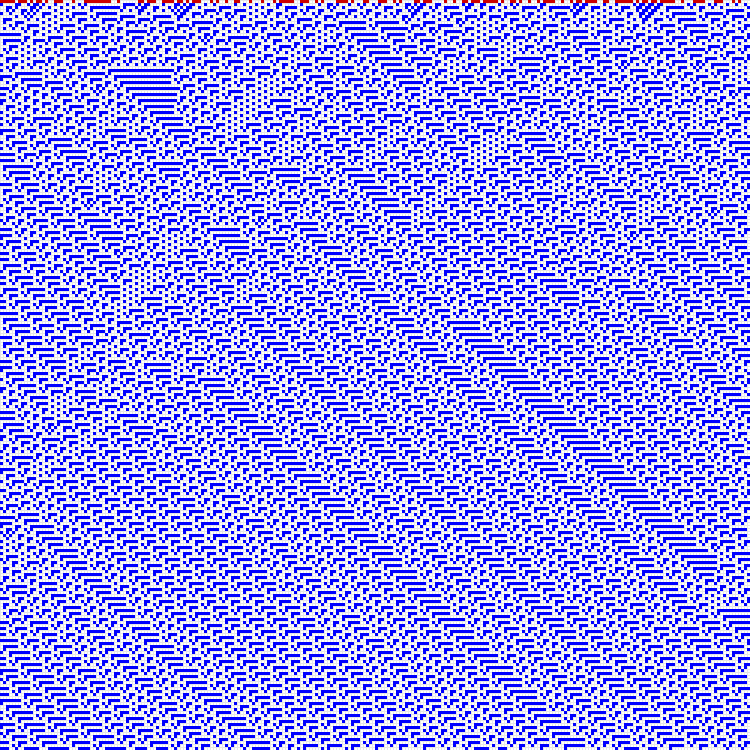} & \includegraphics[width=31mm]{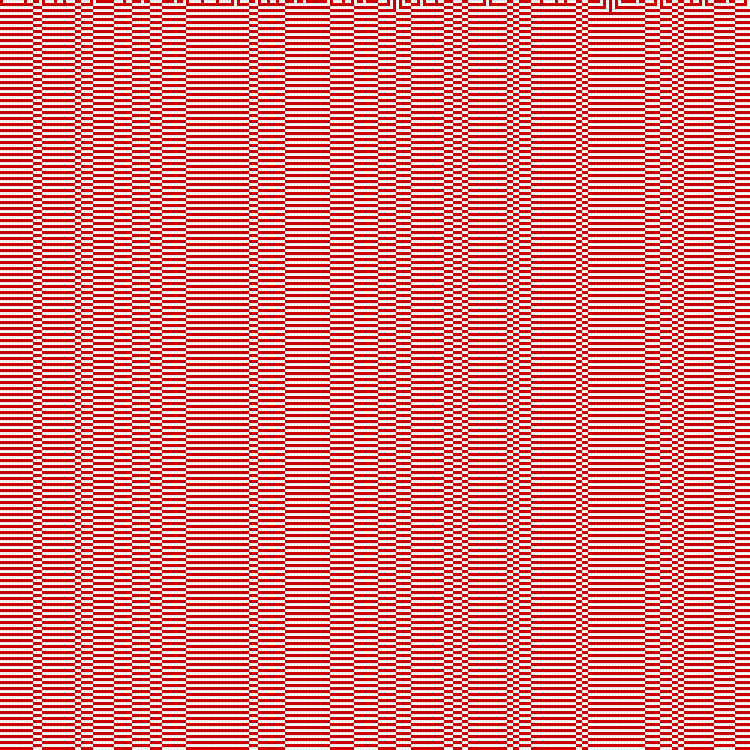} &   \includegraphics[width=31mm]{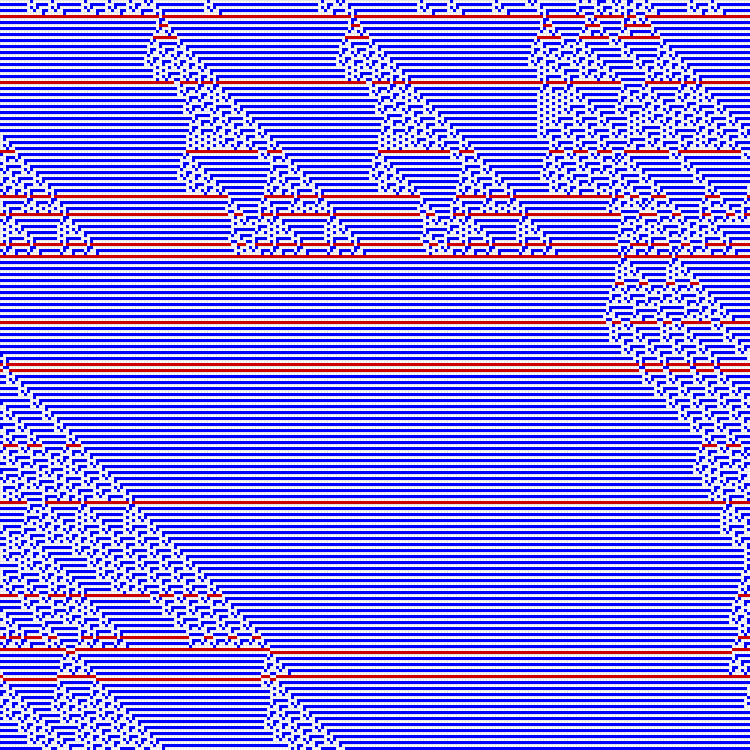} &   \includegraphics[width=31mm]{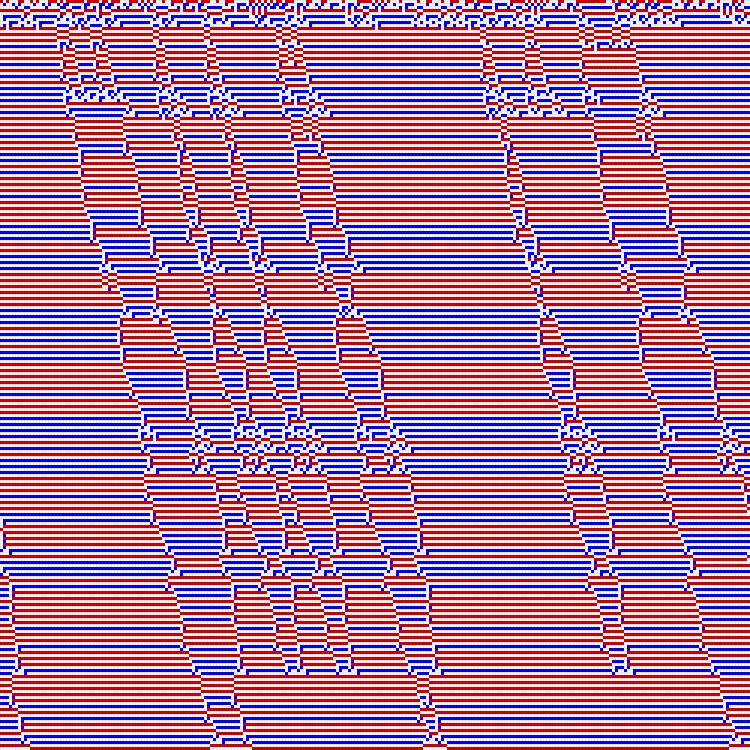} &   \includegraphics[width=31mm]{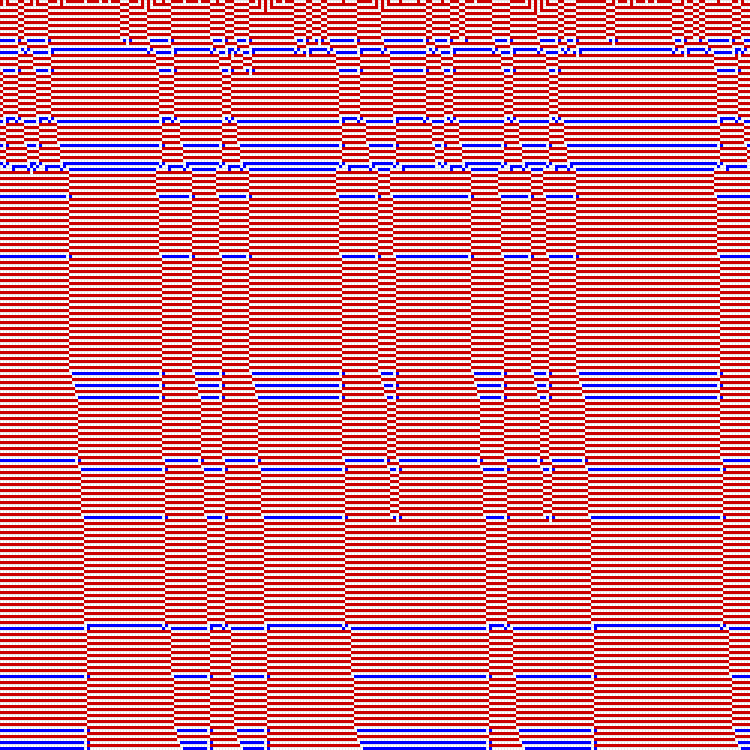} \\
				ECA 105 & ECA 72 & (105,72)[0.1] & (105,72)[0.5] & (105,72)[0.9] \\[6pt]
				\includegraphics[width=31mm]{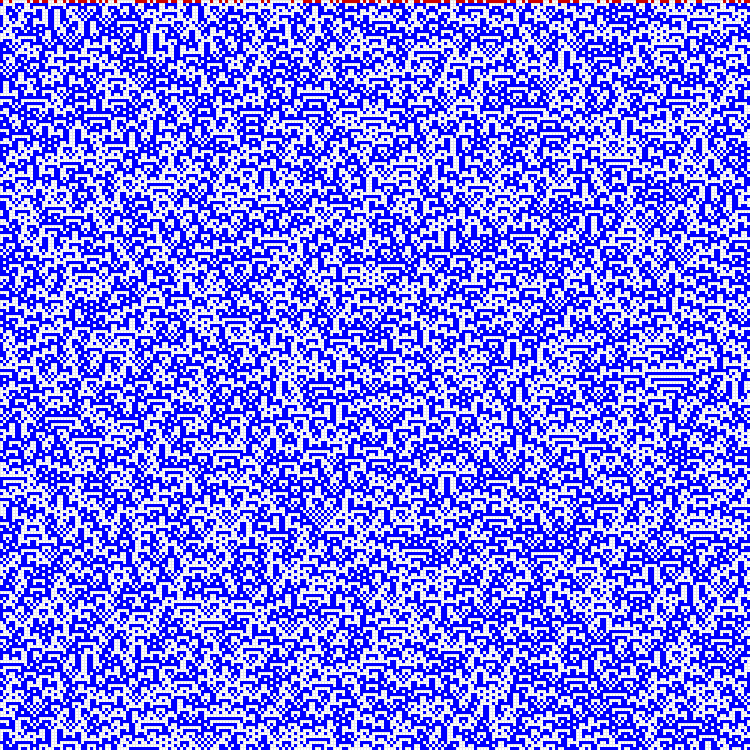} & \includegraphics[width=31mm]{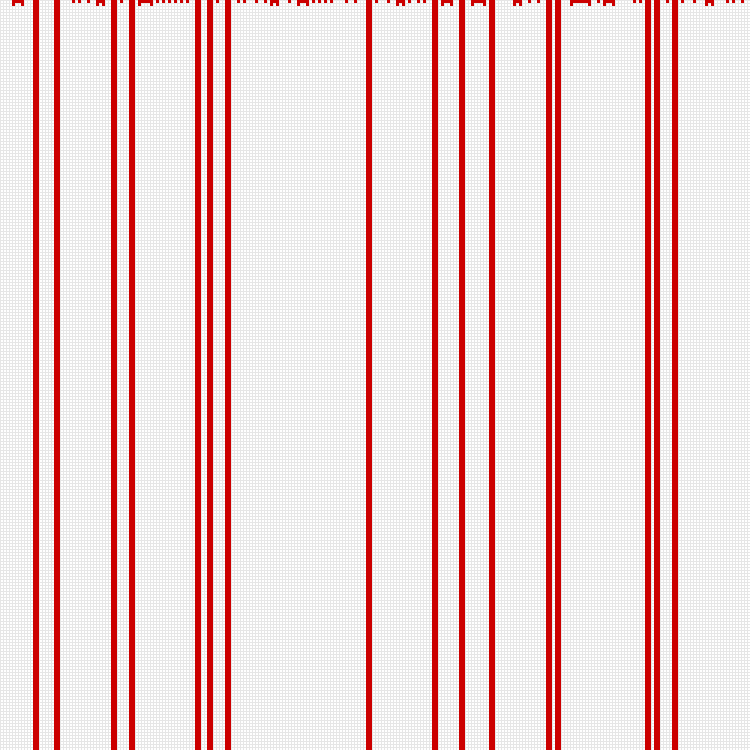} &   \includegraphics[width=31mm]{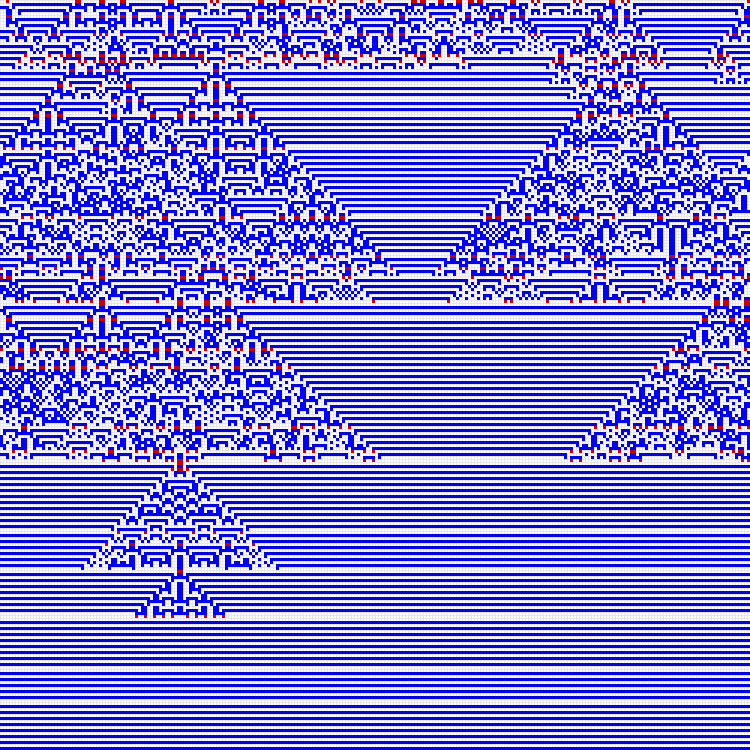} &   \includegraphics[width=31mm]{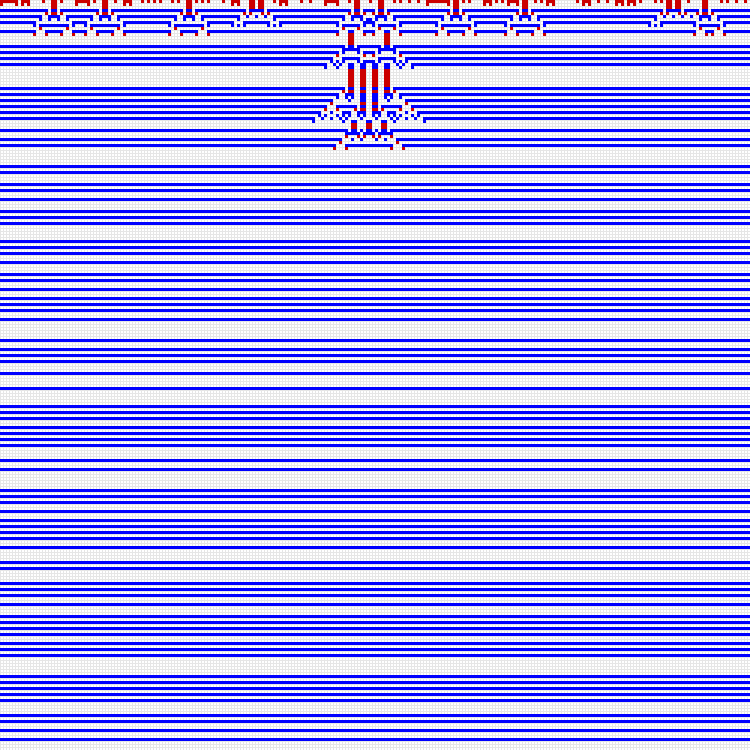} &   \includegraphics[width=31mm]{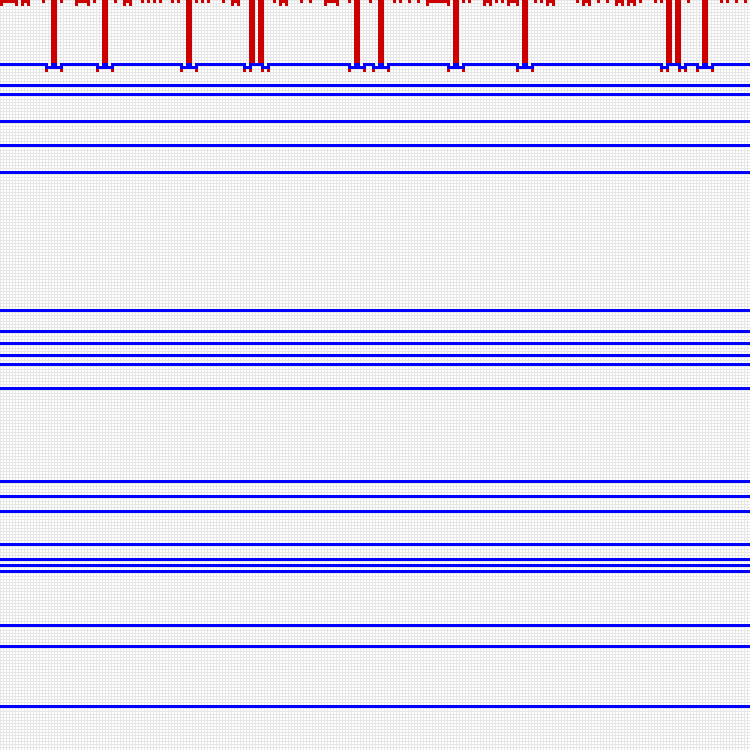} \\
				ECA 45 & ECA 162 & (45,162)[0.1] & (45,162)[0.5] & (45,162)[0.9] \\[6pt]
				\includegraphics[width=31mm]{TEM_IMAGE/Added/rule45.png} & \includegraphics[width=31mm]{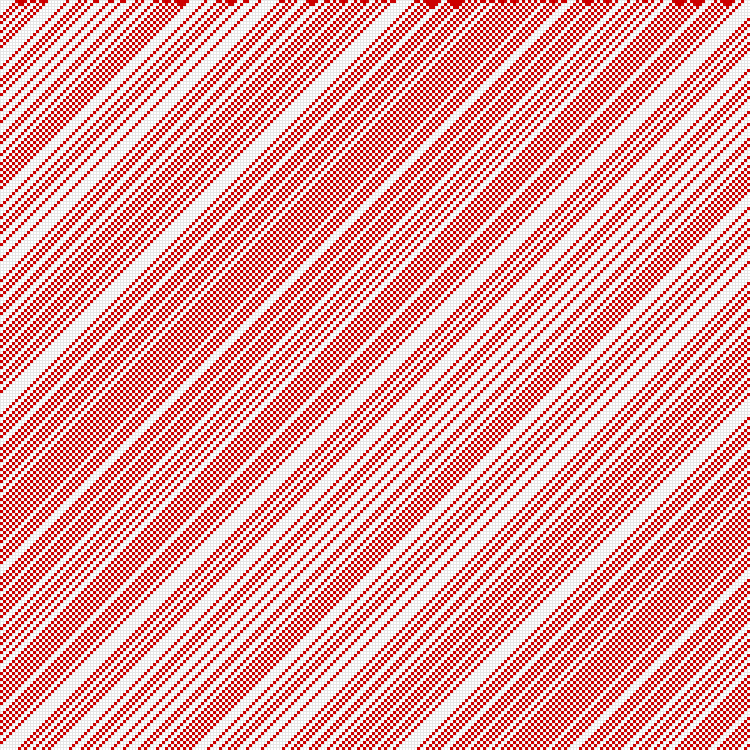} &   \includegraphics[width=31mm]{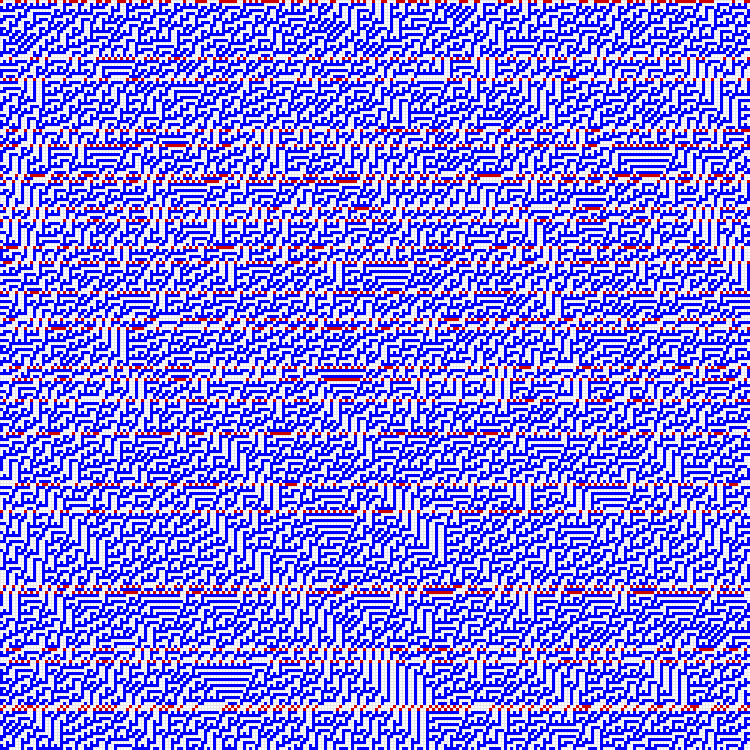} &   \includegraphics[width=31mm]{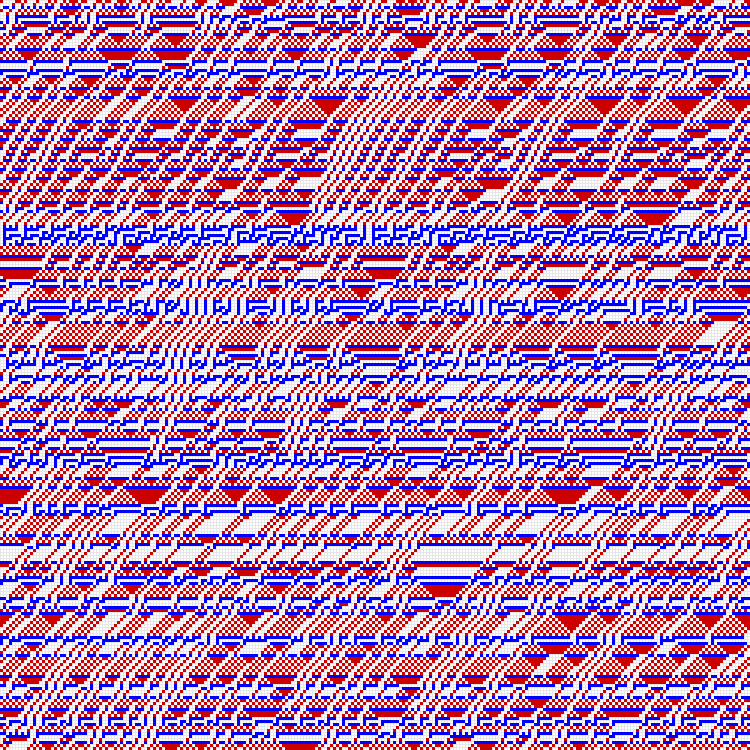} &   \includegraphics[width=31mm]{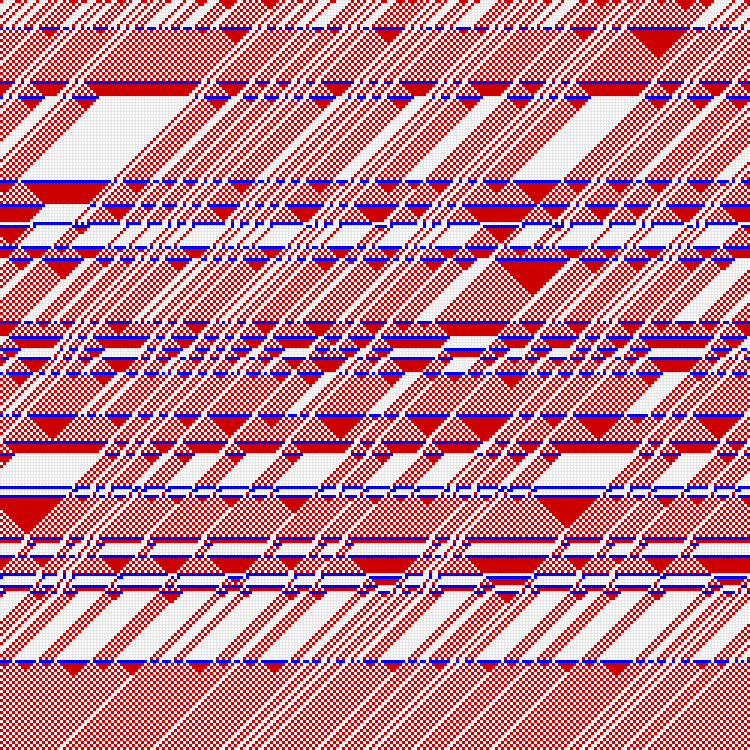} \\
				ECA 126 & ECA 78 & (126,78)[0.1] & (126,78)[0.5] & (126,78)[0.9] \\[6pt]
				\includegraphics[width=31mm]{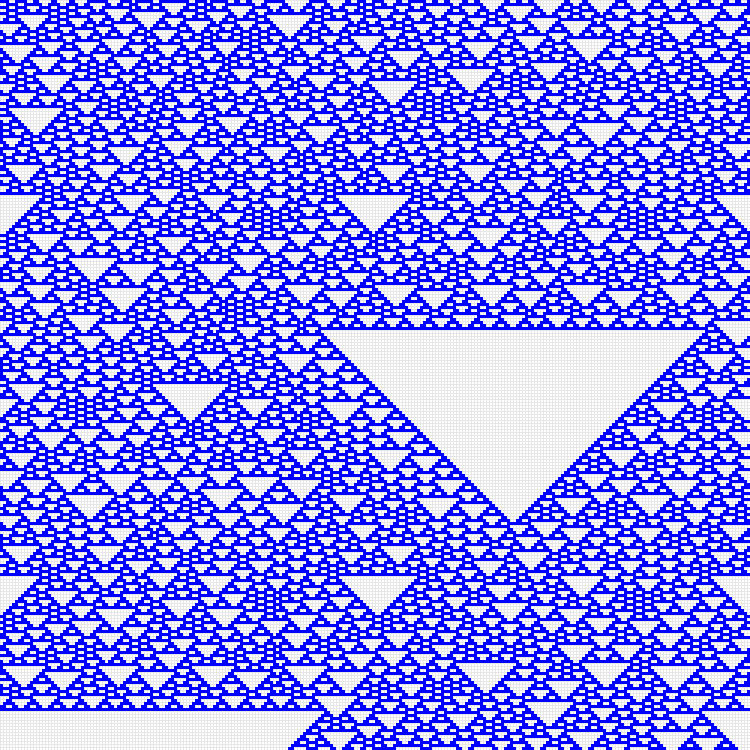} & \includegraphics[width=31mm]{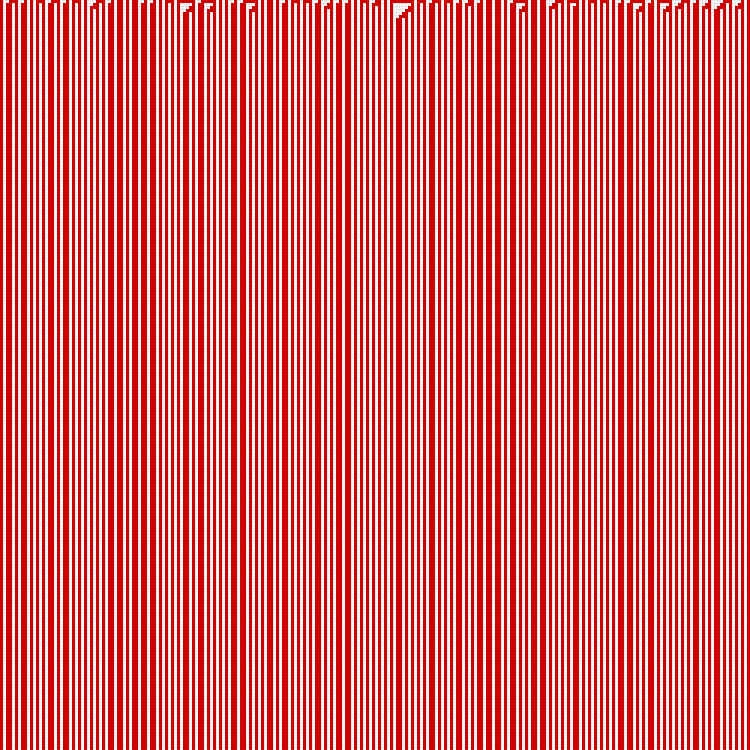} &   \includegraphics[width=31mm]{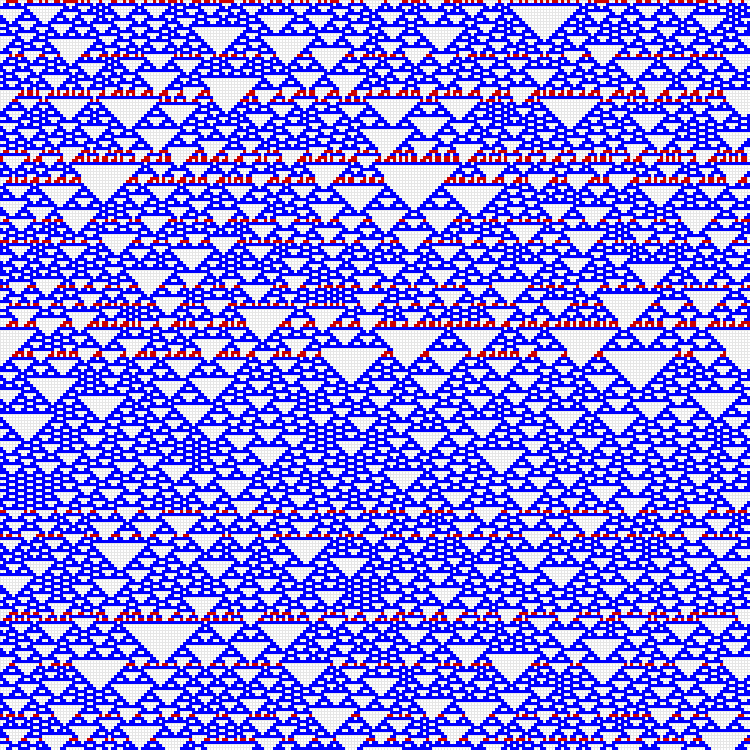} &   \includegraphics[width=31mm]{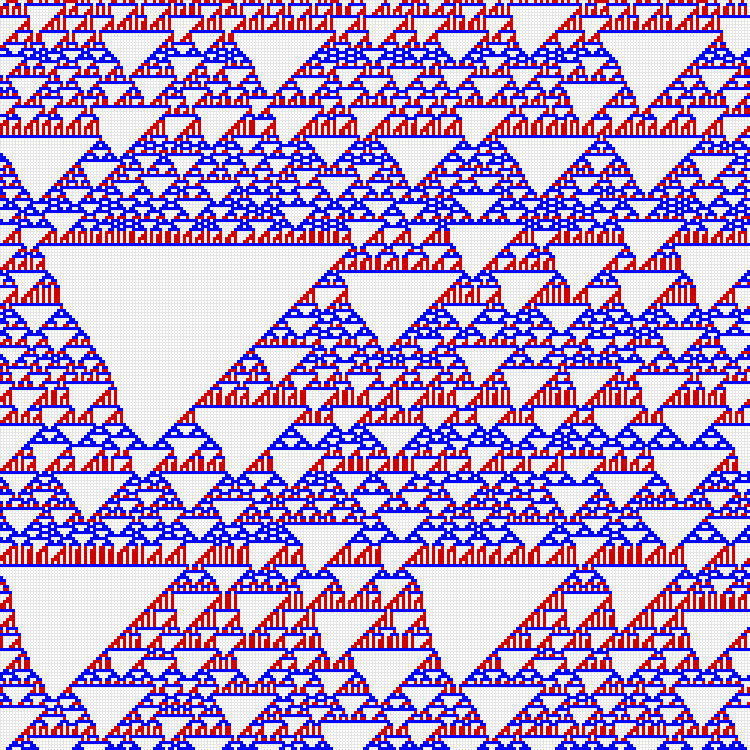} &   \includegraphics[width=31mm]{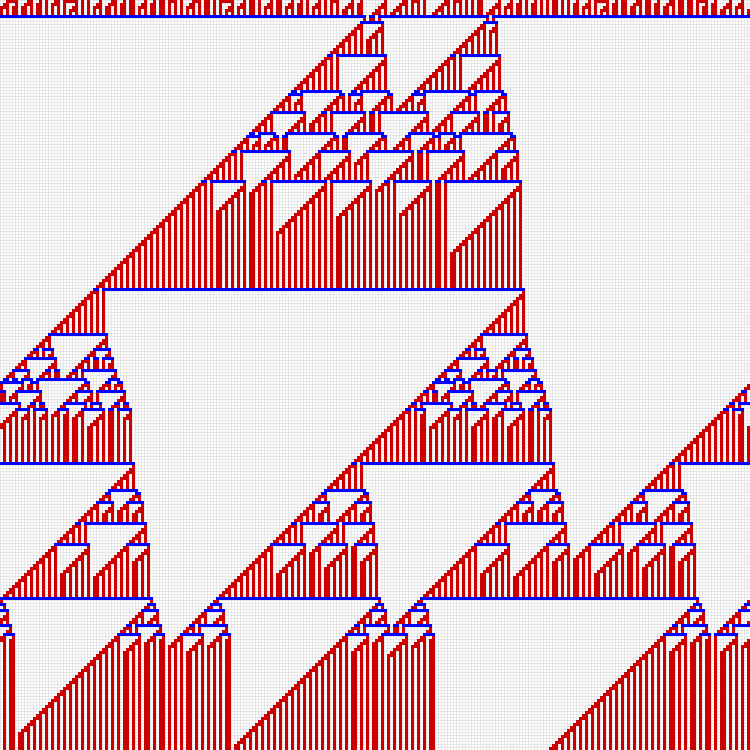} \\
		\end{tabular}}
		\caption{Stochastic CAs ($f,g$) dynamics where either $\mathcal{C}$(($f,g$)) = $\mathcal{C}$($f$) or $\mathcal{C}$(($f,g$)) = $\mathcal{C}$($g$). Here, $\mathcal{C}$($f$) $\neq$ $\mathcal{C}$($g$).}
		\label{TSCA4}
	\end{center}
\end{figure*}

\begin{figure*}[hbt!]
	\begin{center}
		\scalebox{0.8}{
			\begin{tabular}{ccccc}
				ECA 13 & ECA 32 & (13,32)[0.1] & (13,32)[0.5] & (13,32)[0.9] \\[6pt]
				\includegraphics[width=31mm]{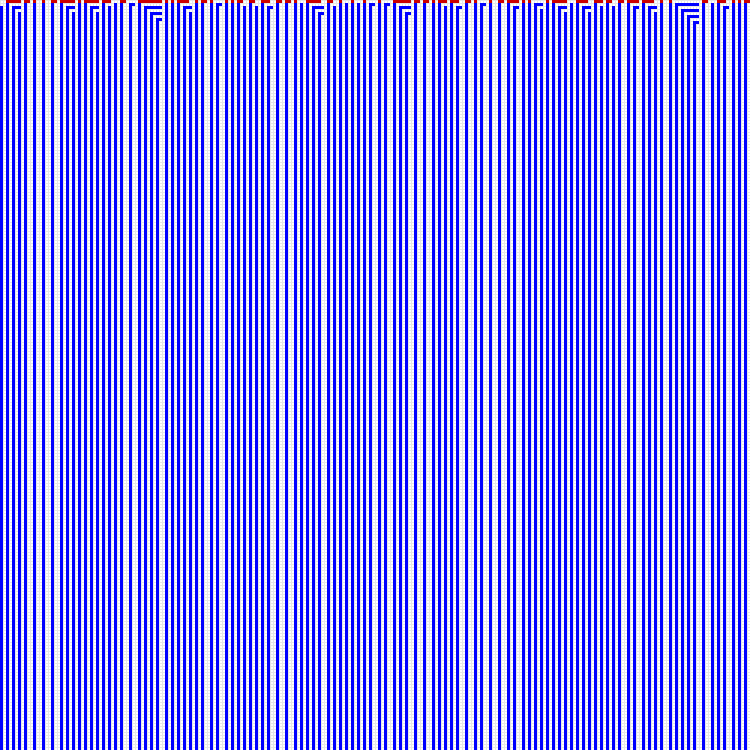} & \includegraphics[width=31mm]{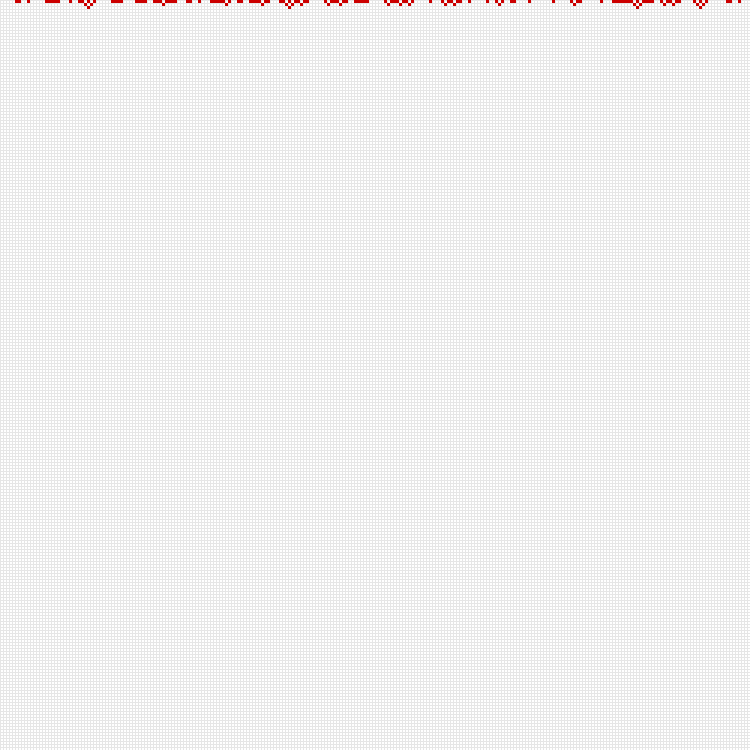} &   \includegraphics[width=31mm]{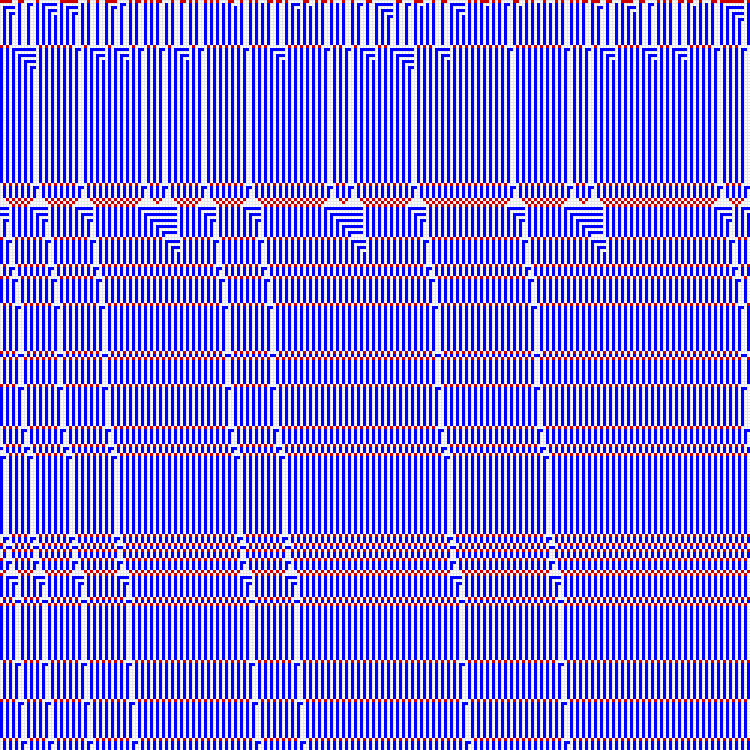} &   \includegraphics[width=31mm]{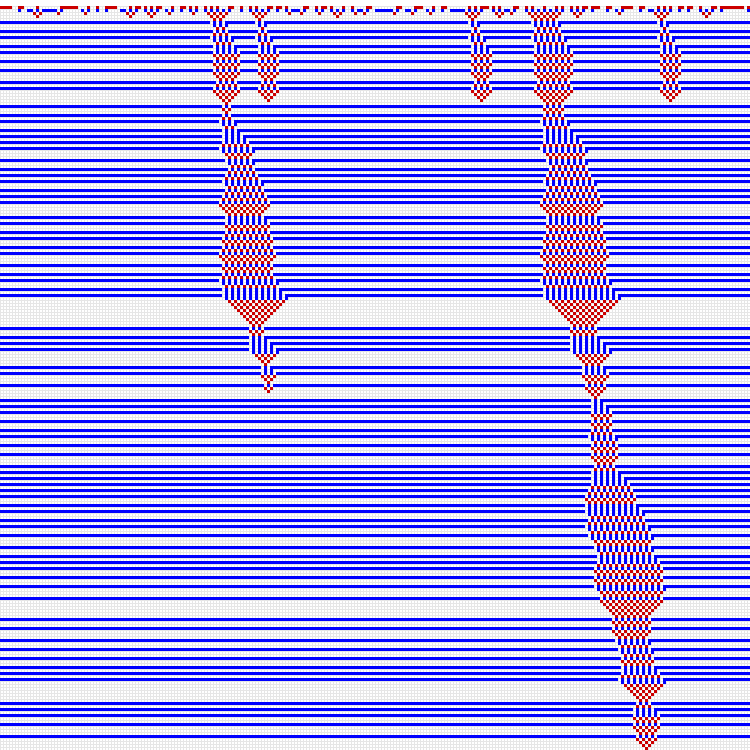} &   \includegraphics[width=31mm]{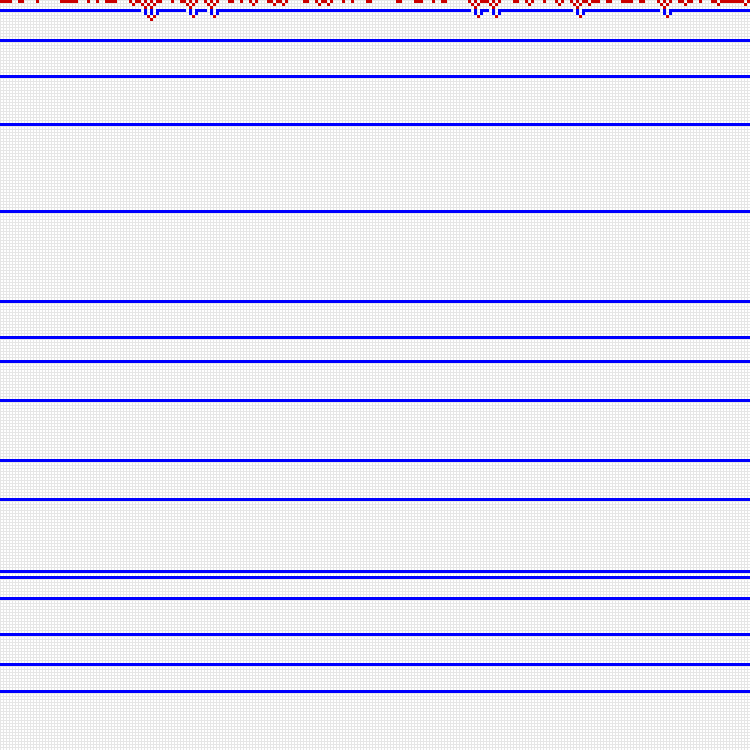} \\
				ECA 146 & ECA 40 & (146,40)[0.1] & (146,40)[0.5] & (146,40)[0.9] \\[6pt]
				\includegraphics[width=31mm]{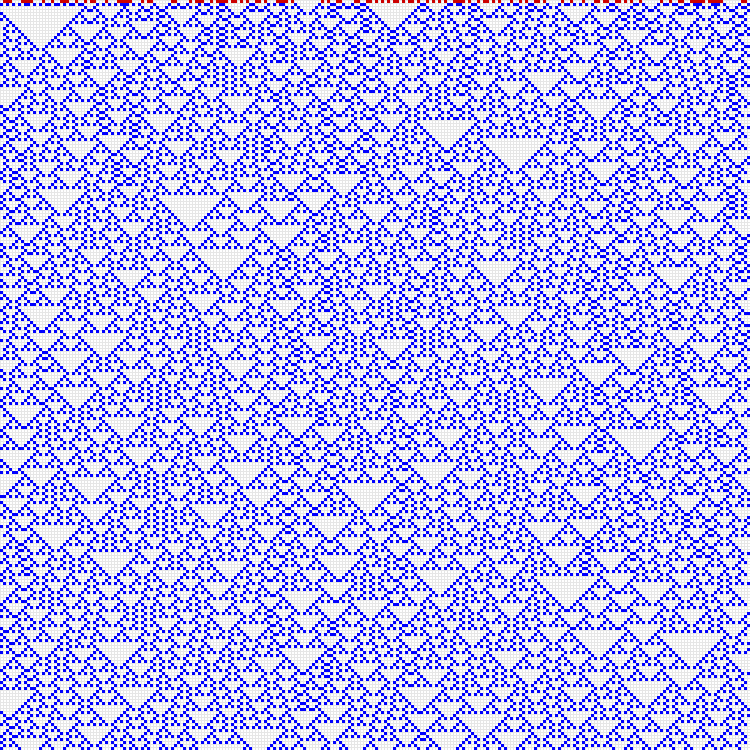} & \includegraphics[width=31mm]{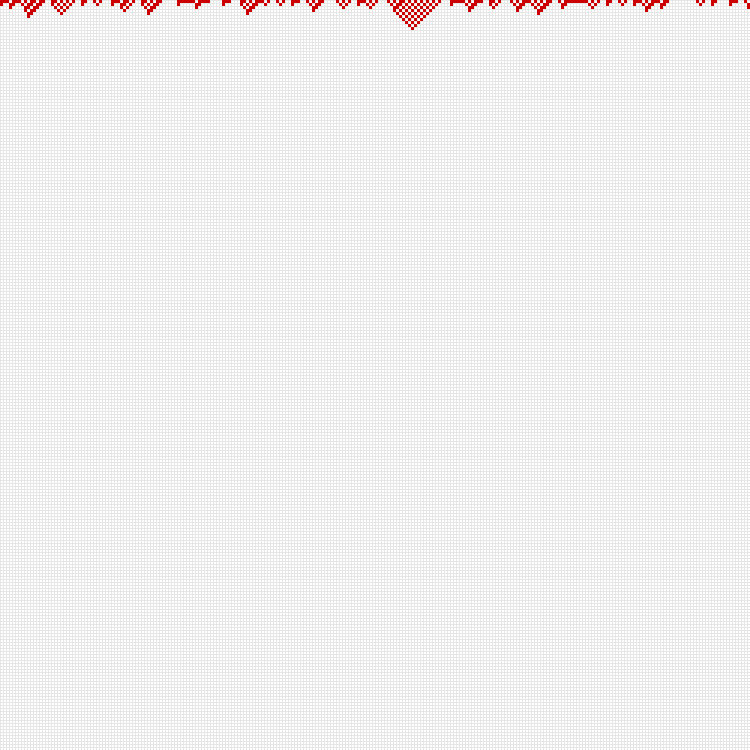} &   \includegraphics[width=31mm]{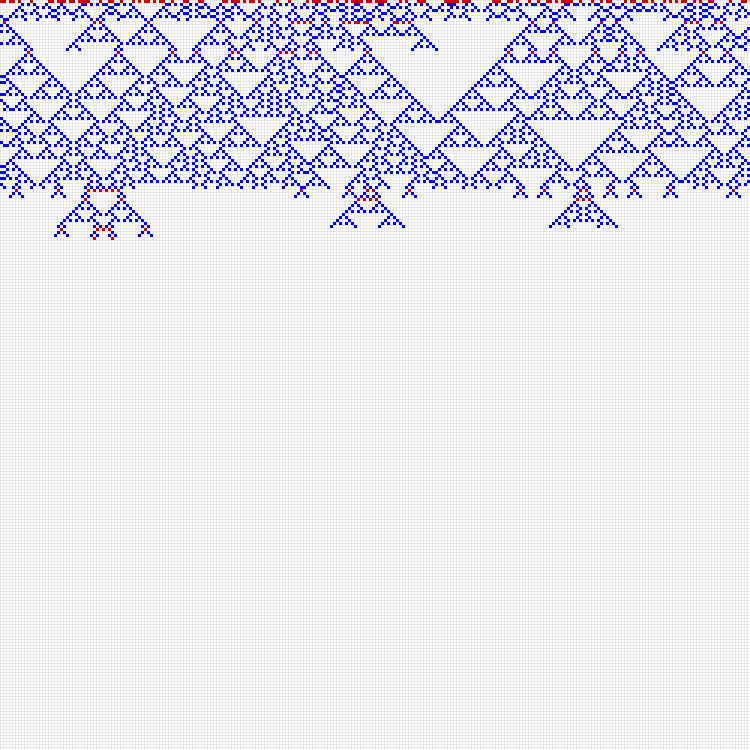} &   \includegraphics[width=31mm]{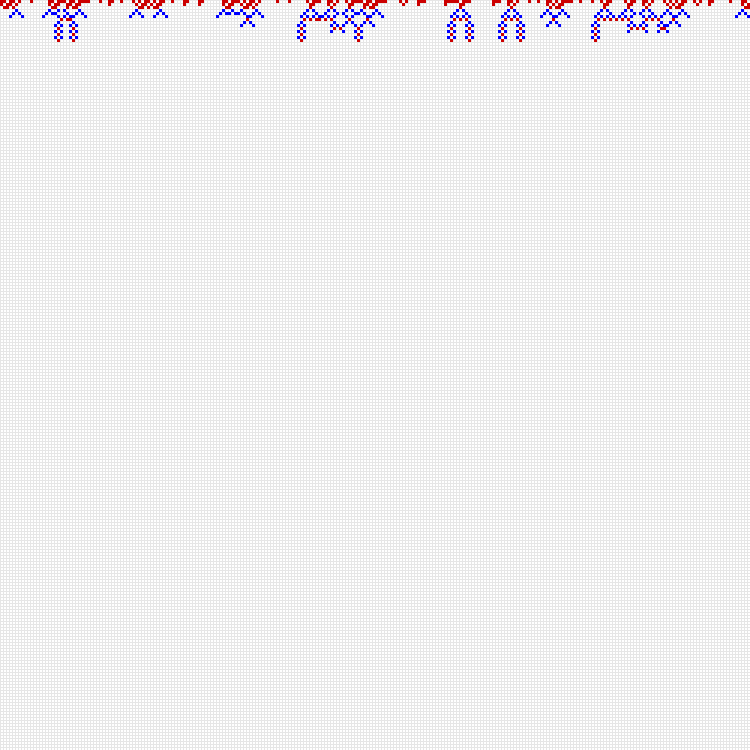} &   \includegraphics[width=31mm]{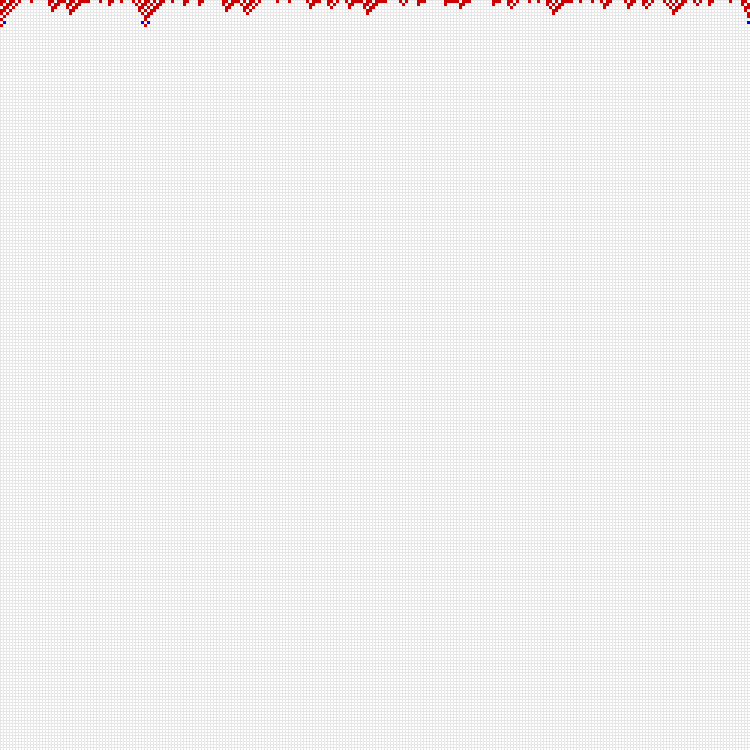} \\
				ECA 164 & ECA 136 & (164,136)[0.1] & (164,136)[0.5] & (164,136)[0.9] \\[6pt]
				\includegraphics[width=31mm]{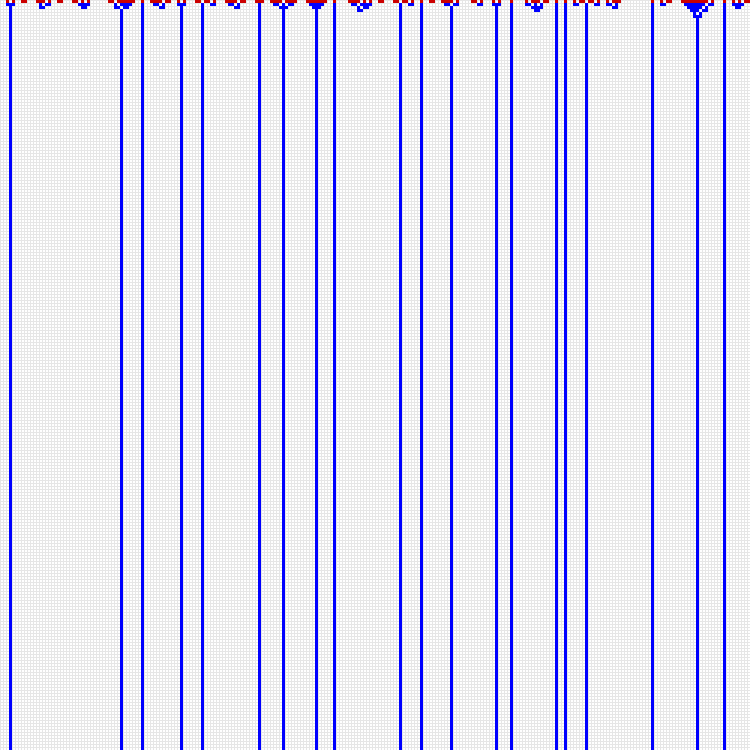} & \includegraphics[width=31mm]{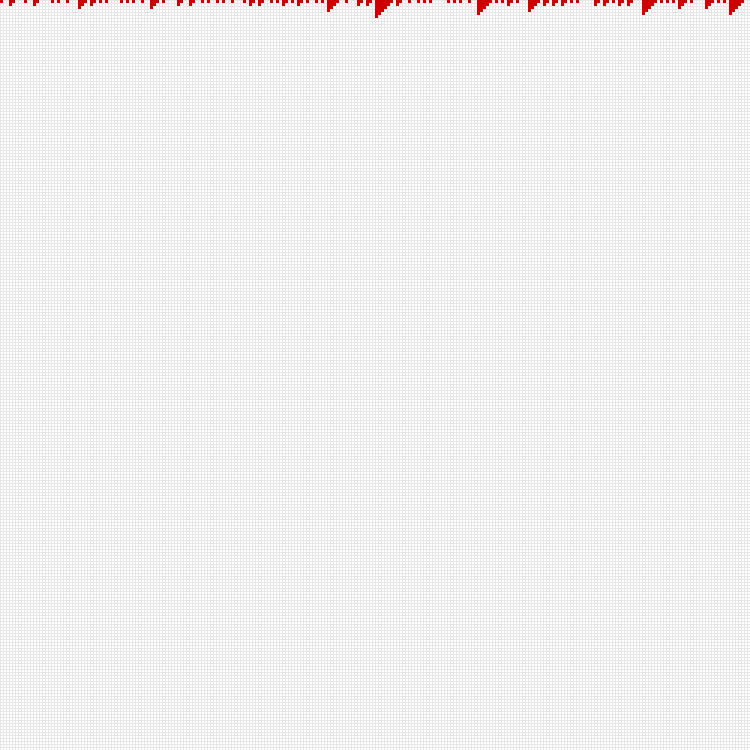} &   \includegraphics[width=31mm]{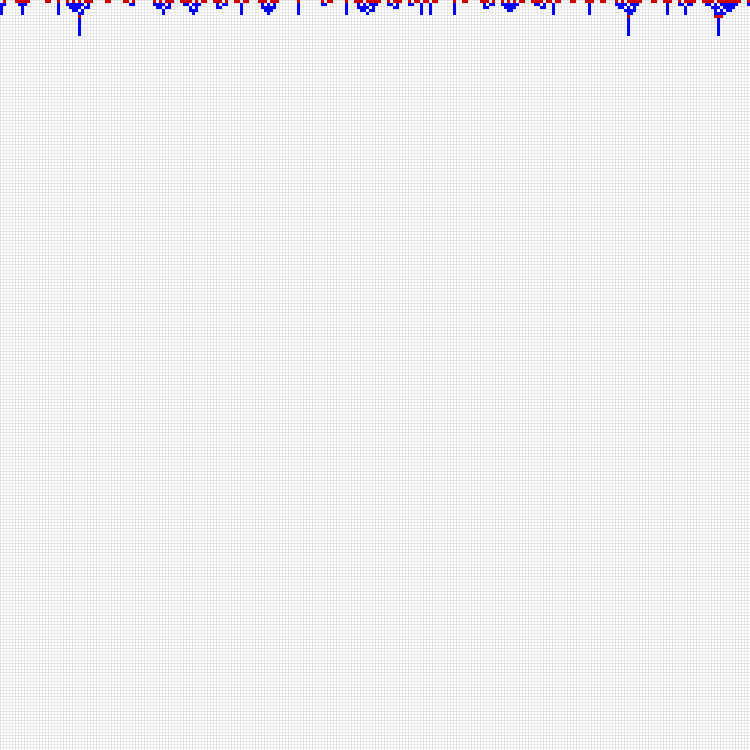} &   \includegraphics[width=31mm]{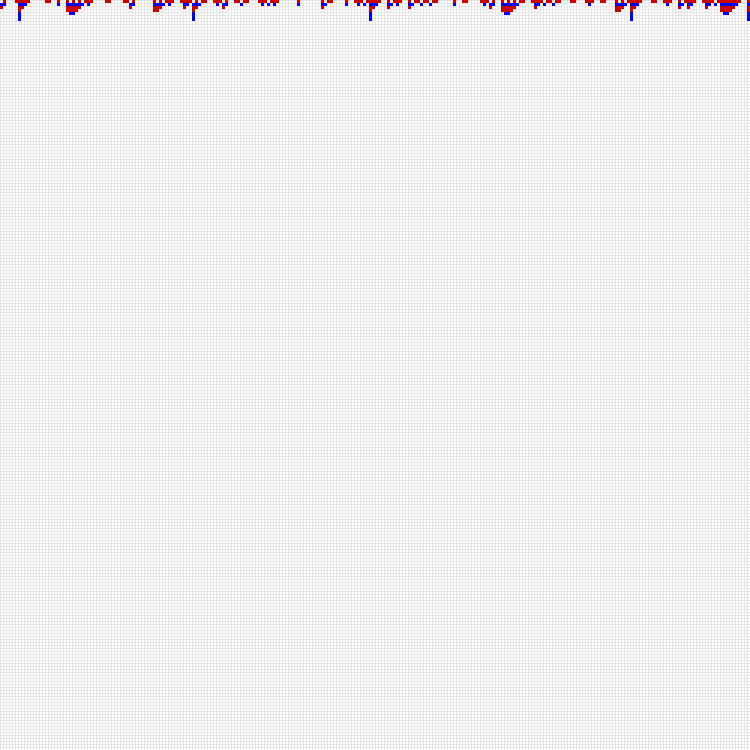} &   \includegraphics[width=31mm]{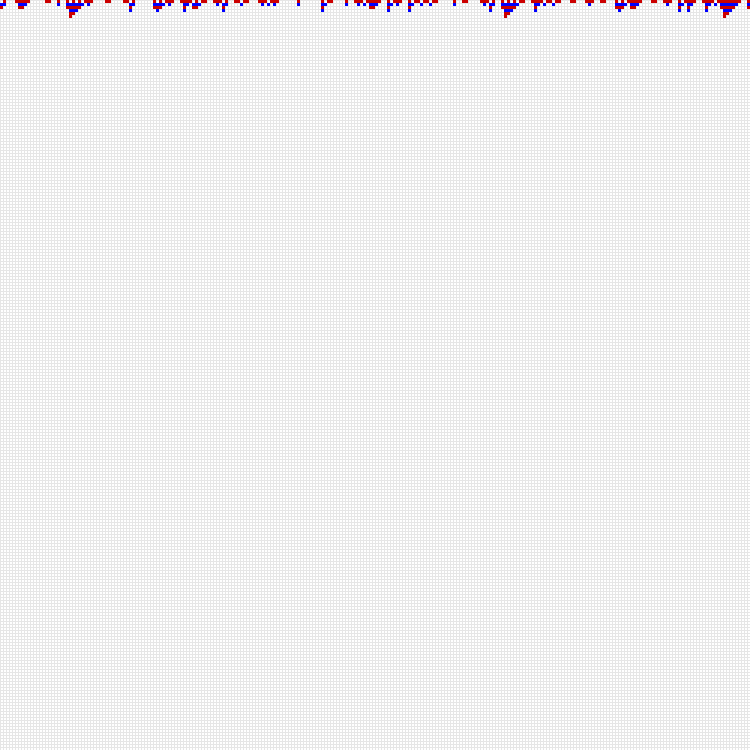} \\
		\end{tabular}}
		\caption{Stochastic CAs ($f,g$) dynamics where either $\mathcal{C}$(($f,g$)) = $\mathcal{C}$($f$) or $\mathcal{C}$(($f,g$)) = $\mathcal{C}$($g$). Here, $\mathcal{C}$($f$) $\neq$ $\mathcal{C}$($g$).}
		\label{TSCA5}
	\end{center}
\end{figure*}

\subsection{Dynamics when $\mathcal{C}$($f$) = $\mathcal{C}$($g$)}

Let us first explore the situation where $\mathcal{C}$($f$) = $\mathcal{C}$($g$). Under this setting, we have observed two kinds of results in our experiments. 

\begin{itemize}
	\item $\mathcal{C}$(($f,g$)) = $\mathcal{C}$($f$); and
	\item $\mathcal{C}$(($f,g$)) $\neq$  $\mathcal{C}$($f)$
\end{itemize}

Whenever $f$ and $g$ both are chosen from Class A, the resultant dynamics remains the same (in class A). On the other hand, only for some of the cases when $f$ and $g$ both are chosen either from class B or class C, the stochastic CA behaves similarly. That is, $\mathcal{C}$(($f,g$)) = $\mathcal{C}$($f$). A minute observation of Fig.~\ref{Fig1} can reveal such $f$ and $g$. Here, we show six such example behavior when ($f,g$) = ($22,18$), ($150,126$), ($43,77$), ($33,5$), ($45,30$) and($172,140$).

Fig.~\ref{Fig2} shows the space-time diagram for ECA $22$ and ECA $18$. Here, both of the ECAs individually show chaotic  behavior (Wolfram's class III), i.e. class C dynamics according to our renaming. Now, if rule $22$ is considered as default rule ($f$) and rule $18$ is added as noise ($g$) at different probability, the resultant dynamics remains chaotic. As sample, we show the space-time diagram of the stochastic CA for $\tau$ $\in$ \{$0.1, 0.5, 0.9$\}. It is also to note that if we progressively change the $\tau$, then the cellular system's dynamics remains unchanged. Similarly, the temporally stochastic CA ($150,126$) and ($45,30$) for $\tau \in$ \{$0.1,0.5,0.9$\} also exhibit chaotic dynamics, and the ECAs $150$, $126$, $45$ and $30$ individually show chaotic dynamics (see Fig.~\ref{Fig2} and Fig.~\ref{TSCA1}). Next, we have chosen both $f$ and $g$ from class B. Fig.~\ref{Fig2} depicts the space-time diagram for ECA $43$ and $77$ where both of the ECAs individually show periodic behavior (Wolfram's class II). Now, if rule $43$ is considered as default rule ($f$) and rule $77$ is added as noise ($g$), the resultant dynamics remains periodic (left shift). As an evidence, Fig.~\ref{Fig2} shows the space time diagram for stochastic CA ($43,77$) for $\tau \in$ \{$0.1,0.5,0.9$\}. Similarly, the temporally stochastic CA ($33,5$) and ($172,140$) for $\tau \in$ \{$0.1,0.5,0.9$\} also exhibit periodic behavior (Wolfram's class II), and the ECAs $33$, $5$, $172$ and $140$ individually show periodic dynamics (see Fig.~\ref{TSCA1}). 

\begin{figure*}[hbt!]
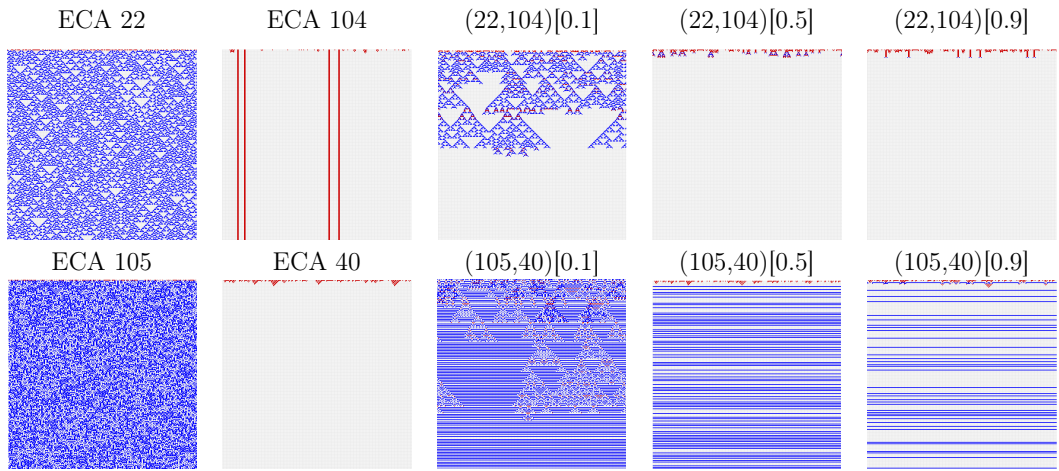

	\begin{center}
		\begin{adjustbox}{width=\columnwidth,center}
			\begin{tabular}{ccccc}
				ECA 22 & ECA 104 & (22,104)[0.1] & (22,104)[0.5] & (22,104)[0.9] \\ [6pt]
				\includegraphics[width=31mm]{TEM_IMAGE/D22-eps-converted-to.pdf} & \includegraphics[width=31mm]{TEM_IMAGE/D104-eps-converted-to.pdf} &   \includegraphics[width=31mm]{TEM_IMAGE/D22_104_1-eps-converted-to.pdf} &   \includegraphics[width=31mm]{TEM_IMAGE/D22_104_5-eps-converted-to.pdf}  &   \includegraphics[width=31mm]{TEM_IMAGE/D22_104_9-eps-converted-to.pdf} \\
				
				ECA 105 & ECA 40 & (105,40)[0.1] & (105,40)[0.5] & (105,40)[0.9] \\
				\includegraphics[width=31mm]{TEM_IMAGE/D105-eps-converted-to.pdf} & \includegraphics[width=31mm]{TEM_IMAGE/D40-eps-converted-to.pdf}  &   \includegraphics[width=31mm]{TEM_IMAGE/D105_40_1-eps-converted-to.pdf} &   \includegraphics[width=31mm]{TEM_IMAGE/D105_40_5-eps-converted-to.pdf} &   \includegraphics[width=31mm]{TEM_IMAGE/D105_40_9-eps-converted-to.pdf} \\

		\end{tabular}
	\end{adjustbox}
		\caption{Stochastic CAs ($f,g$) dynamics where $\mathcal{C}$(($f,g$)) $\neq$ $\mathcal{C}$($f$) and $\mathcal{C}$(($f,g$)) $\neq$ $\mathcal{C}$($g$). Here, $\mathcal{C}$($f$) $\neq$ $\mathcal{C}$($g$).}
		\label{Fig4A}
	\end{center}
\end{figure*}

Let us now present the rest cases with $\mathcal{C}$($f)$ = $\mathcal{C}$($g$) = class B or class C but $\mathcal{C}$(($f,g$)) $\neq$  $\mathcal{C}$($f)$. So we have following four cases: 

\begin{itemize}
	\item[(i)] $\mathcal{C}$($f$) = $\mathcal{C}$($g$) = class B and $\mathcal{C}$(($f,g$)) = class C;
	\item[(ii)] $\mathcal{C}$($f$) = $\mathcal{C}$($g$) = class B and $\mathcal{C}$(($f,g$)) = class A;
	\item[(iii)] $\mathcal{C}$($f$) = $\mathcal{C}$($g$) = class C and $\mathcal{C}$(($f,g$)) = class A; and
	\item[(iv)] $\mathcal{C}$($f$) = $\mathcal{C}$($g$) = class C and $\mathcal{C}$(($f,g$)) = class B.
	
\end{itemize}

Let us start with case (i) where $\mathcal{C}$($f$) = $\mathcal{C}$($g$) = class B and $\mathcal{C}$(($f,g$)) = class C. Here, we show two such example behaviors when ($f,g$) = ($164,131$) and ($164,13$). Observe the class B behavior of ECA $164$ and ECA $131$ in Fig.~\ref{Fig3}. However, in Fig.~\ref{Fig3}, the CA ($164,131$) for $\tau \in$ \{$0.1,0.2,0.3$\} shows chaotic dynamics. The same situation arises for temporally stochastic CA ($164,13$). One can observe the interesting Pascal's triangle patterns in ($164,13$) for $\tau \in$ \{$0.1,0.2,0.3$\}, see Fig.~\ref{Fig3}. Similarly, observe the class B behavior of ECA $9$, ECA $77$, ECA $130$, ECA $13$, ECA $172$, ECA $38$, ECA $232$, ECA $134$ and ECA $164$ in Fig.~\ref{TSCA2} and Fig.~\ref{TSCA3}. However, in Fig.~\ref{TSCA2}, the CA ($9,77$) for $\tau \in$ \{$0.1,0.2,0.3$\} shows chaotic dynamics. The same situation arises for temporally stochastic CA ($130,13$) , ($172,77$), ($130,13$), ($77,130$). This dynamics is a point of interest of this study because couple of two periodic (simple) rules depict kind of chaotic dynamics under temporally stochastic environment. In our previous study \cite{KAMILYA2019116}, we have observed that in spatial mixing environment (i.e. non-uniform CA), two chaotic rules together show periodic (simple) behavior. However, the opposite dynamics has not been observed ever. Hence, this special dynamics is one of the rich assets of this study.

According to case (ii), $\mathcal{C}$($f$) = $\mathcal{C}$($g$) = class B and $\mathcal{C}$(($f,g$)) = class A. As example, if rule $200$ is considered as default rule ($f$) and rule $130$ is added as noise ($g$) at different probabilities, the resultant dynamics shows the evolving to homogeneous configuration, i.e. class A. Note that, ECA $200$ and $130$ individually show periodic behavior (Wolfram's class II). As sample, we show the space-time diagram of the couple ($200,130$) for $\tau$ $\in$ \{$0.1, 0.2, 0.3$\}, see Fig.~\ref{Fig3}. The same situation arises for temporally stochastic CA ($38,232$) and ($134,164$) (see Fig.~\ref{TSCA3}).

In our experiment, case (iii) and case (iv) dynamics have not been observed for any temporally stochastic CA. However, similar {\em class transition} dynamics is observed for case (iv), see Section~\ref{ctd} for details. 
\begin{figure*}[hbt!]
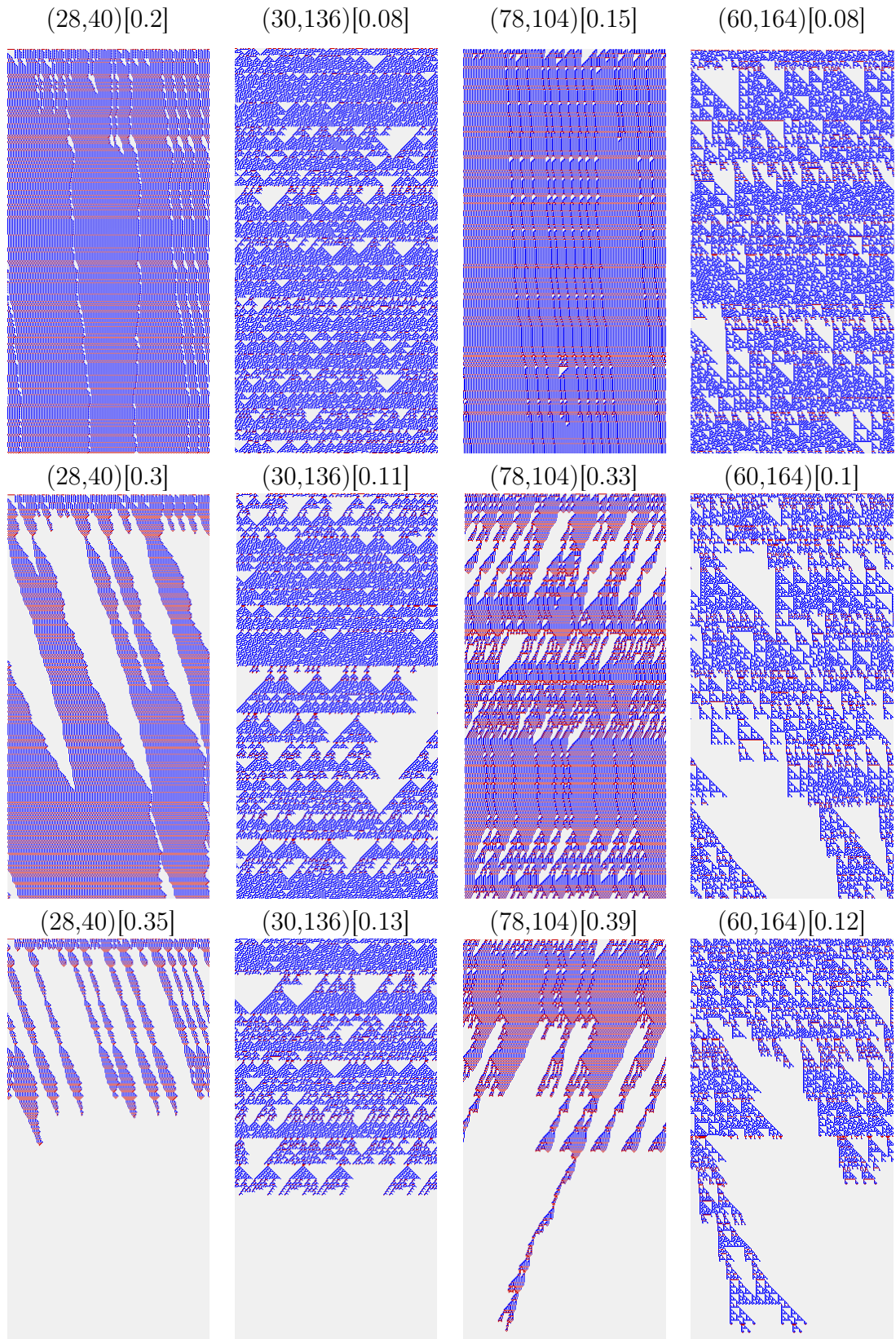

	\begin{center}
		\begin{tabular}{cccc}
			(28,40)[0.2] & (30,136)[0.08] & (78,104)[0.15] & (60,164)[0.08] \\[6pt]
			\includegraphics[width=33mm]{TEM_IMAGE/PT28_40_2-eps-converted-to.pdf} & \includegraphics[width=33mm]{TEM_IMAGE/PT30_136_08-eps-converted-to.pdf}  &   \includegraphics[width=33mm]{TEM_IMAGE/PT78_104_15-eps-converted-to.pdf}  &   \includegraphics[width=33mm]{TEM_IMAGE/PT60_164_08-eps-converted-to.pdf}\\
			(28,40)[0.3] & (30,136)[0.11] & (78,104)[0.33] & (60,164)[0.1] \\
			\includegraphics[width=33mm]{TEM_IMAGE/PT28_40_33-eps-converted-to.pdf} & \includegraphics[width=33mm]{TEM_IMAGE/PT30_136_15-eps-converted-to.pdf}  &   \includegraphics[width=33mm]{TEM_IMAGE/PT78_104_33-eps-converted-to.pdf}  &   \includegraphics[width=33mm]{TEM_IMAGE/PT60_164_1-eps-converted-to.pdf}\\ 
			(28,40)[0.35] & (30,136)[0.13] & (78,104)[0.39] & (60,164)[0.12] \\
			\includegraphics[width=33mm]{TEM_IMAGE/PT28_40_36-eps-converted-to.pdf} & \includegraphics[width=33mm]{TEM_IMAGE/PT30_136_18-eps-converted-to.pdf}  &   \includegraphics[width=33mm]{TEM_IMAGE/PT78_104_39-eps-converted-to.pdf}   &   \includegraphics[width=33mm]{TEM_IMAGE/PT60_164_12-eps-converted-to.pdf}\\
		\end{tabular}
		\caption{Phase transition behavior of stochastic CAs ($28,40$),($30,136$),($78,104$),($60,164$).}
		\label{Fig5}
	\end{center}
\end{figure*}

\begin{figure*}[hbt!]
	\begin{center}
		\scalebox{0.8}{
			\begin{tabular}{cccc}
				(90,104)[0.02] & (90,104)[0.07]  & (90,104)[0.09]  & (90,104)[0.15]\\ [6pt]
				\includegraphics[width=33mm]{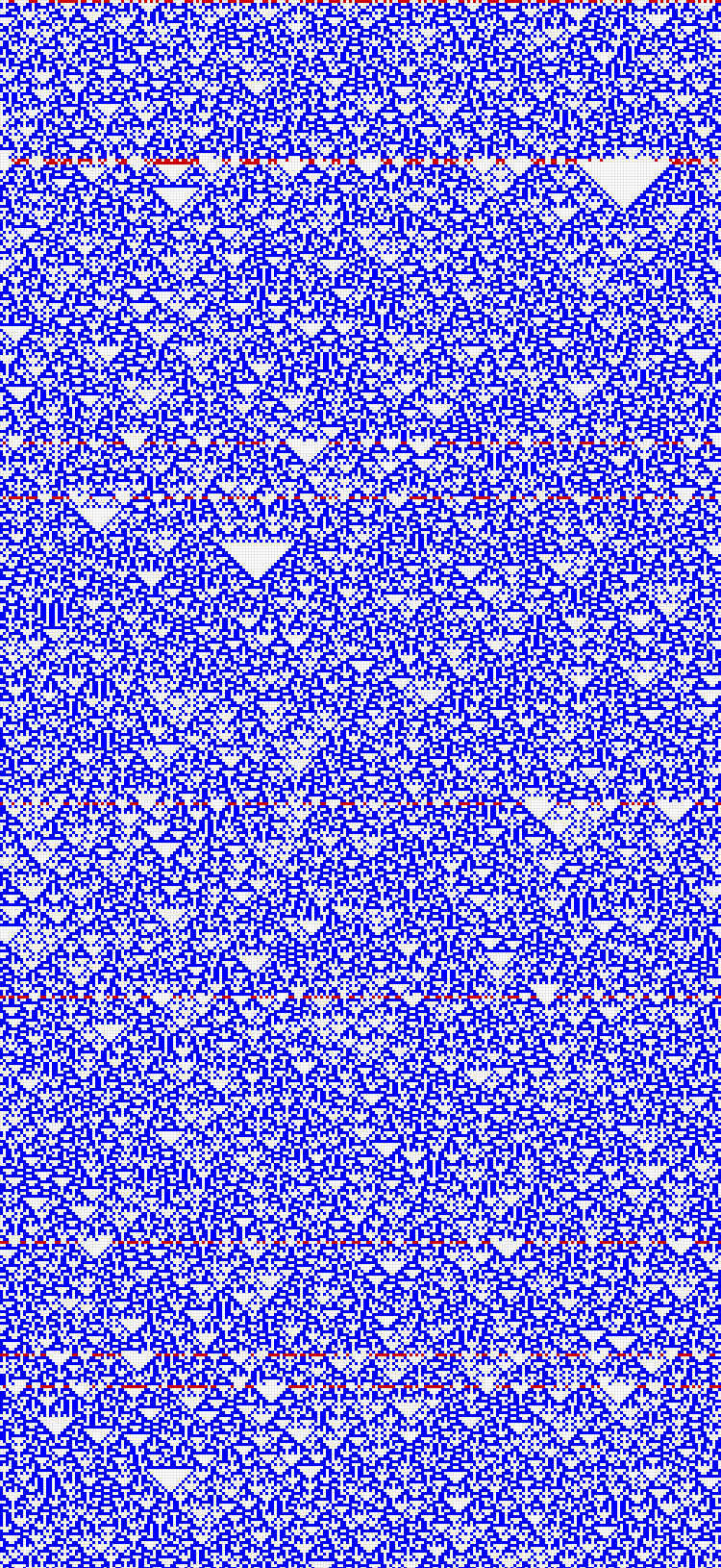} & \includegraphics[width=33mm]{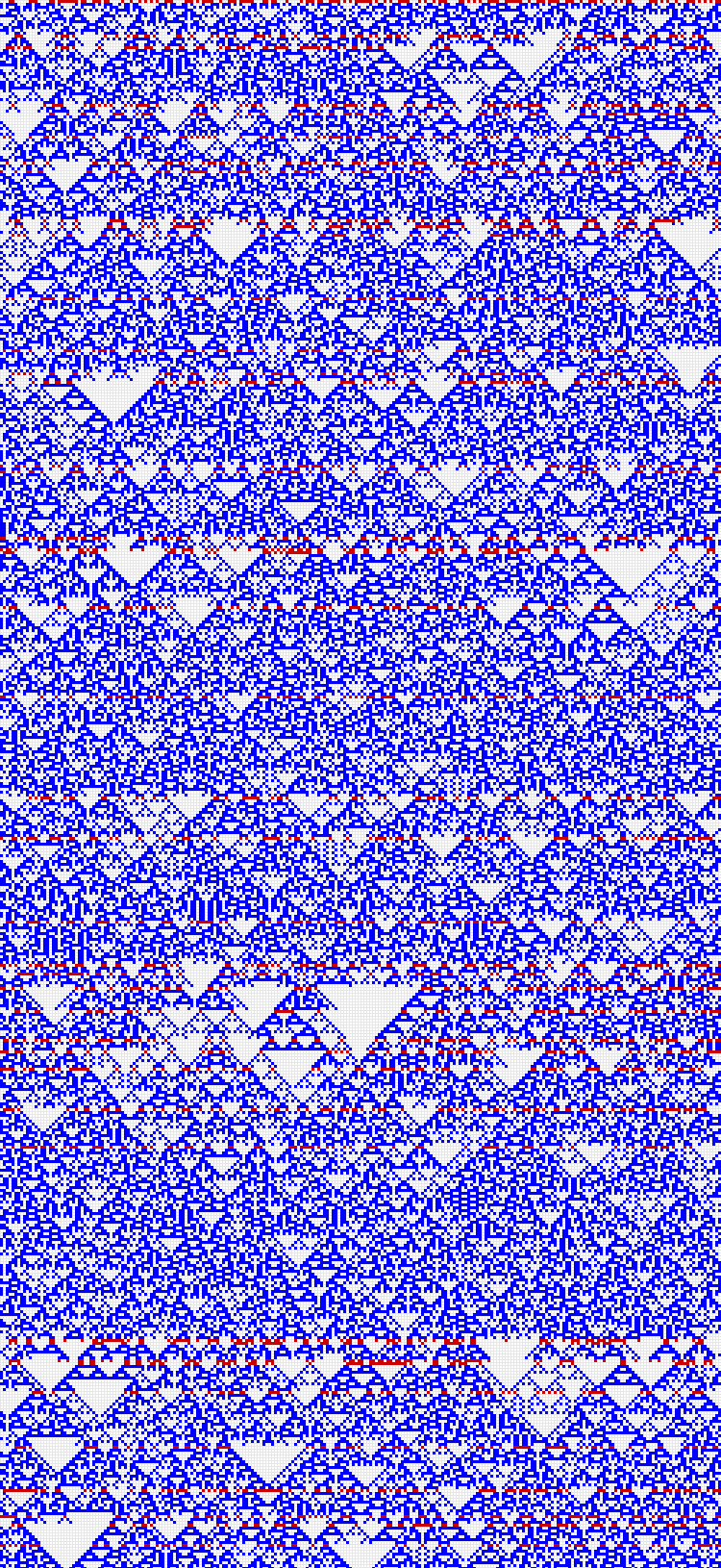} &   \includegraphics[width=33mm]{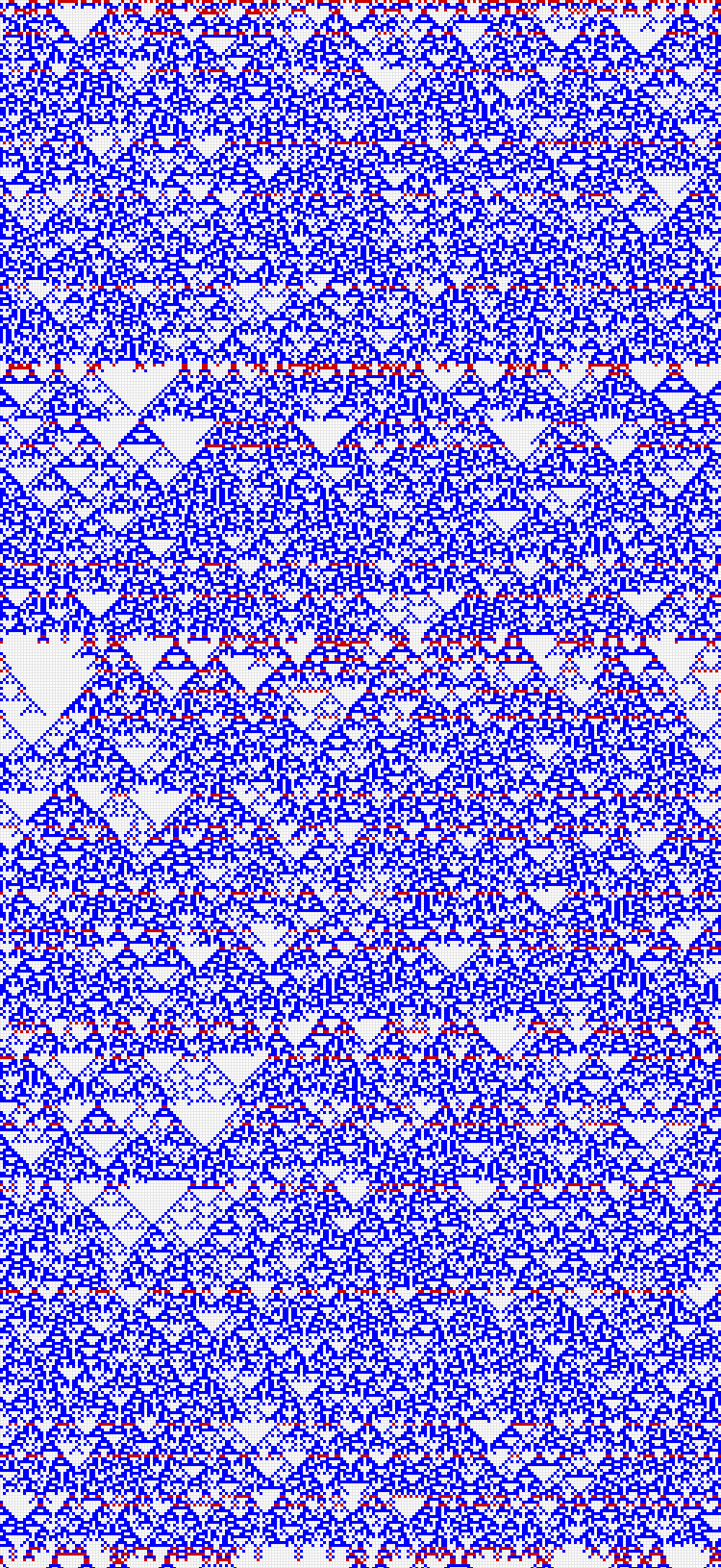} &   \includegraphics[width=33mm]{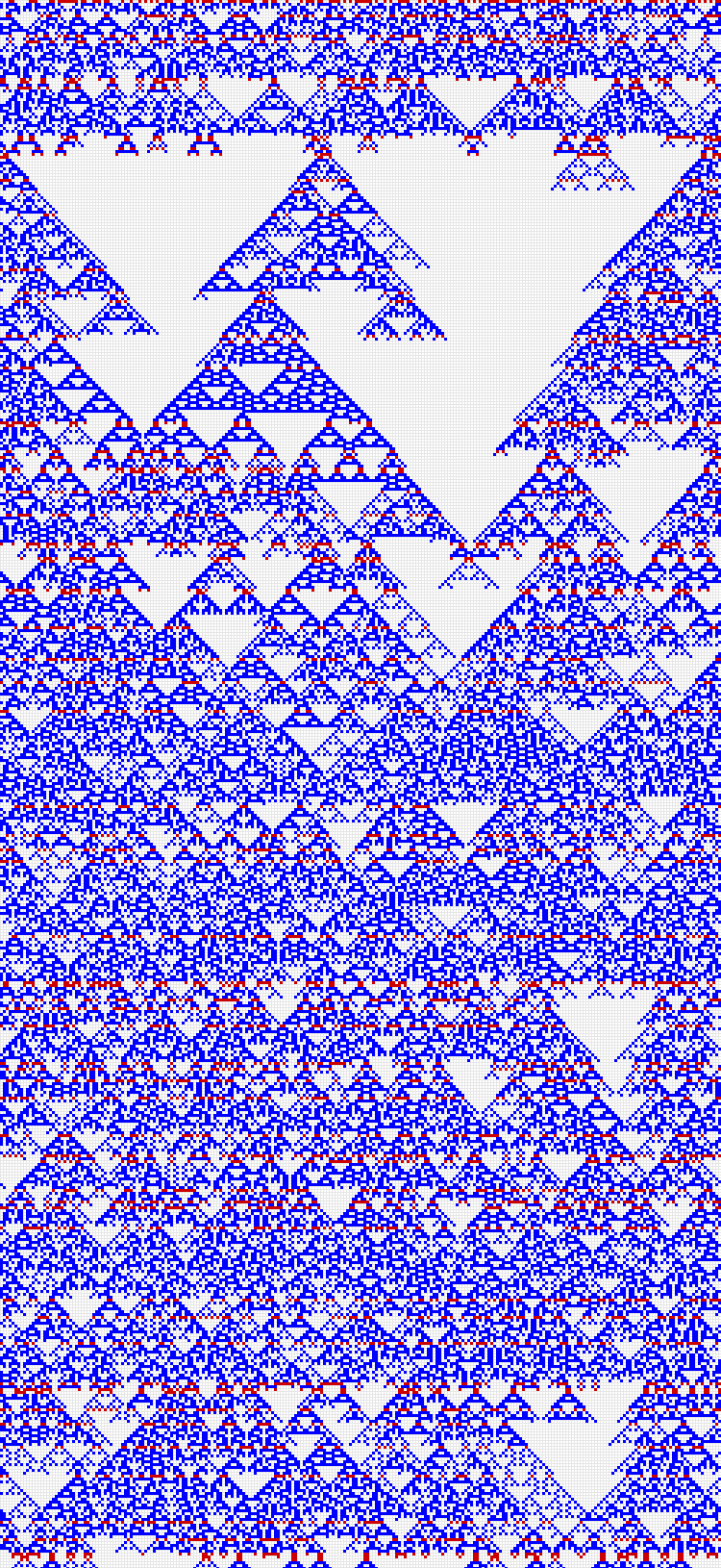}\\ 
				(90,104)[0.16] & (90,104)[0.17]  & (90,104)[0.18]  & (90,104)[0.19]\\ [6pt]
				\includegraphics[width=33mm]{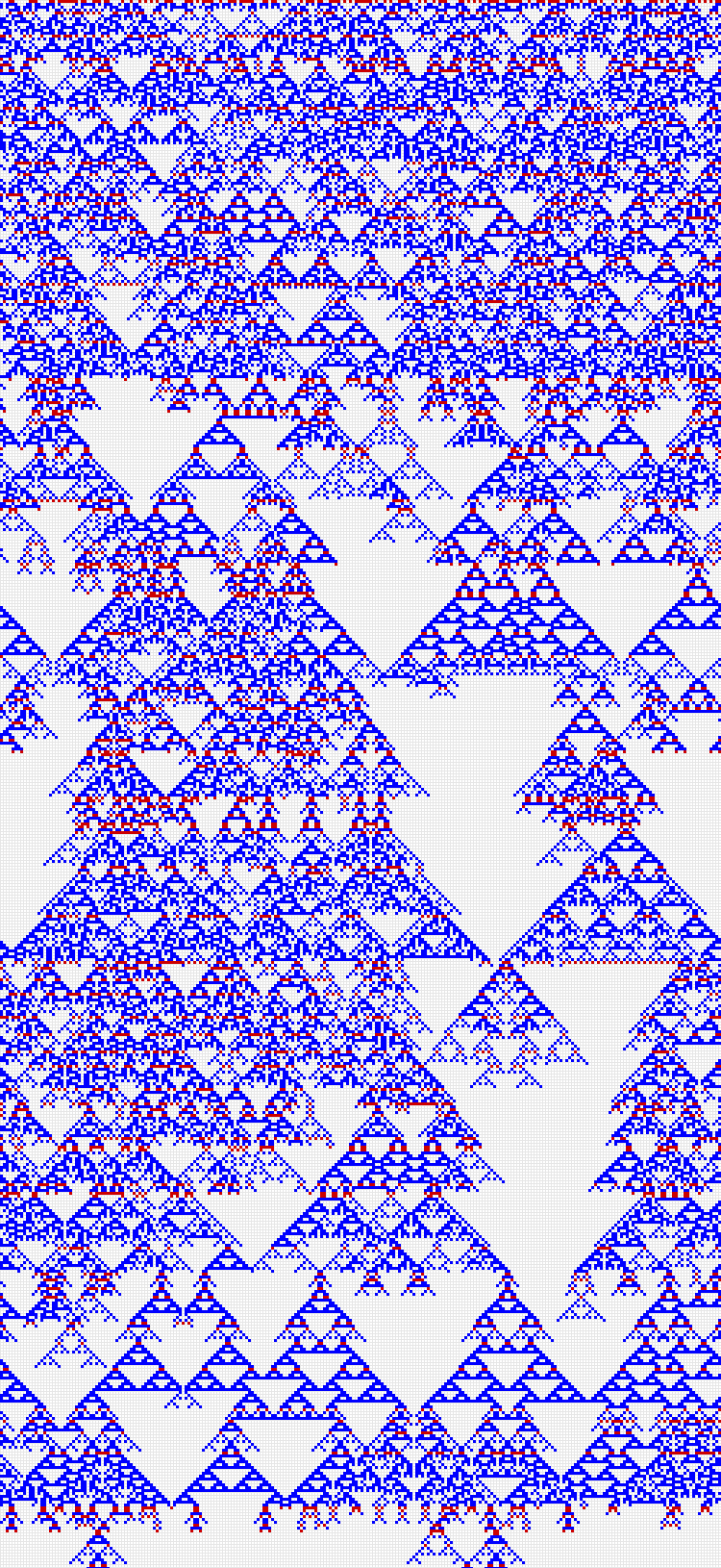} & \includegraphics[width=33mm]{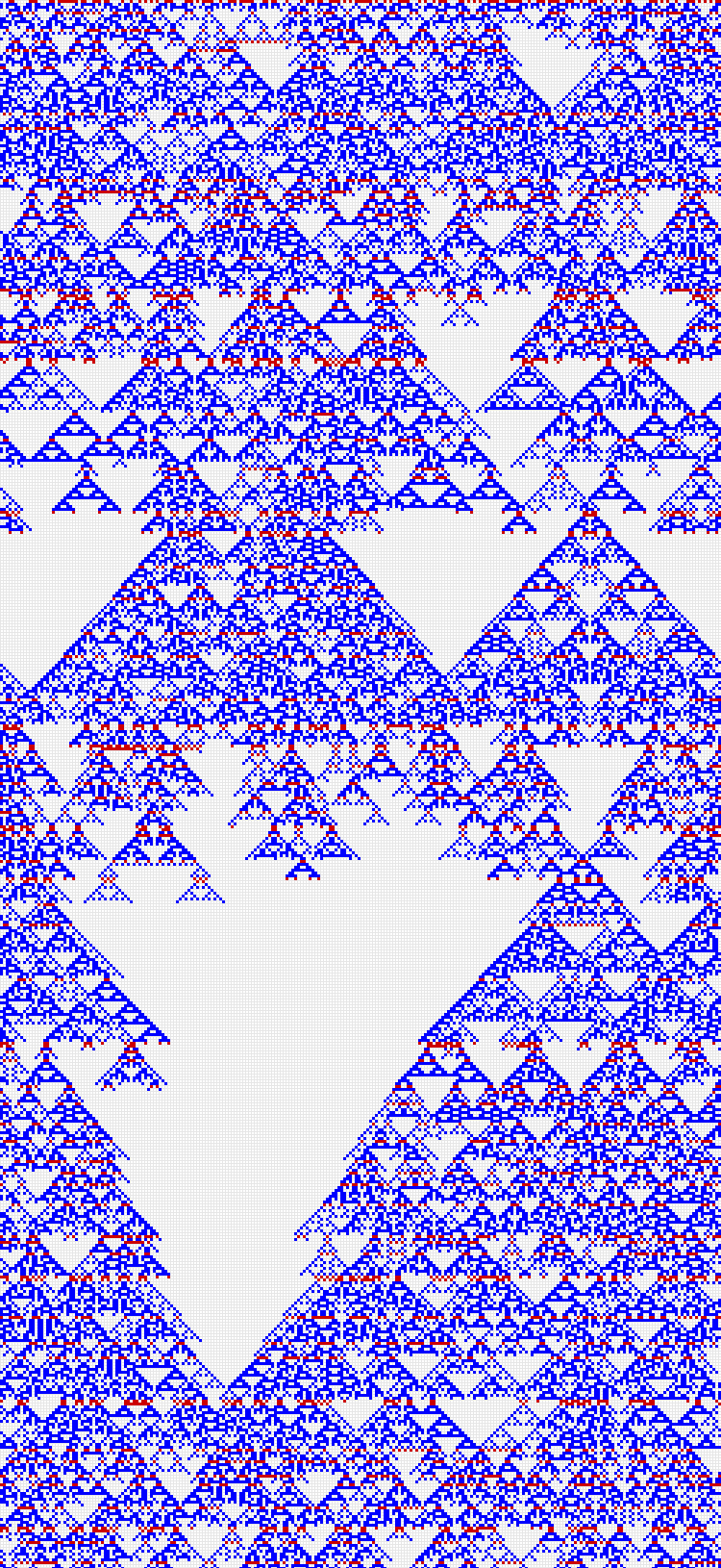} &   \includegraphics[width=33mm]{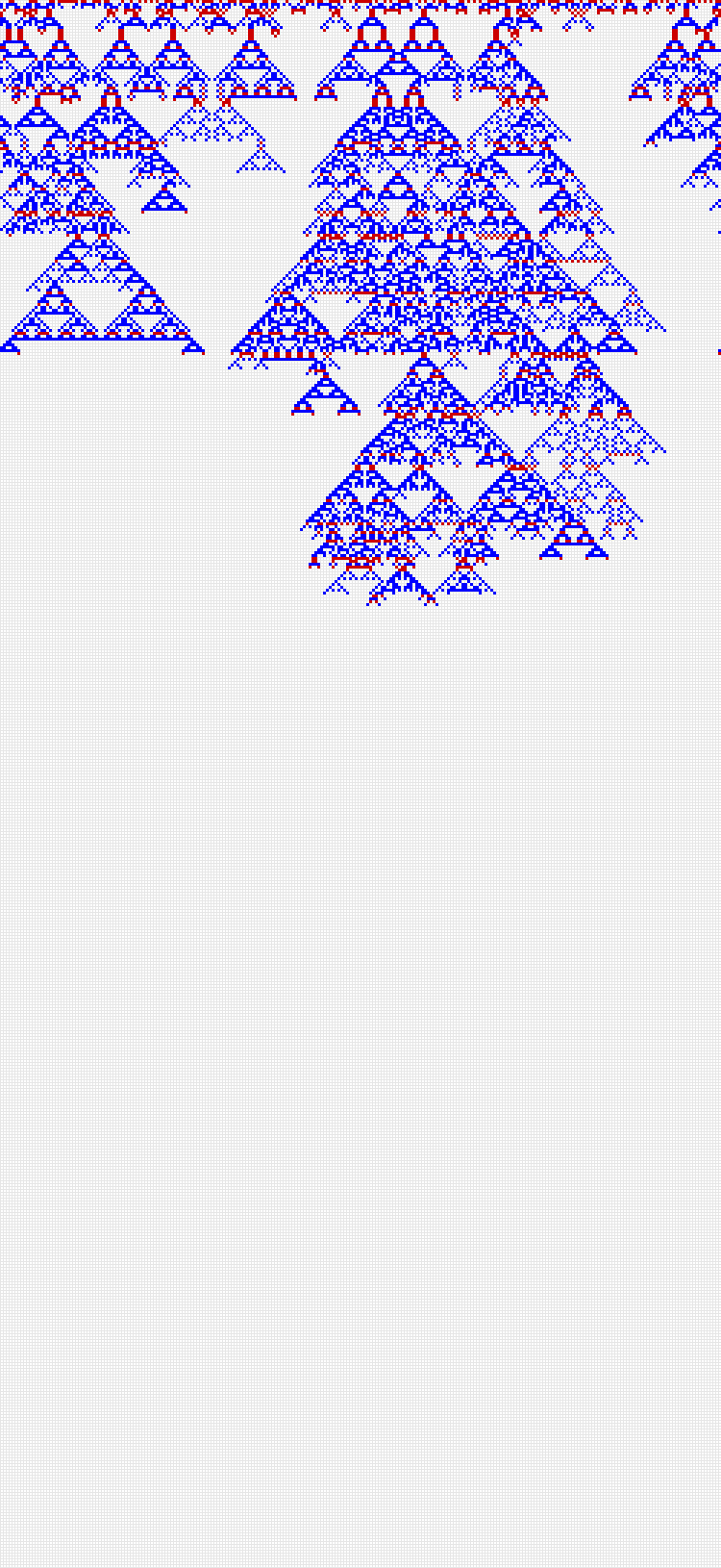} &   \includegraphics[width=33mm]{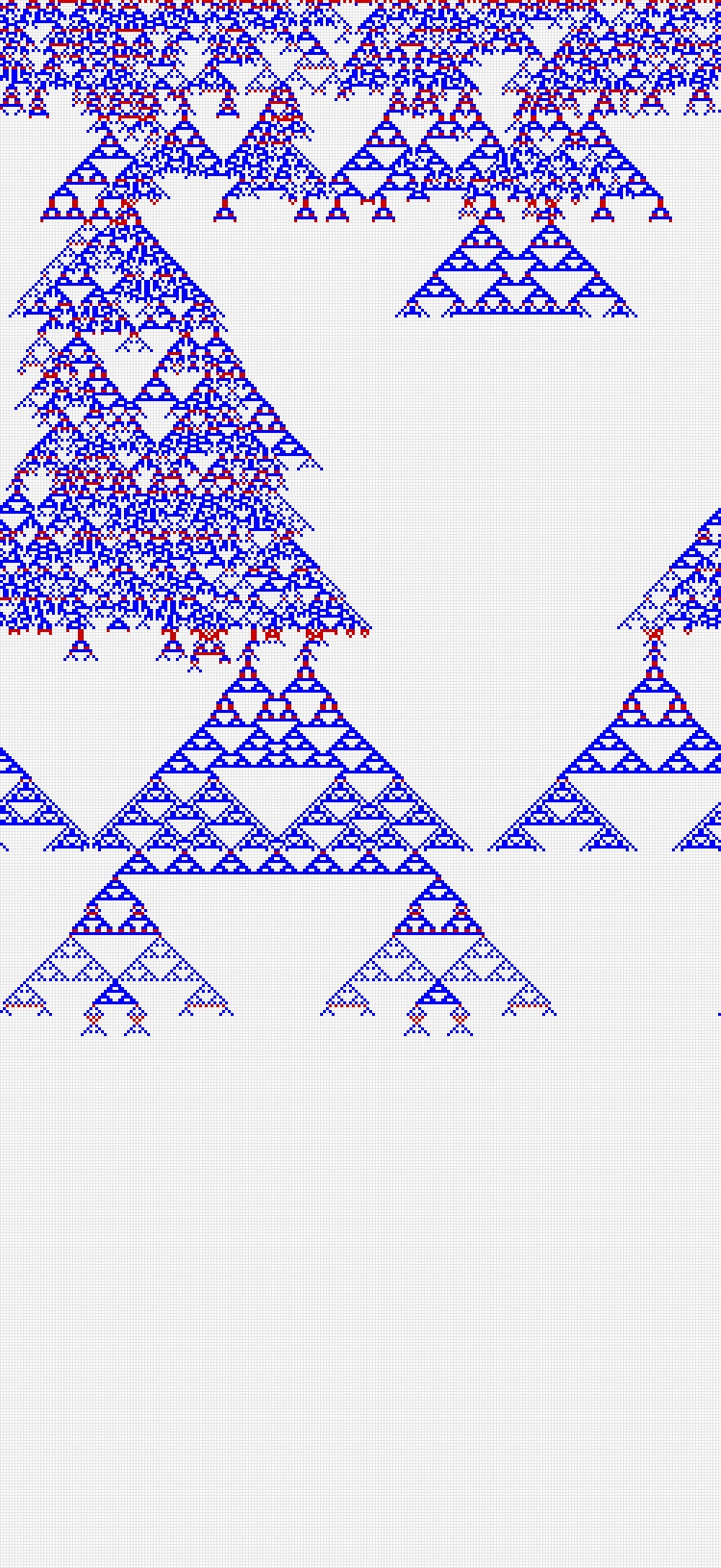}\\ 
				(90,104)[0.20] & (90,104)[0.22]  & (90,104)[0.25]  & (90,104)[0.46]\\ [6pt]
				\includegraphics[width=33mm]{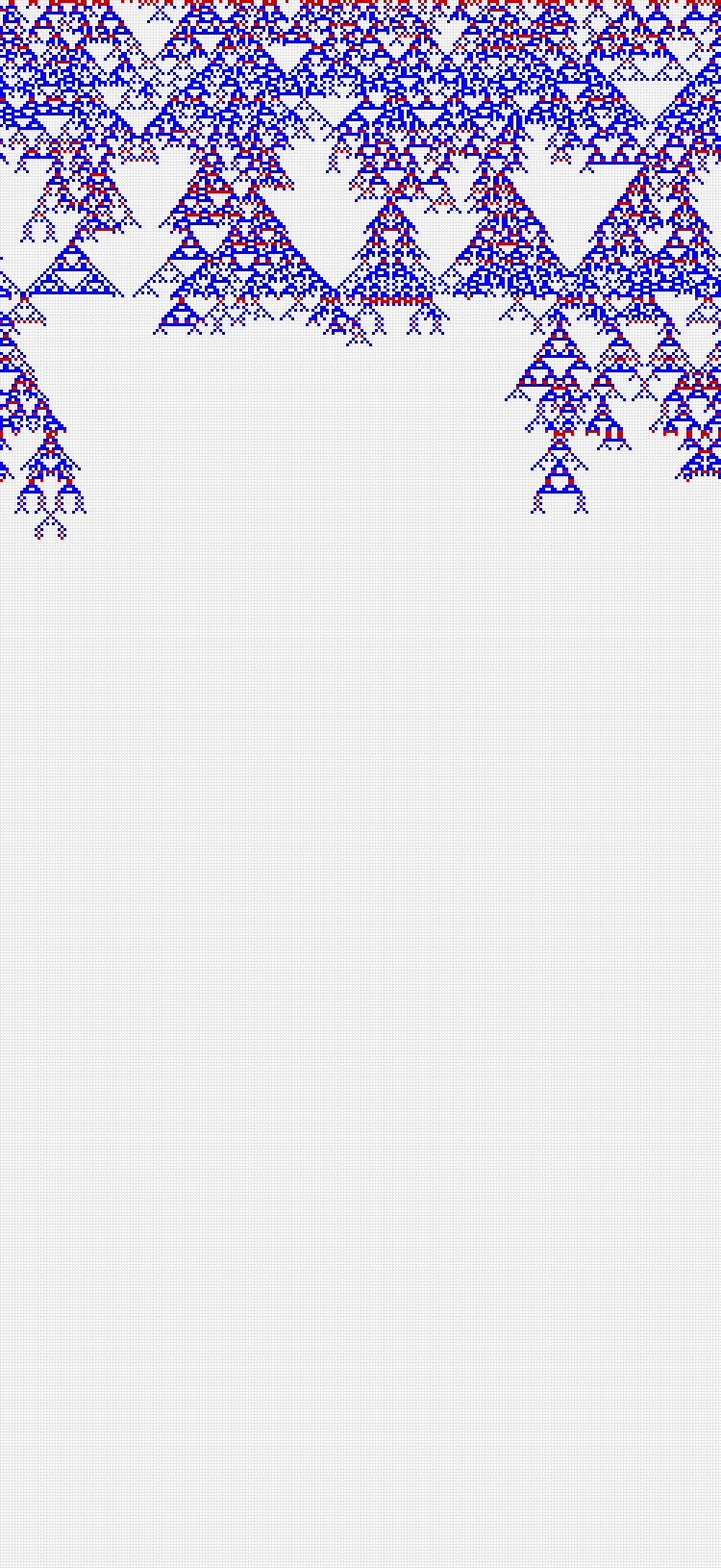} & \includegraphics[width=33mm]{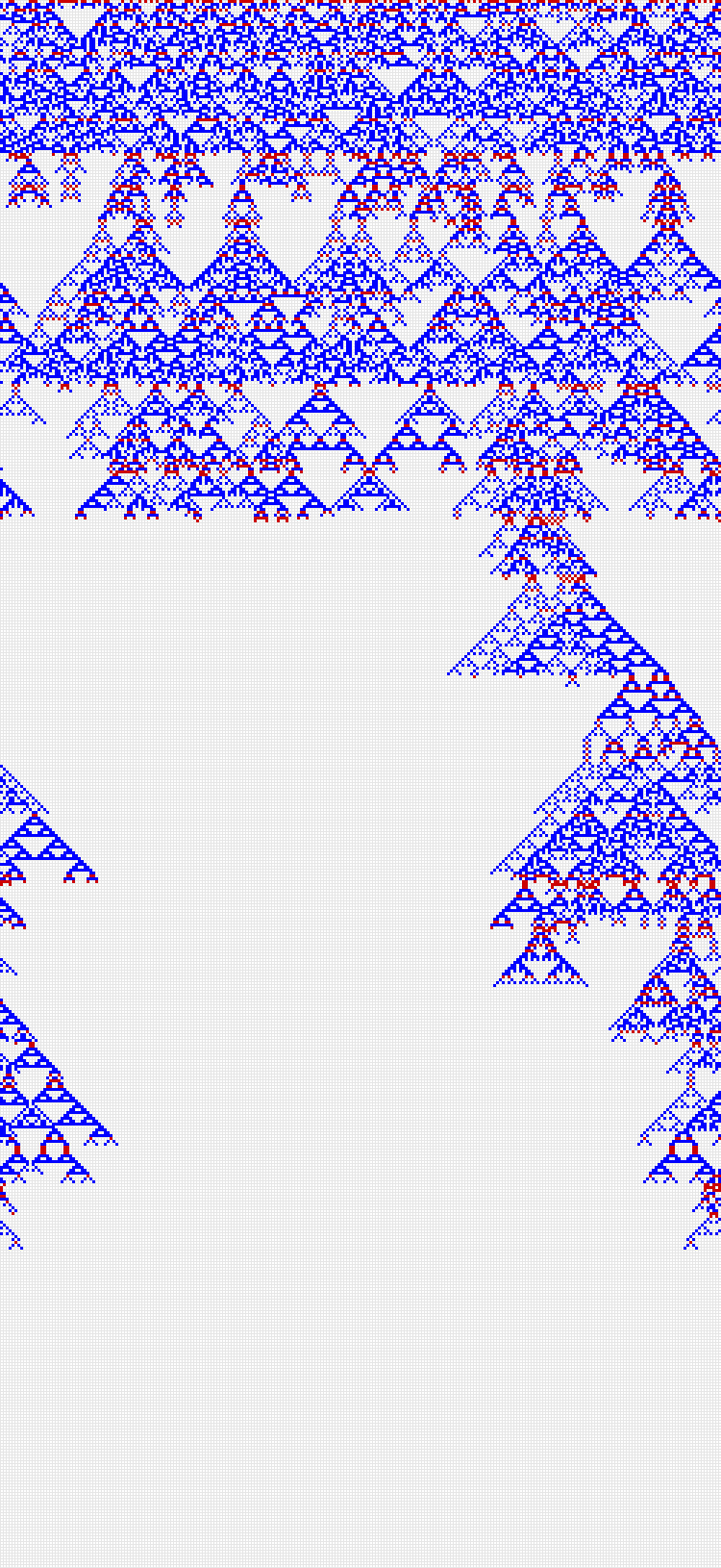} &   \includegraphics[width=33mm]{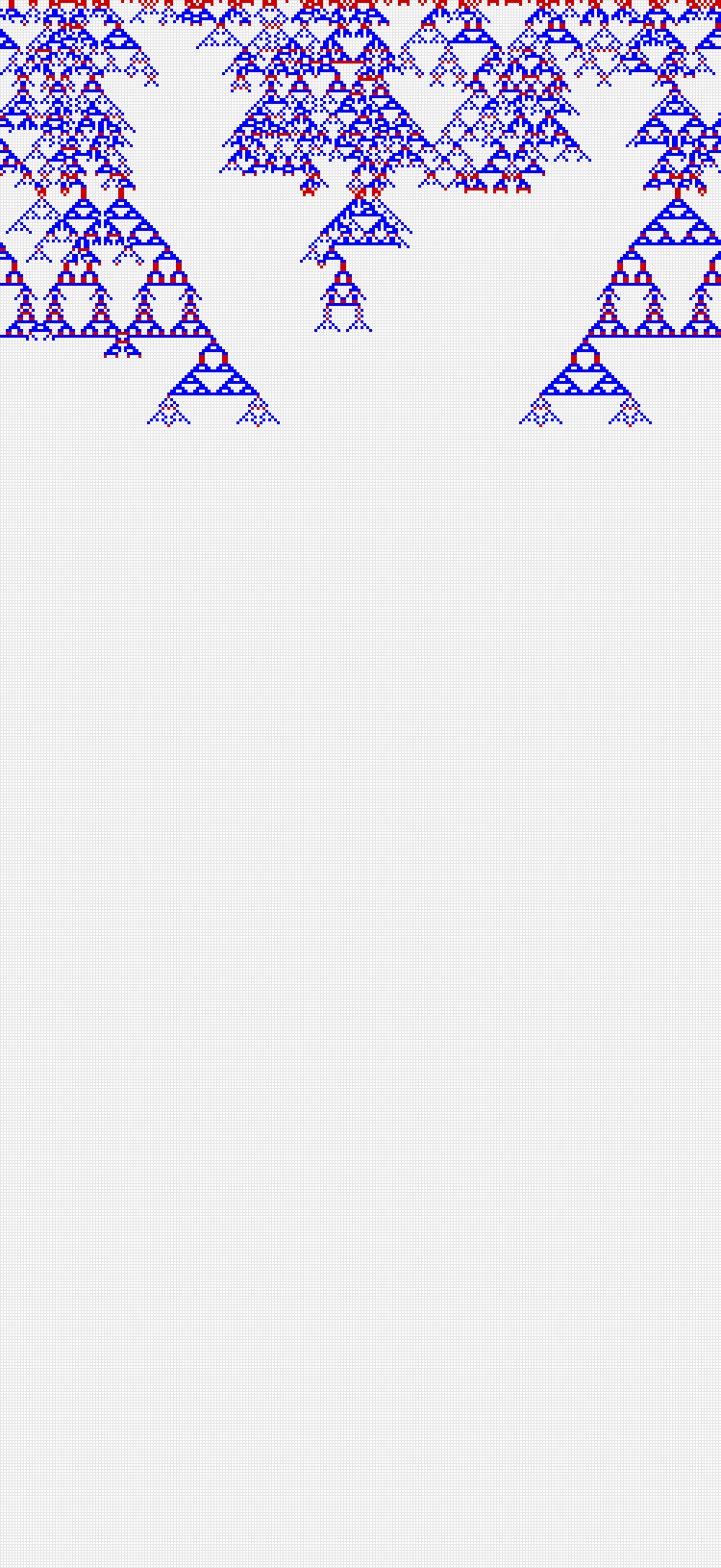} &   \includegraphics[width=33mm]{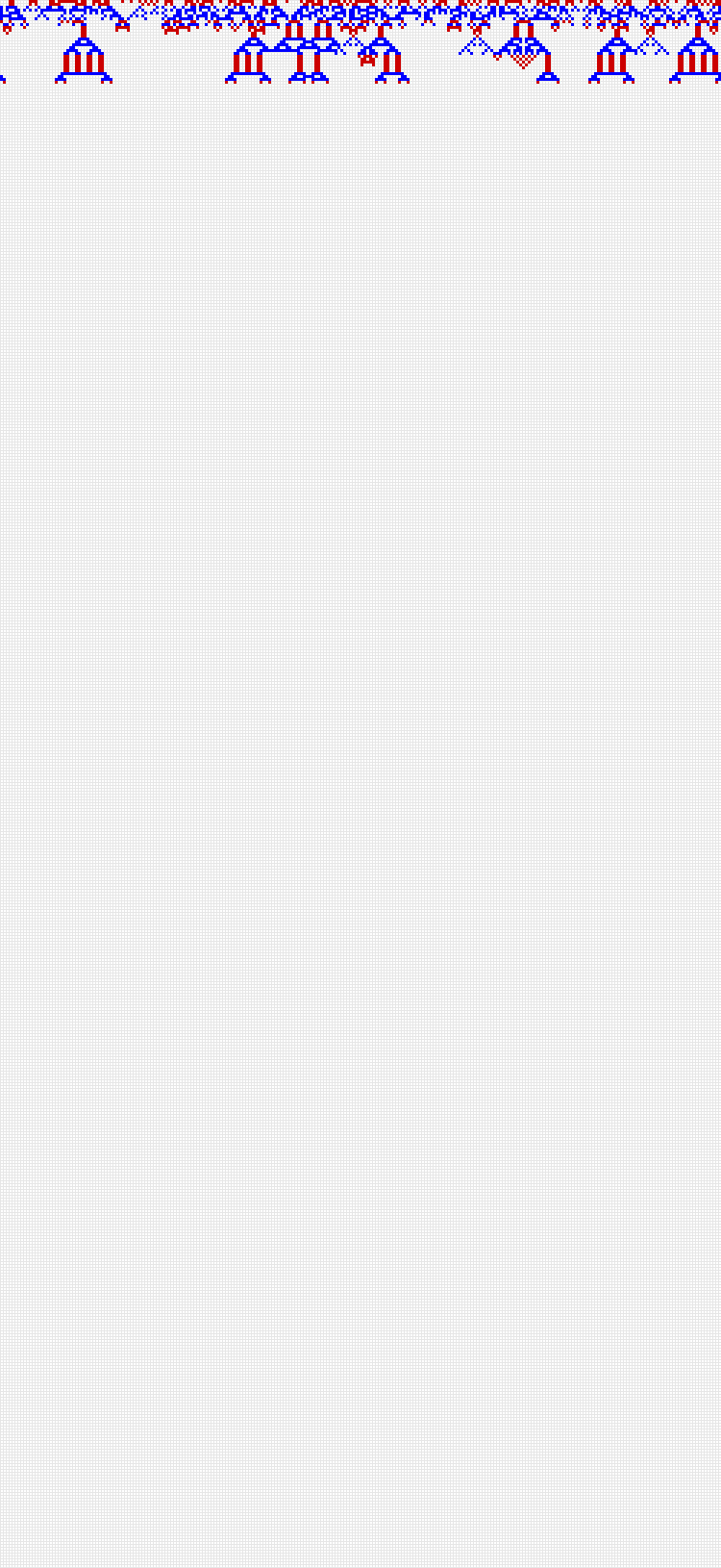}\\ 
t 		\end{tabular}}
		\caption{Phase transition behavior of stochastic CAs ($90,104$) with different $\tau$ values.}
		\label{TSCA6}
	\end{center}
\end{figure*}

\begin{figure*}[hbt!]
	\begin{center}
		\scalebox{0.8}{
			\begin{tabular}{cccc}
				(156,160)[0.09] & (156,160)[0.20] & (156,160)[0.24] & (156,160)[0.27]\\ [6pt]
				\includegraphics[width=33mm]{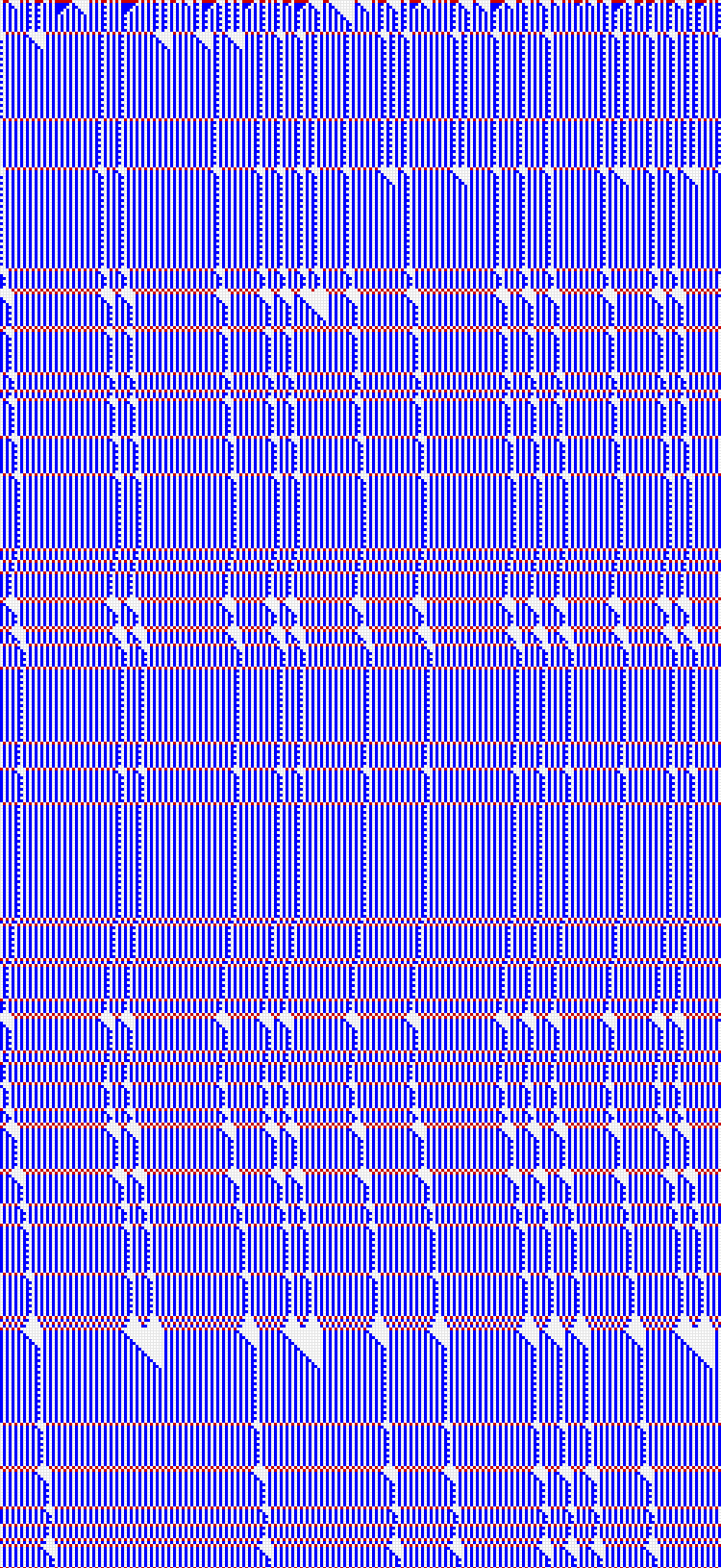} & \includegraphics[width=33mm]{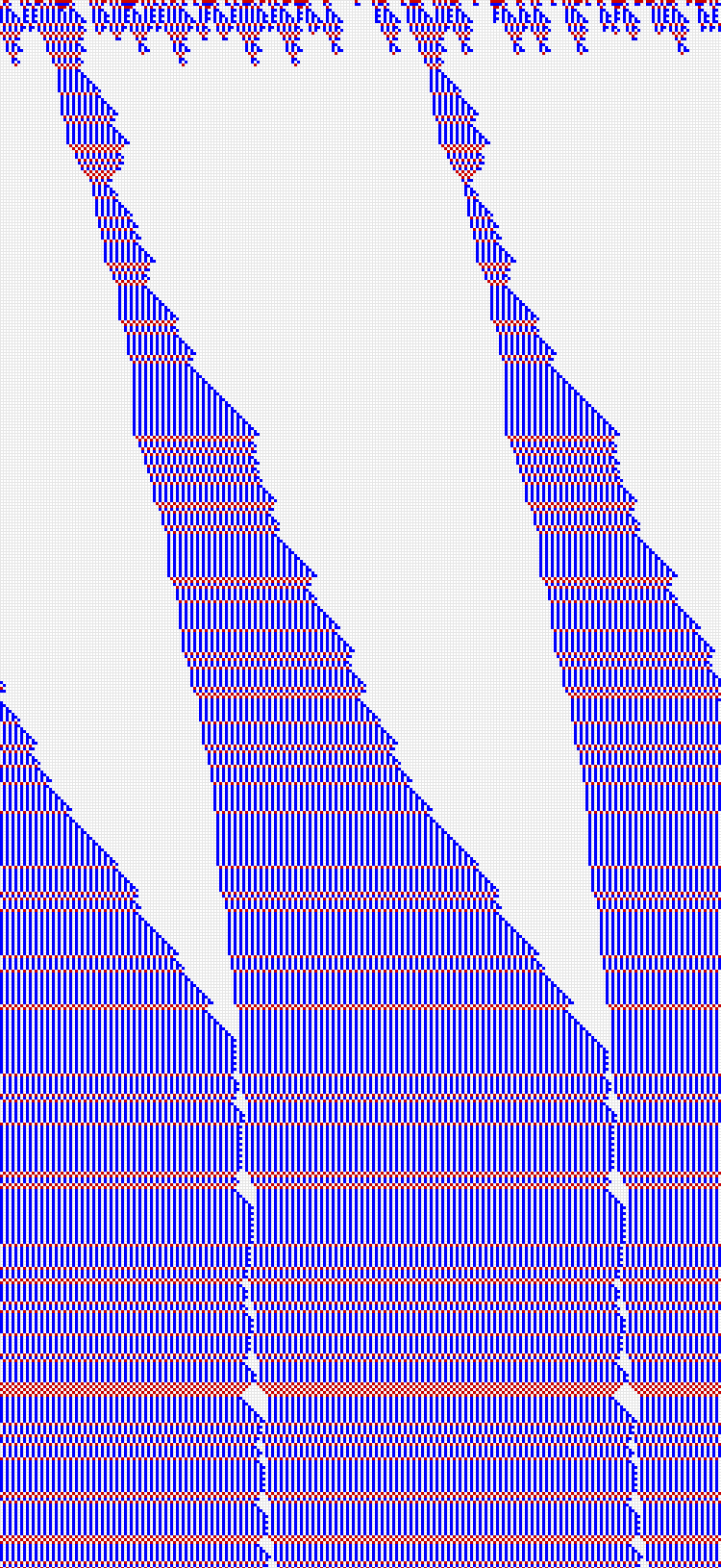} &   \includegraphics[width=33mm]{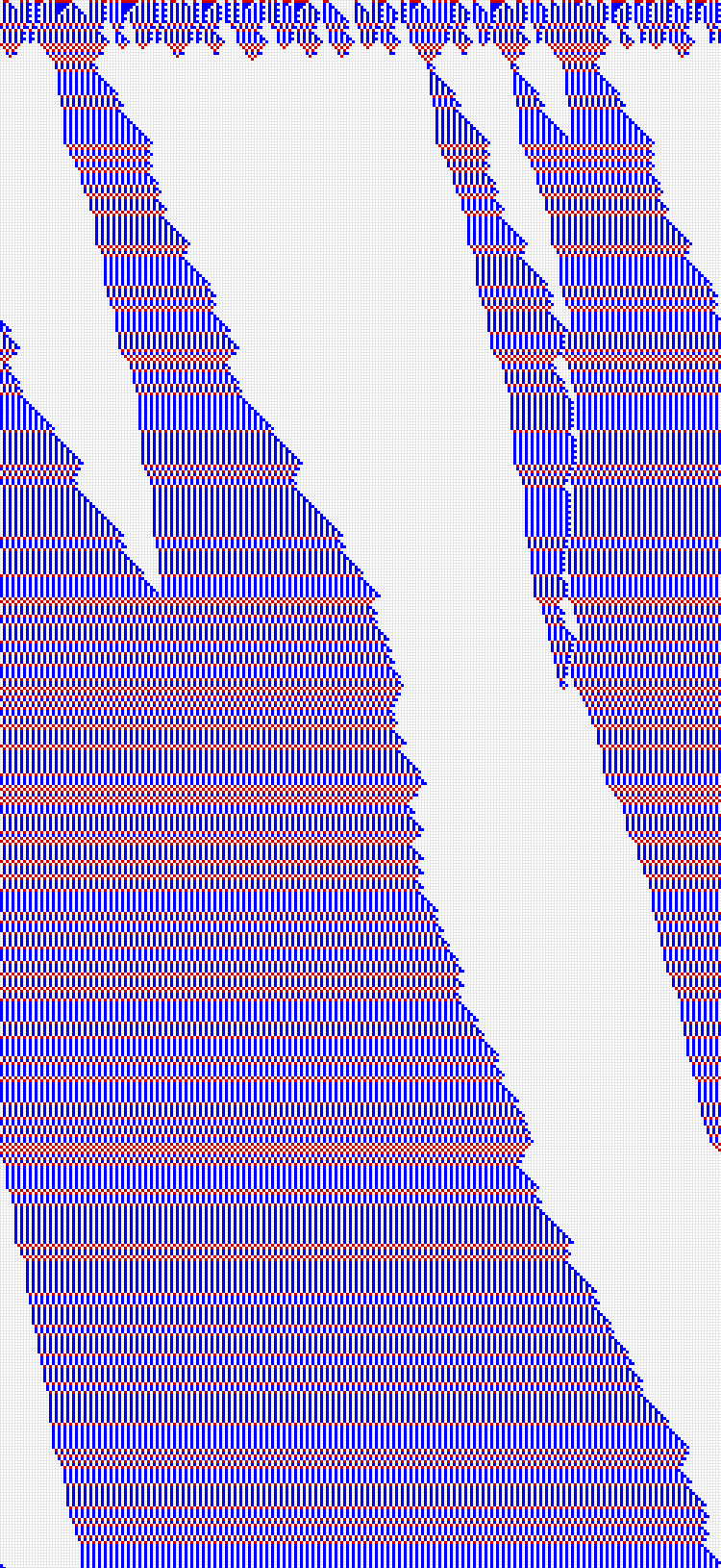} &   \includegraphics[width=33mm]{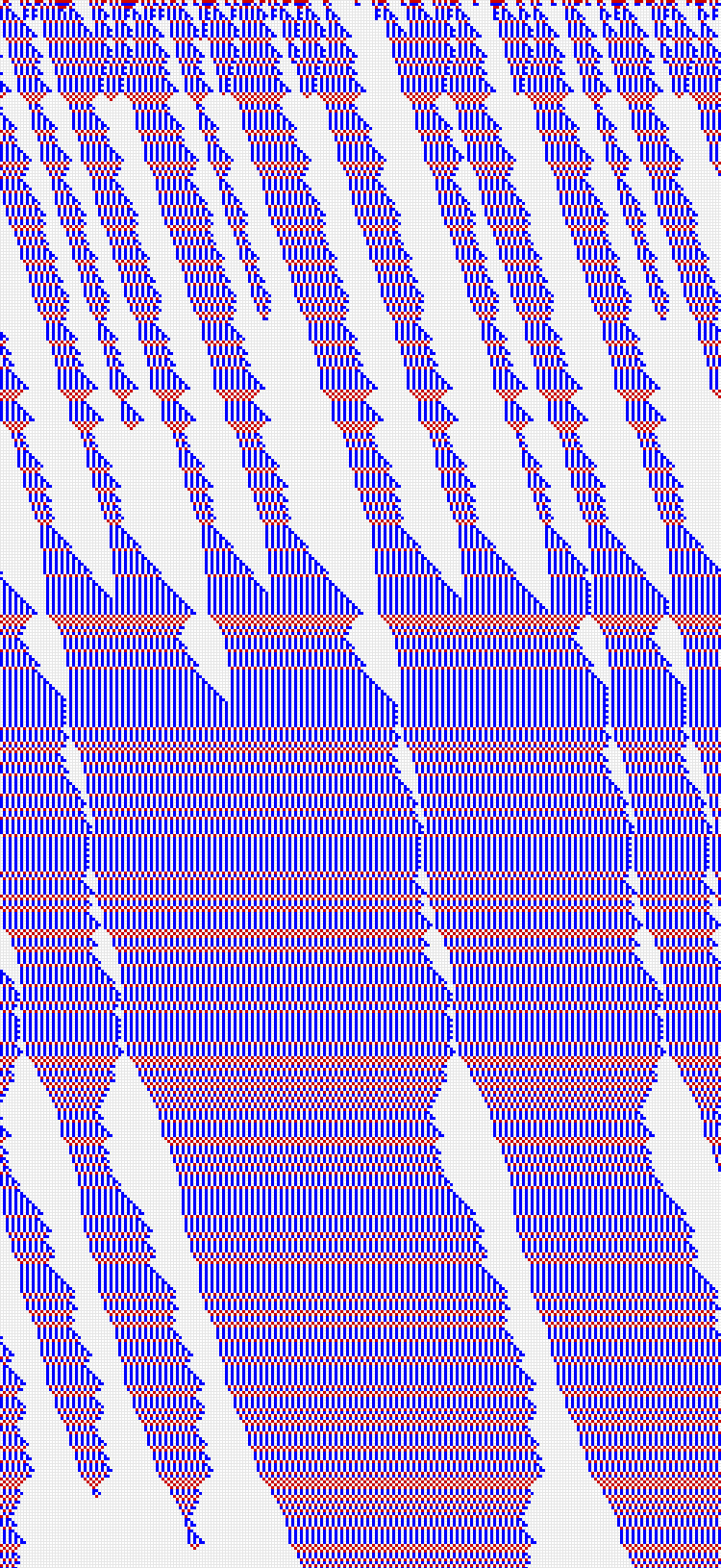}\\ 
				(156,160)[0.28] & (156,160)[0.29] & (156,160)[0.30] & (156,160)[0.31]\\ [6pt]
				\includegraphics[width=33mm]{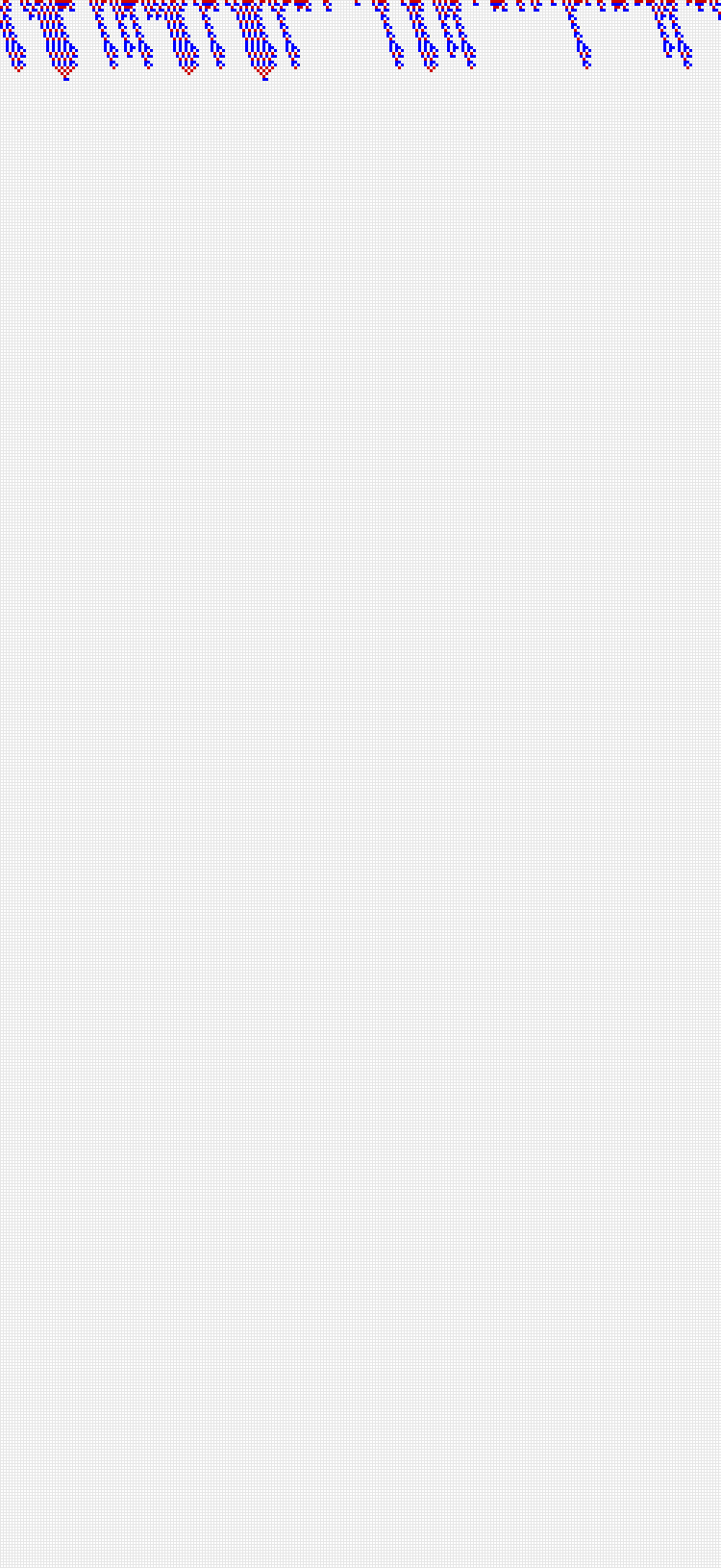} & \includegraphics[width=33mm]{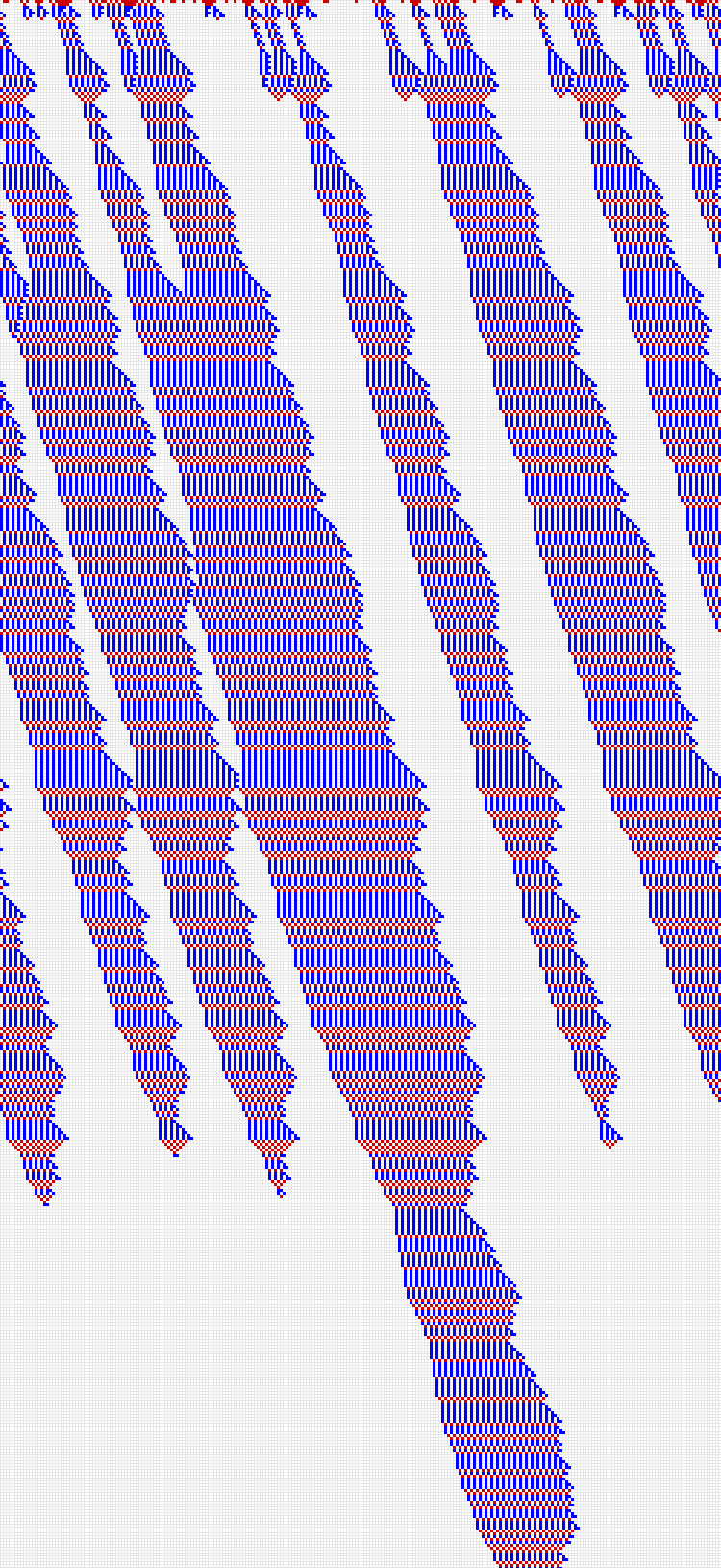} &   \includegraphics[width=33mm]{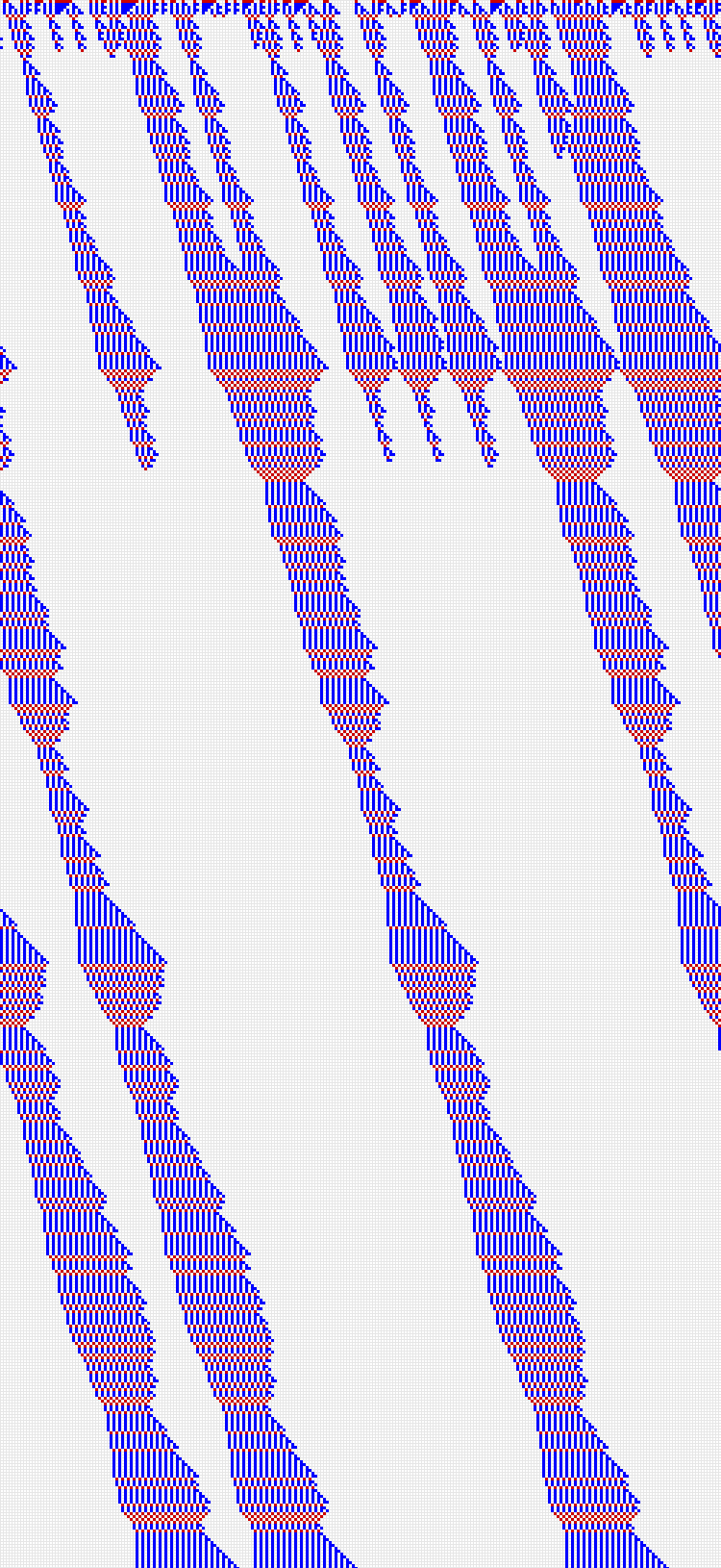} &   \includegraphics[width=33mm]{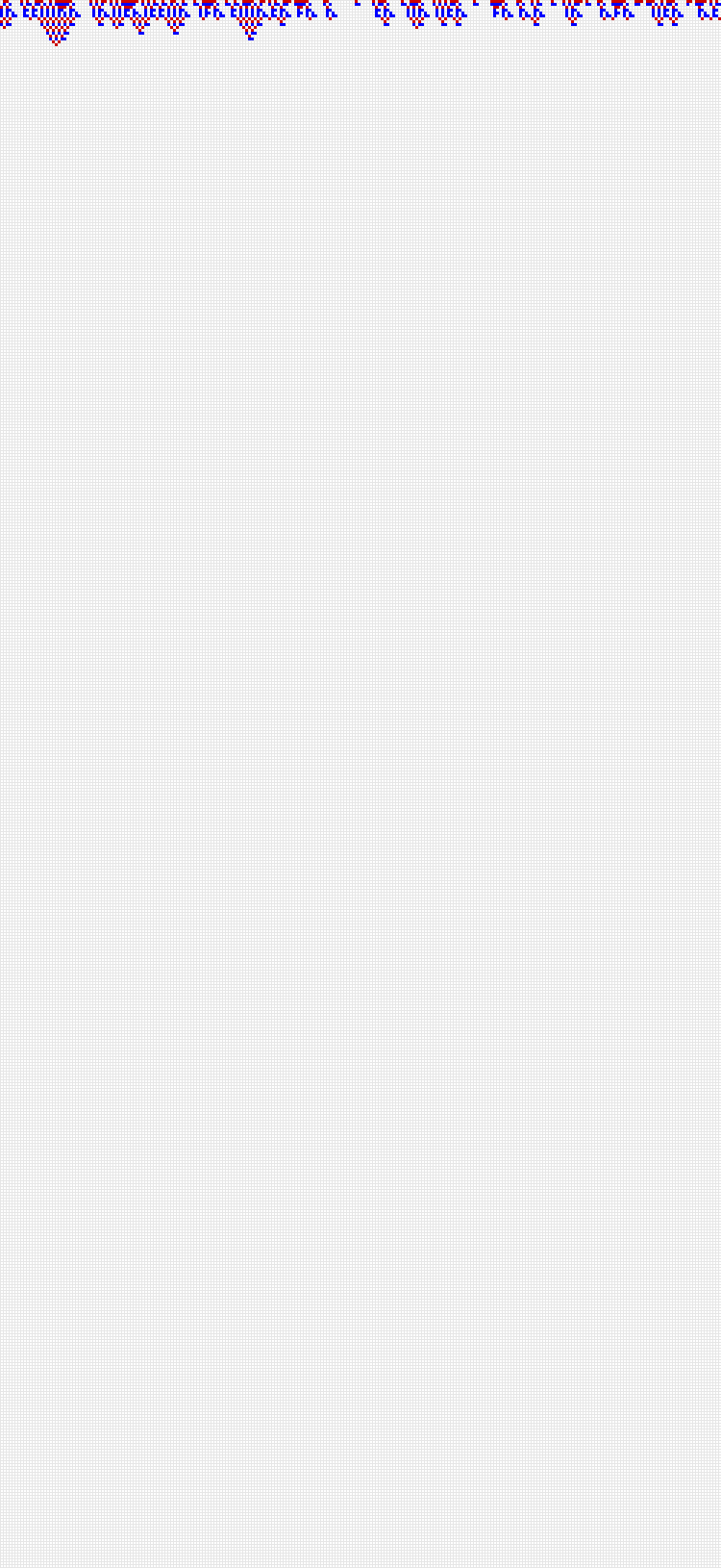}\\ 
				(156,160)[0.32] & (156,160)[0.40] & (156,160)[0.50] & (156,160)[0.60]\\ [6pt]
				\includegraphics[width=33mm]{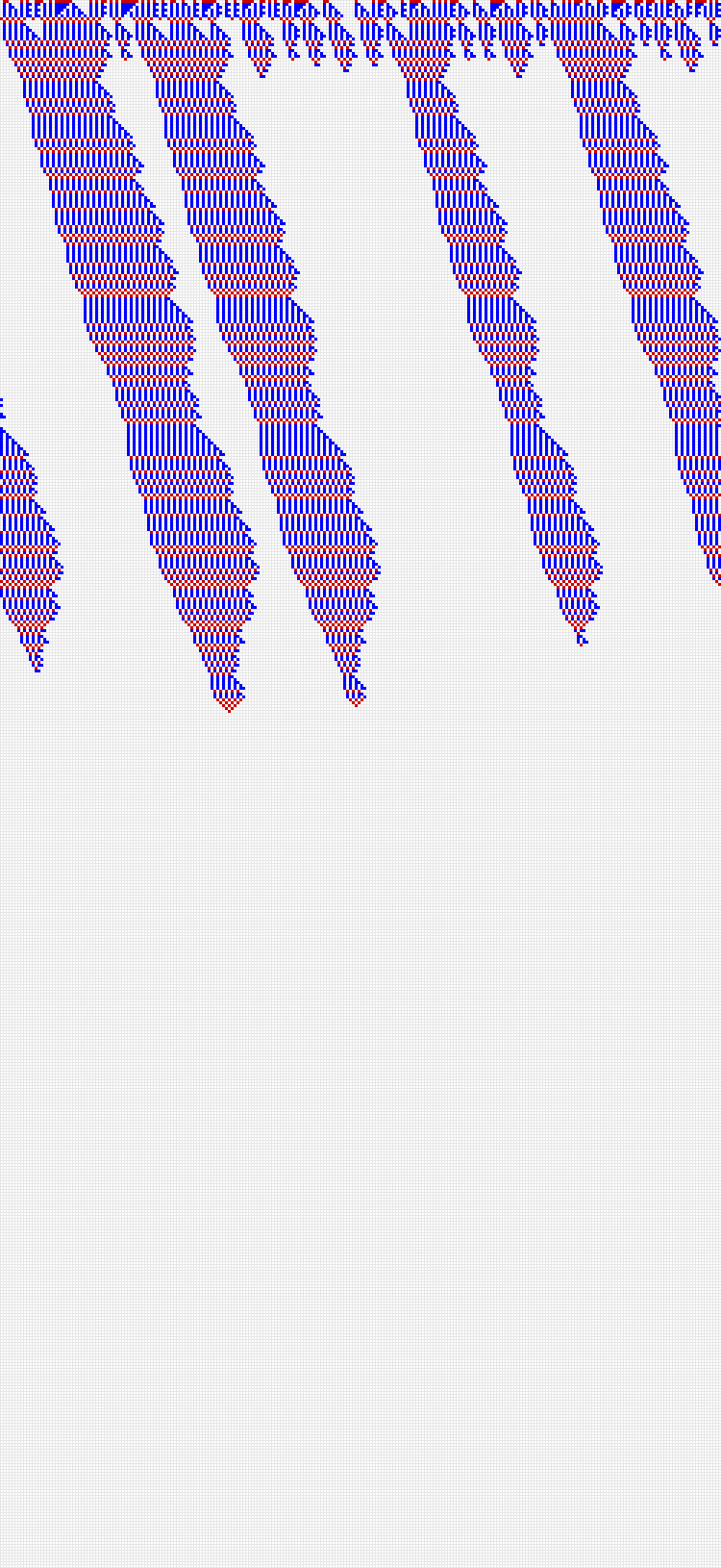} & \includegraphics[width=33mm]{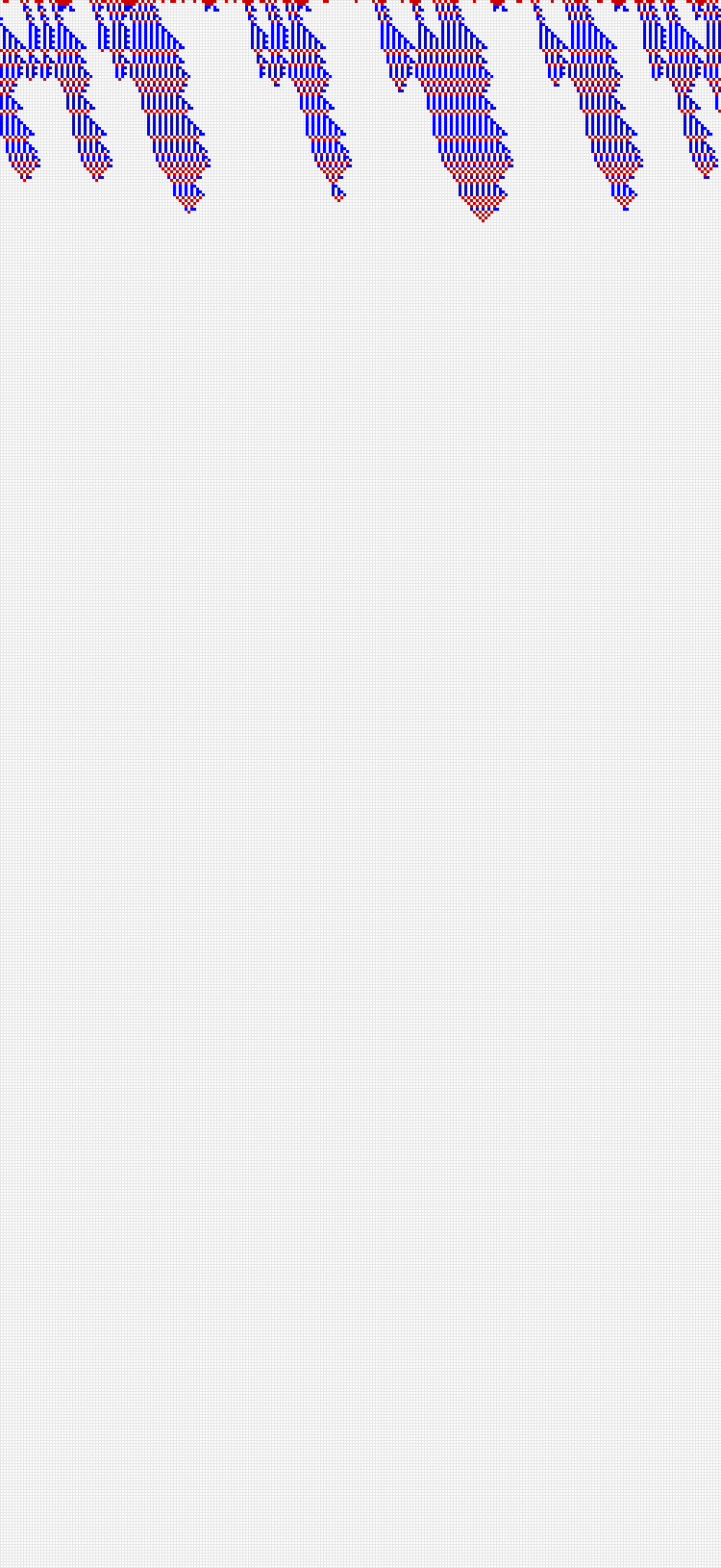} &   \includegraphics[width=33mm]{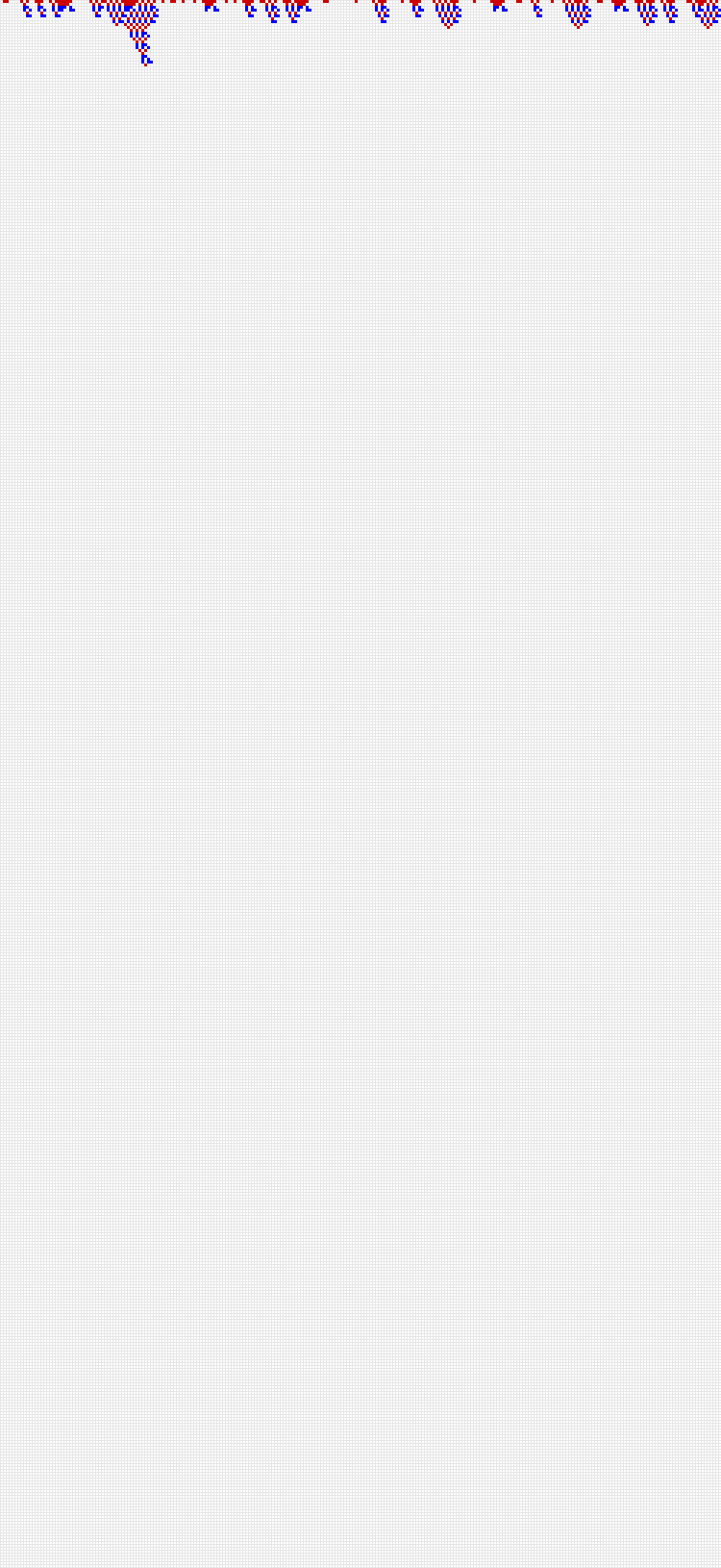} &   \includegraphics[width=33mm]{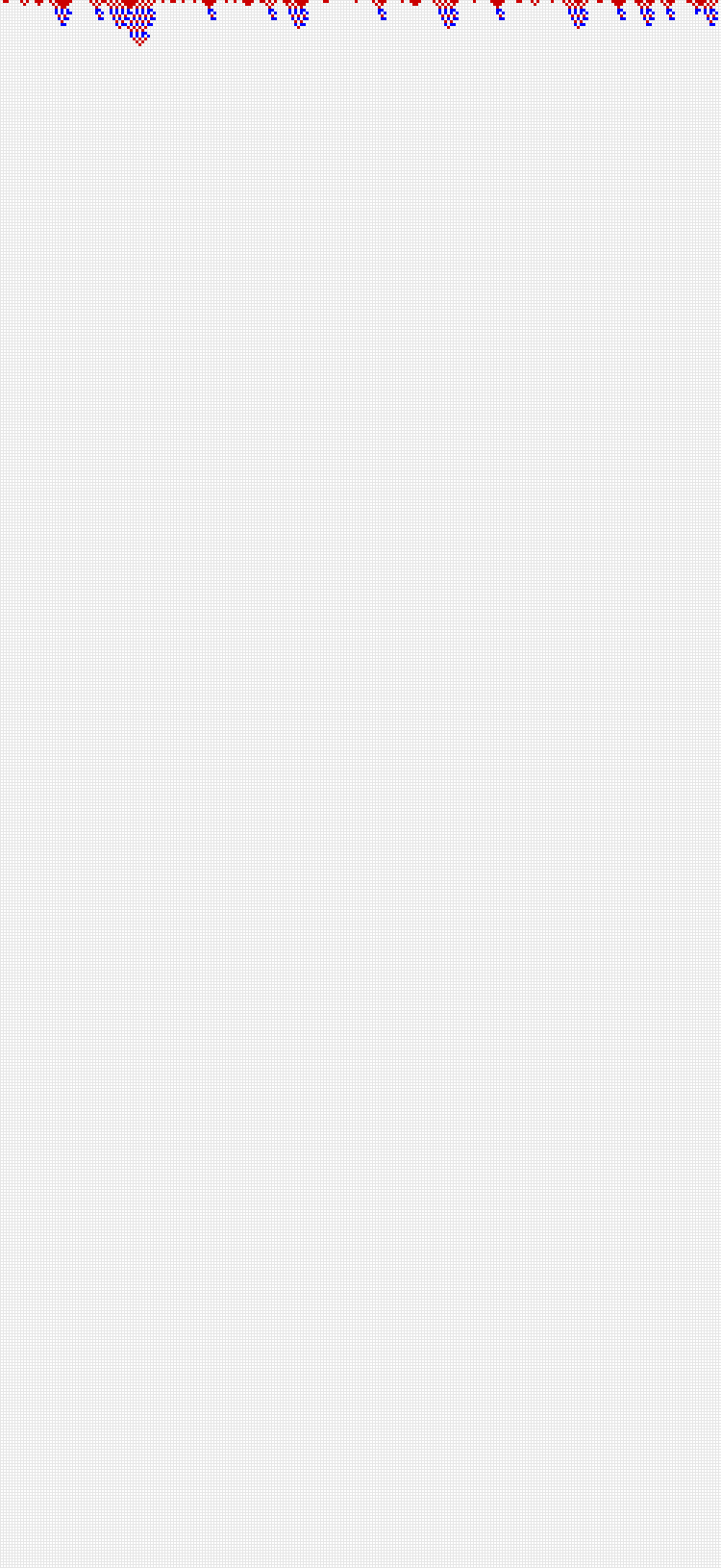}\\ 
				t 		\end{tabular}}
		\caption{Phase transition behavior of stochastic CAs ($156,160$) with different $\tau$ values.}
		\label{TSCA7}
	\end{center}
\end{figure*}

\subsection{Dynamics when $\mathcal{C}$($f$) $\neq$ $\mathcal{C}$($g$)}

Next, we explore the situation where $\mathcal{C}$($f$) $\neq$ $\mathcal{C}$($g$). Under this setting, we have observed two kinds of results in our experiments. 

\begin{itemize}
	\item $\mathcal{C}$(($f,g$)) = $\mathcal{C}$($f$) or $\mathcal{C}$(($f,g$)) = $\mathcal{C}$($g$); and
	\item $\mathcal{C}$(($f,g$)) $\neq$ $\mathcal{C}$($f$) and $\mathcal{C}$(($f,g$)) $\neq$ $\mathcal{C}$($g$)
\end{itemize}

Let us focus on the first possibility, i.e.  $\mathcal{C}$(($f,g$)) = $\mathcal{C}$($f$) or $\mathcal{C}$(($f,g$)) = $\mathcal{C}$($g$). So, here, we have following cases:

\begin{itemize}
	\item[(i)] $\mathcal{C}$($f$) = class C, $\mathcal{C}$($g$) = class B, and $\mathcal{C}$(($f,g$)) = $\mathcal{C}$($f$);
	\item[(ii)] $\mathcal{C}$($f$) = class C, $\mathcal{C}$($g$) = class B, and $\mathcal{C}$(($f,g$)) = $\mathcal{C}$($g$);
	\item[(iii)] $\mathcal{C}$($f$) = class C, $\mathcal{C}$($g$) = class A, and $\mathcal{C}$(($f,g$)) = $\mathcal{C}$($f$);
	\item[(iv)] $\mathcal{C}$($f$) = class C, $\mathcal{C}$($g$) = class A, and $\mathcal{C}$(($f,g$)) = $\mathcal{C}$($g$);
	\item[(v)] $\mathcal{C}$($f$) = class B, $\mathcal{C}$($g$) = class A, and $\mathcal{C}$(($f,g$)) = $\mathcal{C}$($f$); and
	\item[(vi)] $\mathcal{C}$($f$) = class B, $\mathcal{C}$($g$) = class A, and $\mathcal{C}$(($f,g$)) = $\mathcal{C}$($g$).
\end{itemize}

Let us start with case (i) where $\mathcal{C}$($f$) = class C, $\mathcal{C}$($g$) = class B, and $\mathcal{C}$(($f,g$)) = $\mathcal{C}$($f$). Observe the chaotic (class C) and periodic (class B) behavior of ECA $122$ and $37$ respectively. However, in Fig.~\ref{Fig4}, the CA ($122,37$) shows chaotic dynamics for $\tau \in$ \{$0.1,0.5,0.9$\}. That is $\mathcal{C}$(($122,37$)) = $\mathcal{C}$($122$). Next, according to case (ii), $\mathcal{C}$($f$) = class C, $\mathcal{C}$($g$) = class B, and $\mathcal{C}$(($f,g$)) = $\mathcal{C}$($g$) which is the opposite of case (i). Here also, ECAs $22$ and $7$ individually show chaotic and periodic dynamics respectively. To show the second case, if rule $22$ is considered as default rule and rule $7$ is added as noise at different probability, the resultant dynamics shows periodic dynamics. Observe that, the space time diagrams of ($22,7$) for $\tau \in$ \{$0.1,0.5,0.9$\} shows little but chaotic nature in first few time steps. However, when the time progresses, they move towards periodic behavior (see Fig.~\ref{Fig4}). If one progressively changes the temporal noise rate, the same dynamics is observed. 

In our experiment (see Fig.~\ref{Fig1}), none of the stochastic CAs shows the third case where $\mathcal{C}$($f$) = class C, $\mathcal{C}$($g$) = class A, and $\mathcal{C}$(($f,g$)) = $\mathcal{C}$($f$). However, the case (iv) is observed for some of the stochastic CAs. Observe that, ECA $22$ and ECA $128$ respectively show chaotic (class C) and evolving to homogeneous configuration (class A) dynamics in Fig.~\ref{Fig4}. The temporally stochastic CA ($22,128$) for $\tau \in$ \{$0.1,0.5,0.9$\} exhibits homogeneous configurations in Fig.~\ref{Fig4}.

The remaining cases are $\mathcal{C}$($f$) = class B, $\mathcal{C}$($g$) = class A, $\mathcal{C}$(($f,g$)) = $\mathcal{C}$($f$) (case (v)) and $\mathcal{C}$($f$) = class B, $\mathcal{C}$($g$) = class A, $\mathcal{C}$(($f,g$)) = $\mathcal{C}$($g$) (case (vi)). In Fig.~\ref{Fig4}, ECA $11$ and ECA $44$ show class B dynamics (class B). On the other hand, ECAs $8$ and ECA $40$ depict evolving to homogeneous configuration, i.e. class A dynamics. As an evidence of case (v), if rule $11$ is considered as default rule ($f$) and rule $8$ is added as noise ($g$) at different probability, the resultant dynamics shows periodic dynamics. As sample, we show the space-time diagram of the stochastic CA for $\tau$ $\in$ \{$0.1, 0.5, 0.9$\} in Fig.~\ref{Fig4}. In case (vi), the stochastic CA ($44,40$) for $\tau \in$ \{$0.1,0.2,0.3$\} shows class A dynamics.

Let us now present the remaining situation where $\mathcal{C}$($f$) $\neq$ $\mathcal{C}$($g$), $\mathcal{C}$(($f,g$)) $\neq$ $\mathcal{C}$($f$) and $\mathcal{C}$(($f,g$)) $\neq$ $\mathcal{C}$($g$). Here we have following three cases. 

\begin{itemize}
	\item[(i)] $\mathcal{C}$($f$) = class B, $\mathcal{C}$($g$) = class A, and $\mathcal{C}$(($f,g$)) = class C;
	
	\item[(ii)] $\mathcal{C}$($f$) = class C, $\mathcal{C}$($g$) = class B, and $\mathcal{C}$(($f,g$)) = class A; and
	
	\item[(iii)] $\mathcal{C}$($f$) = class C, $\mathcal{C}$($g$) = class A, and $\mathcal{C}$(($f,g$)) = class B.
\end{itemize}

In our experiment, case (i) dynamics has not been observed for any temporally stochastic CA. So, let us start with case (ii) where $\mathcal{C}$($f$) = class C, $\mathcal{C}$($g$) = class B, and $\mathcal{C}$(($f,g$)) = class A. In Fig.~\ref{Fig4A}, observe the chaotic (class C) and periodic (class B) behavior of ECA $22$ and $104$ respectively. However, Fig.~\ref{Fig4A} shows that the CA ($22,104$) evolves to homogeneous configuration (class A) for $\tau \in$ \{$0.1,0.5,0.9$\}. For the third case, if rule $105$ is considered as default rule ($f$) and rule $40$ is added as noise ($g$) at different probability, the resultant dynamics exhibits class B dynamics (see Fig.~\ref{Fig4A} for $\tau \in$ \{$0.1,0.5,0.9$\}). Note that, ECA $105$ belongs to class C and ECA $40$ individually shows the behavior of class A.

\subsection{Phase transition dynamics}
According to the literature of CAs, the occurrence of phase transition is an interesting phenomenon in different non-classical CAs \cite{BoureFC12,jca/Fates14,ALONSOSANZ2005383,doi:10.}. The occurrence of phase transition can be defined as the following way: there exists a critical value of non-uniformity rate \footnote{Here, non-uniformity rate means synchrony rate of a non-classical updating scheme or mixing rate of different CA rule or any other non-uniform scheme mixing rate.} which distinguishes the behavior of the system in two different `phases'- {\em passive phase} (i.e. the system converges to a homogeneous fixed point of all $0$'s) and {\em active phase} (i.e. the system oscillates around a fixed non-zero density). For the $\alpha$-, $\beta$- and $\gamma$-synchronous updating scheme, this brutal change of behavior was noted in \cite{BoureFC12,jca/Fates09}. The phase transition of ECA with memory was identified in \cite{ALONSOSANZ2005383,doi:10.}. Recently, this abrupt change of behavior has been studied by Fat\'es \cite{Fates17} for {\em Diploid}\footnote{The rules of diploid cellular automata are obtained with random mixing of two deterministic ECA rules.} cellular automata. 

According to Fig.~\ref{Fig1} (in blue), some of the temporally stochastic CAs show this {\em phase transition} behavior. Here, we broadly distinguish these CAs into following two categories.
\begin{itemize}
	\item[(i)] The stochastic CAs which are associated with at least one class C rule; and
	\item[(ii)] The stochastic CAs which are not associated with any class C rule.
\end{itemize} 

\begin{figure*}[hbt!]
	\begin{center}
		\begin{tabular}{cc}
			(28,40),$\tau_c = 0.34$ & (30,136),$\tau_c = 0.131$\\[6pt]  
			\includegraphics[width=.45\textwidth]{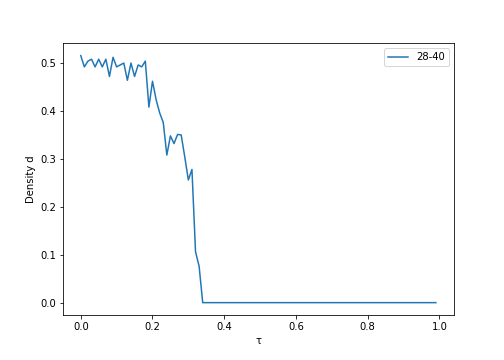} & \includegraphics[width=.45\textwidth]{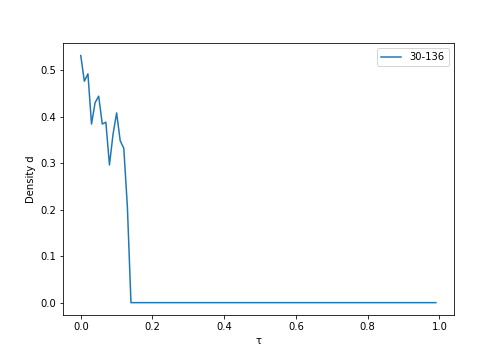} \\ 
			(78,104),$\tau_c = 0.39$ & (60,164),$\tau_c = 0.124$ \\[6pt]  \includegraphics[width=.45\textwidth]{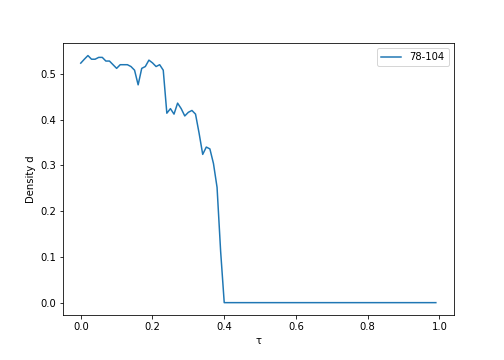}  &   \includegraphics[width=.45\textwidth]{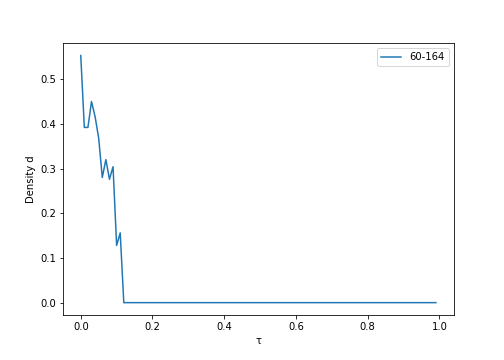}\\
			(90,104),$\tau_c = 0.18$ & (156,160),$\tau_c = 0.28$ \\[6pt]
			\includegraphics[width=.45\textwidth]{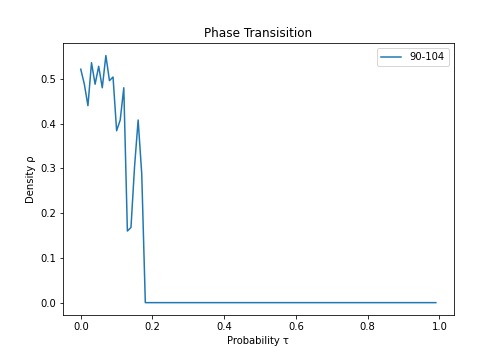}  &   \includegraphics[width=.45\textwidth]{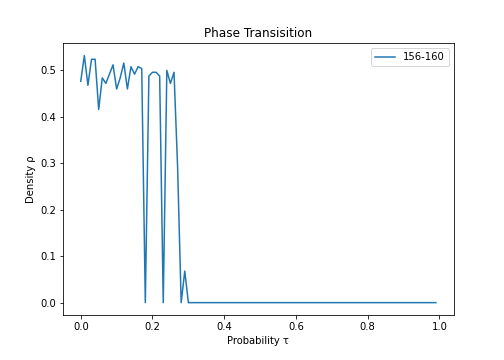}\\
		\end{tabular}
		\caption{The plot shows the profile of density parameter as a function of temporal noise rate ($\tau$) for CAs (28,40), (30,136), (78,104), (60,164), (90,104) and (156,160) .}
		\label{Fig6}
	\end{center}
\end{figure*}

For the first category, we show two such example behavior when ($f,g$) = ($30,136$) and ($60,164$). ECAs $30$ and $60$ individually show (chaotic) class C dynamics (see Fig.~\ref{Fig5}). Observe that, stochastic CAs ($30,136$) and ($60,164$) depict phase transition for small value of $\tau$ where $\tau_c = 0.131$ for CA ($30,136$) and $\tau_c = 0.124$ for CA ($60,164$). Here, for both of the cases (CAs ($30,136$) and ($60,164$)), the cellular systems show chaotic dynamics, however, due to small amount of noise the system converges to homogeneous fixed point of all $0$'s. Fig~\ref{Fig6} shows the profile of density parameter as a function of $\tau$ for the above mentioned CAs.

On the other hand, for the second category, we again show two such example behaviors when ($f,g$) =  ($28,40$) and ($78,104$). Here, none of the ECAs ($28$,~$40$,~$78$,~$104$) belongs to the class C. In Fig~\ref{Fig5}, stochastic CAs ($28,40$) and ($78,104$) show phase transition for relatively large value of $\tau$ : $\tau_c = 0.34$ for CA ($28,40$) and $\tau_c = 0.39$ for CA ($78,104$). As an evidence, see Fig~\ref{Fig6} for the profile of density parameter.

Therefore, the stochastic CAs with one class C rules show phase transition for small value of noise, on a contrary, stochastic CAs, which are not associated with any class C rules, show phase transition for relatively large value of noise. As more evidence, $\tau_c = 0.08$ for the CA ($60,168$), $\tau_c = 0.09$ for the CA ($30,160$), $\tau_c = 0.09$ for the CA ($150,200$), $\tau_c = 0.109$ for the CA ($90,168$) where one rule of each couple is from class C. On the other hand, $\tau_c = 0.49$ for the CA ($50,40$), $\tau_c = 0.48$ for the CA ($178,160$), $\tau_c = 0.44$ for the CA ($58,32$), $\tau_c = 0.33$ for the CA ($156,168$) where none of stochastic CAs are associated with class C rules. Of course, there are exceptions. However, in general, we can argue that the stochastic CAs with class C rules show less resistance against {\em effective noise}. Here, the word `effective noise' indicates the noise which can able to (at least) do the phase transition. If the noise is not associated with any impact on the stochastic CA, then the situation is different (we have seen these kind of example earlier).

\begin{figure*}[hbt!]
	\begin{center}
		\begin{adjustbox}{width=\columnwidth,center}
		\begin{tabular}{ccccc}
			ECA 45 & ECA 18 & (45,18)[0.1] & (45,18)[0.2] & (45,18)[0.3] \\[6pt]
			\includegraphics[width=31mm]{TEM_IMAGE/T45-eps-converted-to.pdf} & \includegraphics[width=31mm]{TEM_IMAGE/NR18-eps-converted-to.pdf} &  \includegraphics[width=31mm]{TEM_IMAGE/T45_18_1-eps-converted-to.pdf} & \includegraphics[width=31mm]{TEM_IMAGE/T45_18_2-eps-converted-to.pdf}  &   \includegraphics[width=31mm]{TEM_IMAGE/T45_18_3-eps-converted-to.pdf} \\
			ECA 106 & ECA 18 & (106,18)[0.1] & (106,18)[0.3] & (106,18)[0.5] \\
			\includegraphics[width=31mm]{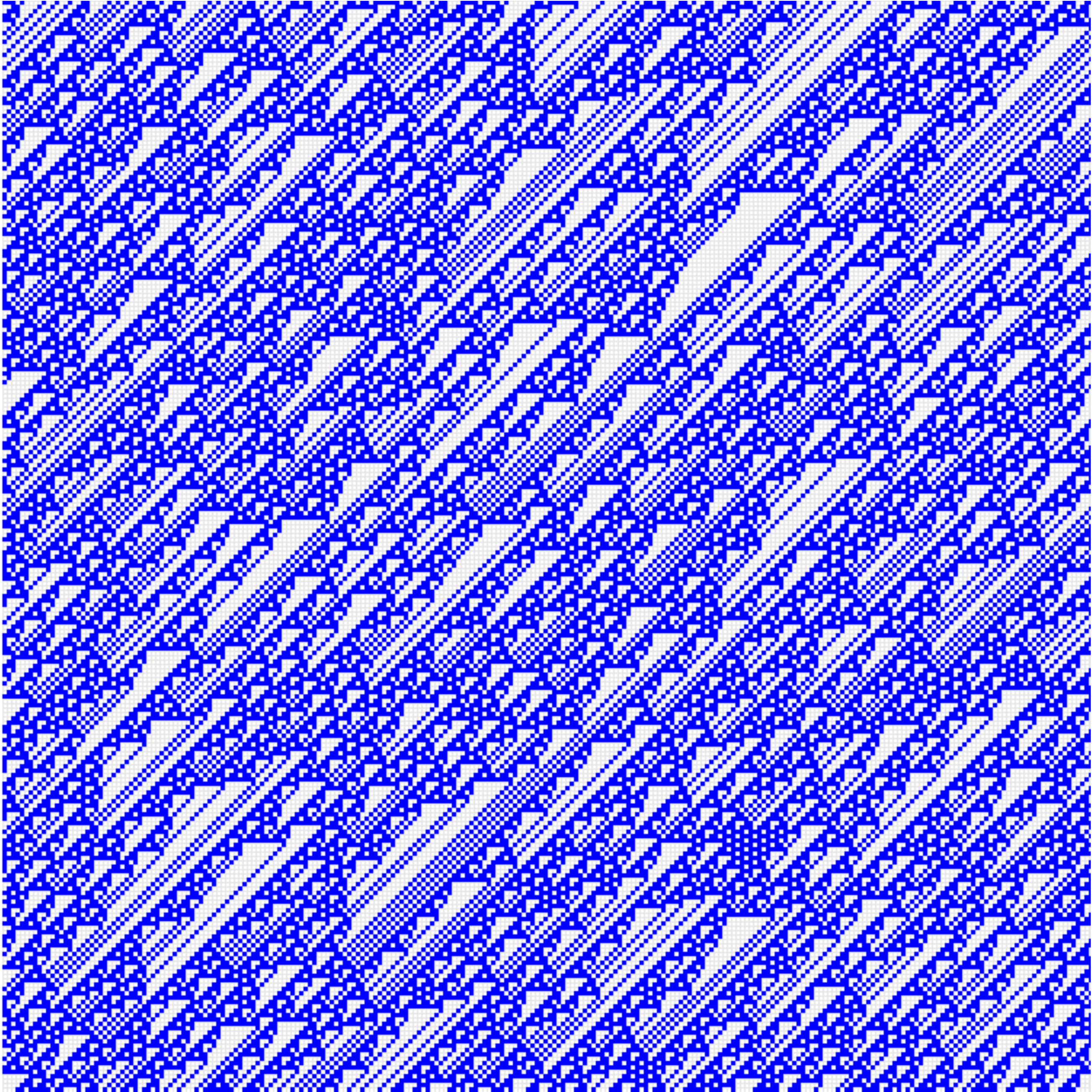} & \includegraphics[width=31mm]{TEM_IMAGE/NR18-eps-converted-to.pdf} &\includegraphics[width=31mm]{TEM_IMAGE/T106_18_1-eps-converted-to.pdf} & \includegraphics[width=31mm]{TEM_IMAGE/T106_18_3-eps-converted-to.pdf}  &   \includegraphics[width=31mm]{TEM_IMAGE/T106_18_5-eps-converted-to.pdf} \\ 
			ECA 77 & ECA 44 & (77,44)[0.1] & (77,44)[0.3] & (77,44)[0.5] \\
			\includegraphics[width=31mm]{TEM_IMAGE/T77-eps-converted-to.pdf} & \includegraphics[width=31mm]{TEM_IMAGE/T44-eps-converted-to.pdf} & \includegraphics[width=31mm]{TEM_IMAGE/T77_44_1-eps-converted-to.pdf} & \includegraphics[width=31mm]{TEM_IMAGE/T77_44_3-eps-converted-to.pdf}  &   \includegraphics[width=31mm]{TEM_IMAGE/T77_44_5-eps-converted-to.pdf} \\ 		
			
			ECA 9 & ECA 58 & (9,58)[0.1] & (9,58)[0.3] & (9,58)[0.7] \\
			\includegraphics[width=31mm]{TEM_IMAGE/T9-eps-converted-to.pdf} & \includegraphics[width=31mm]{TEM_IMAGE/T58-eps-converted-to.pdf} &\includegraphics[width=31mm]{TEM_IMAGE/T9_58_1-eps-converted-to.pdf} & \includegraphics[width=31mm]{TEM_IMAGE/T9_58_3-eps-converted-to.pdf}  &   \includegraphics[width=31mm]{TEM_IMAGE/T9_58_7-eps-converted-to.pdf} \\ 
		\end{tabular}
		\end{adjustbox}
		\caption{Class transition dynamics of stochastic CAs (45,18), (106,18), (77,44), (9,58) where $\mathcal{C}$($f$) = $\mathcal{C}$($g$).}
		\label{Fig7}
	\end{center}
\end{figure*}

\subsection{Class transition dynamics}
\label{ctd}

Next, {\em class transition} is another $\tau$ sensitive dynamics where the cellular systems change their class dynamics for a critical value of $\tau_t$. Formally, we write $\mathcal{C}$(($f,g$))[$\tau$] $\neq$ $\mathcal{C}$(($f,g$))[$\tau'$] where $\tau \in$ [0,$\tau_t$] and $\tau' \in$ ($\tau_t$,1]. Under this umbrella of dynamics, we observe two kinds of results in our experiments.
\begin{itemize}
	\item $\mathcal{C}$($f$) = $\mathcal{C}$($g$); and
	\item $\mathcal{C}$($f$) $\neq$ $\mathcal{C}$($g$).
\end{itemize}
In the first case, $f$ and $g$ are chosen from the same class. So we have following two cases:
\begin{itemize}
	\item[(i)] $\mathcal{C}$($f$) = $\mathcal{C}$($g$) = class C; and
	\item[(ii)] $\mathcal{C}$($f$) = $\mathcal{C}$($g$) = class B.
\end{itemize}

Let us start with case (i) where $\mathcal{C}$($f$) = $\mathcal{C}$($g$) = class C. Here, we show two such example behavior when ($f,g$) = ($45,18$) and ($106,18$). Fig~\ref{Fig7} depicts these interesting phenomenon where two chaotic/complex rules are involved in the stochastic CA. Observe the (chaotic) class C behavior of ECA $45$ and ECA $18$ in Fig.~\ref{Fig7}. Here, the stochastic CA ($45,18$) shows (chaotic) class C dynamics for $\tau = 0.1$. However, the cellular system ($45,18$) shows periodic class B dynamics for $\tau = 0.3$. Fig.~\ref{Fig7} shows the class transition space time diagram of CA ($45,18$) for $\tau \in$ \{$0.1,0.2,0.3$\}. We denote this dynamics in the following way -- \{chaotic $\rightsquigarrow$ periodic\}. Now, if we consider the CA ($18,45$) which is exchange symmetric to the CA ($45,18$), it shows the \{periodic $\rightsquigarrow$ chaotic\} dynamics. Hence, we can write that \{chaotic $\leftrightsquigarrow$  periodic\} dynamics is observed for this cellular system. Similarly, the stochastic CA ($106,18$) exhibits class B dynamics for $\tau = 0.1$, however the CA shows chaotic class C dynamics for $\tau = 0.5$. Note that, here also, ECA $106$ and $18$ individually show complex/chaotic class C dynamics. Interesting to note that, here, couple of two chaotic rules show periodic dynamics for a range of noise rate.

\begin{figure*}[hbt!]
	\begin{center}
		\begin{adjustbox}{width=\columnwidth,center}
		\begin{tabular}{ccccc}
			ECA 122 & ECA 14 & (122,14)[0.1] & (122,14)[0.4] & (122,14)[0.7] \\[6pt]
			\includegraphics[width=31mm]{TEM_IMAGE/T122-eps-converted-to.pdf} & \includegraphics[width=31mm]{TEM_IMAGE/T14-eps-converted-to.pdf} & \includegraphics[width=31mm]{TEM_IMAGE/T122_14_1-eps-converted-to.pdf} & \includegraphics[width=31mm]{TEM_IMAGE/T122_14_4-eps-converted-to.pdf}  &   \includegraphics[width=31mm]{TEM_IMAGE/T122_14_7-eps-converted-to.pdf} \\
			ECA45 & ECA 136 & (45,136)[0.1] & (45,136)[0.3] & (45,136)[0.5] \\
			\includegraphics[width=31mm]{TEM_IMAGE/T45-eps-converted-to.pdf} & \includegraphics[width=31mm]{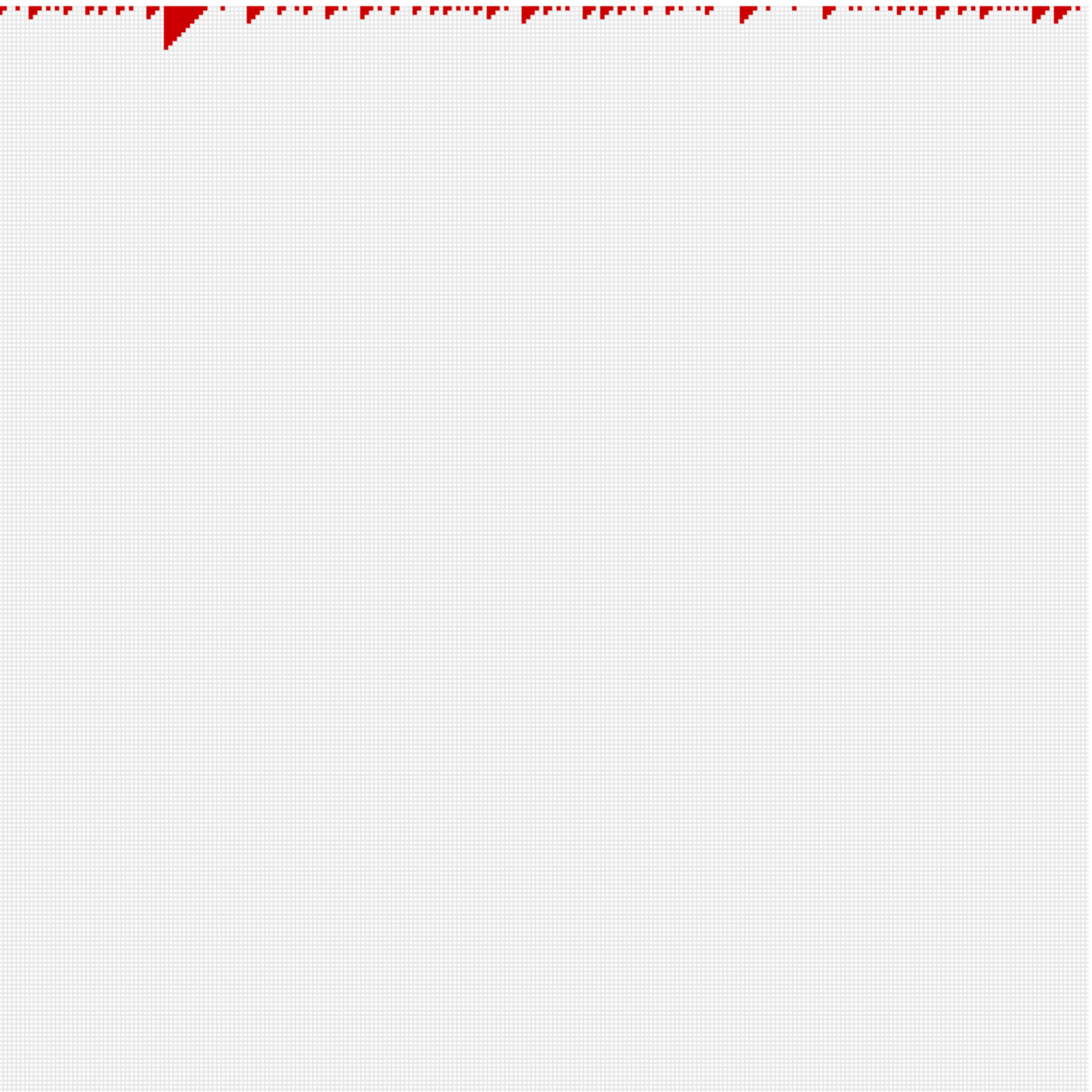} & \includegraphics[width=31mm]{TEM_IMAGE/T45_136_1-eps-converted-to.pdf} & \includegraphics[width=31mm]{TEM_IMAGE/T45_136_3-eps-converted-to.pdf}  &   \includegraphics[width=31mm]{TEM_IMAGE/T45_136_5-eps-converted-to.pdf} \\ 
			
			ECA131 & ECA 136 & (131,136)[0.1] & (131,136)[0.5] & (131,136)[0.9] \\
			\includegraphics[width=31mm]{TEM_IMAGE/T131-eps-converted-to.pdf} & \includegraphics[width=31mm]{TEM_IMAGE/T136-eps-converted-to.pdf} & \includegraphics[width=31mm]{TEM_IMAGE/T131_136_1-eps-converted-to.pdf} & \includegraphics[width=31mm]{TEM_IMAGE/T131_136_5-eps-converted-to.pdf}  &   \includegraphics[width=31mm]{TEM_IMAGE/T131_136_9-eps-converted-to.pdf} \\ 
		\end{tabular}
		\end{adjustbox}
		\caption{Class transition dynamics of stochastic CAs (122,14), (45,136), (131,136) where $\mathcal{C}$($f$) $\neq$ $\mathcal{C}$($g$).}
		\label{Fig8}
	\end{center}
\end{figure*}

Let us now discuss case (ii) : $\mathcal{C}$($f$) = $\mathcal{C}$($g$) = class B. Here, we also show two such example behavior when ($f,g$) = ($77,44$) and ($9,58$). Observe the periodic class B behavior of ECA $77$ and ECA $44$ in Fig.~\ref{Fig7}. Here, the stochastic CA ($77,18$) shows class C dynamics for $\tau = 0.5$, however the cellular system shows periodic dynamics for $\tau = 0.1$. Fig.~\ref{Fig7} shows the class transition space-time diagram of CA ($77,44$) for $\tau \in$ \{$0.1,0.3,0.5$\}. Similarly, CA ($9,58$) shows class C dynamics for high value of $\tau$ (see ($9,58$)[$0.7$], in Fig.~\ref{Fig8}). On the other hand, ($9,58$)[$0.1$] depicts periodic dynamics. Here also, both the ECAs $9$ and $58$ individually depicts periodic dynamics (class B).

Next we discuss the class transformation dynamics for $\mathcal{C}$($f$) $\neq$ $\mathcal{C}$($g$). So we have following three situations:
\begin{itemize}
	\item[(i)] $\mathcal{C}$($f$) = class C and $\mathcal{C}$($g$) = class B;
	\item[(ii)] $\mathcal{C}$($f$) = class C and $\mathcal{C}$($g$) = class A; and
	\item[(iii)] $\mathcal{C}$($f$) = class B and $\mathcal{C}$($g$) = class A;
\end{itemize}

Let us start with case (i) where $\mathcal{C}$($f$) = class C and $\mathcal{C}$($g$) = class B. For this case, we show the dynamics of stochastic CA ($122,14$). Observe that, ECA 122 shows chaotic class C dynamics, on the other hand, ECA $14$ shows class B dynamics. Here, the CA ($122,14$) exhibits chaotic dynamics for $\tau = 0.1$, see Fig~\ref{Fig8}. However, if we progressively increase the noise rate $\tau$, then the stochastic CA shows periodic dynamics (see Fig.~\ref{Fig8} for $\tau \in$ \{$0.4,0.7$\}).

In the next case (case (ii)), $\mathcal{C}$($f$) = class C and $\mathcal{C}$($g$) = class A. In Fig.~\ref{Fig8}, ECA $45$ individually depicts chaotic class C dynamics, and  ECA $136$ belongs to class A, i.e. evolving to homogeneous configuration. Here, the stochastic CA ($45,136$) shows periodic dynamics for high value of $\tau$ ($\tau = 0.5$). However, if we decrease the $\tau$ value, the cellular system shows chaotic dynamics. Space-time diagrams of ($45,136$) for $\tau \in$ \{$0.1,0.3,0.5$\} are shown in Fig.~\ref{Fig8}. 

For case (iii), where $\mathcal{C}$($f$) = class B and $\mathcal{C}$($g$) = class A, we consider the example of stochastic CA ($131,136$). According to the classification, ECA $131$ belongs to class B. Whereas, ECA $136$ evolves to homogeneous configuration dynamics (class A). Now, if rule $131$ is considered as default rule ($f$) and rule $136$ is added as noise ($g$) at low probability, the resultant dynamics shows chaotic behavior. However, if we progressively increase the noise rate, the cellular system shows class A dynamics. As an evidence, Fig.~\ref{Fig8} depicts class transition dynamics of CA ($131,136$) for $\tau \in$ \{$0.1,0.5,0.9$\}. Here also, couple of two periodic and homogeneous rules show chaotic dynamics for a range of noise rate.

\section{Summary}
\label{S5}

This chapter constitutes a first step in the exploration of the space of the temporally stochastic CAs. We have identified that some of the stochastic CAs are affected by the temporal noise, but they are not sensitive to temporal noise rate (i.e. if we progressively vary the temporal noise rate, the cellular system's dynamical behavior remains unchanged). These CAs have shown diverse set of results. We should mention some richest evidence, where
\begin{itemize}
	\item[] $\mathcal{C}$(periodic,periodic) = chaotic;
	\item[] $\mathcal{C}$(periodic,periodic) = homogeneous;
	\item[] $\mathcal{C}$(chaotic,periodic) = homogeneous; and
	\item[] $\mathcal{C}$(chaotic,homogeneous) = periodic.
\end{itemize}
That is, these stochastic CAs ($f,g$) are totally destructed from the individual class of $f$ and $g$. However, when $f$ and $g$ are from different class, there are interesting situations where one of their class dominates in the stochastic CA. Interestingly, there are many cases where chaotic behavior is dominated by temporal noise of type periodic and homogeneous behavior. 

On the other hand, temporal noise sensitive stochastic CAs have shown phenomenon, like phase transition and class transition. It is interesting to note that, in general, stochastic CAs with (at least one) chaotic rules have shown less resistance during phase transition (i.e. critical value of noise rate is low). However, the stochastic CAs without any chaotic rule have exhibited more resistance during phase transition (i.e. critical value of noise rate is high). This is also one of the exiting observation of this study. 

The rarest phenomenon shown by these CAs is class transition. We should mention the following worthy examples,
\begin{itemize}
	\item[] $\mathcal{C}$(chaotic,chaotic) = \{chaotic $\leftrightsquigarrow$ periodic\};
	\item[] $\mathcal{C}$(periodic,periodic) = \{periodic $\leftrightsquigarrow$ chaotic\}; and
	\item[] $\mathcal{C}$(periodic,homogeneous) = \{chaotic $\leftrightsquigarrow$ homogeneous\}.
\end{itemize}

In terms of numbers, we have performed a large number of experiments on $3828$ temporally stochastic CAs. Table~\ref{Table:classes_dynamics} depicts different situations (as mentioned in Section~\ref{S4}) with number of temporally stochastic CAs that show the mentioned behavior.

\begin{table}[ht]
	\centering
	\begin{adjustbox}{width=\columnwidth,center}
		\begin{tabular}{ccc|c} \hline 
			Cases & & & No. of CAs \\ \hline \hline
			$\mathcal{C}$($f$) =  $\mathcal{C}$($g$) & & & \\
			& $\mathcal{C}$(($f,g$)) =  $\mathcal{C}$($f$) & &  \\
			& & $\mathcal{C}$($f$) =  $\mathcal{C}$($g$) = $\mathcal{C}$(($f,g$)) = class A & 28\\
			& & $\mathcal{C}$($f$) =  $\mathcal{C}$($g$) = $\mathcal{C}$(($f,g$)) = class B & 1436\\
			& & $\mathcal{C}$($f$) =  $\mathcal{C}$($g$) = $\mathcal{C}$(($f,g$)) = class C & 149\\
			& $\mathcal{C}$(($f,g$)) $\neq$  $\mathcal{C}$($f$) & &  \\
			& & $\mathcal{C}$($f$) =  $\mathcal{C}$($g$) = class A, $\mathcal{C}$(($f,g$)) = class B & 0\\
			& & $\mathcal{C}$($f$) =  $\mathcal{C}$($g$) = class A, $\mathcal{C}$(($f,g$)) = class C & 0\\
			& & $\mathcal{C}$($f$) =  $\mathcal{C}$($g$) = class B, $\mathcal{C}$(($f,g$)) = class A & 113\\
			& & $\mathcal{C}$($f$) =  $\mathcal{C}$($g$) = class B, $\mathcal{C}$(($f,g$)) = class C & 18\\
			& & $\mathcal{C}$($f$) =  $\mathcal{C}$($g$) = class C, $\mathcal{C}$(($f,g$)) = class A & 0\\
			& & $\mathcal{C}$($f$) =  $\mathcal{C}$($g$) = class C, $\mathcal{C}$(($f,g$)) = class B & 0\\
			$\mathcal{C}$($f$) $\neq$  $\mathcal{C}$($g$) & & & \\ \hline
			& $\mathcal{C}$(($f,g$)) =  $\mathcal{C}$($f$) or $\mathcal{C}$(($f,g$)) =  $\mathcal{C}$($g$) & &  \\
			& & $\mathcal{C}$($f$) = class B,  $\mathcal{C}$($g$) = class A, $\mathcal{C}$(($f,g$)) = $\mathcal{C}$($f$) & 167\\
			& & $\mathcal{C}$($f$) = class B,  $\mathcal{C}$($g$) = class A, $\mathcal{C}$(($f,g$)) = $\mathcal{C}$($g$) & 297\\
			& & $\mathcal{C}$($f$) = class C,  $\mathcal{C}$($g$) = class A, $\mathcal{C}$(($f,g$)) = $\mathcal{C}$($f$) & 0\\
			& & $\mathcal{C}$($f$) = class C,  $\mathcal{C}$($g$) = class A, $\mathcal{C}$(($f,g$)) = $\mathcal{C}$($g$) & 98\\
			& & $\mathcal{C}$($f$) = class C,  $\mathcal{C}$($g$) = class B, $\mathcal{C}$(($f,g$)) = $\mathcal{C}$($f$) & 89\\
			& & $\mathcal{C}$($f$) = class C,  $\mathcal{C}$($g$) = class B, $\mathcal{C}$(($f,g$)) = $\mathcal{C}$($g$) & 67\\
			& $\mathcal{C}$(($f,g$)) $\neq$  $\mathcal{C}$($f$) and $\mathcal{C}$(($f,g$)) $\neq$  $\mathcal{C}$($g$) & &  \\
			& & $\mathcal{C}$($f$) = class B,  $\mathcal{C}$($g$) = class A, $\mathcal{C}$(($f,g$)) = class C & 0\\
			& & $\mathcal{C}$($f$) = class C,  $\mathcal{C}$($g$) = class A, $\mathcal{C}$(($f,g$)) = class B & 20\\
			& & $\mathcal{C}$($f$) = class C,  $\mathcal{C}$($g$) = class B, $\mathcal{C}$(($f,g$)) = class A & 57\\ \hline
			Class & & & \\
			Transition(CT) & $\mathcal{C}$($f$) =  $\mathcal{C}$($g$)  & & \\
			& & $\mathcal{C}$($f$) =  $\mathcal{C}$($g$) = class A, $\mathcal{C}$(($f,g$)) = CT & 0\\
			& & $\mathcal{C}$($f$) =  $\mathcal{C}$($g$) = class B, $\mathcal{C}$(($f,g$)) = CT & 297\\
			& & $\mathcal{C}$($f$) =  $\mathcal{C}$($g$) = class C, $\mathcal{C}$(($f,g$)) = CT & 4\\
		 	& $\mathcal{C}$($f$) $\neq$  $\mathcal{C}$($g$)  & & \\
			& & $\mathcal{C}$($f$) = class B,  $\mathcal{C}$($g$) = class A, $\mathcal{C}$(($f,g$)) = CT & 4\\
			& & $\mathcal{C}$($f$) = class C,  $\mathcal{C}$($g$) = class A, $\mathcal{C}$(($f,g$)) = CT & 12\\
			& & $\mathcal{C}$($f$) = class C,  $\mathcal{C}$($g$) = class B, $\mathcal{C}$(($f,g$)) = CT & 873\\ \hline
			Phase & & & 99\\
			Transition & & & \\ \hline
	\end{tabular}
	\end{adjustbox}
	\caption{Different situations according to Section~\ref{S4} and number of temporally stochastic CAs, out of $3828$, corresponding to every situation.}
	\label{Table:classes_dynamics}
\end{table}

To sum up, this chapter has indicated a rich possibilities of this temporally stochastic CAs. However, in this chapter, we have explored these CAs primarily through experiment. Therefore, the proper theoretical understanding regarding the temporally stochastic CA, is still an open question to further explore.

\chapter{Pattern Classification with Temporally Stochastic Cellular Automata}
\label{chap4}

\section{Introduction}
In a classical cellular automaton (CA), a rule ($f$) is applied to each and every cell of the lattice to evolve the CA from one configuration to its next configuration \cite{BhattacharjeeNR16}. In this work, we deviate from the classical CA, and introduce another rule, say $g$, in the cellular structure which is applied to all the cells in a time step with probability $\tau$. The rule $g$ may be considered as noise of the cellular structure and $\tau$ as the noise rate. The rule $f$ can be called as default rule, which is applied to all cells in a time step with probability $1-\tau$. We name  these cellular automata (CAs) as \emph{Temporally Stochastic Cellular Automata} (TSCAs).

In this work, we take only ECAs rules as our default rule and noise to study this class of automata. We further consider the CAs as finite, which use periodic boundary condition. We first study the dynamical behavior of these TSCAs through an extensive experiment, and classify them as Class A, Class B and Class C by observing their behavior following the Wolfram’s~\cite{Wolfram94} and Li \& Packard’s~\cite{LangtonII,Genaro13} classification. Then we identify a set of TSCAs that converge to fixed points from any initial configuration.

The CAs that converge to fixed point from any seed have been widely employed for the design of pattern classifier~\cite{DasMNS09,Sethi2016,RAGHAVAN1993145}. In this work we also utilize the convergent TSCAs to develop two-class pattern classifier. However, there are some convergent TSCAs which are having a single fixed point (attractor). These CAs cannot act as a two-class pattern classifier. Similarly, a convergent TSCA having enormous number of fixed points (attractor) is not a good classifier. Using these criteria, we identify a set of convergent TSCAs that can act as a good classifier. To evaluate the performance of the proposed classifier, we choose standard data sets which are taken from \url{http://www.ics.uci.edu/~mlearn/MLRepository.html}. It is observed that the proposed classifier performs nicely in \emph{training phase} as well as \emph{testing phase}. Finally we compare the performance of the proposed classifier with that of well-known classifiers. It is found that the proposed classifier is very much competitive with the best-performing classifiers.

\subsection{Dynamical behavior}\label{dynamics}
Stephen Wolfram \cite{Wolfram94} introduced following general classification of the ECAs (defined over $\mathbb{Z}$) depending on their dynamical behavior:
\begin{itemize}
	\item[] Class I: evolving to a homogeneous configuration;
	\item[] Class II: evolving periodically;
	\item[] Class III: evolving chaotically;
	\item[] Class IV: class of complex rules.
\end{itemize}
Later, Li and Packard have identified some periodic rules (Class II) as locally chaotic \cite{LangtonII,Genaro13}. For TSCAs, we target to identify their dynamical behavior and to classify the TSCAs as above. We take $f$ and $g$ from $88$ minimum representative ECA rules, and then consider all possible combinations of these $88$ ECAs rules. Here, total $\frac{88\times87}{2}=3828$ couple of $(f, g)$ are sufficient because, the rest are exchange symmetry of $f$ and $g$. 

We arrange a large number of experiments to understand dynamical behavior of TSCAs. We map the dynamics of TSCAs into following three classes $-$
\begin{itemize}
	\item[] Class A: which is similar to Wolfram's Class I.
	\item[] Class B: which is similar to Wolfram's Class II except locally chaotic rules.
	\item[] Class C: similar to Wolfram's Class III and Class IV, including three locally chaotic rules.
\end{itemize}

\begin{figure}[!ht]
	\subfloat[ECA 164]{
		\begin{minipage}[c][1\width]{
				0.3\textwidth}
			\label{fig1:ex1:a}
			\centering
			\includegraphics[width=1\textwidth]{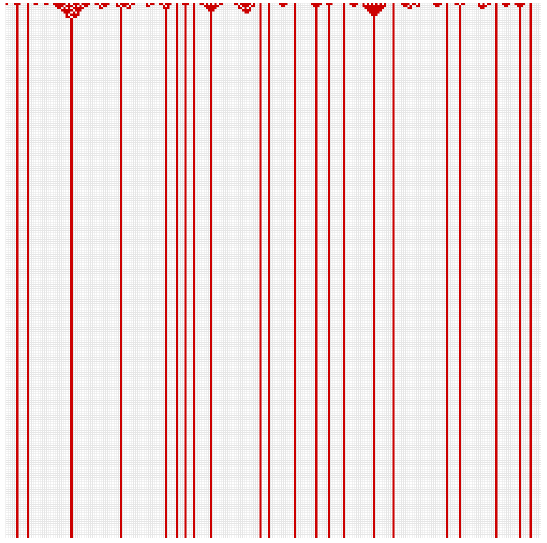}
	\end{minipage}}
	\hfill 	
	\subfloat[ECA 131]{
		\begin{minipage}[c][1\width]{
				0.3\textwidth}
			\label{fig1:ex1:b}
			\centering
			\includegraphics[width=1\textwidth]{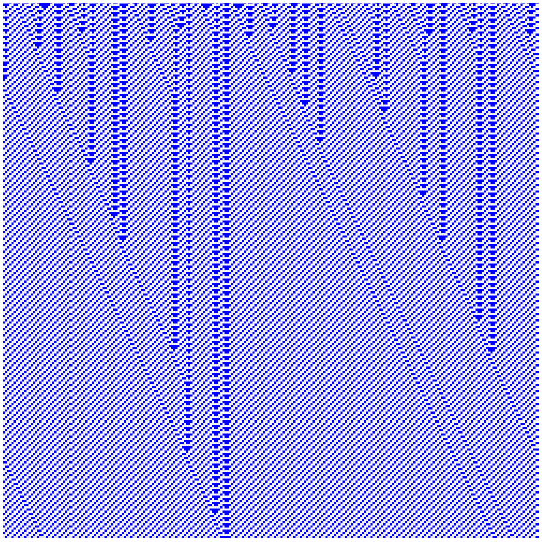}
	\end{minipage}}
	\hfill	
	\subfloat[$(164, 131);\: \tau=0.1$ ]{
		\begin{minipage}[c][1\width]{
				0.33\textwidth}
			\label{fig1:ex1:c}
			\centering
			\includegraphics[width=1.05\textwidth]{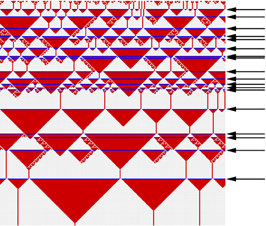}
	\end{minipage}}
	\caption{Stochastic CAs ($f$, $g$) dynamics, when $C((f, g)) \neq C(f)$ and $C(f) = C(g)$.
	}
	\label{fig1:ex1}
\end{figure}

Now, there are two possibilities for a couple ($f$, $g$) $(i)$ $f$ and $g$ belong to the same class; $(ii)$ $f$ and $g$ are from different class. We denote the class of $f$ and $g$ as $C(f)$ and $C(g)$ respectively and class of ($f$, $g$) is denoted by $C((f,\; $g$))$. We find amazing experimental outcomes:

\begin{itemize}
	\item If $C(f) = C(g)$ in a TSCA, one option is $C((f, g)) = C(f)$, which has been seen in a significant number of TSCAs. The TSCA ($f$, $g$), where $C(f) = C(g)$ = Class $A$, approaches to a homogeneous configuration, much like ECA $f$ and ECA $g$. On the other hand, $C((f, g)) \neq C(f)$ could be conceivable. That is, the noise has a significant impact on these TSCAs($f$, $g$) (as an evidence, see Fig~\ref{fig1:ex1}).
	
	\item  Next case where $C(f)\neq C(g)$, shows the dynamics where one of the rule's class dominates, i.e. $C((f, g)) = C(f)$ or $C((f, g)) = C(g)$. Under this case, the TSCA ($f$, $g$) with $C(f) =$ Class C and $C(g) =$ Class B shows the dynamics as $C((f, g)) = C(g)$ (see Example~\ref{ex3}). Fig~\ref{fig3} shows an evidence of this situation. Here, ECA $22$ and ECA $7$ respectively belong to class III and II and, the CA $(22, 7)$ shows periodic behavior (like Wolfram’s Class II), see Fig~\ref{fig3:e}, where the class of ECA	$7$ dominates.
	
	\item  A TSCA ($f$, $g$) with $C(f) \neq C(g)$, on the other hand, depicts dynamics in which none of the rule's classes dominates, i.e. $C((f, g)) \neq C(f)$ and $C((f, g)) \neq C(g)$. The TSCA ($f$, $g$) with $C(f) =$ Class C and $C(g) =$ Class A displays $C((f, g)) =$ Class B, with none of the rule's classes dominating (see Fig~\ref{fig4}).  In Fig~\ref{fig4:c}, ECA $105$ belongs to Class III and ECA $40$ belongs to Class I. However, the TSCA $(105,40)[0.2]$ shows periodic behavior (like Wolfram’s Class II).
\end{itemize}
\begin{table}[!htbp] 
	\centering
	\scriptsize
	
	\begin{adjustbox}{width=\columnwidth,center}
		\begin{tabular}{c|c|c|c} \hline 
			Class & Conditions & Number of TSCAs& $\tau$ \\ \hline
			&&&\\
			&  $\mathcal{C}$($f$) =  $\mathcal{C}$($g$) = $\mathcal{C}$(($f,g$)) = Class A & 28&\\
			&  $\mathcal{C}$($f$) =  $\mathcal{C}$($g$) = Class B, $\mathcal{C}$(($f,g$)) = Class A & 113&\\
			A&  $\mathcal{C}$($f$) =  $\mathcal{C}$($g$) = Class C, $\mathcal{C}$(($f,g$)) = Class A & 0&\\
			
			& $\mathcal{C}$($f$) = Class B,  $\mathcal{C}$($g$) = Class A, $\mathcal{C}$(($f,g$)) = $\mathcal{C}$($g$) & 297&\\
			
			& $\mathcal{C}$($f$) = Class C,  $\mathcal{C}$($g$) = Class A, $\mathcal{C}$(($f,g$)) = $\mathcal{C}$($g$) & 98&\\
			& $\mathcal{C}$($f$) = Class C,  $\mathcal{C}$($g$) = Class B, $\mathcal{C}$(($f,g$)) = Class A & 57&\\
			&&&\\
			&&&\\
			&&&\\
			&  $\mathcal{C}$($f$) =  $\mathcal{C}$($g$) = $\mathcal{C}$(($f,g$)) = Class B & 1436& \\	
			& $\mathcal{C}$($f$) =  $\mathcal{C}$($g$) = Class A, $\mathcal{C}$(($f,g$)) = Class B & 0&\\
			B  & $\mathcal{C}$($f$) =  $\mathcal{C}$($g$) = Class C, $\mathcal{C}$(($f,g$)) = Class B & 0&\\
			& $\mathcal{C}$($f$) = Class B,  $\mathcal{C}$($g$) = Class A, $\mathcal{C}$(($f,g$)) = $\mathcal{C}$($f$) & 167& Insensitive\\
			& $\mathcal{C}$($f$) = Class C,  $\mathcal{C}$($g$) = Class B, $\mathcal{C}$(($f,g$)) = $\mathcal{C}$($g$) & 67&\\
			& $\mathcal{C}$($f$) = Class C,  $\mathcal{C}$($g$) = Class A, $\mathcal{C}$(($f,g$)) = Class B & 20&\\
			&&&\\
			&&&\\
			&&&\\
			
			&  $\mathcal{C}$($f$) =  $\mathcal{C}$($g$) = $\mathcal{C}$(($f,g$)) = Class C & 149&\\
			& $\mathcal{C}$($f$) =  $\mathcal{C}$($g$) = Class A, $\mathcal{C}$(($f,g$)) = Class C & 0&\\
			C & $\mathcal{C}$($f$) =  $\mathcal{C}$($g$) = Class B, $\mathcal{C}$(($f,g$)) = Class C & 18&\\
			& $\mathcal{C}$($f$) = Class C,  $\mathcal{C}$($g$) = Class A, $\mathcal{C}$(($f,g$)) = $\mathcal{C}$($f$) & 0&\\
			& $\mathcal{C}$($f$) = Class C,  $\mathcal{C}$($g$) = Class B, $\mathcal{C}$(($f,g$)) = $\mathcal{C}$($f$) & 89&\\
			& $\mathcal{C}$($f$) = Class B,  $\mathcal{C}$($g$) = Class A, $\mathcal{C}$(($f,g$)) = Class C & 0&\\
			&&&\\
			
			\hline
			&&&\\
			$*$&$-$&1289& Sensitive\\
			&&&\\
			\hline
			
		\end{tabular}
	\end{adjustbox}	
	\caption{Distribution of TSCAs under different classes ( $*$ indicates $-$ No specific class can be obtained)}
	\label{Table4}
\end{table}
\begin{example}\label{ex1}
	Let us consider a TSCA($164, 131$)[$0.1$], where $f=164$ and $g=131$ are applied with probability $0.1$ and $0.9$ at each step of the evolution. Fig~\ref{fig1:ex1:a} and Fig~\ref{fig1:ex1:b} show the space-time diagrams of ECA $164$ and ECA $131$ with random initial configuration. Whereas, Fig~\ref{fig1:ex1:c} shows the space-time diagram of TSCA($164, 131$)[$0.1$]. Rule $131$ is applied at the time step marked by $\leftarrow$ (arrow) in Fig~\ref{fig1:ex1:c}. It is interesting to note here that the dynamical behavior of TSCA can widely vary from that of the CAs with default rule and noise.
\end{example}
\begin{example}\label{ex3}
	Fig~\ref{fig3} shows the dynamics where one of the rule's class dominates. Rule $22$ belongs to Class III and rule $7$ belongs to Class II, and the CA shows periodic behavior (like Wolfram's Class II)(see Fig~\ref{fig3:e}), where the class of rule $7$ dominates . 
	\begin{figure}[!ht]
		\subfloat[ECA 22]{
			\begin{minipage}[c][1.2\width]{
					0.23\textwidth}
				\label{fig3:a}
				\centering
				\includegraphics[width=1\textwidth]{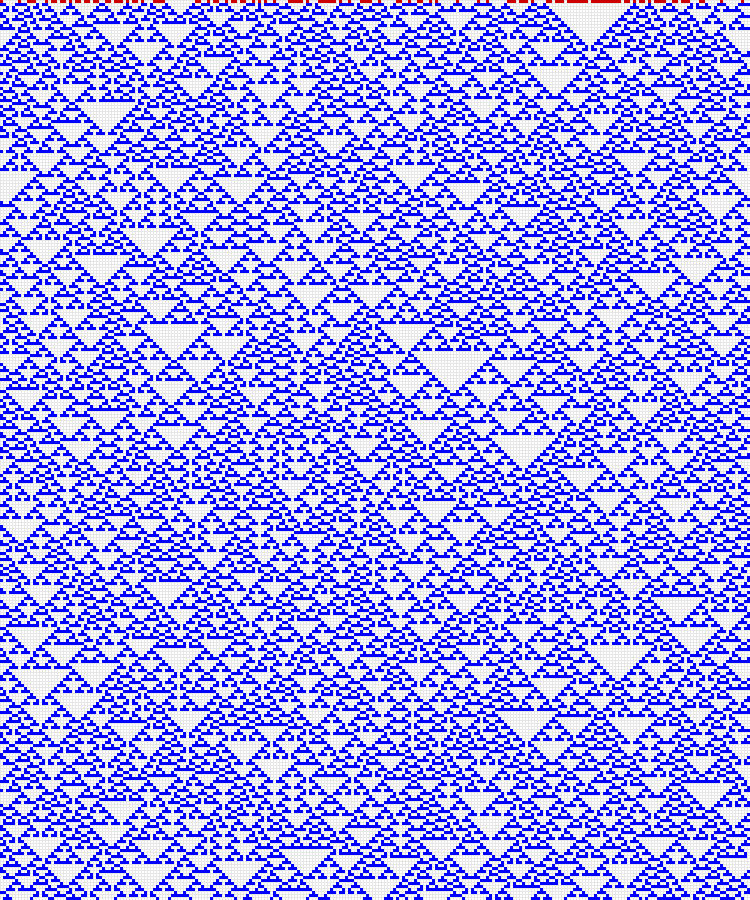}
		\end{minipage}}
		\hfill 	
		\subfloat[ECA 7]{
			\begin{minipage}[c][1.2\width]{
					0.23\textwidth}
				\label{fig3:b}
				\centering
				\includegraphics[width=1\textwidth]{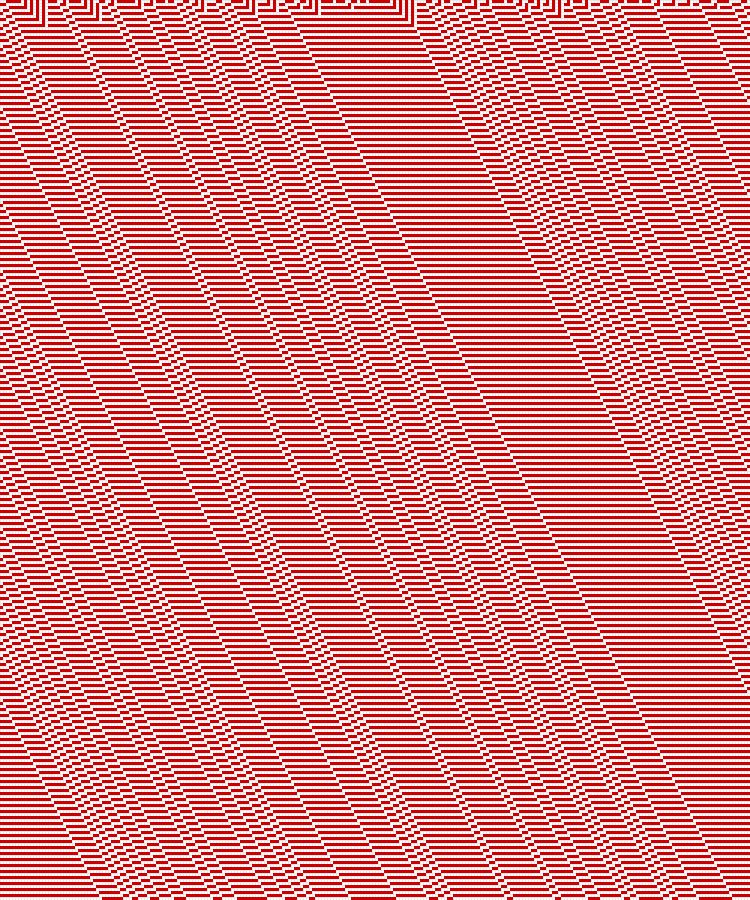}
		\end{minipage}}
		\hfill	
		\subfloat[$\tau=0.1$ ]{
			\begin{minipage}[c][1.2\width]{
					0.23\textwidth}
				\label{fig3:c}
				\centering
				\includegraphics[width=1\textwidth]{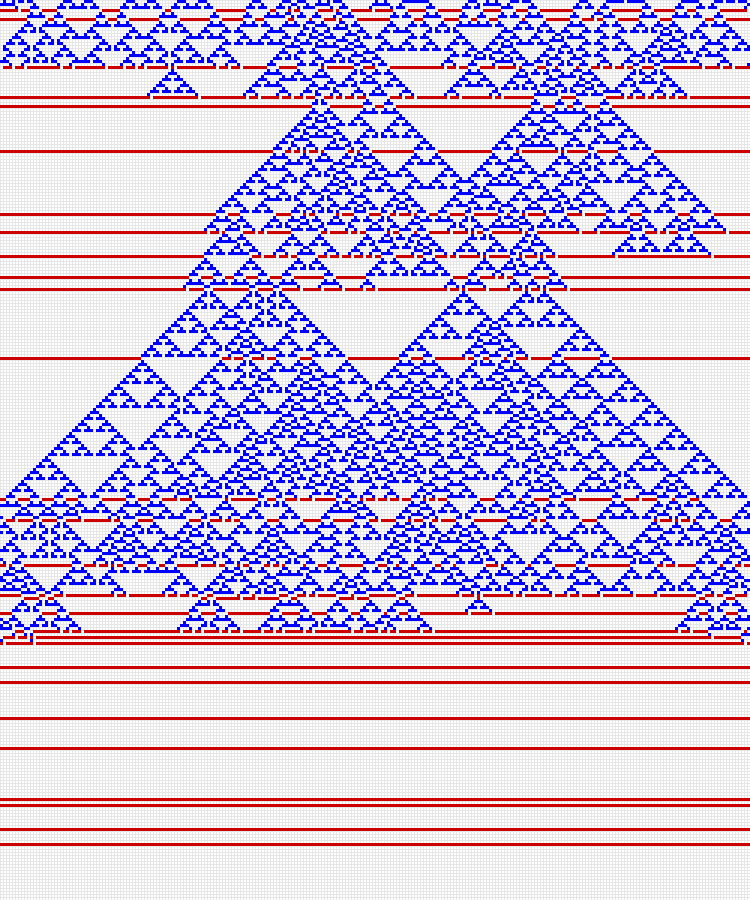}
		\end{minipage}}
		\hfill		
		\subfloat[$ \tau=0.8$ ]{
			\begin{minipage}[c][1.2\width]{
					0.23\textwidth}
				\label{fig3:e}
				\centering
				\includegraphics[width=1\textwidth]{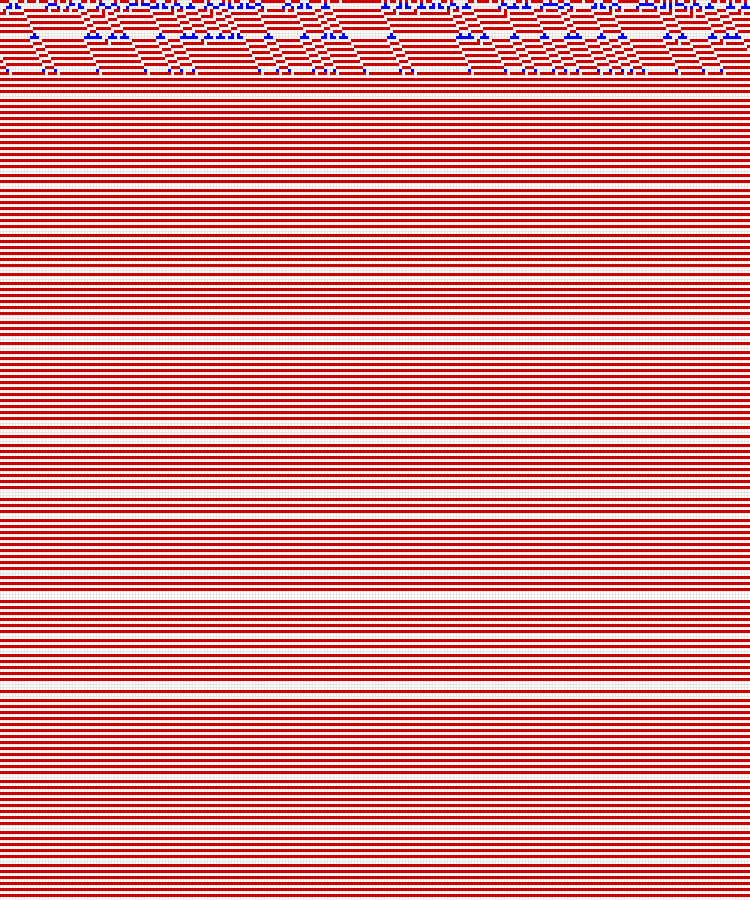}
		\end{minipage}}
		\caption{Dynamics of CA $(22, 7)[0.1]$ and CA $(22, 7)[0.8]$ $(C((f, g)) = C(g)$ and $C(f) \neq C(g))$. 
		}
		\label{fig3}
	\end{figure}
	
\end{example}
\begin{example}\label{ex5}
	Fig~\ref{fig4} shows the dynamics where none of the rule's class dominates. Rule $105$ belongs to Class III and rule $40$ belongs to Class I. The TSCA($105$,$40$)[$0.6$] shows periodic behavior (like Wolfram's Class II) (see Fig~\ref{fig4:e}).
\end{example}
\begin{figure}[!ht]
	\subfloat[ECA 105]{
		\begin{minipage}[c][1.2\width]{
				0.23\textwidth}
			\label{fig4:a}
			\centering
			\includegraphics[width=1\textwidth]{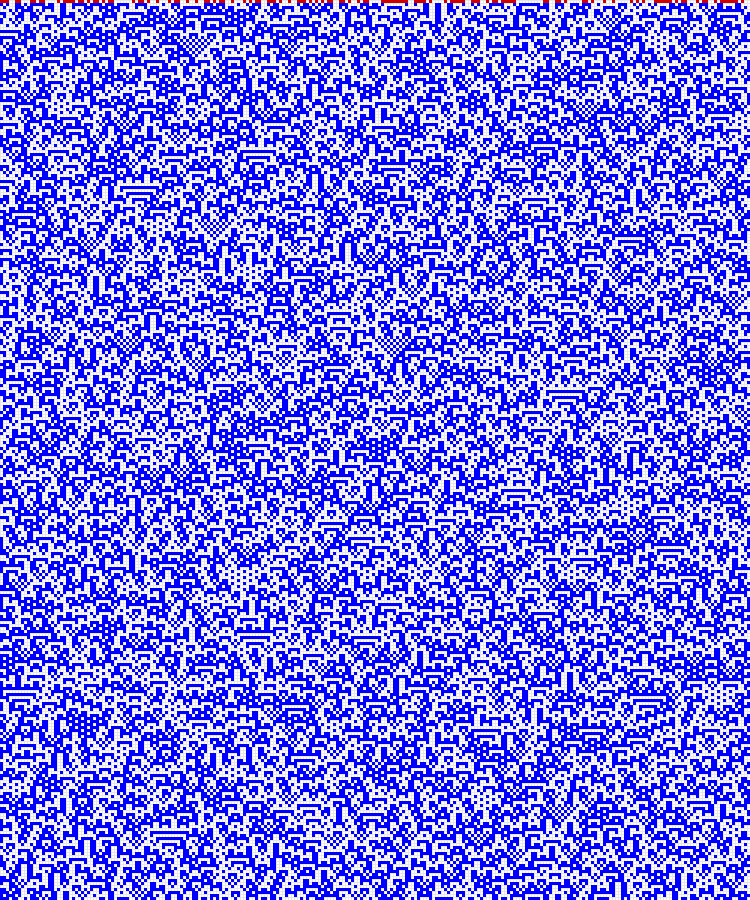}
	\end{minipage}}
	\hfill 	
	\subfloat[ECA 40]{
		\begin{minipage}[c][1.2\width]{
				0.23\textwidth}
			\label{fig4:b}
			\centering
			\includegraphics[width=1\textwidth]{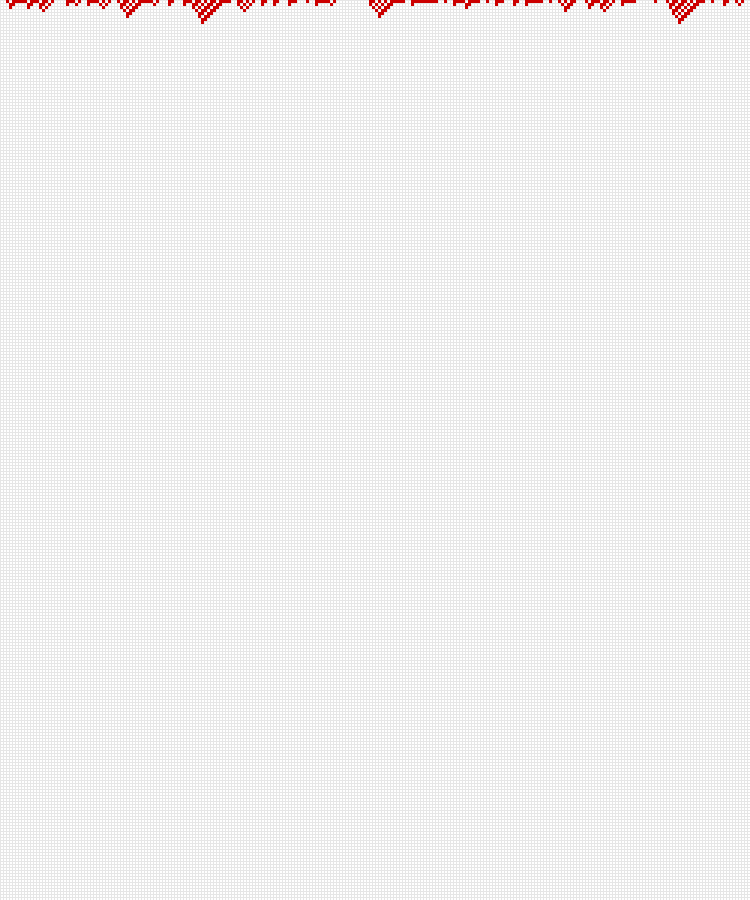}
	\end{minipage}}
	\hfill	
	\subfloat[$\tau=0.2$ ]{
		\begin{minipage}[c][1.2\width]{
				0.23\textwidth}
			\label{fig4:c}
			\centering
			\includegraphics[width=1\textwidth]{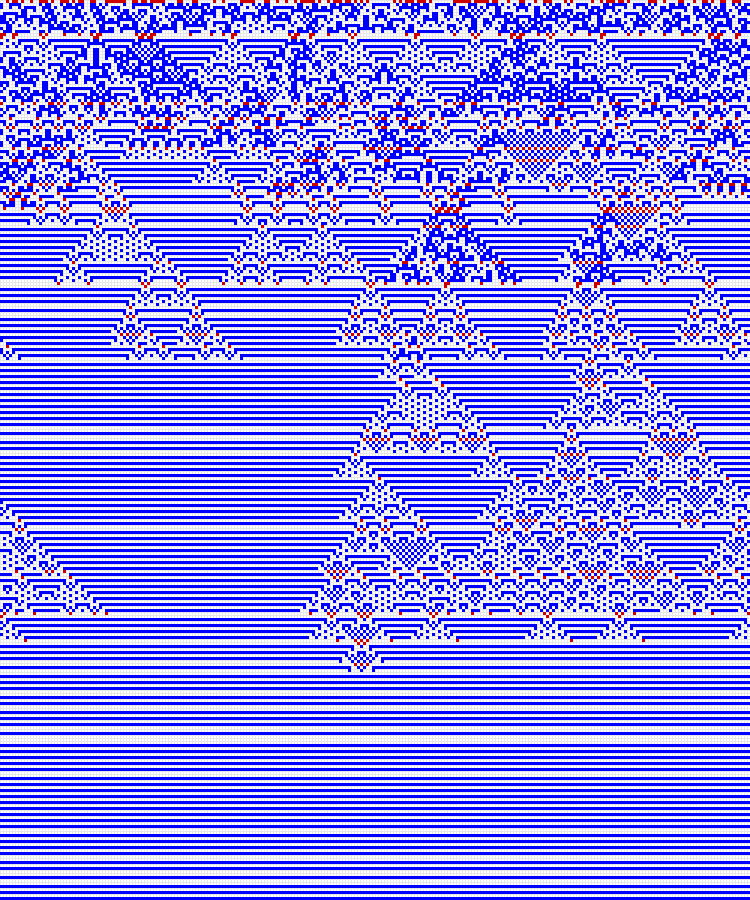}
	\end{minipage}}
	\hfill		
	\subfloat[$ \tau=0.6$ ]{
		\begin{minipage}[c][1.2\width]{
				0.23\textwidth}
			\label{fig4:e}
			\centering
			\includegraphics[width=1\textwidth]{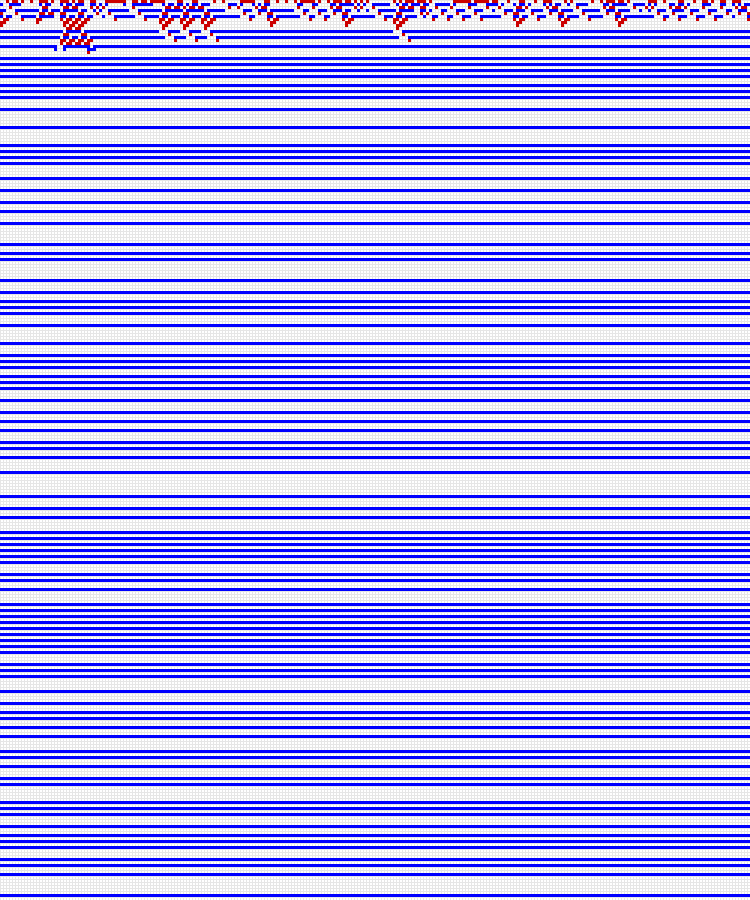}
	\end{minipage}}
	\caption{Dynamics of CA $(105, 40)[0.2]$ and CA $(105, 40)[0.6]$ $(C((f, g)) \neq C(f)$, $(C((f, g)) \neq C(g)$ and $C(f) \neq C(g))$..
	}
	\label{fig4}
\end{figure}

We have found that out of $3828$ TSCAs, $593$, $1690$ and $256$ TSCAs belong to Class A, Class B and Class C respectively. Table~\ref{table4} shows the summary of the outcome. It is also found that for a number of TSCAs, the noise rate ($\tau$) has not been playing a significant role. That is, for these cases, if we progressively vary the temporal noise rate, the cellular system's dynamics remains unchanged. Therefore, these TSCAs are $\tau$-insensitive. 

However, there are $1289$ cases, out of $3828$ TSCAs, where the noise rate $(\tau)$ has been playing a significant role, i.e. $\tau-$sensitive. These TSCAs show phase
transition~\footnote{Some TSCAs show a discontinuity after a critical value of temporal noise rate which brutal change of behavior is well known as phase transition.} and class transition~\footnote{For a set of TSCAs, the class dynamics of the system changes after a critical value of $\tau$.} dynamics. Fig~\ref{TSCA7} phase transition behavior andFig~\ref{Fig8} class transition behavior. However, the goal of the current work is to explore the pattern classification capability of TSCAs for which $\tau-$sensitive CAs are not appropriate. Therefore, we next deal with the CAs, dynamical behavior of which
are independent of $\tau$, i.e. $\tau-$insensitive. For our next purpose, we identify the $\tau-$insensitive TSCAs which converge to fixed point from any initial configuration.
\subsection{Convergence}\label{convergence}
During evolution, a CA approaches to a set of configurations which form an attractor. If the set is a singleton, we call the attractor as fixed point. Whenever all the attractors are fixed points, we call the CAs as convergent.
\begin{definition}\label{def0}
	A TSCA($f$, $g$)[$\tau$] is called as convergent TSCA if the CA converges to a fixed point from any initial configuration and for any $\tau$ and $n$, where $n$ is the number of cells of the TSCA.
\end{definition}
In other words, for a given seed, a TSCA converges to a fixed point, if both $f$ and $g$ may converge to a fixed point separately for the same seed. Following a large number of experiments, we identify the set of TSCAs that converge to fixed points. Here, the experimental study shows that $424$ couple of CAs converge to fixed point starting from any initial configuration and for any $\tau$ and $n$, see Table~\ref{pattern1}.

However, there are some cases where the convergence feature of various TSCAs changes depending on the size n and $\tau$ . As an evidence, Fig~\ref{Fig5} depicts the dynamics of TSCA $(30, 136)$ which converges to all$-0$ configuration after a critical value of $\tau$ (here, $\tau = 0.13$). However, $(30, 136)$ oscillates around a fixed non-zero density for $\tau = 0.08$ and $\tau = 0.11$, in Fig~\ref{Fig5}. Earlier, we have mentioned that this type of brutal change of behavior is well known as second-order phase transition~\cite{Sethi2016}. Similarly, the couple $(131, 136)$ converges to all$-1$ configuration for $\tau$ value $0.5$ and $0.9$, see Fig~\ref{Fig8}. On the other hand, it shows chaotic dynamics for $\tau = 0.1$. Here, the class dynamics of the system changes after a critical value of $\tau$ , i.e. class transition. However, for the current study, we exclude these TSCAs.

Although $424$ convergent TSCAs are identified to design pattern classifier, the general demand is multiple attractor TSCA.



\begin{definition}\label{def3}
	If a convergent TSCA is having more than one fixed point, the TSCA is called as Multiple Attractor TSCA.
\end{definition}
\begin{example}
	The TSCA ($12,4$)[$\tau$] with $4$ cells (and any $\tau$) is a convergent TSCA and is having seven fixed points $-$ $0000$, $0001$, $0010$, $0100$, $1000$, $0101$ and $1010$. Hence, it is a multiple attractor TSCA.
\end{example}

Previously, we have mentioned that to designing a pattern classifier, we exclude the TSCAs which
\begin{itemize}
	\item are associated with single fixed point (see Table~\ref{pattern1}, in black); and
	\item are associated with large number of attractors, specifically, couples with ECA
	$204$ (see Table~\ref{pattern1}, in blue).
\end{itemize}
Therefore, to design a pattern classifier, we need to pick a few couples from the Table~\ref{pattern1} after excluding the above situations. Finally, we find a few of $114$
couples (see Table~\ref{pattern1}, in bold) which are the candidate of the proposed pattern
classifier.

\begin{table}[htbp]
	\begin{center}
		
		\begin{adjustbox}{width=\columnwidth,center}
			\begin{tabular}{|cccccccccccc|} \hline\hline
				(2, 0) & (4, 0) & (6, 0) & (6, 4) & (8, 0) & (8, 2) & (8, 4) & (8, 6) & (10, 0)& (10, 8) & (12, 0) & 
				(12, 8) \\ (14, 0) & (14, 4) & (18, 0) & (18, 8) & (22, 0) & (22, 4) & (22, 8) & (24, 0) & (24, 8) & (26, 0) & (26, 8) &
				(28, 0) \\ (28, 8) & (30, 0) & (30, 4) & (32, 0) & (32, 2) & (32, 8) & (32, 10) & (32, 18) &(32, 24) & (32, 26) & (34, 0) &
				(34, 8) \\ (36, 0) & (36, 6) & (36, 8) & (36, 32 & (38, 0) & (38, 4) & (38, 8) & (38, 32) & (40, 0) & (40, 2) & (40, 8) & (40, 10) \\ (40, 24) & (40, 36) & (40, 38) & (42, 0) & (42, 8) & (44, 0) &
				(44, 8) & (44, 32) & (46, 0) & (46, 4) & (46, 32) & (50, 0) \\ (50, 8) & (54, 0) & (54, 4) & (54, 8) & (54, 32) & (54, 40) & (56, 0) & (56, 8) & (58, 0) & (60, 0) & (60, 4) & (60, 8) \\ (60, 32) & (72, 0) & (72, 2) & (72, 4) & 
				(72, 6) & (72, 8) & (72, 12) & (72, 24) & (72, 28) & (72, 32) & (72, 34) & (72, 36) \\ (72, 38) & (74, 0) & (74, 8) & (74, 32) & (74, 72) & (76, 0) & (76, 8) & (78, 0) & (78, 18) & (90, 0) & (90, 8) & (90, 32) \\ (94, 0) & (104, 0) & (104, 2) & (104, 8) & (104, 24) & (104, 36) & (104, 38) & (104, 44) & (104, 74) & (106, 0) & (106, 8) & (106, 72) \\ (108, 0) & (108, 8) & (108, 32) & (108, 40) & (110, 0) & (110, 4) & (110, 32) & (122, 0) & (122, 8) & (122, 36) & (126, 0) & (126, 4) \\(126, 32) & (128, 0) & (128, 2) & (128, 4) & (128, 6) & (128, 8) & (128, 10) & (128, 12) & (128, 14) & (128, 18) & (128, 22) & (128, 24) \\ (128, 26) & (128, 28) & (128, 32) & (128, 34) & (128, 36) & (128, 38) & (128, 40) & (128, 42) & (128, 44) & (128, 46) & (128, 50) & (128, 54) \\ (128, 56) & (128, 58) & (128, 60) & (128, 72) & (128, 74) & (128, 76) & (128, 78) & (128, 94) & (128, 104) & (128, 106) & (128, 108)& (128, 110) \\ (130, 0) & (130, 8) & (130, 32) & (130, 40) & (130, 72) & (130, 104) & (131, 128) & (132, 0) & (132, 6) & (132, 8) & (132, 14) & (132, 38) \\ (132, 46) & (132, 72) & (134, 0) & (134, 4) & (134, 8) & (134, 36) & (134, 72) & (136, 0) & (136, 2) & (136, 4) & (136, 6)& (136, 8) \\ (136, 10) & (136, 12) & (136, 18) & (136, 22) & (136, 24) & (136, 26) & (136, 28) & (136, 32) & (136, 34) & (136, 36) & (136, 38) & (136, 40) \\ (136, 42) & (136, 44) & (136, 50) & (136, 54) & (136, 56) & (136, 72) & (136, 74) & (136, 76) & (136, 90) & (136, 104) & (136, 106) & (136, 108) \\ (138, 0) & (138, 8) & (138, 32)&\ (138, 40) &	(140, 0) & (140, 8) & (140, 72) & (142, 0) & (142, 4) & (146, 0) & (146, 8) & (146, 32) \\ (146, 78) & (150, 0) & (150, 4) & (150, 8) & (152, 0) & (152, 8) & (152, 32) & (152, 40) & (152, 72) & (152, 104) & (154, 0) & (154, 8) \\ (154, 32) & (154, 40) & (156, 0) & (156, 8) & (156, 72) & (156, 126) & (156, 131) & (160, 0) & (160, 2) & (160, 8) & (160, 10) & (160, 18) \\ (160, 24) & (160, 26) & (160, 36) & (160, 38) & (160, 44) & (160, 54) & (160, 72) & (160, 74) & (160, 108) & (160, 131) & (162, 0) & (162, 8) \\	(162, 72) & (164, 0) & (164, 6) & (164, 8) & (164, 32) & (164, 40) & (164, 72) & (164, 104) & (168, 0) & (168, 2) & (168, 8) & (168, 10) \\ (168, 24) & (168, 36) & (168, 38) & (168, 54) & (170, 0) & (170, 8) & (172, 0) & (172, 8) & (172, 32) & (178, 0) & (178, 8) & (178, 131) \\ (184, 0) & (184, 8) & (200, 0) & (200, 2) & (200, 4) & (200, 6) & (200, 8) & (200, 12) & (200, 24) &	(200, 28) & (200, 32) & (200, 34) \\ (200, 36) & (200, 38) & (204, 0) & (204, 8) & (232, 0) & (232, 2) & (232, 8) & (232, 24) &(232, 36) & (232, 38) & (232, 44) &
				
				\textbf{(12, 4)} \\ \textbf{(36, 4)} & \textbf{(36, 12)} & \textbf{(44, 4)} &\textbf{ (44, 12)} & \textbf{(44, 36)} & \textbf{(76, 4) }&\textbf{ (76, 12)} &\textbf{(76, 72)} & \textbf{(78, 76) }&\textbf{ (94, 78)} &\textbf{ (104, 72)} &\textbf{ (108, 4)} \\ \textbf{(108, 12)} & \textbf{(108, 36)} & \textbf{(108, 44)}  &\textbf{(108, 72)} & \textbf{(130, 128)} &\textbf{(132, 4)}&\textbf{ (132, 12)} & \textbf{(132, 36)} & \textbf{(132, 44)} & \textbf{(132, 76) }&\textbf{ (132, 108)}  &\textbf{(132, 128)} \\\textbf{ (134, 128)} &\textbf{ (134, 132)} & \textbf{(136, 128)} & \textbf{(136, 130)} &\textbf{ (136, 132)} & \textbf{(136, 134)} &\textbf{ (138, 128)}  &\textbf{(138, 136)} & \textbf{(140, 4)} & \textbf{(140, 12)} & \textbf{(140, 36)} & \textbf{(140, 44)} \\ \textbf{(140, 76)} &\textbf{ (140, 108)} & \textbf{(140, 128)}&\textbf{(140, 132) }&\textbf{ (140, 136)} &\textbf{ (142, 128)} & \textbf{(142, 132)} &\textbf{ (146, 128)} &\textbf{ (146, 136)} & \textbf{(150, 128)} & \textbf{(150, 136) } &\textbf{(152, 128)} \\\textbf{ (152, 136)} & \textbf{(154, 128)}&\textbf{ (154, 136)} &\textbf{ (156, 128)} & \textbf{(156, 136)} &\textbf{ (160, 128) }& \textbf{(160, 130)} &\textbf{(160, 136)} & \textbf{(160, 138)} & \textbf{(160, 146)} & \textbf{(160, 152)} & \textbf{(160, 154)} \\ \textbf{(162, 128)} & \textbf{(162, 136)} & \textbf{(164, 4)}  &\textbf{(164, 12)} &\textbf{(164, 36)} & \textbf{(164, 44)} & \textbf{(164, 108) }&\textbf{ (164, 128)} &\textbf{ (164, 132) }& \textbf{(164, 134)}&\textbf{ (164, 136) } &\textbf{(164, 140)}\\\textbf{ (164, 160)} & \textbf{(168, 128)} & \textbf{(168, 130)} &\textbf{ (168, 136)} &\textbf{ (168, 138)} & \textbf{(168, 152)} &\textbf{ (170, 128)} & \textbf{ (170, 136)} & \textbf{(172, 4)} &\textbf{ (172, 12)} &\textbf{ (172, 36)} &\textbf{ (172, 128)} \\\textbf{ (172, 132) }& \textbf{(172, 136)} &\textbf{ (172, 140)} & \textbf{(178, 128)} & \textbf{(178, 136)} &\textbf{ (184, 128) }& \textbf{(184, 136)}&\textbf{ (200, 72)} & \textbf{(200, 76)} &\textbf{ (200, 128)} &\textbf{ (200, 130)} &\textbf{ (200, 132)} \\ \textbf{(200, 134)} & \textbf{(200, 136)} & \textbf{(200, 140)} & \textbf{(200, 152)} & \textbf{(200, 156) }& \textbf{(200, 160)} &\textbf{ (200, 162)}  &\textbf{(200, 164) }& \textbf{(232, 72)} & \textbf{(232, 108)} & \textbf{(232, 128)} &\textbf{ (232, 130)} \\ \textbf{(232, 136)} & \textbf{(232, 154)} & \textbf{(232, 164)}  &\textbf{(232, 172)} & \textbf{(232, 200)}&

				\textcolor{blue}{(204, 4)} & \textcolor{blue}{(204, 12)} & \textcolor{blue}{(204, 36)} & \textcolor{blue}{(204, 72)} & \textcolor{blue}{(204, 76)} & \textcolor{blue}{(204, 78)} & \textcolor{blue}{(204, 128)} \\ \textcolor{blue}{(204, 132)}& \textcolor{blue}{(204, 136)} & \textcolor{blue}{(204, 140)} & \textcolor{blue}{(204, 200)}&&&&&&&&\\\hline		\hline	
			\end{tabular}
		\end{adjustbox}
	\caption{ Couples of TSCAs that converge to Fixed Points}
	\label{pattern1}
	\end{center}
\end{table}  

\section{Multiple Attractor TSCA as Pattern classifier}\label{pattern_classifier}
A $n$-cell TSCA with $k$ fixed points can act as $k$-class classifier. Each class contains a set of configurations that converge to a single fixed point. Hence the fixed point can act as representative of the set. Now to design a two-class classifier, a set of fixed points, out of $k$ fixed points, needs to represent one class whereas the rest fixed points shall represent the other class. From implementation  point of view, all the fixed points along with their class interaction are to be stored in memory. Whenever class of an input pattern ($P$) is to be found out, the TSCA runs with the pattern as seed. Based on the fixed point, where the TSCA settles down, the class of $P$ is declared.

\begin{figure}
	\centering
	\includegraphics[width=0.6\textwidth]{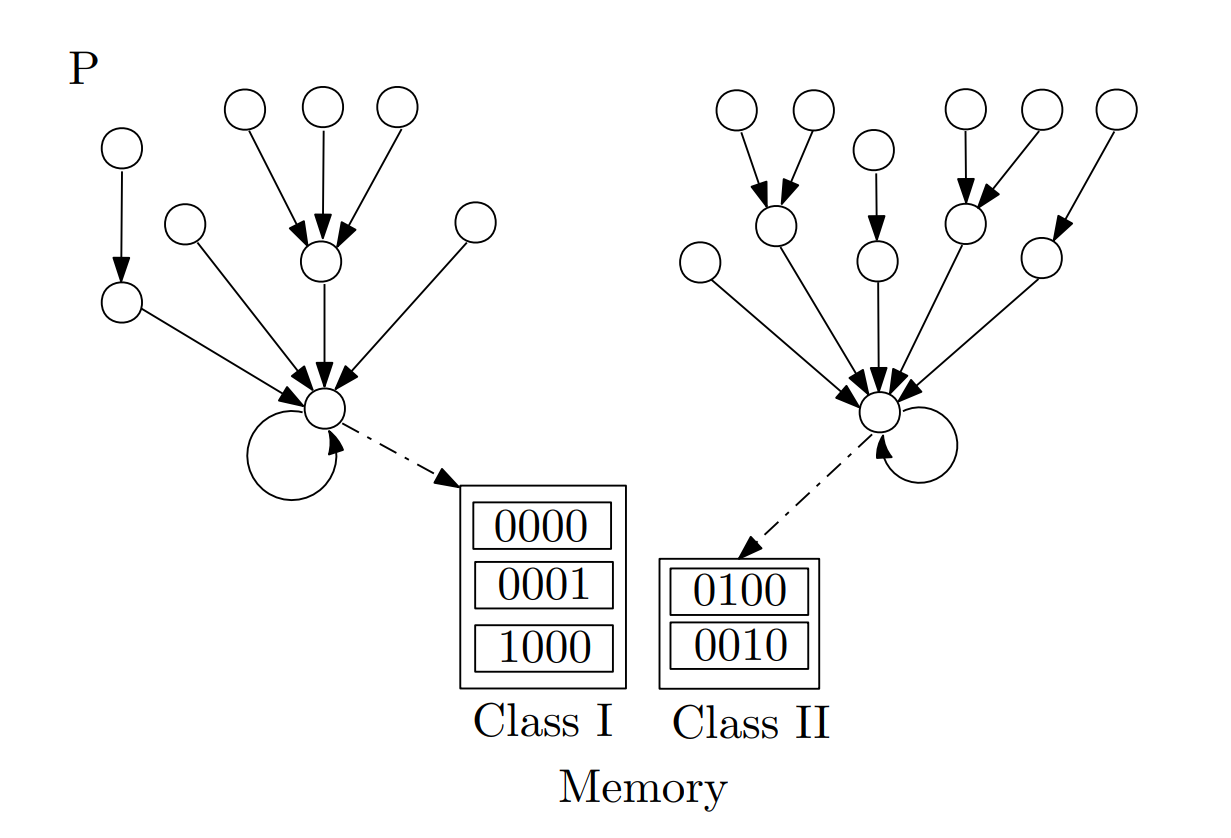}
	\caption{The TSCA ($108$, $44$) was classified using multiple fixed points.}
	\label{fig:pattern1}
\end{figure}

As an example, the $4$-cell convergent TSCA($108,44$)[$0.1$] which has five attractors may be used as a two-class pattern classifier. Assume that the fixed points $0000$, $0001$ and $1000$ represent Class I, and the rest fixed points $0010$ and $0100$ represents Class II. Whenever a pattern, say $1101$ is given, the TSCA is run with $1101$ as seed. After some time, the CA reaches to a fixed point, say $1000$. Since $1000$ represents Class I, class of $1101$ is declared as I. Hence this multiple attractor TSCA can act as two-class pattern classifier, see Fig~\ref{fig:pattern1}.

For good classifiers, the patterns are to be distributed evenly throughout the attractor basins. In real-world datasets, however, the attractor basins may mix up the patterns of two classes. As a result, we evaluate the classifier's performance in terms of classification accuracy, which is defined as the ratio of properly classified patterns to total patterns. The formula for calculating efficiency is as follows:
\begin{align}
	Efficiency &= \frac{\text{No. of properly classified patterns}}{\text{Total no. of patterns }}\times100\%
\end{align}

However, a multiple-attractor TSCA may not be a good pattern classifier. To measure the performance and effectiveness of a TSCA, we pass it through the \emph{training phase} and \emph{testing phase}. 
\subsection{Training phase}\label{training}
As shown in Table~\ref{pattern1}, there are $114$ multiple fixed attractor TSCAs. These TSCAs can act as potential candidates for pattern classification. However, in order to find the most effective classifier, we train all the candidates using patterns of two disjoint datasets, say $P_1$ and $P_2$. A TSCA from the set of candidates is loaded first with patterns of $P_1$ and $P_2$, and constantly updated until the TSCA reaches to a fixed point. We keep track of all the attractors and the number of patterns converge on them. If more patterns from pattern set $P_1$ converge to the attractor than the patterns from pattern set $P_2$, the attractor is declared to be of Class I and stored in \emph{attractorset-$1$}; otherwise, the attractor is of Class II and stored in \emph{attractorset-$2$}. At the end, we have two sets of attractors. The following formula is used to determine the efficiency of a TSCA:

\begin{align}
	Efficiency &= \frac{\sum_{i=1}^{m} max(n_1^i, n_2^i)}{\big|P_1\big|+\big|P_2\big|}
\end{align}
Here, $n^i_1$ and $n^i_2$ are the maximum number of patterns converged to the $i^{th}$ fixed point attractor of a TSCA from dataset $P_1$ and $P_2$, respectively. $\big|P_1\big|$ and $\big|P_2\big|$ are the number of patterns of two datasets used for pattern classification. The \emph{training phase} produces a TSCA with highest efficiency, \emph{attractorset-$1$}, and \emph{attractorset-$2$} as output. Note that, the output of this phase is used as input of the \emph{testing phase}(see Section~\ref{testing}).

As an example, let us consider Monk$-1$ dataset ($11$-bit data) for classification. Let us take the TSCA $(76, 72)[0.2]$ as two-class pattern classifier (similar to Fig~\ref{fig:pattern1}), with two pattern set $P_1$ and $P_2$ loaded to the TSCA as Class I and Class II respectively. There are total $169$ patterns including $P_1$ and $P_2$, out of which $2$ patterns of $P_2$ and $4$ pattern of $P_1$ are wrongly identified as in Class I and Class II respectively. Hence, $163$ patterns are properly classified, which gives training efficiency as $96.4497\%$. 

To get the best candidate TSCA, we train all the $114$ TSCAs of Table~\ref{pattern1} by Monk$-1$ dataset. The result of the training is noted in Table~\ref{table4}. We find that the  TSCA($76,72$)[$0.1$] with training efficiency $97.54\%$, as the best performing TSCA. This TSCA acts as our desired classifier.

\subsection{Testing phase}\label{testing}
In this phase, a new collection of patterns are used to find the efficacy of the designed classifier. The attractor sets \emph{attractorset-1} and \emph{attractorset-2} with a TSCA (output of training phase) and the pattern sets ($P_1$ and $P_2$) are taken as input in this phase. The TSCA is loaded with the patterns of $P_1$ and $P_2$ and updated till all the patterns converge to any fixed point attractor. The number of patterns successfully detected by the classifier is used to measure the TSCA's efficiency. For example, if an attractor is present in \emph{attractorset-1} then count only the number of patterns from dataset $P_1$ converge to the attractor as correctly identified patterns, similarly, if an attractor is present in \emph{attractorset-2}, count only the number of patterns from dataset $P_2$ that converge to the attractor as correctly identified patterns. TSCAs with their training and testing efficiencies for different datasets are reported in Table~\ref{table5}.

\begin{table}[htbp]
	\begin{center}
		
		\begin{adjustbox}{width=\columnwidth,center}
			\begin{tabular}{|ccccccccc|} \hline 
				TSCAs & \makecell{ Efficiency\\ (in \%)} & \makecell{ Number of\\ Attractor} &TSCAs & \makecell{ Efficiency\\ (in \%)} & \makecell{ Number of\\ Attractor}&TSCAs & \makecell{ Efficiency\\ (in \%)} & \makecell{ Number of\\ Attractor}\\\hline
				(12,4)[0.1]  & 86.066 & 199 & (36,4)[0.9]  & 68.033 & 67 & (36,12)[0.9]  & 84.426 & 67\\(44,4)[0.1]  & 83.607 & 67 & (44,36)[0.1]  & 83.607 & 67 & (44,12)[0.1]  & 84.426 & 67\\(76,4)[0.1]  & 96.721 & 199 & (76,12)[0.1]  & 98.361 & 199 & (76,72)[0.1]  & 97.541 & 67\\(78,76)[0.1]  & 65.574 & 23 & (94,78)[0.1]  & 73.77 & 23 & (104,72)[0.9]  & 64.754 & 34\\(108,4)[0.1]  & 81.967 & 67 & (108,36)[0.1]  & 85.246 & 67 & (108,72)[0.4]  & 80.328 & 34\\(108,12)[0.1]  & 88.525 & 67 & (108,44)[0.1]  & 89.344 & 67 & (130,128)[0.1]  & 50.0 & 2\\(132,128)[0.1]  & 72.951 & 2 & (132,36)[0.1]  & 75.41 & 67 & (132,4)[0.1]  & 74.59 & 199\\(132,12)[0.9]  & 87.705 & 199 & (132,44)[0.9]  & 86.885 & 67 & (132,76)[0.9]  & 88.525 & 199\\(132,108)[0.9]  & 91.803 & 67 & (134,128)[0.1]  & 50.0 & 2 & (134,132)[0.1]  & 50.0 & 2\\(136,128)[0.1]  & 50.0 & 2 & (136,130)[0.1]  & 50.0 & 2 & (136,132)[0.1]  & 50.0 & 2\\(136,134)[0.1]  & 50.0 & 2 & (138,128)[0.1]  & 50.0 & 2 & (138,136)[0.1]  & 50.0 & 2\\(140,128)[0.1]  & 88.525 & 2 & (140,36)[0.1]  & 86.885 & 67 & (140,4)[0.1]  & 85.246 & 199\\(140,132)[0.1]  & 88.525 & 200 & (140,136)[0.1]  & 86.066 & 2 & (140,108)[0.9]  & 94.262 & 67\\(140,44)[0.2]  & 88.525 & 67 & (140,12)[0.1]  & 84.426 & 199 & (140,76)[0.9]  & 94.262 & 199\\(142,128)[0.1]  & 50.0 & 2 & (142,132)[0.1]  & 50.0 & 2 & (146,128)[0.1]  & 50.0 & 2\\(146,136)[0.1]  & 50.0 & 2 & (150,128)[0.1]  & 50.0 & 2 & (150,136)[0.1]  & 50.0 & 2\\(152,128)[0.1]  & 50.0 & 2 & (152,136)[0.1]  & 50.0 & 2 & (154,128)[0.1]  & 50.0 & 2\\(154,136)[0.1]  & 50.0 & 2 & (156,128)[0.1]  & 50.0 & 2 & (156,136)[0.1]  & 50.0 & 2\\(160,128)[0.1]  & 50.0 & 2 & (160,130)[0.1]  & 50.0 & 2 & (160,136)[0.1]  & 50.0 & 2\\(160,138)[0.1]  & 50.0 & 2 & (160,146)[0.1]  & 50.0 & 2 & (160,152)[0.1]  & 50.0 & 2\\(160,154)[0.1]  & 50.0 & 2 & (162,128)[0.1]  & 50.0 & 2 & (162,136)[0.1]  & 50.0 & 2\\(164,128)[0.1]  & 68.033 & 2 & (164,4)[0.1]  & 70.492 & 67 & (164,36)[0.2]  & 70.492 & 67\\(164,132)[0.9]  & 72.951 & 68 & (164,108)[0.9]  & 87.705 & 67 & (164,134)[0.7]  & 73.77 & 2\\(164,160)[0.9]  & 73.77 & 2 & (164,136)[0.9]  & 77.869 & 2 & (164,44)[0.9]  & 86.066 & 67\\(164,140)[0.8]  & 86.066 & 68 & (164,12)[0.9]  & 85.246 & 67 & (168,128)[0.1]  & 50.0 & 2\\(168,130)[0.1]  & 50.0 & 2 & (168,136)[0.1]  & 50.0 & 2 & (168,138)[0.1]  & 50.0 & 2\\(168,152)[0.1]  & 50.0 & 2 & (170,128)[0.1]  & 50.0 & 2 & (170,136)[0.1]  & 50.0 & 2\\(172,128)[0.1]  & 81.148 & 2 & (172,4)[0.1]  & 77.869 & 67 & (172,36)[0.1]  & 81.967 & 67\\(172,132)[0.1]  & 82.787 & 68 & (172,136)[0.1]  & 83.607 & 2 & (172,12)[0.4]  & 86.066 & 67\\(172,140)[0.7]  & 85.246 & 68 & (178,128)[0.1]  & 50.0 & 2 & (178,136)[0.1]  & 50.0 & 2\\(184,128)[0.1]  & 50.0 & 2 & (184,136)[0.1]  & 50.0 & 2 & (200,160)[0.1]  & 87.705 & 2\\(200,162)[0.1]  & 86.885 & 2 & (200,130)[0.1]  & 86.885 & 2 & (200,132)[0.1]  & 86.885 & 2\\(200,152)[0.1]  & 86.885 & 2 & (200,128)[0.1]  & 87.705 & 2 & (200,140)[0.1]  & 86.885 & 2\\(200,76)[0.1]  & 86.066 & 67 & (200,156)[0.1]  & 88.525 & 2 & (200,72)[0.1]  & 86.066 & 67\\(200,136)[0.2]  & 87.705 & 2 & (200,164)[0.8]  & 89.344 & 2 & (200,134)[0.4]  & 90.984 & 2\\(232,130)[0.1]  & 81.967 & 2 & (232,72)[0.2]  & 83.607 & 34 & (232,128)[0.1]  & 83.607 & 2\\(232,108)[0.1]  & 84.426 & 34 & (232,164)[0.2]  & 85.246 & 2 & (232,136)[0.1]  & 82.787 & 2\\(232,154)[0.7]  & 90.984 & 2 & (232,200)[0.1]  & 81.967 & 200 & (232,172)[0.7]  & 88.525 & 2\\\hline
			\end{tabular}
		\end{adjustbox}
	\caption{Effectiveness of TSCAs During Training of Monk-1 Dataset}
	\label{table4}
	\end{center}
\end{table}
\begin{table}[htbp]
	\begin{center}
		
		\begin{adjustbox}{width=\columnwidth,center}
			\begin{tabular}{|ccccccc|} \hline 
				Datasets & \makecell{TSCA\\ Size} & \makecell{Training \\Efficiency} &\makecell {Margin of Error\\ in Training} & \makecell{Testing\\ Efficiency} & \makecell{Margin of Error\\ in Testing} & \makecell{Proposed\\ TSCAs}\\
				\hline
				Monk-1 & 11 & 97.54 & 0.223 & 86.08 & 0.3112 & (76, 72)[0.1]\\
				Monk-2 & 11 & 96.45 & 0.2012 & 88.22 & 0.2068 & (76, 72)[0.1]\\
				Monk-3 & 11 & 98.36 & 0.123 & 94.21 & 0.2406 & (76, 72)[0.1]\\
				Haber man & 9 & 80.27 & 0.4321 & 80.76& 0.6730 & (132, 108)[0.4]\\
				Heart-statlog & 16 & 99.26 & 0.541 & 92.59 & 0.7679 & (232, 154)[0.8]\\
				Tic-Tac-Toe & 18 & 100 & 0 & 99.48 & 0.1679 & (140, 12)[0.9]\\
				Hepatitis & 19 & 100 & 0.6089 & 97.3 & 0.7303 & (232, 172)[0.6]\\
				Spect Heart & 22 & 97.33 & 0.4326 & 95.699 & 0.4133 & (172, 140)[0.3]\\
				Appendicitis  & 28 & 97.56 & 0.1921 & 95 & 0.7874 & (76, 72)[0.4]\\
				\hline
			\end{tabular}
		\end{adjustbox}
	\caption{Pattern classifiers' performance over various datasets (for proposed classifier)}
	\label{table5}		
	\end{center}
\end{table}

\subsection{Margin of error}
As previously stated, the classifier is a \emph{Temporally Stochastic CA}-based classifier, in which the noise rule $g$ is applied with a probability $\tau$ and the cells are stochastically updated. This might happen in different ways for different runs, resulting in varying efficiency. As a result, these categorization differences must be recorded. To obtain these information, we determine the margin of error for both the training and testing phases. A margin of error expresses as a variation of small amount in case of change of circumstances, i.e. the maximum expected difference between the true parameter and a sample estimation of that parameter \cite{Cochran1977}. We estimate the margin of error for sample size $m$ using Equation~\ref{mr}  \cite{Cochran1977}. 
\begin{align}\label{mr}
	Marginal\; Error &= Z_{n/2}(\frac{\sigma}{\sqrt{m}})	
\end{align}
We have considered $m=30$ samples for the experimentation with $\sigma$ as the variance.
\begin{align}\label{var}
	\sigma &= \sqrt{\frac{\sum(x_i-\bar{x})^2}{(m-1)}}
\end{align}
The efficiency of the $i^{th}$ sample is $x_i$ , and the mean of the sample efficiencies is $\bar{x}$. As we consider the confidence level for our sampling experiments is $95$\% percent \cite{Cochran1977}, we set $Z_{n/2} = 1.96$. Table~\ref{table5} displays the margin of error of different classifiers in training and testing phase. It is found that the margin of error is very less in both training and testing phase. Hence, During different runs, the suggested classifier's efficiency fluctuates somewhat.

\subsection{Comparison}
For the study of the efficiency of the proposed two-class pattern classifier, we employed nine datasets: Monk-1, Monk-2, Monk-3, Haber-man, Heart-statlog, Tic-Tac-Toe, Spect heart, Hepatitis and Appendicitis. The datasets are preprocessed suitably to fit the input features of the classifier.

The classification accuracy of the proposed classifier is compared with different existing standard algorithms such as Bayesian, C4.5 \cite{Salzberg1994}, MLP (Multilayer Perceptron), TCC, MTSC, ASVM, LSVM, Sparse grid, Traditional CA \cite{DasMNS09} and Asynchronous CA \cite{Sethi2016}. 	

\begin{table}[htbp]
	\begin{center}
		
		\begin{adjustbox}{width=\columnwidth,center}
		\begin{tabular}{|cccc|} \hline 
			Datasets & \makecell{Algorithm} & \makecell{Efficiency in \%} &\makecell {Efficiency of proposed classifier \\ with Margin of Error} \\
			\hline
			Monk-1 & Bayesian & 99.9&86.08 $\pm$ 0.3112 \\
			&C4.5 & 100 & (TSCA(76, 72)[0.1]) \\
			&TCC & 100 &\\
			&MTSC & 98.65 &\\
			&MLP & 100 &\\
			&Traditional CA & 61.111 &\\
			&Asynchronous CA & 81.519 &\\\hline				
			Monk-2 & Bayesian & 69.4&88.22 $\pm$ 0.2068\\
			&C4.5 & 66.2 &  (TSCA(76, 72)[0.1])\\
			&TCC & 78.16 &\\
			&MTSC & 77.32 &\\
			&MLP & 75.16 &\\
			&Traditional CA & 67.129 &\\
			&Asynchronous CA & 73.410 &\\\hline				
			Monk-3 & Bayesian & 92.12&94.21 $\pm$ 0.2406\\
			&C4.5 & 96.3 &  (TSCA(76, 72)[0.1])\\
			&TCC & 76.58 &\\
			&MTSC & 97.17 &\\
			&MLP & 98.10 &\\
			&Traditional CA & 80.645 &\\
			&Asynchronous CA & 83.749 &\\\hline
			Haber-man & Traditional CA  & 73.499&80.76 $\pm$ 0.6730 \\
			&Asynchronous CA & 77.493 & (TSCA(132, 108)[0.4])\\\hline
			Spect Heart & Traditional CA  & 91.978 & 95.699 $\pm$ 0.4133 \\
			&Asynchronous CA & 100 & (TSCA(172, 140)[0.3])\\\hline
			Tic-Tac-Toe & Sparce grid  & 98.33 & 99.48 $\pm$ 0.1679 \\
			&ASVM & 70.00 & (TSCA(140, 12)[0.9])\\
			&LSVM & 93.330 &\\
			&Traditional CA & 93.330 &\\
			&Asynchronous CA & 99.721 &\\\hline
			Heart-statlog&Bayesian&82.56&92.59$\pm$0.7679\\
			&C4.5&80.59&(TSCA(232,154)[0.8])\\
			&Logit-boost DS &82.22&\\ \hline
			Hepatitis&Bayesian&84.18&97.3 $\pm$ 0.7303\\
			&C4.5&82.38&(TSCA(232,172)[0.6])\\
			&Logit-boost DS &81.58&\\ \hline
			Appendicitis&-&-&95$\pm$ 0.7874\\
			&&&(TSCA(76,72)[0.4])\\
			
			\hline
		\end{tabular}
	\end{adjustbox}
	\caption{Classification accuracy compared to other well-known classifiers}
	\label{table6}
	\end{center}
	
\end{table}

The performance of our proposed TSCA-based classifier is compared to that of other well-known classifiers, as shown in Table~\ref{table6}. We observed that our proposed TSCA-based two-class pattern classifier performs much better than traditional CA-based classifier and it becomes more competitive and performs reliably better than other well known classifier algorithms.
\section{Summary}

In this chapter, we have suggested a variant of cellular automata, termed as \emph{Temporally Stochastic CA} (TSCA), in which, instead of one local rule, two rules ($f$ and $g$) have been utilized. Where, rule $f$ acts as a default rule and $g$ acts as a noise rule, applied with probability $\tau$ (noise rate). After analyzing their dynamics, we have identified the convergent TSCAs that have been used to design two-class pattern classifier.

A two-class pattern classifier was developed using these TSCAs ($114$ in number)(see Table~\ref{pattern1}). For a given dataset, we have chosen a TSCA with the highest efficiency to build a classifier. In comparison to existing common algorithms, our suggested design of TSCA-based two-class pattern classifier offers competitive performance. As a two-class pattern classifier, one can employ TSCAs with optimal number of fixed point attractors to improve performance. This will be the focus of our future works.

%




\chapter{Affinity Classification Problem by Stochastic Cellular Automata}
\label{chap5}
\section{Introduction}
\noindent The density classification problem is a well-known problem in cellular automata (CAs). Given an initial configuration, this problem asks to find a binary cellular automaton (CA) that converges to all-$0$ (resp. all-$1$) configuration, a fixed point, if number of $0$’s (resp. $1$s) in the initial configuration in higher than the number of $1$s (resp. $0$s). That is, the CA reaches all-$1$ configuration if it has an \emph{affinity} towards $1$ in its initial configuration with respect to the density of $1$ in it and reaches all-$0$ otherwise. However, sometimes, the requirement of many applications is that, this density itself is to be treated as a variable -- still a binary CA is required that can converge to the all-$1$ (resp. all-$0$) configuration. In this chapter, we introduce this problem as a generalization of the density classification problem. Formally the problem can be stated as:\\\\
\textbf{Problem Statement:} \emph{Given an initial configuration, find a binary cellular automaton that converges to all-$1$ configuration if density of $1$s is more than $\rho$. Otherwise, it converges to all-$0$ configuration.}\\\\
Here, $\rho$ is calculated as the density of $1$s in the initial configuration and all-$0$ and all-$1$ are the only fixed points of the CA. We name this problem as \emph{Affinity Classification Problem} as the CA has an affection towards the all-$1$ configuration. When $\rho = 0.5$, the problem is reduced to the classical density classification problem.

In literature, several attempts have been taken to solve the density classification problem. However, in \cite{PhysRevLett.74.5148}, it is proved that it is impossible to solve this problem with $100\%$ accuracy using classical CAs. Because of this, research efforts have been shifted towards finding the non-classical CAs which can solve the problem {\em almost} perfectly.
In \cite{Fuk05},  it is shown that the density classification task is solvable by running in sequence the trivial combination of elementary rules $184$ and $232$. This solution is extended for two-dimension using a stochastic component into each of these two rules in \cite{fuks2015solving}. In \cite{fates13}, a stochastic CA is used to solve the problem with an arbitrary precision. In this solution, the cells of 1-dimensional CA stochastically choose a rule in each step from a set of rules to evolve. These non-classical CAs can be named as \emph{spatially} stochastic CAs. Target has also been taken to tackle this problem with non-uniform CA where the cells can use different rules to evolve. A non-uniform CA that does the best density classification task is identified in \cite{NazmaTh}. However, neither (spatial) stochastic CA nor non-uniform CA can perfectly solve the density classification problem. Whereas, the non-classical CA of Ref.\cite{Fuk05} which may be called as \emph{temporally} non-uniform CA, can do it perfectly. 

As the affinity classification problem is an extension of the density classification problem, it is most likely to be unsolvable using classical CAs. We may need non-classical CA with temporal non-uniformity and stochastic component for this. Hence, to solve this problem, in this work, we introduce \emph{temporally stochastic} CAs. We define our problem over two dimensional binary CAs and use two different CA rules uniformly over the grid. The default rule is deterministic, whereas, another rule is stochastic whose application time is dependent on some probability. Section~\ref{model} describes the proposed model. The simulation and convergence to the solution for different density is shown in Section~\ref{simulation}. It is shown that our model is not blind as it \emph{intelligently} decides and converges to its \emph{point of attraction}.
Finally, we show that this model has several applications including as model for \emph{self-healing systems} (Section~\ref{application}).

\section{The Model}\label{model}

\noindent The proposed cellular automaton is defined over two-dimensional square grid which uses periodic boundary condition. The CA is binary and considers Moore neighborhood dependency; that is, a cell takes any of the two states $0$ or $1$ and depends on itself and it's eight nearest neighbors. At a time stamp $t$, a cell can be updated using one of the two rules $f$ and $g$. Here, $f$ is deterministic and the default rule for the grid, whereas, $g$ is stochastic and is applied with some probability. As the CA is defined over Moore neighborhood, both $f$ and $g$ are having the same domain and range:
\[f:\{0,1\}^9 \rightarrow \{0,1\} \text{            and            } g:\{0,1\}^9 \rightarrow \{0,1\} \]

Let us now first discuss about the default rule $f$. This rule is spatially deterministic -- at any time, it is applied over all cells uniformly. At each time step $t+1$, this rule updates the state of cell ${(i,j)}$ depending on the present states of its neighboring cells:
\begin{scriptsize}
	$${(i-1,j)}, {(i-1,j-1)}, {(i,j-1)},  {(i+1,j-1)}, {(i+1,j)}, {(i+1,j+1)}, {(i,j+1)}, {(i-1,j+1)}$$
\end{scriptsize}
Let $s^t_{i,j}$ be the present state of cell ${(i,j)}$ and $\mathscr{C}^d_{(i,j)}$ represents for the cell $(i,j)$, $s^t_{i,j}=d$ where $d \in \{0, 1\}$.
Then $f$ works in the following way:
\begin{align*}
	s^{t+1}_{i,j} & = f(s^t_{i-1,j}, s^t_{i-1,j-1}, s^t_{i,j-1} ,  s^t_{i,j} , s^t_{i+1,j-1}, s^t_{i+1,j}, s^t_{i+1,j+1} ,  s^t_{i,j+1} , s^t_{i-1,j+1})\\
	& = \begin{cases}
		0 & \text{ if } s^t_{i,j}=1 \text{ and } \sum\limits_{\substack{i-1 \le l \le i+1,\\ j-1 \le m \le j+1}}\mathscr{C}^0_{(l,m)}> K  \\
		1 & \text{ if } s^t_{i,j}=0 \text{ and } \sum\limits_{\substack{i-1 \le l \le i+1,\\ j-1 \le m \le j+1}}\mathscr{C}^1_{(l,m)}=8-K\\
		s^t_{i,j} & \text{ otherwise }
	\end{cases}
\end{align*}
where $K$ is a constant and $0\le K \le 8$. 
That means, if a cell is $1$ and it has more than $K$ neighbors with state 0, it becomes 0 in next step; whereas, a cell of state 0 with $(8-K)$ or more neighbors with state 1 becomes 1 in the next step. This number of neighbors required for state transition ($K$) is the \emph{first parameter} of the model.

The most significant characteristics of our model comes from the second rule $g$. As already mentioned, $g$ is a stochastic rule, that is, it is applied to each cell with some probabilities. Moreover, at which time step this rule is to be applied that is also stochastically decided. Hence, we call the CA as a temporally stochastic CA. However, when selected, this rule is also applied uniformly over all cells. Following is the definition of this rule:
\begin{align*}
	s^{t+1}_{i,j} & = g(s^t_{i-1,j}, s^t_{i-1,j-1}, s^t_{i,j-1} ,  s^t_{i,j} , s^t_{i+1,j-1}, s^t_{i+1,j}, s^t_{i+1,j+1} ,  s^t_{i,j+1} , s^t_{i-1,j+1})\\
	& = \begin{cases}
		0 \text{ with probability } \phi(x)& \text{ if } s^t_{i,j}=1 \text{ and } \sum\limits_{\substack{i-1 \le l \le i+1,\\ j-1 \le m \le j+1}}\mathscr{C}^0_{(l,m)}=x \\
		1 \text{ with probability } \psi(x) & \text{ if } s^t_{i,j}=0 \text{ and } \sum\limits_{\substack{i-1 \le l \le i+1,\\ j-1 \le m \le j+1}}\mathscr{C}^1_{(l,m)}=x\\
		s^t_{i,j} & \text{ otherwise }
	\end{cases}
\end{align*}
Here, $\phi(x)$, $\psi(x): \{0,1,\cdots, K\}\rightarrow [0,1]$ are two probability distribution functions.
We denote this $x$ as the number of \textit{supporting neighbors} or simply \emph{support}.

This rule implies, if a cell is at state $1$ and it has $x$ number of neighbors with state $0$, it updates its value to $0$ with some probability $\phi(x)$. Similarly, if a cell is at state $0$ and it has $x$ number of neighbors with state $1$, it updates its value to $1$ with some probability $\psi(x)$. We name $\phi(x)$ as the \emph{affection probability} and $\psi(x)$ as the \emph{repulsion probability} function. These two probability distribution functions are the \emph{second} and \emph{third parameters} of our model.

However this stochastic rule $g$ does not act in each step. When it is to be applied is decided by another probability $p$, which we name as the \emph{upgrade probability}. This $p$ is the \emph{fourth} and final \emph{parameter} of our model. Hence, the parameters required by the model are --
\begin{itemize}
	\item $K$ = number of neighbors required to change from one state to another
	\item $\phi(x)$= affection probability function
	\item $\psi(x)$ = repulsion probability function
	\item $p$ = upgrade probability
\end{itemize}

Observe that, in our model the role of $g$ is to give the cells an extra chance to change their status. During evolution of the CA by $f$ if some cells are left out which are \emph{eager} to update their states but can not do so because of the surrounding neighbors (\emph{hostile environment}), they get a \emph{booster} to upgrade their current status through $g$. This $g$ helps them achieve their desired status even if they have less number of neighboring cells to their \emph{support} (as $x\le K$). But whether the cell will be updated or not, is dependent on the probability value. Both cells with state $0$ and $1$ get this advantage uniformly in terms of the two probability distribution functions $\phi(x)$ and $\psi(x)$. As $g$ gives precedence towards some cells, it is to be applied with a caution -- so there is the \emph{upgrade} probability value $p$ which works as a controlling measure. Therefore, when $K=4$, $f$ works as a simple majority rule and depending on $g$ the system can be inclined towards a specific state. 

Note that, the parameters give us flexibility to design the model according to the need of an application. For example, for the model to have affection to converge to all-$0$ as a fixed point, we can set $\phi(x)$ and $\psi(x)$ accordingly. Similarly, we can change value of our parameter(s) to get different versions of the model which can be used for a specific purpose. In fact, we may also consider that in our model rule $g$ is applied with probability $p$ whereas the rule $f$ is applied with probability $(1-p)$ with $p$ being any probability value. This way of looking at these rules makes both of them \emph{temporally stochastic}. 
The next section shows some simulation results of our model to solve the affinity classification problem taking some specific value of the parameters.

%

\section{Solving Affinity Classification Problem: A Simulation}\label{simulation} 
\noindent We now simulate our proposed model to understand its efficacy in solving the affinity classification problem. As mentioned before, the model is a 2-dimensional finite CA that uses periodic boundary condition. For the simulation purpose, we consider here the grid size as $10^3\times10^3$, that is total number of cells = $10^6$. Further, our model is characterized by four parameters -- $K$, $\phi(x)$, $\psi(x)$ and $p$. In our simulation, we have assumed the following values for the parameters:
\begin{align*}
	K&=4\\
	\phi(x)&=\begin{cases}
		0  &\text{if }  x \le 1\\
		log_K(x) &{2\le x \le K }
	\end{cases}\\
	\psi(x)&=\begin{cases}
		0  &\text{if }  x = 0\\
		e^{x-K} &{1\le x \le K }
	\end{cases}\\
	p&=0.2
\end{align*}

\begin{figure}[h]
	\vspace{-1.5em}
	\subfloat[ \label{fig:12}]{%
		\includegraphics[scale = 0.20]{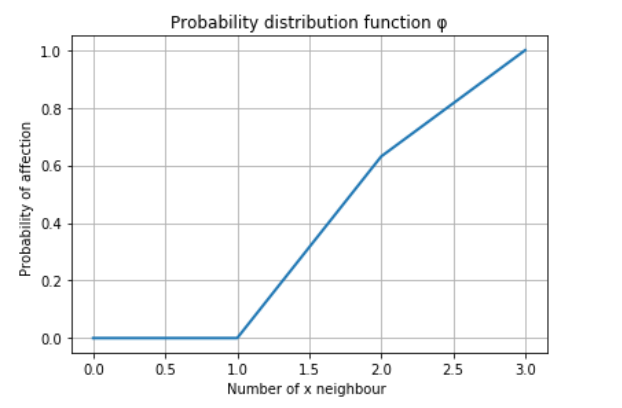}
	}
	\hfill
	\subfloat[ \label{fig:13}]{%
		\includegraphics[scale = 0.20]{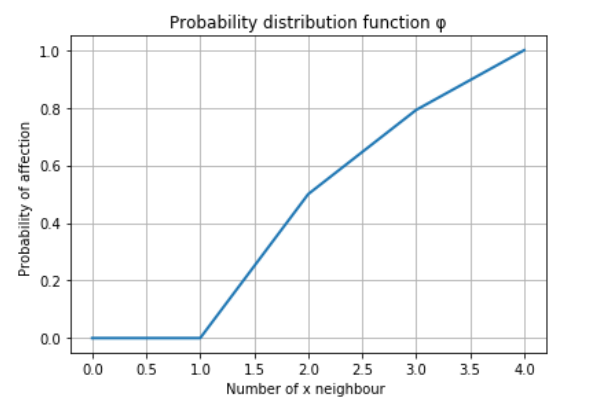}
	}
	\hfill  
	\subfloat[ \label{fig:14}]{%
		\includegraphics[scale = 0.20]{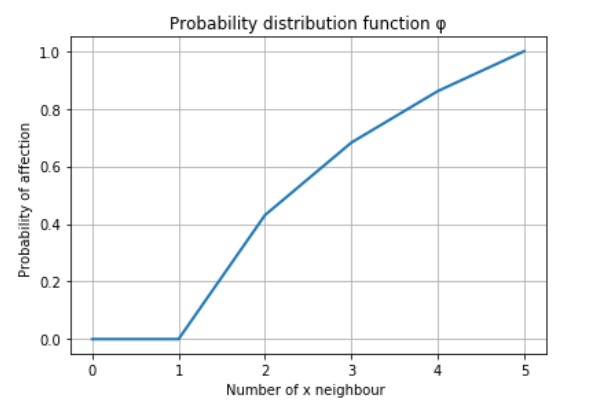}
	}
	\hfill
	\subfloat[ \label{fig:15}]{%
		\includegraphics[scale = 0.20]{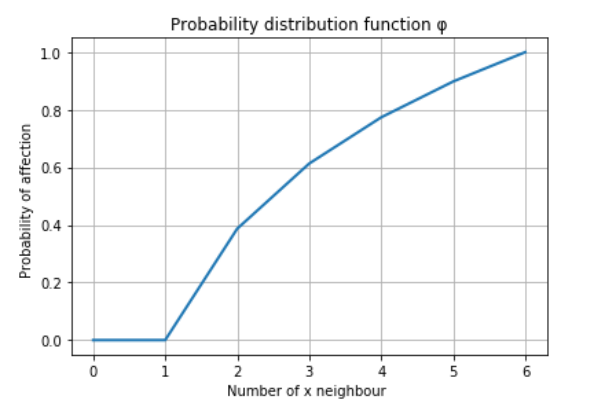}
	}     

	\caption{Graph of $\phi(x)$ for different $K$: a) $K=3$ ; b) $K=4$ ; c) $K=5$ ; d) $K=6$}
	\label{fig:16}
	
\end{figure}

\begin{figure}[!htbp]
	\vspace{-1.5em}
	\subfloat[ \label{fig:23}]{%
		\includegraphics[scale = 0.20]{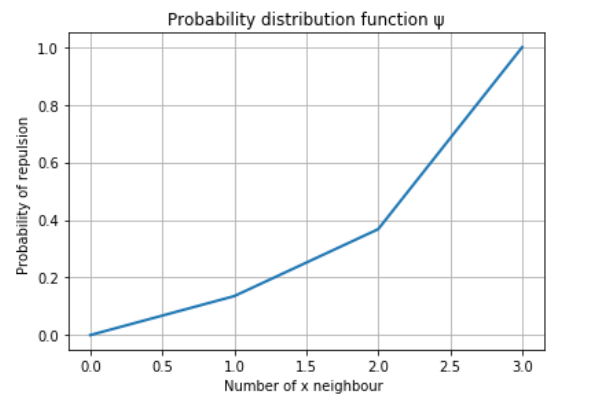}
	}
	\hfill
	\subfloat[ \label{fig:24}]{%
		\includegraphics[scale = 0.20]{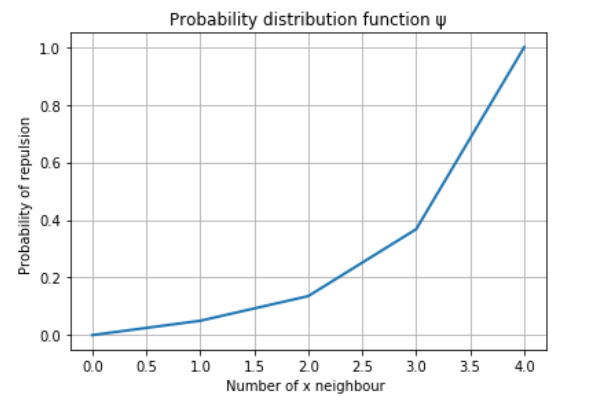}
	}
	\hfill  
	\subfloat[ \label{fig:25}]{%
		\includegraphics[scale = 0.20]{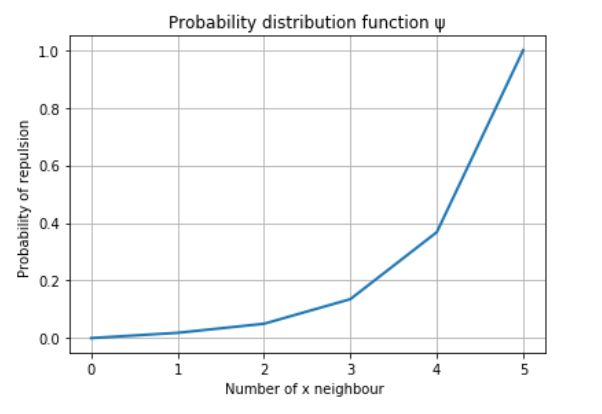}
	}
	\hfill
	\subfloat[ \label{fig:26}]{%
		\includegraphics[scale = 0.20]{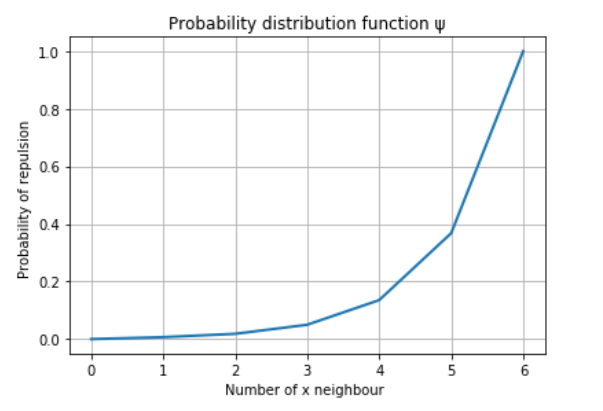}
	}     

	\caption{Graph of $\psi(x)$ for different $K$: a) $K=3$ ; b) $K=4$ ; c) $K=5$ ; d) $K=6$}
	\label{fig:27}
	
\end{figure}

As our model uses Moore neighborhood dependency on 2-D grid, $K$ is very small ($0\le K \le 8$). In this small range of $K$, logarithmic function grows faster than exponential function. Therefore, since we want to observe the affinity of the model towards all-$0$ configuration, we take $\phi(x)$ as a logarithmic function and $\psi(x)$ as an exponential function. As per our model, we use $x=0,1,\cdots,K$ to get the probability values for $\phi(x)$ and $\psi(x)$. We have plotted $\phi(x)$ and $\psi(x)$ for different $x$ to see their behavior at $K=4$ (see Figure \ref{fig:13} and Figure \ref{fig:24} respectively). We can observe that, at $K=4$, $\phi(1)=0.0$, $\phi(2)=0.5$, $\phi(3)=0.79248$, $\phi(4)=1.0$, whereas, $\psi(1)=0.0497$, $\psi(2)=0.1353$, $\psi(3)=0.3679$, $\psi(4)=1.0$. 
%
We observe that, if the count of supporting neighbors $x$ is increased then the probability of changing state from $1$ to $0$ is also increased (Figure~\ref{fig:13}); but, if $x$ is decreased then the probability of changing state from $0$ to $1$ is increased with the growth of the first function being faster than the latter (Figure~\ref{fig:24}).


\subsection{Random Initial Configuration} 
We have experimented our model with huge number of random initial configurations having various $\rho$ where 
$$\rho= \frac{\text{Number of 1s}}{\text{Total number of cells}} $$ 
Following are some sample results from our experiment when $K=4$. Here, $0$ is represented in color \emph{yellow} and $1$ is represented by color \emph{red}.

\begin{figure}[!htbp]
	\vspace{-1.5em}
	\subfloat[ \label{fig:28}]{%
		\includegraphics[scale = 0.20]{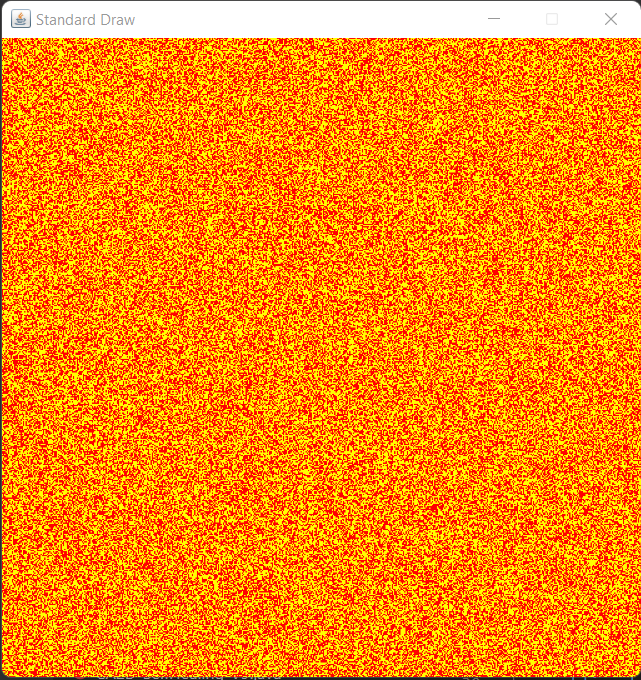}
	}
	\hfill
	\subfloat[ \label{fig:29}]{%
		\includegraphics[scale = 0.20]{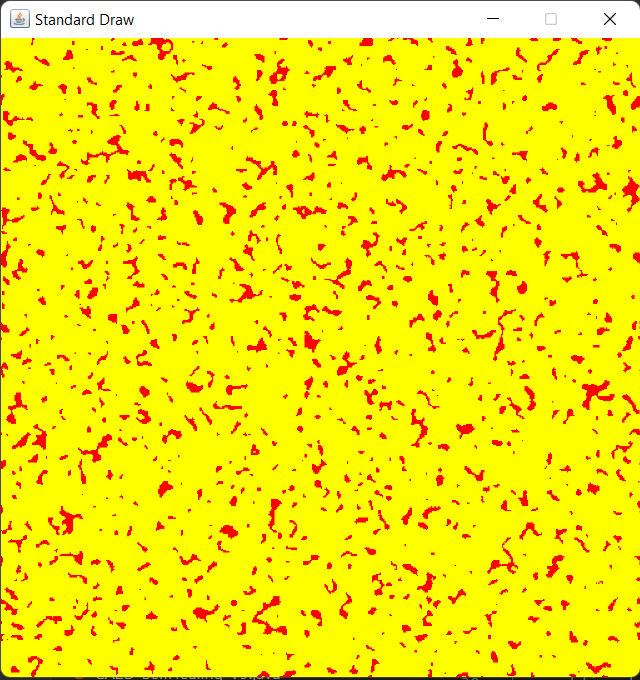}
	}
	\hfill  
	\subfloat[ \label{fig:30}]{%
		\includegraphics[scale = 0.20]{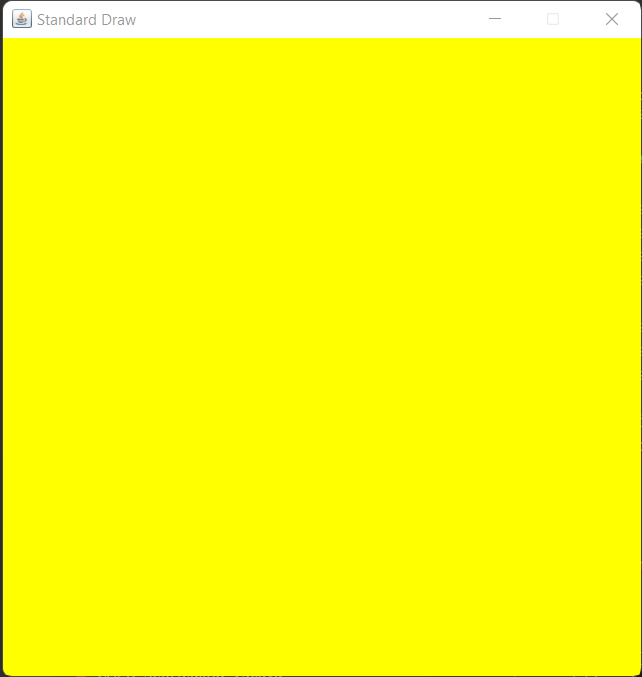}
	}
	
	\caption{For $K=4$ and $\rho=0.475$, the model converge after $150$ iterations: (a) Initial configuration, (b) An intermediate Configuration, (c)Final configuration(all-0)}
	\label{fig:31}
\end{figure}
\noindent Figure~\ref{fig:31} shows that, at $\rho=0.475$, for a random initial configuration, all the cells become yellow after $150$ iterations, that means, the model converge to it's converging point (all-$0$). We have experimented with large number of random initial configurations and seen that, in our experiments, when the initial configuration has $\rho \le 0.675$, the model is converging to all-$0$, otherwise it converges to all-$1$.
\begin{figure}[!htbp]
	\vspace{-1.5em}
	\subfloat[ \label{fig:41}]{%
		\includegraphics[scale = 0.20]{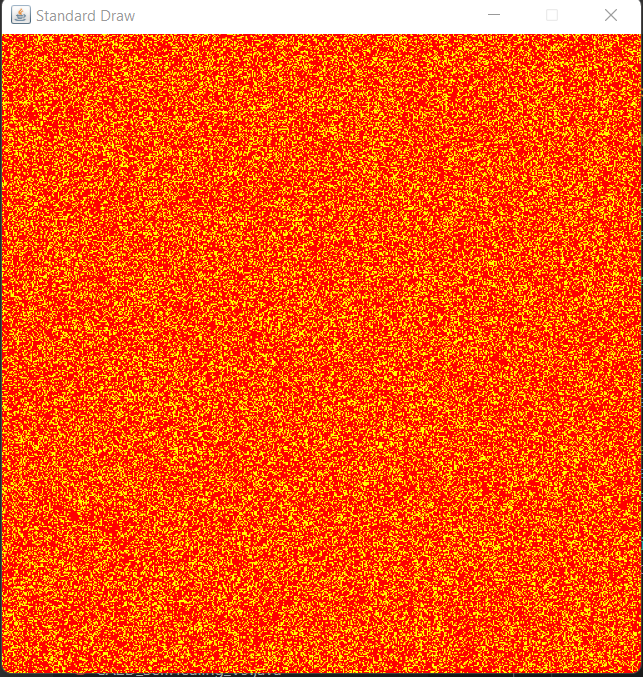}
	}
	\hfill
	\subfloat[ \label{fig:42}]{%
		\includegraphics[scale = 0.20]{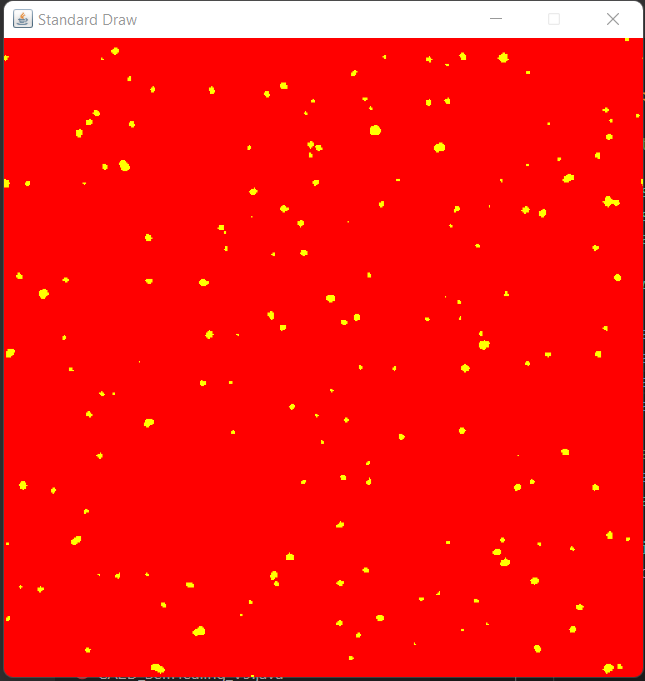}
	}
	\hfill  
	\subfloat[ \label{fig:43}]{%
		\includegraphics[scale = 0.20]{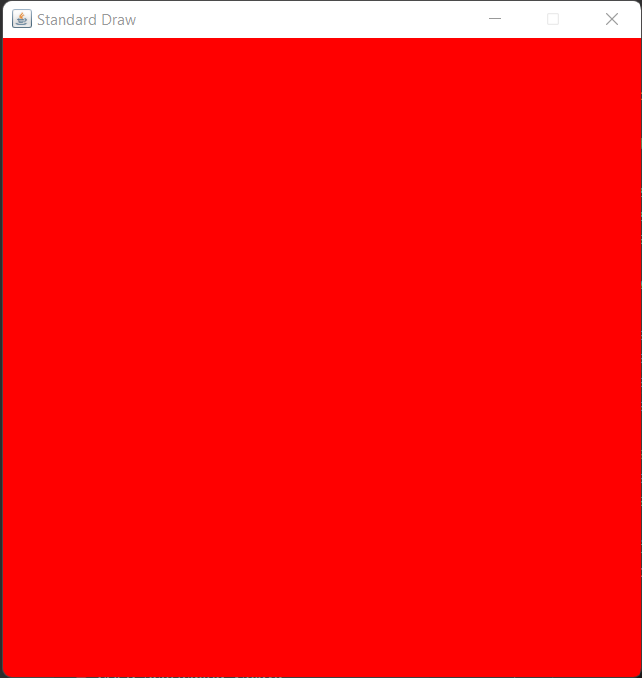}
	}
	
	\caption{For $K=4$ and $\rho=0.6989$: the model converge to all-$1$ after $137$ iterations: (a) Initial configuration, (b)An intermediate Configuration, (c)Final configuration(all-1)}
	\label{fig:4}
\end{figure}

Figure \ref{fig:4} shows another sample random initial configuration with an arbitrary $\rho > 0.675$ (here $\rho = 0.6989$). Here, the model converges to all-$1$ after 137 iterations.
\begin{figure}[!htbp]
	\vspace{-1.5em}
	\subfloat[ \label{fig:51}]{%
		\includegraphics[scale = 0.20]{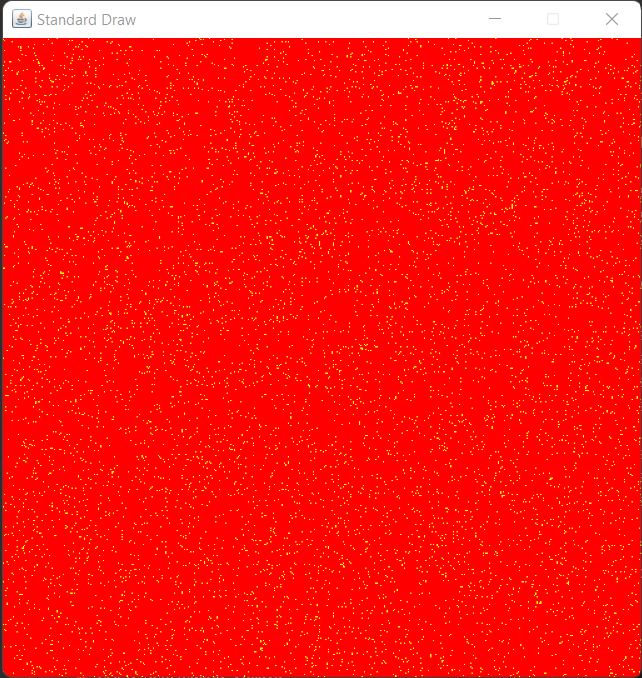}
	}
	\hfill
	\subfloat[ \label{fig:52}]{%
		\includegraphics[scale = 0.20]{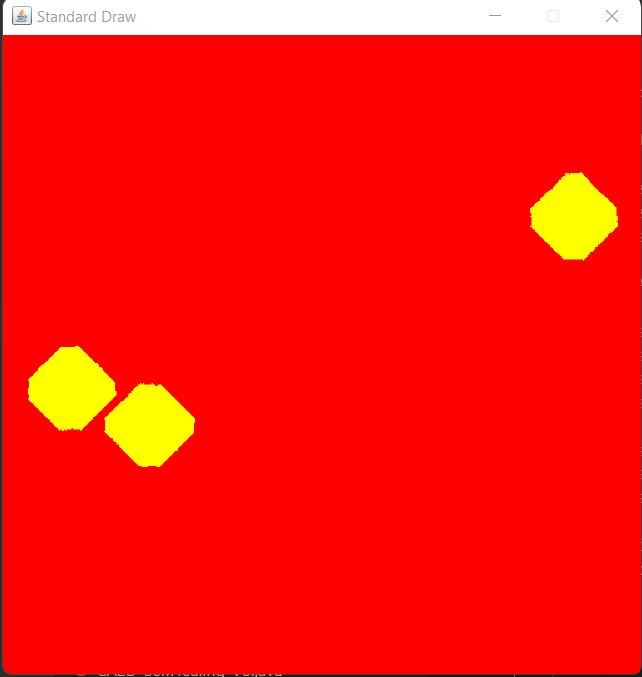}
	}
	\hfill  
	\subfloat[ \label{fig:53}]{%
		\includegraphics[scale = 0.20]{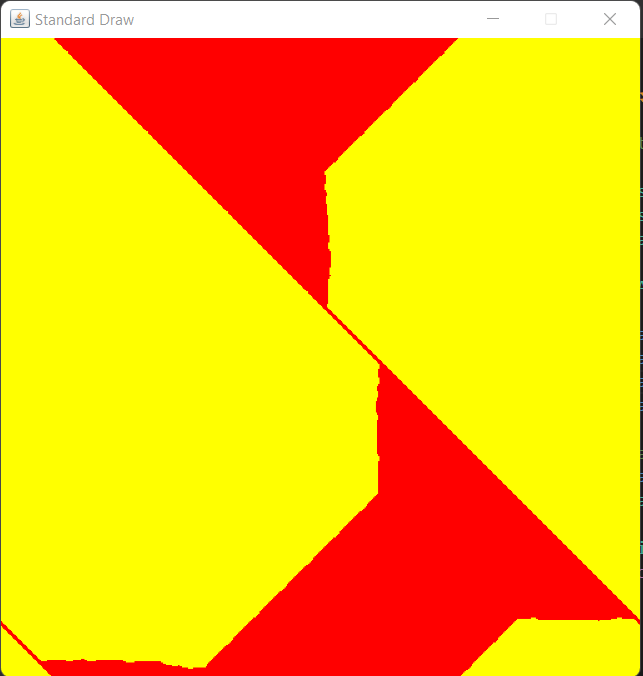}
	}
	\hfill  
	\subfloat[ \label{fig:54}]{%
		\includegraphics[scale = 0.20]{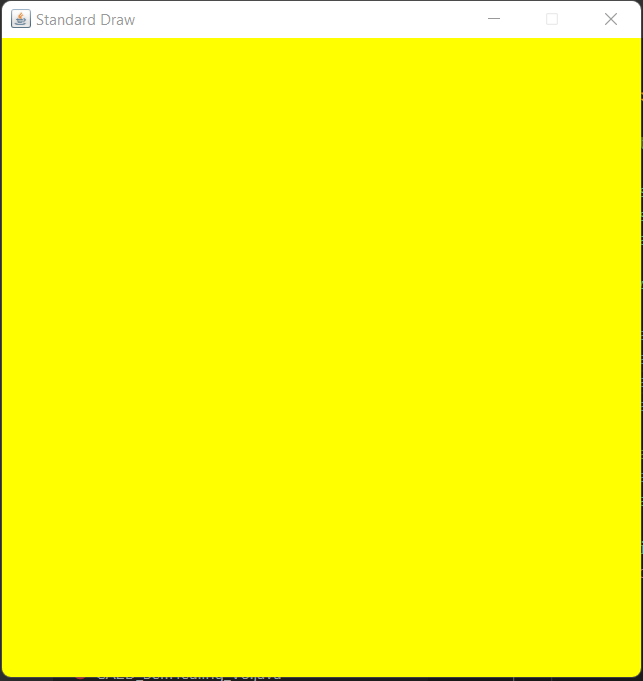}
	}
	
	\caption{For $K=3$ and $\rho=0.969852$, after 3132 iterations, the model converge to all-$0$'s: (a) Initial configuration, (b) An intermediate Configuration, (c) Another intermediate configuration, (d) Final configuration (all-$0$)}
	\label{fig:5}
\end{figure}

Naturally, the question comes, \emph{``Can we increase the affection probability so that even if we take a lot of $1$'s in the initial configuration then also the model converges to all-$0$?''}. To search for this answer, we have again done large number of experiments by varying the value of $K$. In our experiments, we have observed that when we decrease the value of $K$, the model is converging to all-$0$ even though $\rho > 0.68$. For example, for the initial configuration of Figure~\ref{fig:5}, if $K=3$, then although $\rho \le 0.96$, the model is converged to all-$0$s. By further experimentation, we observe that, if the value of $K$ is decreased to $2$ then $\rho$ can be as high as $0.99$ but the model may still converges to all-$0$. Similarly, when we increase the value of $K$ then the value of $\rho$ is to be decreased for converging to all-$0$. 

\begin{table}[!h]
	\centering
	\vspace{-1.5em}
	
	\resizebox{0.99\textwidth}{!}{
		\begin{tabular}{cc}
			\begin{tabular}{|c|c|c|c|}
				\hline\hline
				{\bfseries K } & {\bfseries $\rho$} & {\bfseries  Number of iterations } & {\bfseries Converge to }\\
				\hline
				1 &  0.000002 & 1 &all-0\\
				1 &  0.0002 & 2 &all-0\\
				1 &  0.0051 & 2 &all-0\\
				1 &  0.3 & 3 &all-0\\
				1 &  0.55 & 4 &all-0\\
				1 &  0.67 & 6 &all-0\\
				1 &  0.6864 & 8 &all-0\\
				1 &  0.943 & 21 &all-0\\
				1 &  0.991 & 76 &all-0\\
				1 &  0.997 & 170 &all-0\\
				1 &  0.9995 & 489 &all-0\\
				\hline
				2 &  0.1 & 3 &all-0\\
				2 &  0.3 & 4 &all-0\\
				2 &  0.4 & 5 &all-0\\
				2 &  0.61 & 7 &all-0\\
				2 &  0.74 & 14 &all-0\\
				2 &  0.8 & 16 &all-0\\
				2 &  0.9536 & 90 &all-0\\
				2 &  0.965 & 151 &all-0\\
				2 &  0.982 & 323 &all-0\\
				2 &  0.993 & 759 &all-0\\
				2 &  0.995 & 3 &all-1\\
				2 &  0.9995 & 2 &all-1\\
				\hline
				3 &  0.1 & 3 &all-0\\
				3 &  0.3 & 6 &all-0\\
				3 &  0.4 & 9 &all-0\\
				3 &  0.55 & 20 &all-0\\
				3 &  0.61 & 22 &all-0\\
				3 &  0.6864 & 44 &all-0\\
				3 &  0.8 & 109&all-0\\
				3 &  0.943 & 518 &all-0\\
				3 &  0.953 & 2042 &all-0\\
				3 &  0.96 & 2372 &all-0\\
				3 &  0.965 & 3220 &all-0\\
				3 &  0.982 & 3 &all-1\\
				3 &  0.991 & 3 &all-1\\
				3 &  0.995 & 2 &all-1\\
				
				\hline\hline
			\end{tabular}
			&
			\begin{tabular}{|c|c|c|c|}
				\hline\hline
				{\bfseries K } & {\bfseries $\rho$} & {\bfseries  Number of iterations } & {\bfseries Converge to }\\
				\hline
				
				4 &  0.1 & 4 &all-0\\
				4 &  0.4 & 66 &all-0\\
				4 &  0.55 & 805 &all-0\\
				4 &  0.61 & 3074 &all-0\\
				4 &  0.65 & 7019 &all-0\\
				4 &  0.67 & 12385 &all-0\\
				4 &  0.675 & 16186 &all-0\\
				4 &  0.6864 & 261 &all-1\\
				4 &  0.7 & 159 &all-1\\
				4 &  0.74 & 69 &all-1\\
				4 &  0.8 & 8 &all-1\\
				4 &  0.943 & 4 &all-1\\
				4 &  0.965 & 4 &all-1\\
				4 &  0.991 & 2 &all-1\\
				
				\hline
				5 &  0.06 & 8 &all-0\\
				5 &  0.08 & 14 &all-0\\
				5 &  0.09512 & 8 &all-0\\
				5 &  0.1  & 14 &all-0\\
				5 &  0.3 & 304 &all-1\\
				5 &  0.4  & 91 &all-1\\
				5 &  0.55 & 12 &all-1\\
				5 &  0.61  & 10 &all-1\\
				5 &  0.686 & 6 &all-1\\
				\hline
				6 &  0.001 & 2 &all-0\\
				6 &  0.0051 & 3 &all-0\\
				6 &  0.00994 & 8 &all-0\\
				6 &  0.03 & 638&all-1\\
				6 &  0.0629 & 230 &all-1\\
				6 &  0.076 & 194 &all-1\\
				6 &  0.08 & 98 &all-1\\
				6 &  0.1 & 58 &all-1\\
				\hline
				7 &  0.000002 & 2 &all-0\\
				7 &  0.0002 & 2 &all-0\\
				7 &  0.0004 & 2 &all-0\\
				7 &  0.0005 & 976 &all-1\\
				7 &  0.0009 & 375 &all-1\\
				7 &  0.001 & 417 &all-1\\
				7 &  0.00499 & 136 &all-1\\
				7 &  0.00994 & 83 &all-1\\
				7 &  0.0676 & 25 &all-1\\
				7 &  0.1 & 12 &all-1\\
				7 &  0.55 & 4 &all-1\\
				7 &  0.95 & 2 &all-1\\
				\hline\hline
			\end{tabular}
	\end{tabular}}
	\caption{Relationship between the values of $K$ and $\rho$ where the model converges to all-$0$ or all-$1$}\label{tab1}
\end{table}

Figure \ref{fig:16}and \ref{fig:27} show the variation of the probability distribution functions for different $K$ values. {If $K$ is changed then the growth of the probability distribution functions $\phi(x)$ and $\psi(x)$ are also changed with respect to $K$.} Table~\ref{tab1} gives some of our experimental results. In each of the subtables of this table, column $1$ and $2$ describe the initial configurations in the form of $K$ and $\rho$, whereas, column $3$ and $4$ show experimental outcomes.

\begin{table}[!h]
	\centering
	
	\resizebox{0.6\textwidth}{!}{
		\begin{tabular}{|c|c|c|c|}
			\hline\hline
			{\bfseries $K$} & {\bfseries Number of $0s$ } & {\bfseries Iterations (Time steps) }& {\bfseries Converges to }\\
			\hline
			1 &  1 & 1 &all-1\\
			1 &  2 & 998 &all-0\\
			\hline
			2 &  2 & 1 &all-1\\
			2 &  3 & 1152 &all-0\\
			2 &  4 & 1148 &all-0\\
			\hline
			3 &  3 & 1 &all-1\\
			3 &  4 & 3874 & all-0\\
			3 &  25 & 4093 & all-0\\
			\hline
			4 &  49 & 64 &all-1\\
			4 &  64 & 92 &all-1\\
			4 &  70 & 492 &all-1\\
			4 &  81 & 576 &all-1\\
			4 &  100 & 15915 &all-0\\
			4 &  144 & 16138 &all-0\\
			\hline\hline
	\end{tabular}}
	\caption{Relationship between the values of $K$ and the block of $0$s where the model converges to all-$0$ or all-$1$ considering number of $0$s are placed sequentially in the grid}\label{tab2}
\end{table}

\subsection{Initial Configuration with Block of $0$s and $1$s}\label{sec:block}
Previous subsection shows the results when $0$ and $1$ in the initial configuration are randomly organized. Now, we experiment with initial configurations where block of cells are set to have same value. Table~\ref{tab2} and \ref{tab3} depict our sample results. In Table \ref{tab2}, we consider initial configurations with a small number of consecutive cells at state $0$ and the remaining cells in state $1$. For every value of $K$ ($ 1 \le K \le 7$), column 2 shows the number of consecutive cells having same value in our experiments so that the model converges to all-$0$ . For instance, when $K=4$ then an initial configuration having a block of $100$ consecutive $0$s converges the model to all-$0$. These consecutive $0$s form a cluster and it grows in size to converge the model to all-$0$.
\begin{figure}[!h]
	\vspace{-1.5em}
	\subfloat[ \label{fig:17}]{%
		\includegraphics[scale = 0.20]{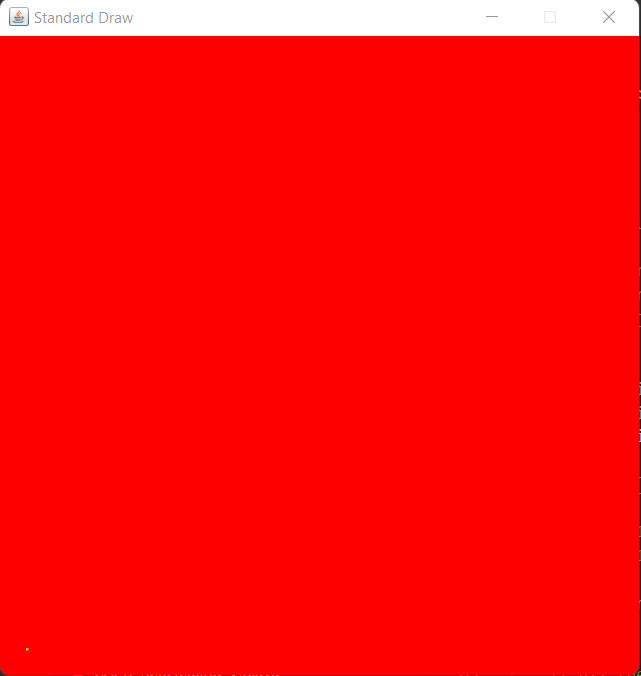}
	}
	\hfill
	\subfloat[ \label{fig:18}]{%
		\includegraphics[scale = 0.2]{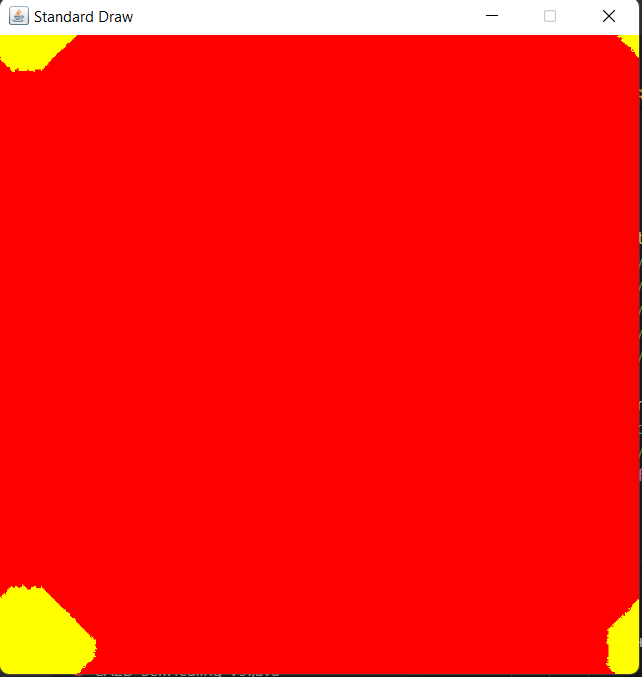}
	}
	\hfill  
	\subfloat[ \label{fig:19}]{%
		\includegraphics[scale = 0.2]{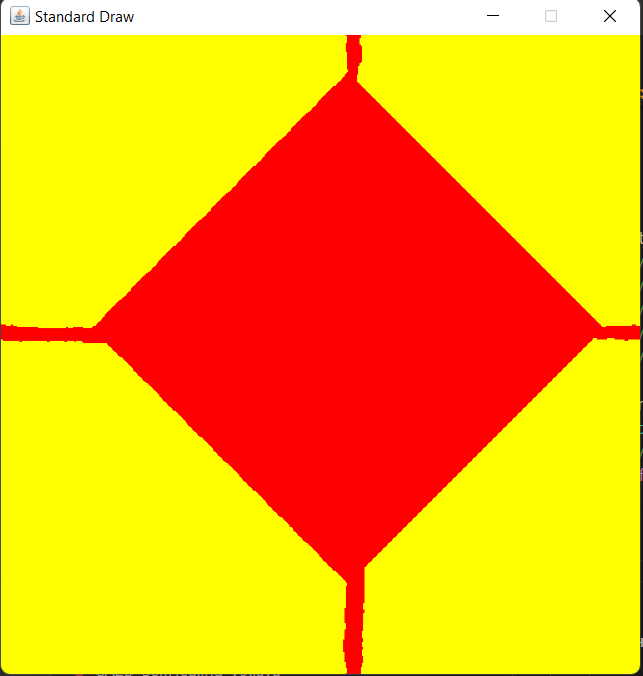}
	}
	\hfill
	\subfloat[ \label{fig:20}]{%
		\includegraphics[scale = 0.2]{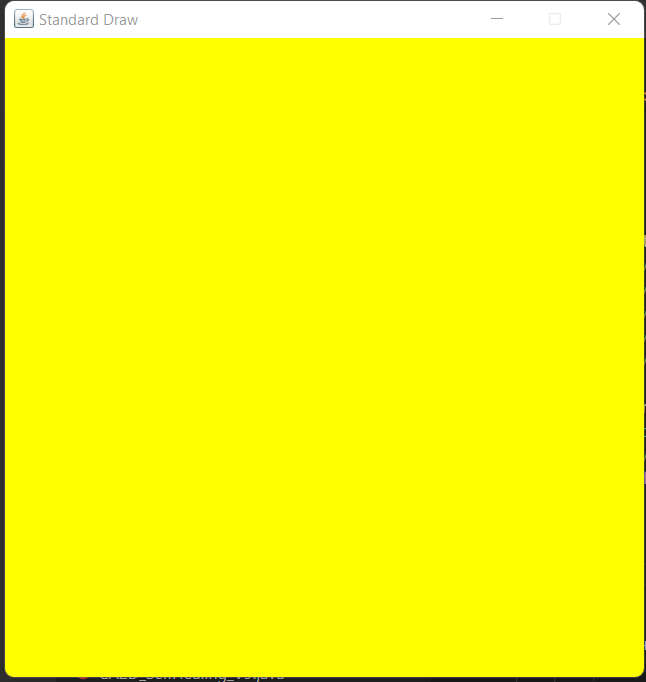}
	}     

	\caption{$K=3$ and the block of $25$ $0s$, the model converge to all $0s$ after 4093 iterations: a) Initial configuration, b)Intermediate Configuration, c)Another intermediate configuration, d) Final Configuration converge to all-$0$}
	\label{fig:21}
	
\end{figure}
Figure~\ref{fig:21} shows a random initial configuration with $10^6$ cells, where only $25$ consecutive cells are in state $0$ and the value of $K$ is $3$. We can observe that, although the number of $0s$ is very less, still the model converges to all-$0$ after $4093$ iterations (see Figure~\ref{fig:20}). {Therefore, our model has affinity to converge to all-$0$ even if at initial configuration the number of $0$s is very less in comparison to the grid size.} 

However, the model does not always converge to the desired fixed point. For example, at $K=3$, for a random initial configuration having block of $0$s of size $81$, the model converges to all-$1$ (Table~\ref{tab2}).
\begin{table}[!htbp]
	\centering
	
	\resizebox{0.6\textwidth}{!}{
		\begin{tabular}{|l|l|l|l|}
			\hline\hline
			{\bfseries $K$ } & {\bfseries Number of $1$'s }& {\bfseries Iterations (Time steps) } & {\bfseries Attractor }\\
			\hline
			5 &  25 & 24 &all-0\\
			5 &  225 & 212 &all-0\\
			5 &  256 & 483 &all-0\\
			5 &  324 & 12148 &all-1\\
			5 &  400 & 11870 &all-1\\
			\hline
			6 &  2 & 1 &all-0\\
			6 &  4 & 13 &all-0\\
			6 &  5 & 1519 &all-1\\
			6 &  6 & 1510 &all-1\\
			6 &  8 & 1507 &all-1\\
			6 &  9 & 1493 &all-1\\
			
			\hline
			7 &  1 & 1 &all-0\\
			7 &  2 & 979 &all-1\\
			
			\hline\hline
			
	\end{tabular}}
	\caption{Relationship between the values of $K$ and the block of $0s$ where the model converges to all-$0$ or all-$1$ considering number of $1s$ are placed sequentially in the grid}\label{tab3}
\end{table}
Table \ref{tab3} depicts some sample results from our experiment where we take small number of $1$s organized in sequential order that is, they make cluster. Here, we can see that, after taking the $K>4$, sometimes the model converges to all-$1$s even the number of $1$'s is very less. 
Therefore, even if the model has an affection to converge to all-$0$, the value of $K$ may take a major role to converge the model in a direction (all-$0$ or all-$1$).

\section{Applications}\label{application}
\noindent As discussed in Section~\ref{model}, the parameters give us flexibility to design our model according to the need of the solution to a particular problem. There are several possible applications of our model. Here we discuss some of them.

\subsection{Modeling Self-healing Systems}\label{sec:application}
Living systems are assumed to be more intelligent than a non-living system. Therefore, to be intelligent, a machine (non-living system) has to emulate the properties of living systems. Among the properties, self-healing is a basic and important biological property which indicates sign of life. 
Self-healing is the ability to reorganize and heal itself. 
If a machine has self-healing ability, it is likely to mimic other properties of living elements like self-replication. Hence, it will be more intelligent just like a living system. We can show that our proposed CA can be used to model any self-healing system where parameters of our abstract model can be interpreted as the characteristics of the self-healing system.

Let us interpret our model as the following. Let the grid of cells embodies a collection of living elements (they can be cells, humans, animals -- anything), where state $0$ means the cell is healthy and $1$ means it is sick. We want to model how much infection the cells can endure and still heal. 
By default, the living system is healthy, that is, all cells are in state $0$. Now, suppose, because of some change in environment, a number of cells get infected and update their states to $1$ (become sick). This is our initial configuration in the model. We start to observe the dynamics of the system from here. Let us consider that, in our model, system's immunity is the immunity of individual cells and as a whole the system's \emph{health} is the \emph{majority} of the \emph{individual} cell's health condition. So, at the initial configuration, if we ask the system, ``\emph{Are you sick?}'', it can answer ``Yes'' or ``No'' depending on density of $1$ ($\rho$). If using this model the system can \emph{heal} itself, that is, comes back to all-$0$, then we can call the model as a model for self-healing systems. At that time, the answer to ``\emph{Are you sick?}'' to the model will always be ``No''. Therefore, our target is to converge the grid to all-$0$ so that we can say there is no infection and ``\emph{The model is {Not Sick}}''. However, if the model converges to all-$1$ then we have to declare, ``\emph{The model is {Sick}}''.

Now, any living body has some inbuilt immunity status. This immunity is represented by the first parameter $K$. Just like immunity is different for different elements, $K$ itself is a variable. When $K=4$, the system can be interpreted as the situation of natural immunity having no prevailing sickness. The deterministic rule $f$ plays the role of natural healing process based on immunity $K$. Results from Section\ref{simulation} show that, if converging point is set to all-0 and $K\le4$ then there is a tendency to converge towards all-$0$ even if in the initial configuration number of 1s $>$ number of $0$s. This indicates, like any living body, our model also wants to become \textit{Not Sick}. 
%
%



\begin{figure}[!h]
	\vspace{-1.5em}
	\subfloat[ \label{fig:h3}]{%
		\includegraphics[scale = 0.2]{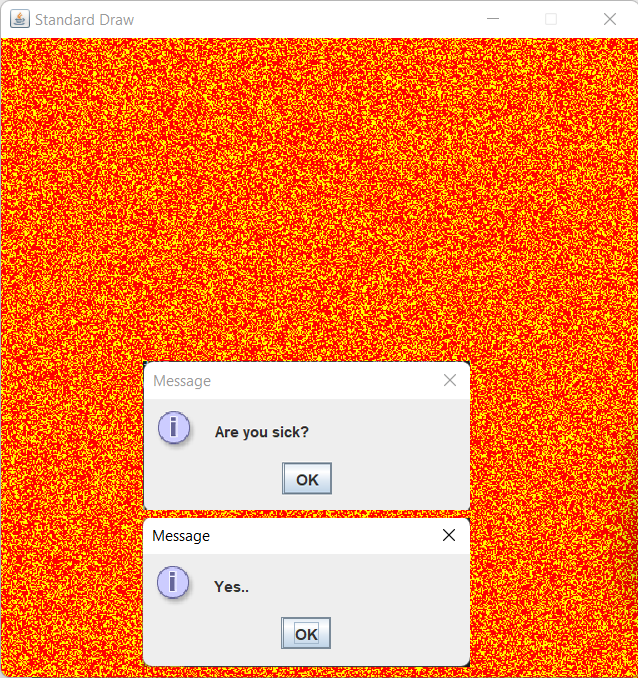}
	}
	\hfill
	\subfloat[ \label{fig:h4}]{%
		\includegraphics[scale = 0.2]{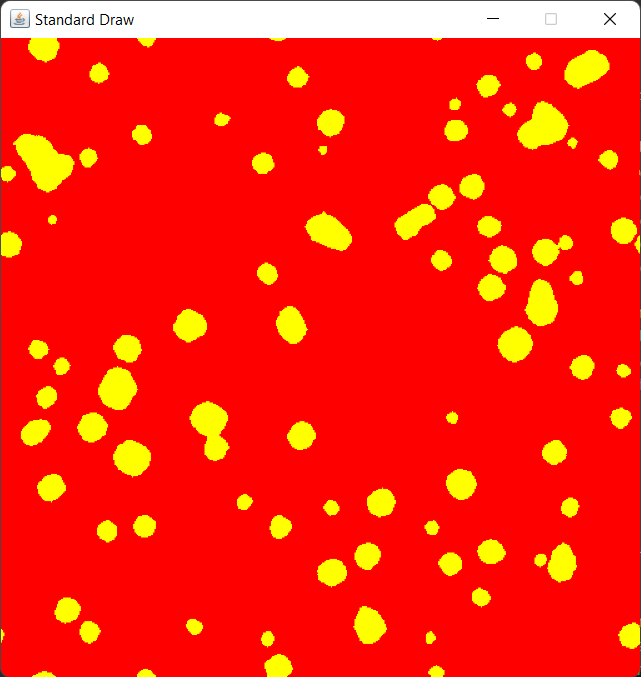}
	}
	\hfill  
	\subfloat[ \label{fig:h5}]{%
		\includegraphics[scale = 0.2]{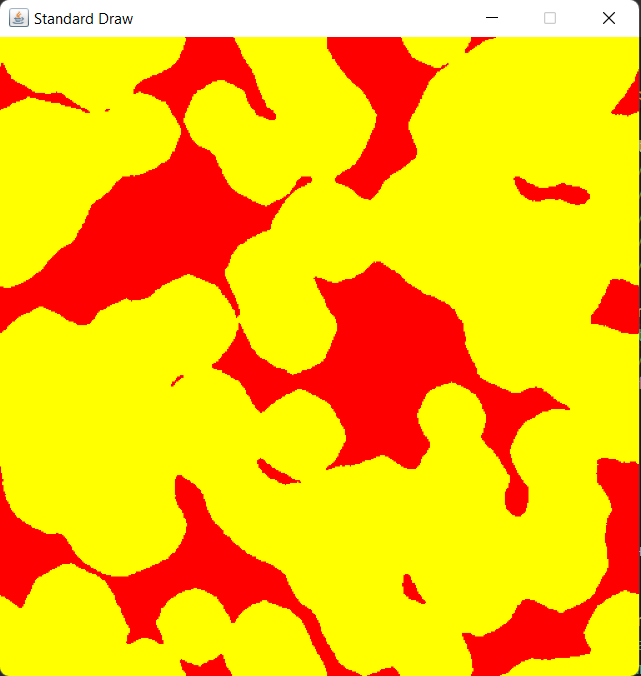}
	}
	\hfill
	\subfloat[ \label{fig:h6}]{%
		\includegraphics[scale = 0.2]{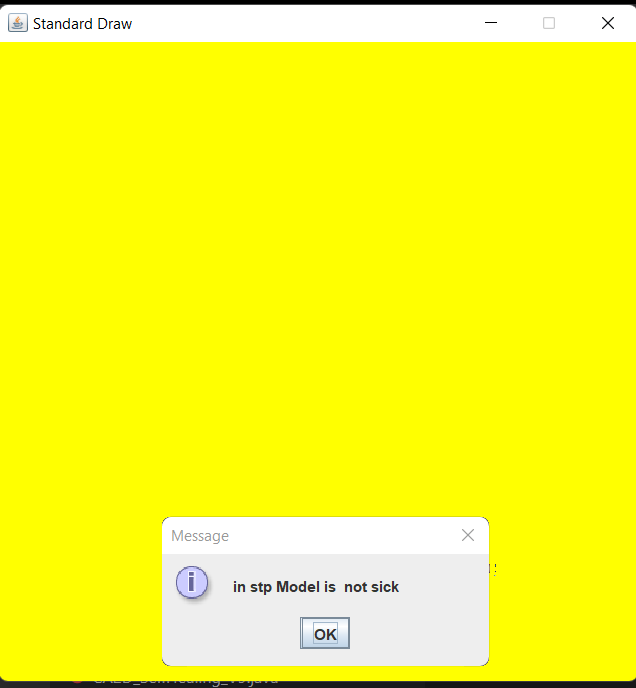}
	}     

	\caption{(a)Initial configuration of a sick model ; (b) and (c) shows two intermediate configurations during evaluation ; (d) The model is healed}
	\label{fig:h7}
	
\end{figure}
However, even in this condition, if number of infected cells become too large ($\rho$ is high), then, according to our rule, the system is \emph{sick}.  So, the inherent immunity is not enough to restore it to its health. For example, if we take a random initial configuration with some infected cells (cell state $1$) where $\rho=0.632275$ ($\rho$ = density of $1$'s) and $K=4$, then, at this stage the model is \textit{Sick} (see Figure \ref{fig:h3}). At this point, the cells are given some \emph{booster} to improve its immunity in terms of $g$. Here, $g$ may be considered as a \emph{vaccine} for the infection. As if, it can bypass the \emph{natural justice} process giving the cells a second chance to live. 
But, whether vaccine will be effective to a cell, is not deterministic (so, $g$ is stochastic). Further, when this vaccine is to be applied to the system is also not pre-determined (temporally stochastic CA with probability $p$). 

Nevertheless, for every cell, the vaccine will not react similarly. A large number of sick cells with favorable environment may become healthy ($\phi(x)$), whereas, some healthy cells with unhealthy environment can become sick ($\psi(x)$). But, if we take $\phi(x)$ as logarithmic and $\psi(x)$) exponential like defined in Section~\ref{simulation}, by choosing $K$ and $x$, we can see that, after some iterations the model converge to all-$0$ (see Table~\ref{tab1}). Then we can say the model is \textit{Not Sick} (Figure~\ref{fig:h6}).

However, if we increase $\rho$ value further (say, from $0.675$ to $0.68$ or more), then for the same $K$ and $x$, the model may be converged to all-$1$ (see Table\ref{tab1}) and the model becomes \textit{Sick}. Therefore, the role of $K$ and $x$ is very important to model self-healing systems. If we want to have our system a larger tendency to heal, then we need to choose the parameters of our model wisely. Moreover, if the affection probability $\phi(x)$ is large, then the system has more tendency to heal.
This probability indicates the ability to repair or heal oneself automatically and evolve oneself according to the demand of the environment.
%

This is how the self-healing is modeled by our CA. It also shows that, our abstract model can be a good interpretation of the role of vaccination in living population. Also, observe that, our proposed model takes the global decision democratically where every single cell take their own decision and the system comes to a consensus. Because of these properties we claim that our proposed model is intelligent.


\subsection{Modeling Transformation Process}\label{sec:intelligence}
In nature and chemical world, we get glimpses of several transformation processes -- water evaporates into vapor, a drop of color in a glass of liquid dissolves giving the whole glass of liquid a lighter shade of that color. All these processes happen to conserve the law of mass and energy. This section shows that our CA can be used to model such transformation processes.

\begin{figure}[!htbp]
	\vspace{-1.5em}
	\subfloat[$t=0$ \label{fig:s1}]{%
		\includegraphics[scale = 0.15]{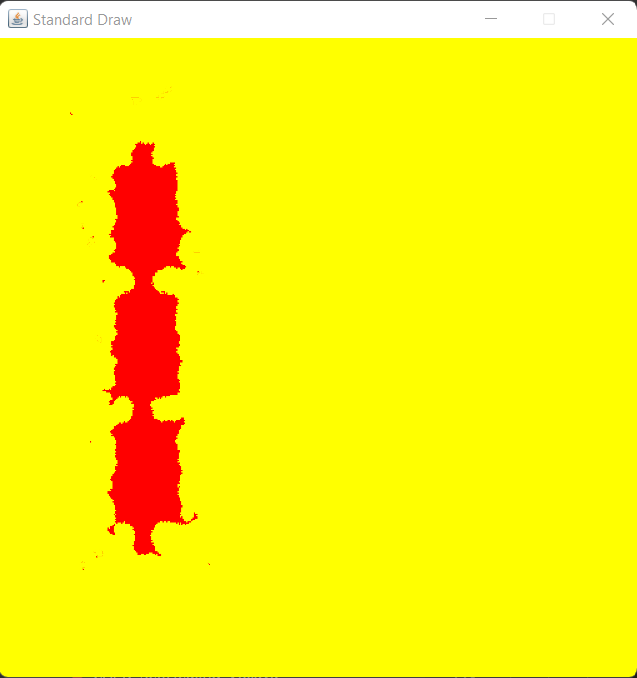} 
	}
	\hfill	
	\subfloat[$t=42$ \label{fig:s2}]{%
		\includegraphics[scale = 0.15]{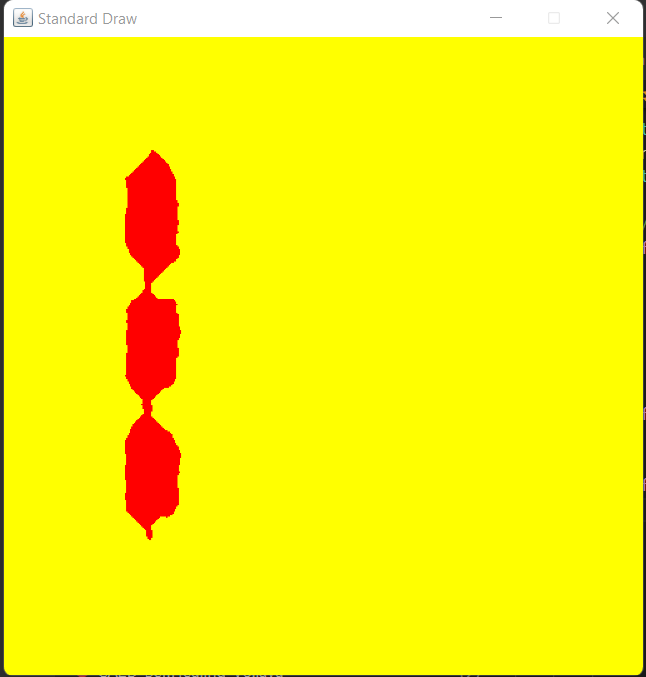} 
	}
	\hfill	
	\subfloat[ $t=78$ \label{fig:s3}]{%
		\includegraphics[scale = 0.15]{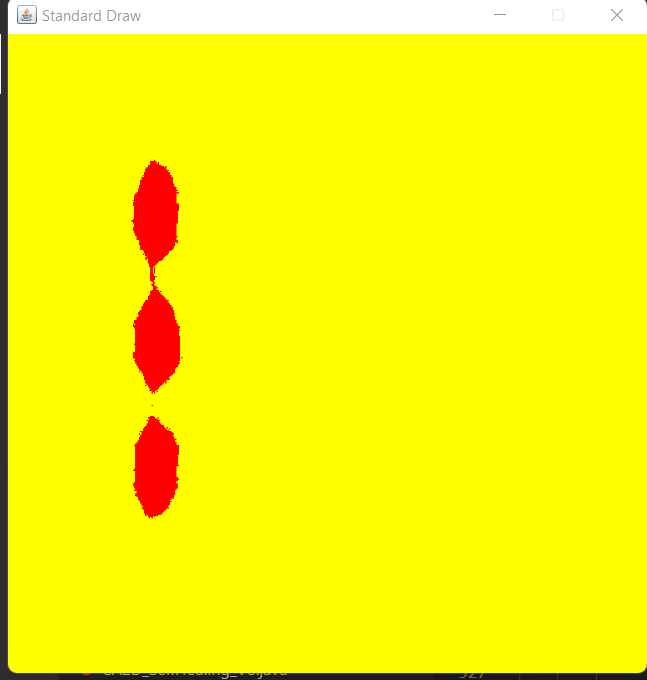}
	}
	\hfill
	\subfloat[$t=108$  \label{fig:s4}]{%
		\includegraphics[scale = 0.15]{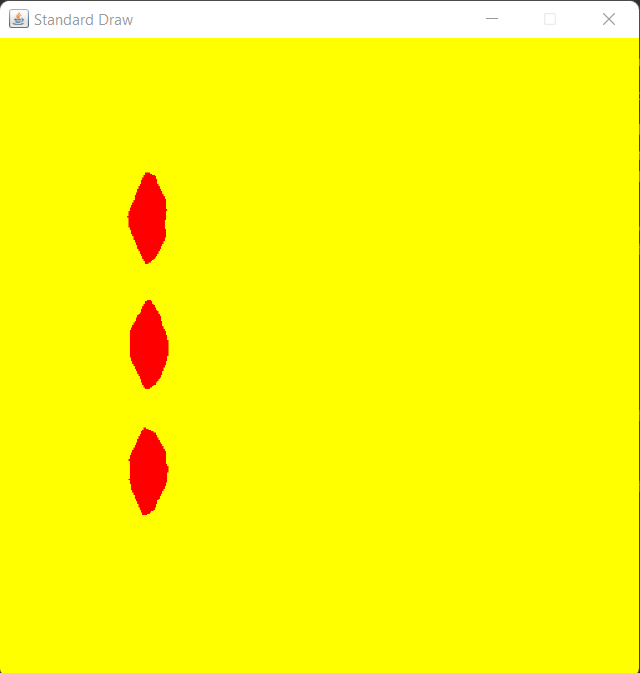}
	}
	\hfill  
	\subfloat[ $t=151$ \label{fig:s5}]{%
		\includegraphics[scale = 0.15]{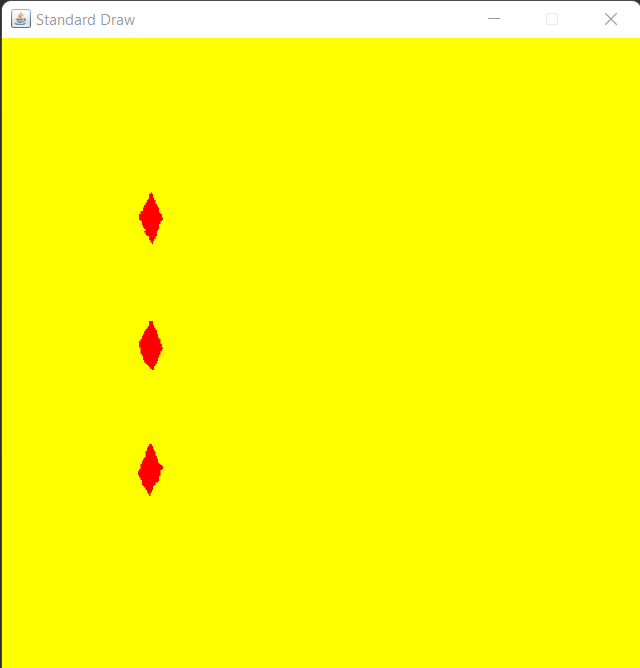}
	}
	\hfill
	\subfloat[$t=180$  \label{fig:s6}]{%
		\includegraphics[scale = 0.15]{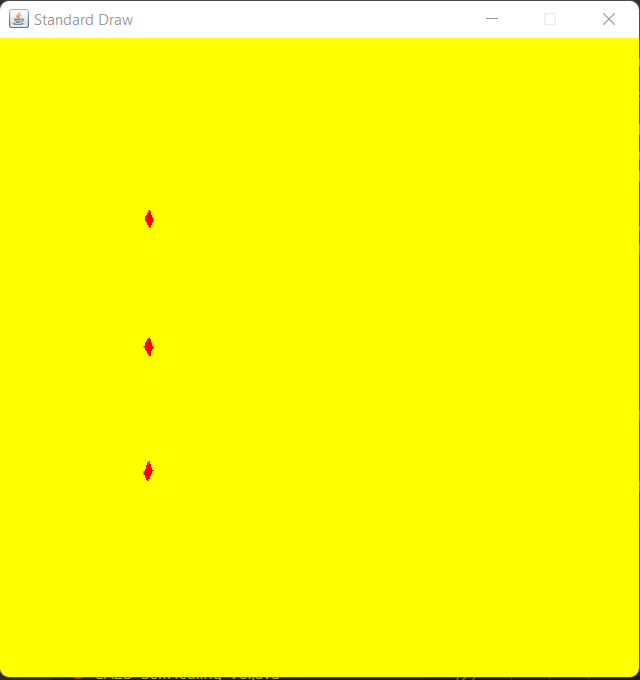}
	}     
	\hfill
	\subfloat[$t=198$  \label{fig:s7}]{%
		\includegraphics[scale = 0.15]{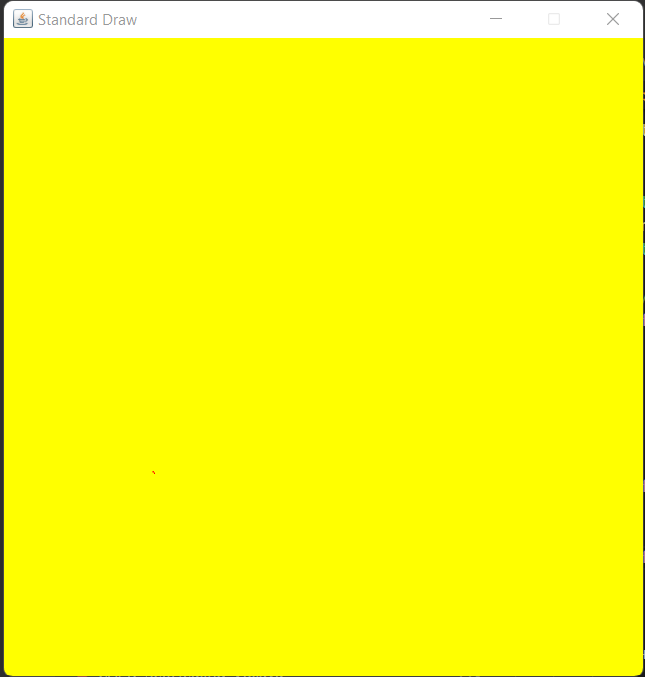}
	}    
	\caption{Simulation of a transformation process}
	\label{fig:7}
	
\end{figure}

During the process of transformation, the particles are divided into smaller sized particles and dissolves until the system comes to an equilibrium. In our model, if we set $K=3$ (and other parameters as same as Section~\ref{simulation}), then, for some special initial configurations, the evolution of the CA looks like transformation processes -- the configuration is divided into two or more smaller configurations. It goes on dividing and dissolving until the system converges to a fixed point which signifies the equilibrium state. For example, in Figure \ref{fig:7}, 
an initial configuration is shown, which, after some iterations, is divided into more than three configurations. It keeps on getting smaller until it converges to the fixed point all-$0$ when the system has reached its equilibrium. Hence, we can say that, by varying the parameters of our model, we can simulate the transformation process from one system to another by our CA.


\subsection{Density Classification Problem}
The \emph{density classification problem} can also be addressed by our model.
According to the definition of \emph{affinity classification problem}, this problem goes down to the former if we take the density of $1$s ($\rho$)=$0.5$. 
However, here, instead of taking $\rho$ as exact $0.5$, we take it as a variable and see how close we can reach to solve this classical problem using our model.

Previous works have established that density classification problem is not solvable by spatially stochastic CA (uniform or non-uniform), but can be solved by using temporally non-uniform CA \cite{Fuk05,fuks2015solving}. So, we also take our CA as temporarily non-uniform with a stochastic component $(g)$ which perfectly fits our model. However, a property of this problem is, there is no affinity towards any state at any time. Hence, to make the system unbiased, we take the number of neighbors required to change from one state to another $(K)$ as $4$. Also, we choose both the probability distribution functions $\phi(x)$ and $\psi(x)$ to be same. That is, if a cell is at state $1$ and it has $x$ number of neighbors with state $0$, it updates its value to $0$ with the same probability distribution function as in case of the cell being at state $0$ with $x$ number of neighbors with state $1$ and gets updated to state $1$. Further, we consider the upgrade probability value $p=0.1$ such that the stochastic component ($g$) is applied with very low probability. 

Here, we show simulation results for two different probability distribution functions -- linear and exponential. 
For the first case, the value of the parameters for the model are:
\begin{align*}
	K&=4\\
	\phi(x)&=\begin{cases}
		\frac{x}{K} &\text{for }  0 \le x \le K\\
	\end{cases}\\
	\psi(x)&=\begin{cases}
		\frac{x}{K} &\text{for }  0 \le x \le K\\
	\end{cases}\\
	p&=0.1
\end{align*}
We have done huge experimentation on random initial configurations over $200 \times 200$ grid based on this model. Some sample simulation results are shown in Table~\ref{tab4}. Here, the first column indicates some $\rho$ values whereas, the third and fourth columns represents that, for each of these $\rho$, out of $100$ experiments how many are converged to all-$0$ and all-$1$ respectively. {In our experiments, we observe that when initial configuration is taken randomly, then for $\rho \le 0.4647$ or $\rho \ge0.54$, our model converges to its fixed point (all-$0$ and all-$1$ respectively). 
	\begin{table}[h]
	
		\centering
		\resizebox{0.9\textwidth}{!}{
			\begin{tabular}{|c|c|c|c|} \hline  \hline   
				$\rho$ (number of $1$'s) & Number of experiments & Converge to all - $0$ & Converge to all - $1$\\  \hline 
				$\le$ 0.4647 & 100&100&0 \\
				0.4779710& 100 & 96&4 \\
				0.49112875& 100 & 73&27 \\
				0.5036035& 100 & 37 & 63  \\
				0.51548925 & 100 & 9 & 91\\
				0.52758225 & 100 & 5 & 95 \\
				0.5394037& 100 & 1 & 99\\
				$\ge$ 0.54 & 100 & 0 & 100 \\
				\hline \hline 
		\end{tabular}}  
		\caption{Taking 2D-square grid ($200 \times 200$) and both $\phi$ and $\psi$ as linear functions}\label{tab4}
	\end{table} 
	
	For the second case, we take both $\phi$ and $\psi$ as exponential functions with $K=4$ and $p=0.1$. Hence, the changed parameters of the model are:
	\begin{align*}
		\phi(x)&=\begin{cases}
			0  &\text{if }  x =0\\
			e^{x-K}  &\text{for }  1 \le x \le K\\
		\end{cases}\\
		\psi(x)&=\begin{cases}
			0  &\text{if }  x =0\\
			e^{x-K}  &\text{for }  1 \le x \le K\\
		\end{cases}
	\end{align*}
	We again repeat our experiments with a large set of random initial configurations over $100 \times 100$ grid. Table~\ref{tab5} shows some sample results of this experiment.
	\begin{table}[h]
	
		\centering
		\resizebox{0.9\textwidth}{!}{
			\begin{tabular}{|c|c|c|c|} \hline  \hline   
				$\rho$ (number of $1$'s) & Number of experiments & Converge to all - $0$ & Converge to all - $1$\\  \hline 
				$\le$ 0.4679  & 100 & 100 & 0  \\
				0.47837 & 100 & 96 & 4  \\
				0.513014 & 100 & 30 & 70  \\
				0.5181367 & 100 & 0 & 100  \\
				$\ge$ 0.520 & 100 & 0 & 100  \\
				\hline \hline 
		\end{tabular} }
		\caption{Taking 2D-square grid with size $100 \times 100$ and both $\phi$ and $\psi$ as exponential functions}\label{tab5}
	\end{table} 
	\begin{figure}[!hbtp]
		\vspace{-1.5em}
		\subfloat[\label{fig:10}]{%
			\includegraphics[scale = 0.2]{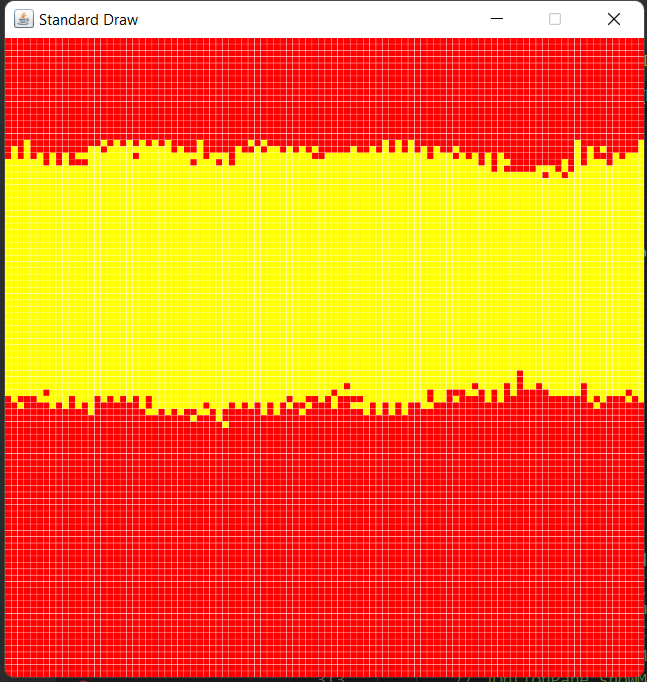}
		}
		\hfill
		\subfloat[\label{fig:11}]{%
			\includegraphics[scale = 0.2]{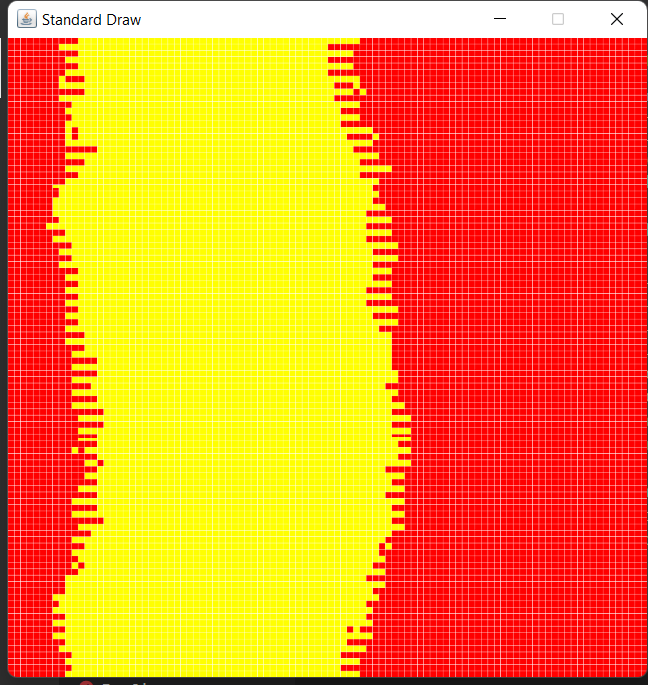}
		}    

		\caption{Some unsolvable configurations for density classification problem in 2D square grid}
		
	\end{figure}
	Here also we observe that, when the initial configuration is random, then for $\rho \le 0.4679$ or $\ge 0.520$, the model reaches its desired fixed point (all-$0$ or all-$1$). However, when the configurations are block of $0$s or $1$s forming a cluster, then it fails to reach the desired fixed point. Figure~\ref{fig:10} shows example of two such patterns where the model can not reach its fixed point (see Section~\ref{sec:block} for more details). 
	\section{Summary}
	\label{future}
	\noindent There are several properties in the living system that make them intelligent -- affection is one of them. In this work, we propose a new problem, named as, \emph{affinity classification problem}. We develop a devoted machine that is embedded in a 2-dimensional cellular automaton having Moore neighborhood dependency and periodic boundary condition. Our model has affection capabilities to a converging point, all-$1$ or all-$0$ and can be characterized by four parameters $K, \phi(x), \psi(x)$ and $p$.
	Using this model, we can develop a self-healing system. We know that, because of self-healing any species can survive in the evolution. As our model has this feature and it takes decision democratically, we can say that the model is acting like a natural living system to some extent and we can conclude that the model become intelligent. 
	
	However, there are some other properties of life which an intelligent machine need to possess; we have to see if our model possess them. Similarly, here we have considered only Moore neighborhood, what kind of behavior might arise if we change the neighborhood dependency for the rules is still not seen. 
	Different other behaviors might emerge by varying the parameters of our model. And, apart from self-healing systems, our model may be useful for other several areas of application. Answers to these questions remain work of the future.
	
\chapter{Searching with Cellular Automata on Cayley Tree}
\label{chap6}
	\section{Introduction}
	In a traditional Cellular Automaton (CA), a lattice of cells and a local rule are the components of a CA. The system operates in discrete time and space, and each cell generates its next state using the same rule. The CA has no memory; a cell switches to its next state depending only on the current states of its neighbors. Cellular Automata have been used in different domains, including biology, image processing, encryption, physics, machine learning~\cite{BhattacharjeeNR16,Sethi2016}, etc.
	
	A variant of Cellular Automata (CAs) in which each cell has an attached memory and an additional processing unit has been proposed in~\cite{Das2022}. Our proposed model, similar to this variant of CA, is developed over Cayley Tree~\cite{CayLeyid01,CayLeyid02} of order $\eta$, where each node represents a cell of the CA. A memory unit is attached to each cell. This is introduced here to efficiently solve the \emph{Searching problem} with In-Memory computation.
	
	Many algorithms have already been developed using CAs~\cite{parallelsorting} to solve computational problems. However, it is critical to emphasize the importance of Cellular Automata as a computational tool for effectively resolving issues involving large amounts of evenly dispersed data. To solve the issue, the components are often exchanged in the form of a cell state. However, in this study, no element exchange is carried out.
	
	We consider a finite CA, where the data elements ($X$) are distributed over the cells' memory. The CA solves the Searching problem, which asks to decide whether a given element (key) exists in a finite set of natural numbers ($X$). If the key $k \in X$, the CA concludes with a positive output as \emph{Found}; otherwise, the CA concludes with a negative output as \emph{Not Found}.
	In Section~\ref{model}, details of the proposed model are reported. The Cayley tree is introduced in Section~\ref{cayley}. The realization of \emph{In-Memory Searching} is reported in Section~\ref{in-memory}. 
	\section{Cellular Automata over Cayley Tree}\label{cayley}
	A Cellular Automaton (CA) is a discrete, abstract computational model that consists of a regular network of finite-state automata, formally known as cells. This section introduces the Cayley tree and the CA is developed over the Cayley Tree, where each cell of the CA is represented as a vertex in the tree.
	\begin{definition}
		A Cayley tree is a tree in which each non-leaf vertex has a constant number of branches $\eta$ (order of the tree). The Cayley tree $\kappa^\eta$ of order $\eta \geq 1$ is an infinite tree, from each vertex of which, there are exactly $\eta+1$ edges. The $\kappa^\eta$ = $(V, E, \nu)$, where $V$ is the set of vertices of $\kappa^n$, $E$ is the set of edges and $\nu$ is the incidence function associating each edge $e\in E$ with its endpoints $v_{neighbors} \in V$.
	\end{definition}
	\begin{figure}[!ht]
		\subfloat[]{
			\begin{minipage}[c][1\width]{
					0.5\textwidth}
				\label{fig1:a}
				\centering
				\includegraphics[width=1\textwidth]{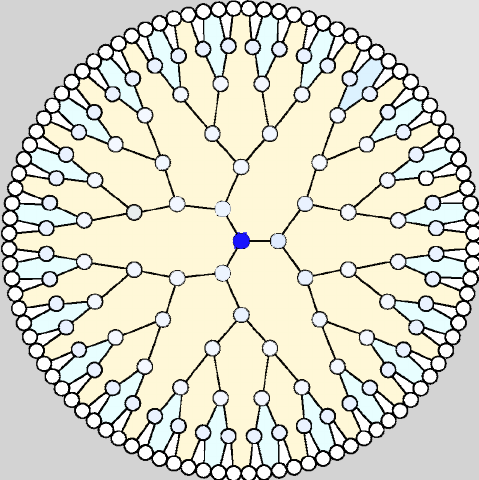}
		\end{minipage}}
		\hfill 	
		\subfloat[]{
			\begin{minipage}[c][1\width]{
					0.5\textwidth}
				\label{fig1:b}
				\centering
				\includegraphics[width=1\textwidth]{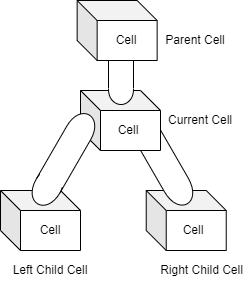}
		\end{minipage}}
		
		\caption{Cayley tree (a) height $h=7$ , (b) Current cell and it's neighbors}
		\label{fig:CayLAyTree}
	\end{figure}
	In the current work, we consider the Cayley tree of height $h$ with a finite number of cells. The distances from the root node to all the leaf nodes (boundary cells) are equal. The number of nodes in the grid is -
	\begin{align}\label{eq1}
		n &=\begin{cases}
			h  &\text{for } h \le 1 \\
			1 + \sum_{i=2}^{h} \eta^{i-2}(\eta+1)&\text{Otherwise}
		\end{cases}
	\end{align}
	In Fig.~\ref{fig:CayLAyTree}, the Cayley tree is of order $\eta=2$ and the height $h=7$, where each of the vertices is denoted as a cell. The total number of cells in the tree is $n=190$ (follow Eq.~\ref{eq1}). The nearest neighborhood comprises three cells. The root node (cell) is marked as \emph{blue}, and is placed in the center of the tree. The neighborhood is depicted in Fig.~\ref{fig1:b}, where each cell has three neighbors (for $\eta=2$). The root has no parent cell as such. Instead of the parent cell, the root has three children. This model follows null boundary conditions. That is, the missing neighbors of a leaf node are assumed to be in state $0$.
	\section{Proposed Computational Model}\label{model}
	The computational model reported in Section~\ref{cayley} is similar to a cellular automaton, where each cell uses a local rule $(f)$ to switch to its next state depending on the present states of its neighbors. Unlike classical CAs, however, a cell uses an additional function, say $g$, that operates on the previous states stored in the memory.
	
	Let us first define $f$. This is the local rule for the CA, which can decide on the next state for the cell. This rule is deterministic and, at any point of time, it is applied over all the cells uniformly. At each time step $t + 1$, this rule updates the state of a cell depending on the present states of its neighboring cells and the internal state - that is, the states of its parent, internal (determined by $g$) and the children.
	The function (rule) $g$ is defined in such a way that it has more computational power than $f$. Rule $g$ generates internal state $s_{internal}$ based on the contents of memory elements and the current state of the cell.
	
	\begin{figure}[h]
		\includegraphics[width=\linewidth]{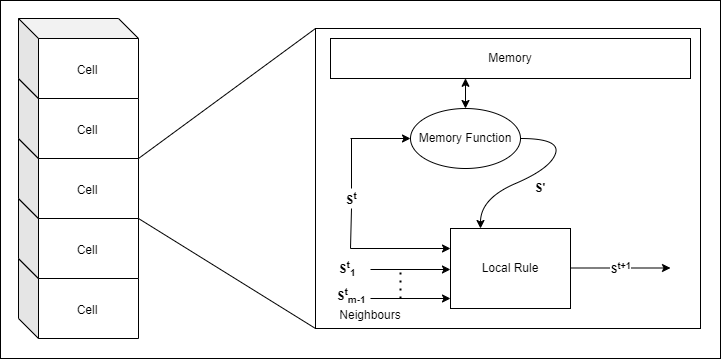}\centering
		\caption{ A typical cell}
		\label{fig:CaModel}
	\end{figure}
	
	The schematic of a typical cell is shown in Fig.~\ref{fig:CaModel} where, $s^t$ is the current state of the considering cell, $s^t_1 ,\cdots, s^t_{m-1}$ are the present sates of neighbors, $s^t_{internal}$ is the internal state and $m$ is the number of neighbors. Since it is desirable that $g$ has to be computationally more powerful than $f$, $g$ is allowed to use the memory of the cell. Let $S$ be the set of states that a cell uses and $\Psi$ be the set of memory symbols, then 
	$g : \mathcal{S} \times \Psi \rightarrow \mathcal{S} \times \Psi$.  
	Here $f$ is the function of present states of $m$ neighbors of the cell and an internal state generated by $g$. Thus, 
	\begin{align}\label{eq4}
		f&:\mathcal{S}^{m+1} \rightarrow \mathcal{S} 
	\end{align}
	Apart from the neighbor’s present states, the internal state is taken into cognizance by $f$ to generate the cells' next state. That is, a cell uses a finite memory which is accessed and modified by $g$. 
	
	The proposed model is defined over the Cayley tree. So, the parent and children of a node are the neighbors of the node. Here, the rule $f$ depends on the parent cell, self, the internal state generated by $g$, and its children ,
	\begin{align}\label{eq5}
		s^{t+1}_{self}  &= f(s^t_{parent}, s^t_{self}, s^t_{internal},s^t_{child_1},s^t_{child_2},\cdots,s^t_{child_\eta})\\
		s^t_{internal}  &= g(s^t_{self})
	\end{align}
	Eq.~\ref{eq5} shows that, for each cell, the next state of the cell at time $t+1$ ($s^{t+1}_{self}$) depends on the cell itself ($s^{t}_{self}$), its corresponding neighbors ($s^t_{parent},s^t_{child_1},s^t_{child_2},\\\cdots,s^t_{child_\eta}$), where $\eta$ is the order of the tree. and an internal state ($s^t_{internal}$) at time $t$. However, based on the cells' updated memory element, the internal state alters. The updating of the memory element depends on the current state of the cell ($s^t_{self}$). Note that, the number of children varies for different orders($\eta$) of the tree. For example, if the order of the tree is $2$ (for $\eta=2$) then, the model considers the state of two children ($s^t_{child_1},s^t_{child_2}$). For the root, there is no parent but the number of children is $\eta+1$. For the leaves, on the other hand, there is no child.
	
	If a cell itself is its neighbour, then the state of the cell and a $f$ is the function of the internal state. To get configurations in this automaton, we need to take the set of memory elements into consideration, along with the set of states. Hence, here a configuration is an assignment $ c : \mathcal{L} \rightarrow \mathcal{S} \times \Psi$. Let us define a observable configuration ($c_\pi$) of a configuration $c$ as $c_\pi\: :\: \mathcal{L} \rightarrow \mathcal{S}$. During computation, observable change occurs at the configuration ($c$). At each time $t \in \mathbb{N}$, a cell is assigned a observable state and we denoted by $ \mathcal{S}=\{0,1,2\}$. Rule $f$ and $g$ are defined as $-$ In traditional CA, the additional function $g$ is void and $f$ takes the role of computation.
	
	\begin{figure}[!htbp]
		\subfloat[]{
			\begin{minipage}[c][1\width]{
					0.24\textwidth}
				\label{sim:a}
				\centering
				\includegraphics[width=1\textwidth]{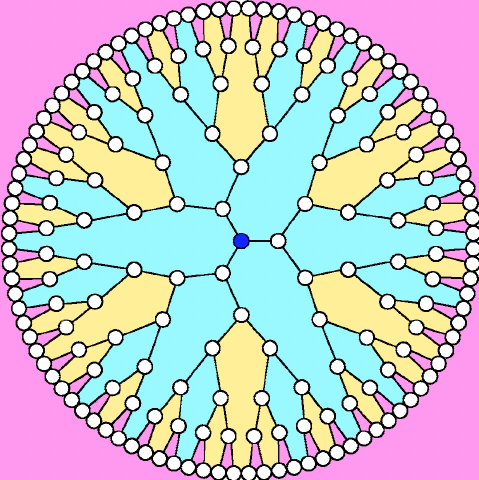}
		\end{minipage}}
		\hfill 	
		\subfloat[]{
			\begin{minipage}[c][1\width]{
					0.24\textwidth}
				\label{sim:b}
				\centering
				\includegraphics[width=1\textwidth]{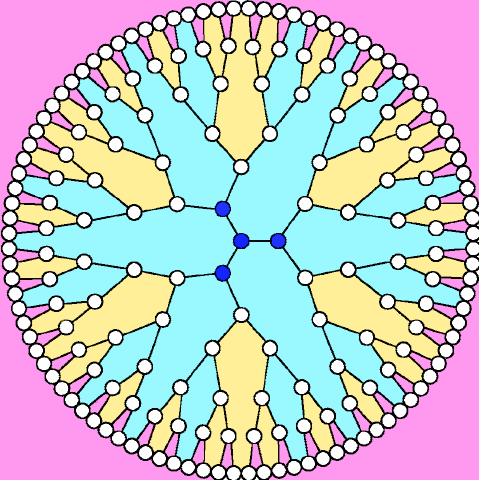}
		\end{minipage}}
		\hfill
		\subfloat[]{
			\begin{minipage}[c][1\width]{
					0.24\textwidth}
				\label{sim:c}
				\centering
				\includegraphics[width=1\textwidth]{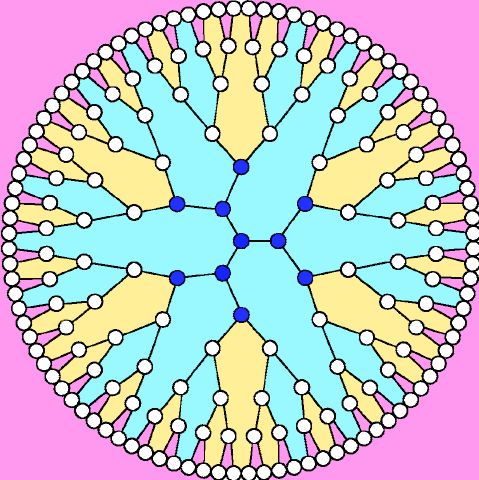}
		\end{minipage}}
		\hfill 	
		\subfloat[]{
			\begin{minipage}[c][1\width]{
					0.24\textwidth}
				\label{sim:d}
				\centering
				\includegraphics[width=1\textwidth]{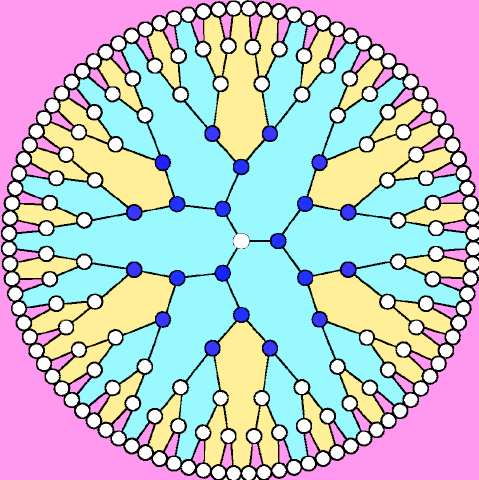}
		\end{minipage}}
		\vfill
		\subfloat[]{
			\begin{minipage}[c][1\width]{
					0.24\textwidth}
				\label{sim:e}
				\centering
				\includegraphics[width=1\textwidth]{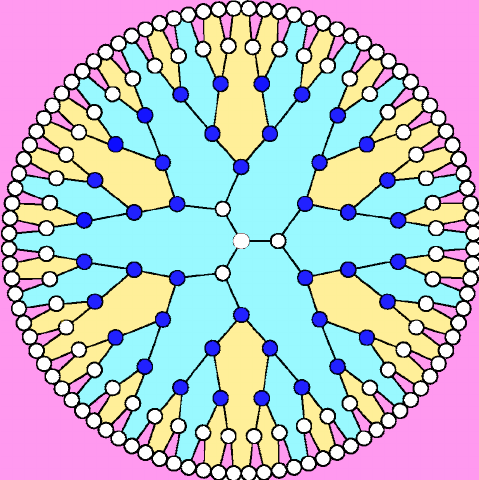}
		\end{minipage}}
		\hfill 	
		\subfloat[]{
			\begin{minipage}[c][1\width]{
					0.24\textwidth}
				\label{sim:f}
				\centering
				\includegraphics[width=1\textwidth]{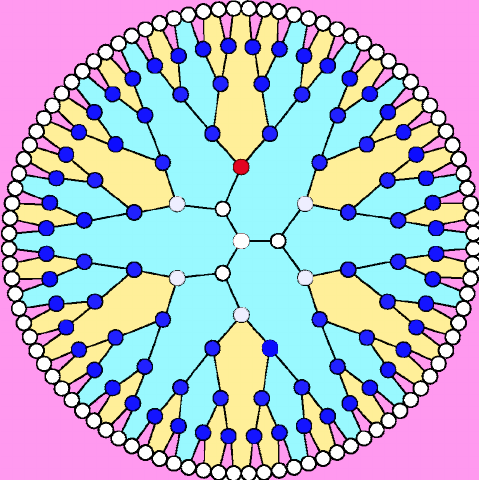}
		\end{minipage}}
		\hfill
		\subfloat[]{
			\begin{minipage}[c][1\width]{
					0.24\textwidth}
				\label{sim:g}
				\centering
				\includegraphics[width=1\textwidth]{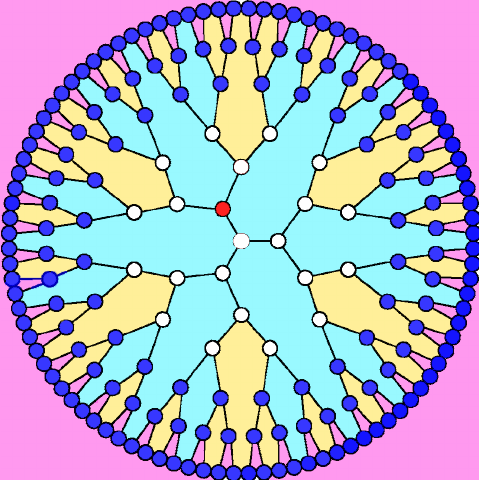}
		\end{minipage}}
		\hfill 	
		\subfloat[]{
			\begin{minipage}[c][1\width]{
					0.24\textwidth}
				\label{sim:h}
				\centering
				\includegraphics[width=1\textwidth]{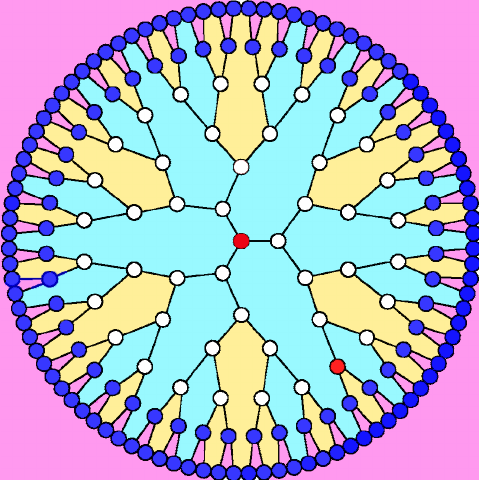}
		\end{minipage}}
		
		\vfill	\subfloat[]{
			\begin{minipage}[c][1\width]{
					0.24\textwidth}
				\label{sim:i}
				\centering
				\includegraphics[width=1\textwidth]{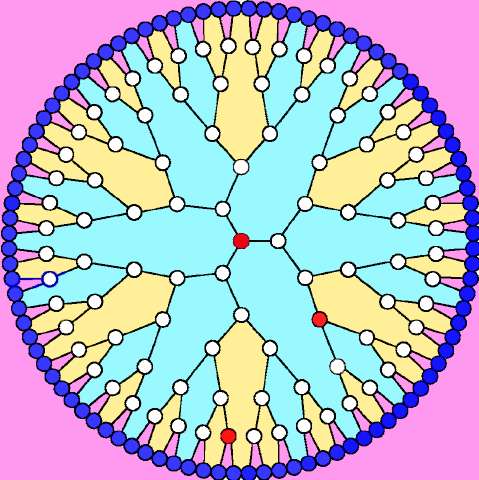}
		\end{minipage}}
		\hfill 	
		\subfloat[]{
			\begin{minipage}[c][1\width]{
					0.24\textwidth}
				\label{sim:j}
				\centering
				\includegraphics[width=1\textwidth]{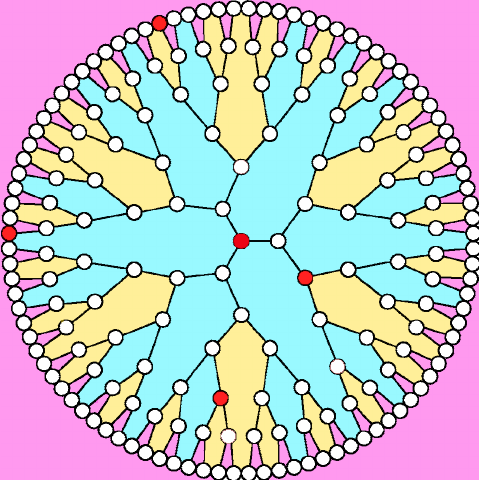}
		\end{minipage}}
		\hfill
		\subfloat[]{
			\begin{minipage}[c][1\width]{
					0.24\textwidth}
				\label{sim:k}
				\centering
				\includegraphics[width=1\textwidth]{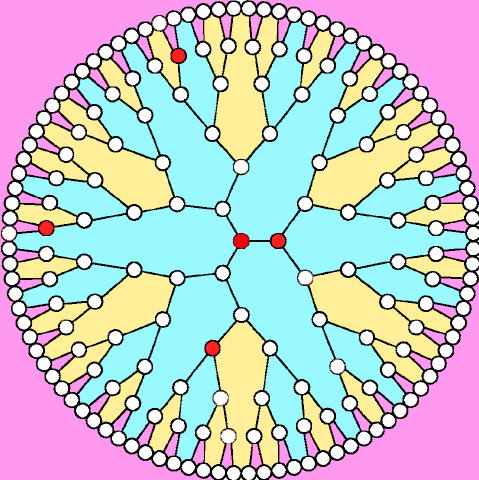}
		\end{minipage}}
		\hfill 	
		\subfloat[]{
			\begin{minipage}[c][1\width]{
					0.24\textwidth}
				\label{sim:l}
				\centering
				\includegraphics[width=1\textwidth]{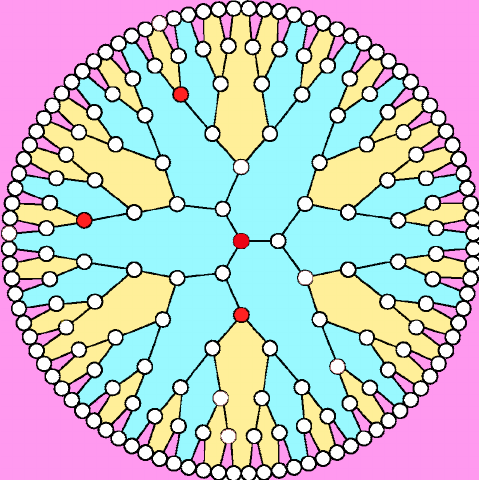}
		\end{minipage}}
		\vfill
		\subfloat[]{
			\begin{minipage}[c][1\width]{
					0.24\textwidth}
				\label{sim:m}
				\centering
				\includegraphics[width=1\textwidth]{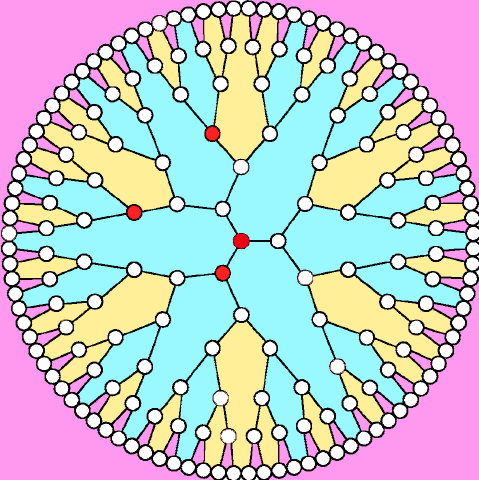}
		\end{minipage}}
		\hfill 	
		\subfloat[]{
			\begin{minipage}[c][1\width]{
					0.24\textwidth}
				\label{sim:n}
				\centering
				\includegraphics[width=1\textwidth]{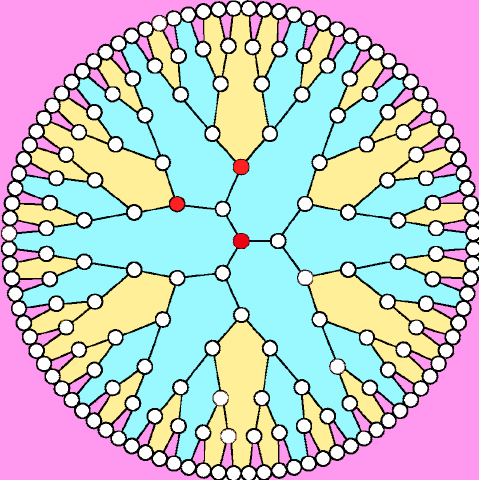}
		\end{minipage}}
		\hfill
		\subfloat[]{
			\begin{minipage}[c][1\width]{
					0.24\textwidth}
				\label{sim:o}
				\centering
				\includegraphics[width=1\textwidth]{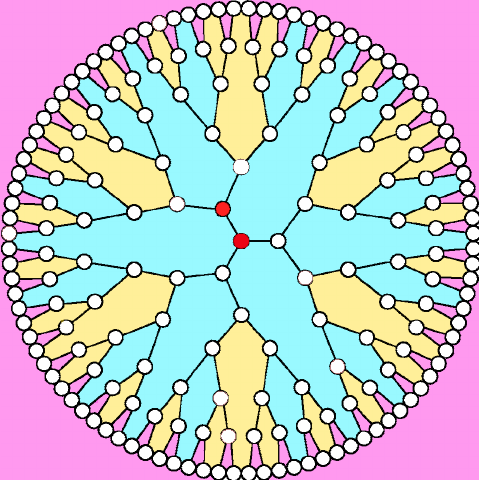}
		\end{minipage}}
		\hfill 	
		\subfloat[]{
			\begin{minipage}[c][1\width]{
					0.24\textwidth}
				\label{sim:p}
				\centering
				\includegraphics[width=1\textwidth]{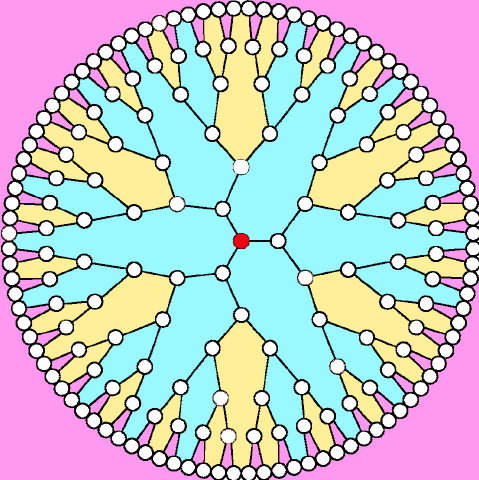}
		\end{minipage}}	
		\caption{Simulation of Searching , where \emph{red} = state 2; \emph{blue} = state $1$; \emph{white} = state $0$. (a) Initial Configuration , (p) Final Configuration.}
		\label{fig:simulation}
	\end{figure}
	
	An initial configuration is given and, after the computation, it converges to some fixed points. The output is taken from that configuration, reachable from the initial configuration. In the initial configuration, the cells are assigned to some states, and the memory of each cell is assigned some memory symbols. Hence, in computation, the rule $f$ of the cells apparently plays the decisive role in computation, but the function $g$ plays a crucial role to realize a fruitful computation on the internal memory. 
	\noindent 
	\section{In-Memory Searching}\label{in-memory}
	\noindent In \emph{In-Memory Computing}, the target is to reduce the workload of a CPU. The data-intensive computations are run in memory. In this section, we report the realization of \emph{In-Memory Computing} around the computing model, introduced in Section~\ref{model}, with Cayley Tree.
	\subsection{Overview}\label{overview}
	For searching, the key $k$ is stored at the root of the Cayley Tree. The elements are distributed over the nodes' memory (CA cells' memory). The tree is constructed in such a way that the distance between the root node and each leaf is equal. There should be more nodes (cells) in the tree than the number of elements to be dealt with. The elements are scattered throughout the cells' memory randomly. The key is stored in the memory of the root. $0$ is placed in the rest of the cells' memory. The memory unit of a node can only retain one element. Initially, the state of the root is set to $1$ and other cells are set to $0$; the internal state of all the cells is set to $0$ if the memory contains $0$, set to $1$ otherwise, (this is considered the \emph{initial configuration} of the model).
	
	The search is composed of two modules-
	\begin{itemize}
		\item[] A. Computation at the root node:
		
		A reduction operation on $k$ by $1$ is done at the root. The root then sends $1$ to its children and sets its next state as $0$ if the updated $k$ is $0$, else it is set to $1$. This process is repeated until the $k$ becomes $0$. However, if the state of any child is $2$ then the next state of the root node switches to $2$; otherwise, it remains $0$.
		\item[] B. Computation at other nodes:
		
		For every time step, when the state of a node is $1$ (respectively $2$) then the node transmits its present state to its children (respectively parent). It then decrements the memory content (element) if the present state is $1$. The next state of the node depends on its nearest neighbor-
		\begin{itemize}
			\item[]Case 1$-$ If the state of any child is $2$ then the next state of the node is set to $2$, otherwise, it follows the parent state (follow \emph{case 2}).
			\item[]Case 2$-$ The next state of the node is set to $1$ if the parent's state is $1$; otherwise, it depends on the memory element (follow\emph{case 3}).
			\item[]Case 3$-$  If the updated memory contains $0$ (internal state is set to $0$), then the next state of the node switches to $2$, otherwise the next state switches to $0$.
			\item[]Case 4$-$ If none of the above criteria match, then the next state remains the same as the present state.
		\end{itemize}
		The CA converges when the states of all the nodes are $0$ and the root node is either $0$ or $2$. If the root node is $2$, the execution concludes with the positive result \emph{"Found"}, but if it is not, the execution ends with the negative output  \emph{"Not Found"}.
	\end{itemize}
	The memory modification is done by $g$ as:
	\begin{align}\label{eq6}
		Memory Element=\begin{cases}
			Memory Element-1 & if \:s^t_{self} = 1   \\
			Memory Element & Otherwise
		\end{cases} 
	\end{align}
	\begin{algorithm}[h]
		\caption{Rule $f$ for the root node:}
		\begin{algorithmic}
			\State Taking states at time $t$ : $s^t_{self}, s^t_{internal}\: and\: s^t_{child_1},s^t_{child_2},\cdots,s^t_{child_\eta}$ as \textbf{input}
			\If{state of any child is $2$} 
			\State $NS \gets 2$
			\ElsIf{$s^t_{self}==2$}
			\State $NS \gets 2$
			\ElsIf{$ s^t_{self}==1$}
			\State Compute $g$
			\If {$s^t_{internal}== 1$}
			\State $NS \gets 1$
			\Else
			\State $NS \gets 0$
			\EndIf
			\Else
			\State $NS \gets s^t_{self}$
			\EndIf
			\State $s^{t+1}_{self} \gets NS$
			\State \textbf{Output : }$s^{t+1}_{self}$ 
		\end{algorithmic}
	\end{algorithm}
	
	The internal state modification is done by $g$ as:
	\begin{align}\label{eq7}
		s^t_{internal}&=\begin{cases}
			0  &\text{if Memory Element}  = 0 \\
			1 &\text{Otherwise}
		\end{cases}
	\end{align}
	
	\begin{algorithm}[h]
		\caption{Rule $f$ for other nodes:}
		\begin{algorithmic}
			\State Taking states at time $t$ : $s^t_{parent}, s^t_{self}, s^t_{internal} \:\: \text{ and } \:\: s^t_{child_1},s^t_{child_2},\cdots,s^t_{child_\eta}$  as \textbf{input}
			\If{state of any child is $2$} 
			\State $NS \gets 2$
			\ElsIf{$s^t_{self}==2$}
			\State $NS \gets 0$
			\ElsIf{$s^t_{parent}==2$}
			\State $NS \gets s^t_{self}$
			\ElsIf{$s^t_{parent}==0 \:and\: s^t_{self}==0$}
			\State $NS \gets 0$
			\ElsIf{$s^t_{parent}==0 \:and\: s^t_{self}==1$}
			\State Compute $g$
			\If {$s^t_{internal}==0$}
			\State $NS \gets 2$
			\Else
			\State $NS \gets 0$
			\EndIf
			\ElsIf{$s^t_{parent}==1 \:and\: s^t_{self}==0$}
			\State $NS \gets 1$
			\ElsIf{$s^t_{parent}==1 \:and\: s^t_{self}==1$}
			\State Compute $g$
			\State $NS \gets 1$
			\EndIf
			\State $s^{t+1}_{self} \gets NS$
			\State \textbf{Output : }$s^{t+1}_{self}$ 
		\end{algorithmic}
	\end{algorithm}
	The leaf nodes have the same $f$ as the other nodes; the only distinction is that because there are no children, the next state is determined by the current state, the state of the parent, and the newly updated internal state.
	
	The computational model is better described with an example run in Fig.~\ref{fig:simulation}. Fig.~\ref{fig:simulation} is a Cayley tree of order two, and the height is $7$. Here, the root node emits consecutive $1$s towards the leaves. The number of $1$s depends on the key element($k$) stored in the root's memory. In this simulation, we assume that the key $k=3$. That is, the root emits three consecutive $1$s and then it goes to state $0$. The same rule is followed when passing these three consecutive $1$s through the nodes. When three $1$s pass through, the memory components on all nodes are reduced by $3$ units. A node then switches to state $2$ if the memory element of the node becomes $0$, and switches to state $0$ otherwise. 
	\begin{figure}[!htbp]
		\subfloat[]{
			\begin{minipage}[c][1\width]{
					0.50\textwidth}
				\label{ex:cell}
				\centering
				\includegraphics[width=1\textwidth]{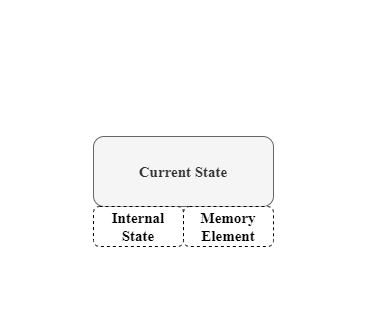}
		\end{minipage}}
		\hfill
		\subfloat[]{
			\begin{minipage}[c][1\width]{
					0.50\textwidth}
				\label{ex:str}
				\centering
				\includegraphics[width=1\textwidth]{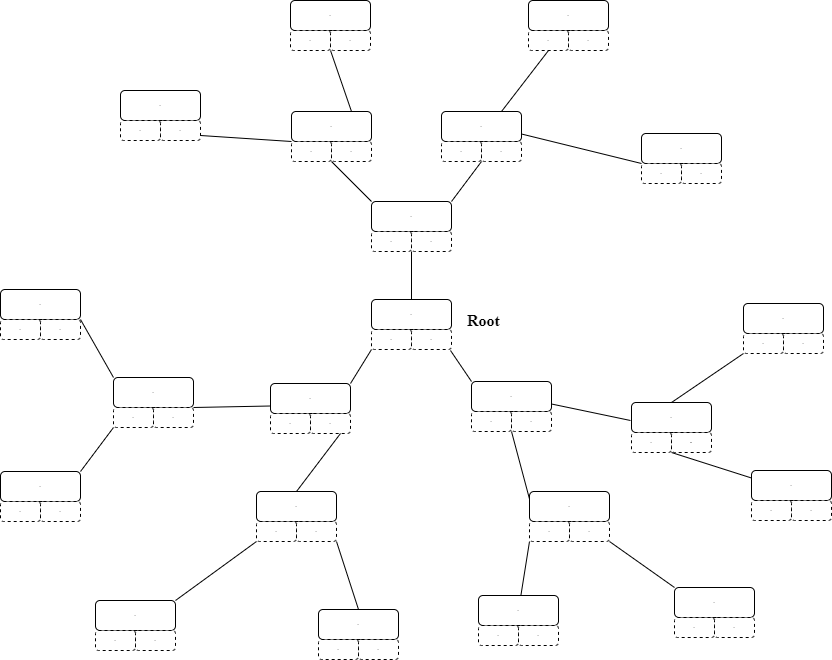}
		\end{minipage}}
		\caption{Cayley Tree of order $\eta=2$ and height $h=4$ (a) Overview of a single node , (b) Structure of the tree.}
		\label{fig:init}
	\end{figure}
	
	If the state of any node changes to $2$ then the next time step, the node sends $2$ to its parent and switches to state $0$ (see Fig.~\ref{sim:g}). In these ways, $2$ is reached to the root. If the root's state switches to $2$ (see Fig.~\ref{sim:h}) then we can consider it as the key \emph{Found}. Note that, after $k+2\times h$ time steps, if the root node's state remains $0$, the key should be regarded as \emph{Not Found}.
	
	Fig~\ref{sim:a} shows the initial configuration with all $0$s nodes and the state of the root is set to $1$. Computation starts from the initial configuration to the final configuration (see  Fig~\ref{sim:p}), where the root is in state $2$ (\emph{red marked}), which means \emph{Found}.
	\begin{example}
		The model is run with $20$ elements ($\Lambda=20$), $X=\{5,4,7,3,\\5,11,3,4,9,3,9,5,2,1,3,7,9,2,5,7\}$ and the key $k$ is $3$. We consider a Cayley tree of height $h=4$ and order $\eta=2$ (see Fig.~\ref{ex:str}), where the number of nodes in the tree is $22$ (see Eq.~\ref{eq1}). The elements are distributed over nodes and key is stored in the root's memory (see Fig.~\ref{ex:a}). The Fig.~\ref{fig:example2} illustrates the detailed execution, where Fig.~\ref{ex:str} shows the structure of the tree and Fig.~\ref{ex:cell} describes each node. Fig.~\ref{ex:a} depicts the initial configuration for the search, Fig.~\ref{ex:j} depicts the configuration where the CA converges. Note that, in Fig.~\ref{ex:h}, the observable state of the root node is $2$, indicates that $k \in X$. In other words, this configuration (see Fig.~\ref{ex:h}) concludes that the key is \emph{Found}.
	\end{example}
	
	\begin{figure}[!htbp]
		\subfloat[]{
			\begin{minipage}[c][1\width]{
					0.50\textwidth}
				\label{ex:a}
				\centering
				\includegraphics[width=1\textwidth]{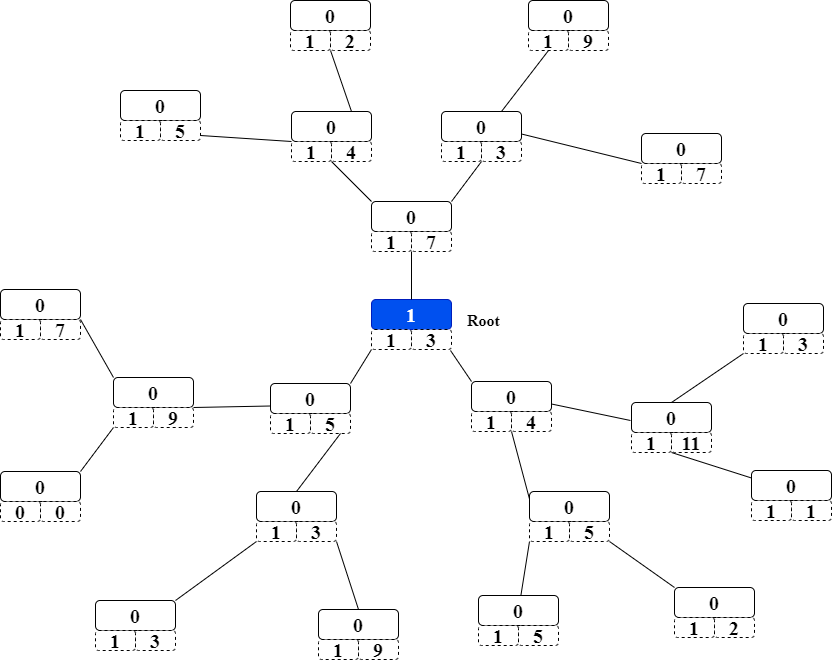}
		\end{minipage}}
		\hfill 	
		\subfloat[]{
			\begin{minipage}[c][1\width]{
					0.50\textwidth}
				\label{ex:b}
				\centering
				\includegraphics[width=1\textwidth]{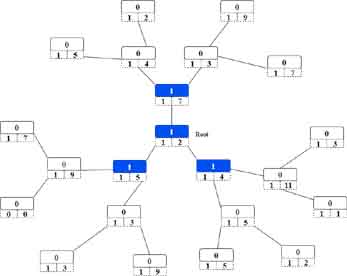}
		\end{minipage}}
	\vfill
		\subfloat[]{
			\begin{minipage}[c][1\width]{
					0.50\textwidth}
				\label{ex:c}
				\centering
				\includegraphics[width=1\textwidth]{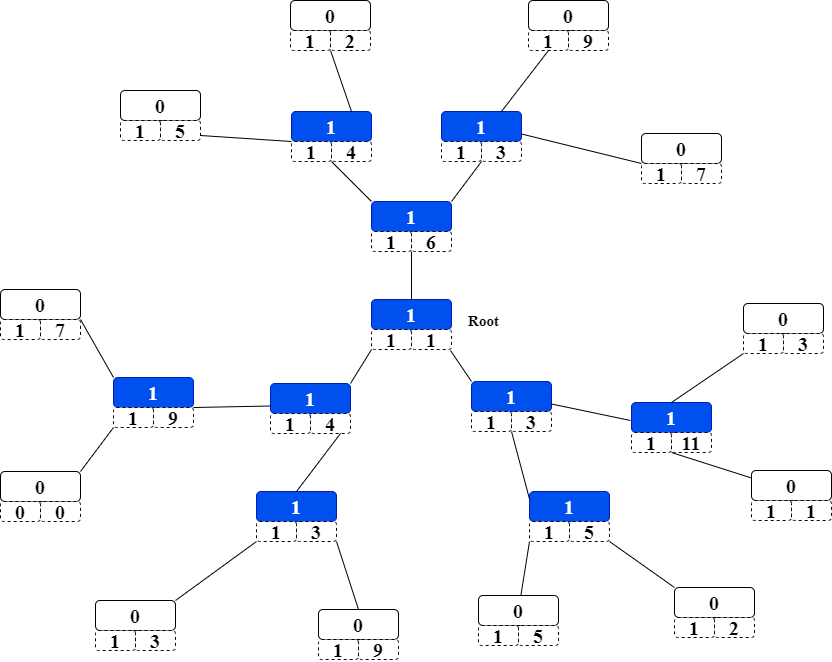}
		\end{minipage}}
		\hfill 	
		\subfloat[]{
			\begin{minipage}[c][1\width]{
					0.50\textwidth}
				\label{ex:d}
				\centering
				\includegraphics[width=1\textwidth]{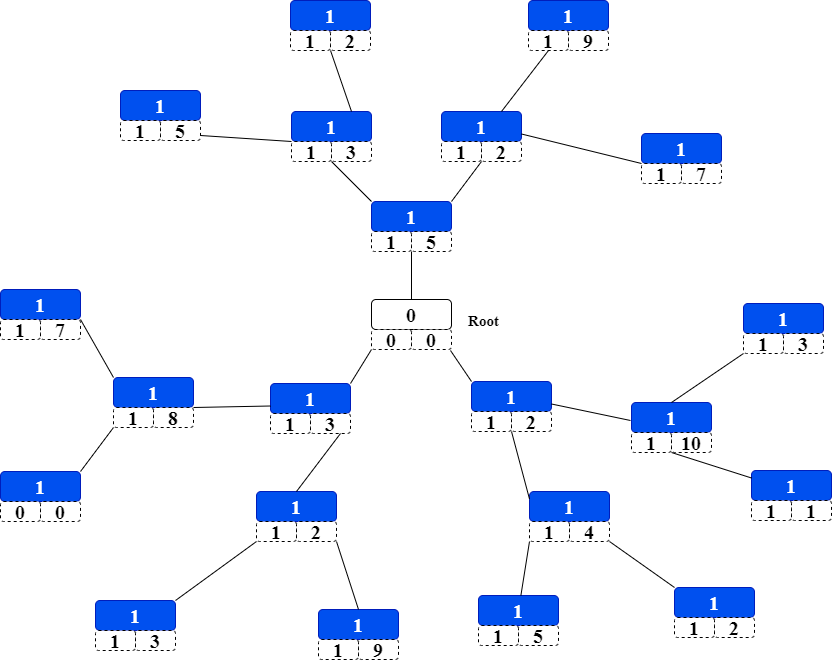}
		\end{minipage}}
		\caption{In-Memory Searching; (a) Initial Configuration.}
	\label{fig:example0}
\end{figure}
\begin{figure}[!htbp]\ContinuedFloat
		\subfloat[]{
			\begin{minipage}[c][1\width]{
					0.50\textwidth}
				\label{ex:e}
				\centering
				\includegraphics[width=1\textwidth]{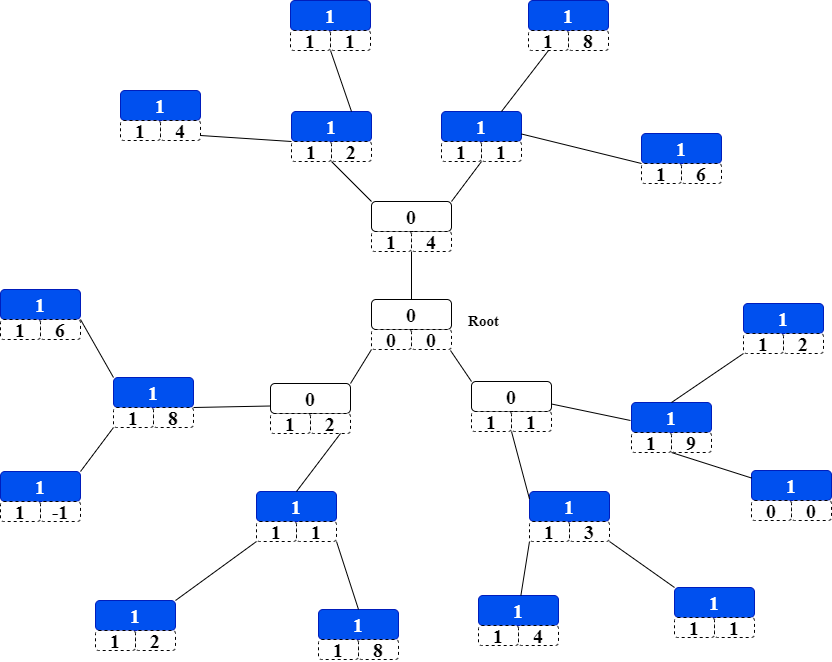}
		\end{minipage}}
		\hfill 	
		\subfloat[]{
			\begin{minipage}[c][1\width]{
					0.50\textwidth}
				\label{ex:f}
				\centering
				\includegraphics[width=1\textwidth]{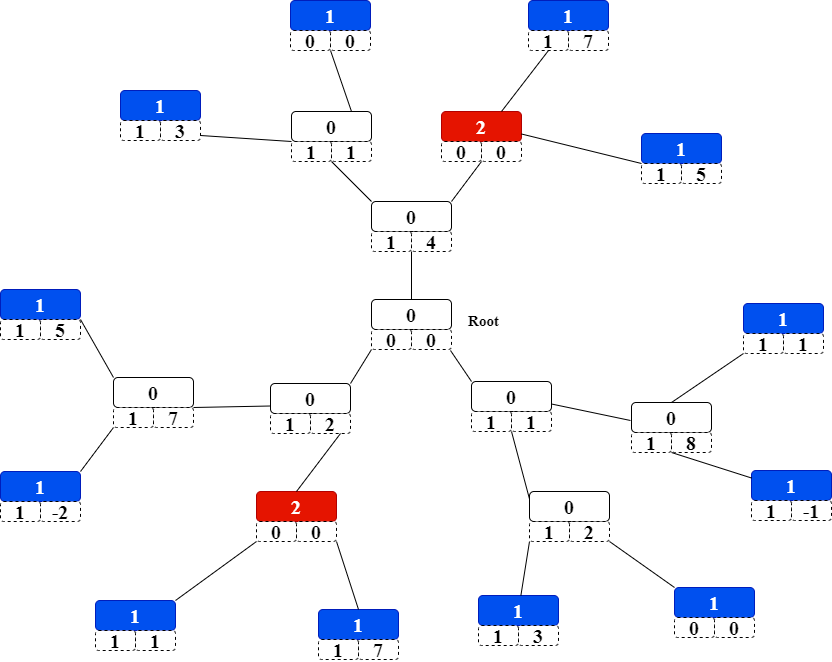}
		\end{minipage}}
		\vfill
		\subfloat[]{
			\begin{minipage}[c][1\width]{
					0.50\textwidth}
				\label{ex:g}
				\centering
				\includegraphics[width=1\textwidth]{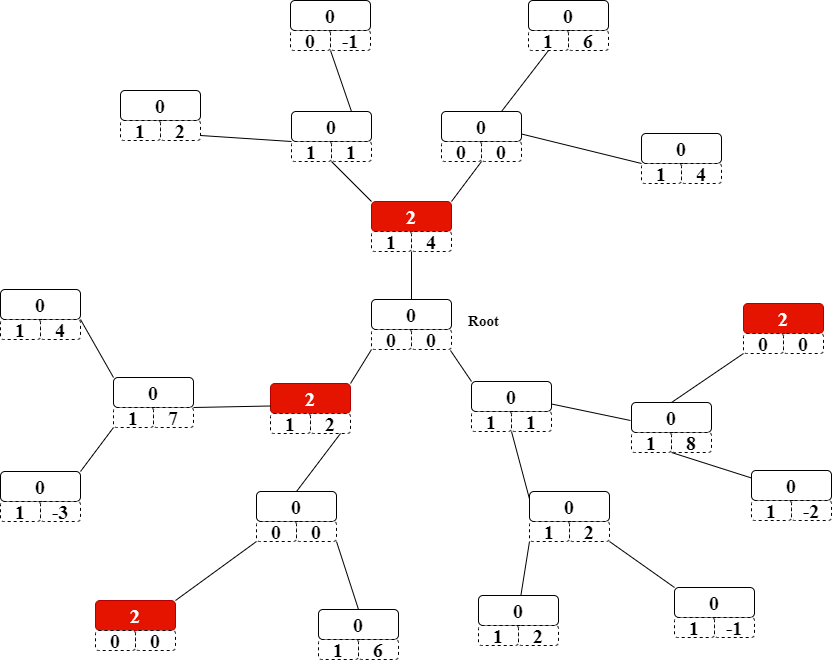}
		\end{minipage}}
		\hfill 	
		\subfloat[]{
			\begin{minipage}[c][1\width]{
					0.50\textwidth}
				\label{ex:h}
				\centering
				\includegraphics[width=1\textwidth]{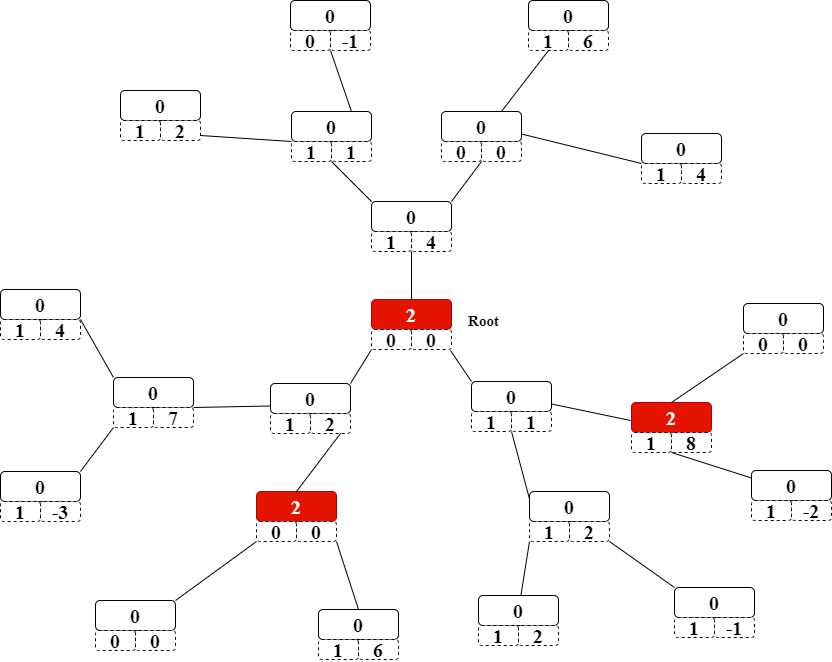}
		\end{minipage}}
		\caption{In-Memory Searching; (h) The root node switches to state $2$.}
		\label{fig:example1}
	\end{figure}
	\begin{figure}[!htbp]\ContinuedFloat
		\subfloat[]{
			\begin{minipage}[c][1\width]{
					0.50\textwidth}
				\label{ex:i}
				\centering
				\includegraphics[width=1\textwidth]{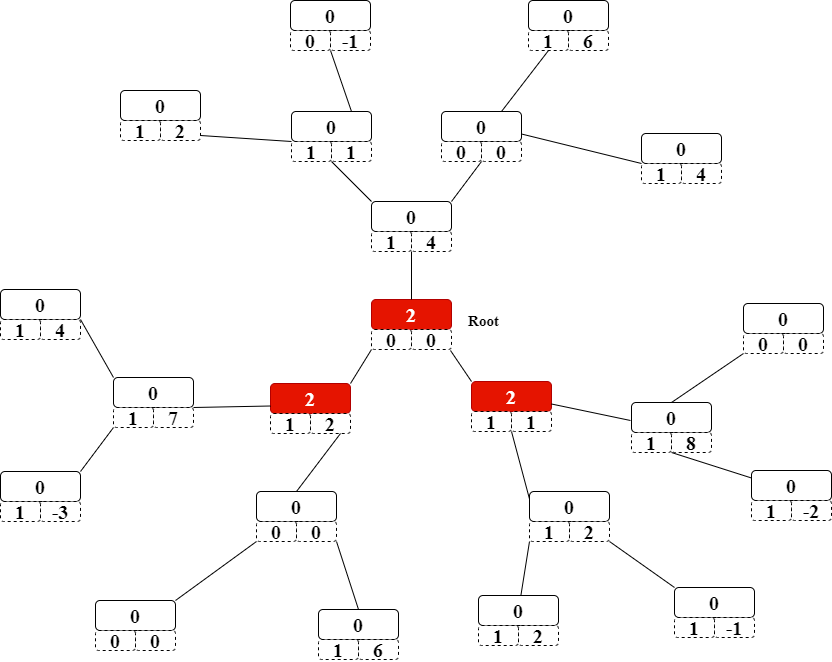}
		\end{minipage}}
		\hfill 	
		\subfloat[]{
			\begin{minipage}[c][1\width]{
					0.50\textwidth}
				\label{ex:j}
				\centering
				\includegraphics[width=1\textwidth]{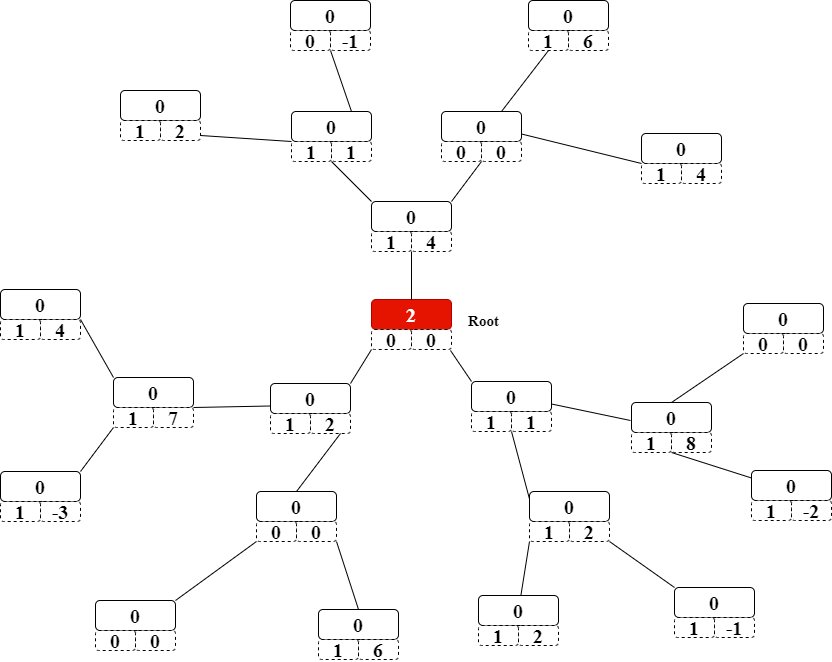}
		\end{minipage}}
		\caption{In-Memory Searching; (j) Final Configuration (where the CA converges).}
		\label{fig:example2}
	\end{figure}
	This searching scheme implements parallelism in the computation. The nodes' local interactions carry out the entire process. If $n$ is the number of elements and $k$ is the searching key, then the basic intuition behind finding the complexity can be that, as the root decrements the key for $k$ times and then wait for the response from the children for $log(n)$ time, thus the total time complexity becomes $O(k+log(n))$.
	
	The arrangement of elements in the array does not affect the flow of the scheme. No matter if the elements in the array are already sorted, reverse sorted, or randomly placed, the scheme works the same for all these cases, and thus the time complexity for all such cases is the same, $O(k+log(n))$.
	
	The most important issue is by sensing only the root node, the outcome of a search is determined. That is, if the root is in state $2$, then \emph{Found}, otherwise \emph{Not Found}.
	\section{Summary}
	In this study, a new type of cellular automata model has been proposed to reduce the workload of the CPU. Each cell of the CA has a memory and an additional processing unit attached to it. This is the initial step in the introduction of a new model capable of resolving computational issues. In this work, a CA was implemented over a Cayley tree, with each node denoting a cell. Each CA cell has some additional processing capability, which aids in resolving the given situation.
	
	We've shown that the model can deal with the \emph{Searching Problem}. By sensing only one node, the model can decide the outcome of the search. It demonstrates the model's efficacy. The focus of our next study is to find the solutions to other computational problems, i.e., the sorting problem, finding the greatest number, etc. 


\chapter{Conclusion}
\label{chap7}
This chapter aims to summarize the thesis' important contributions and to discuss the potential future directions for study in the subject of cellular automata, which has the potential to bring scientists from many academic fields from across the world.
\section{Main Contribution}
The primary goal of this research work has been to study the dynamics of temporally stochastic cellular automata (TSCAs). In this thesis, the TSCAs have been utilized to classify patterns and used to define affinity classification problem. Moreover, the thesis has explored a new kind of CA that has been proposed to reduce the workload of the CPU. Each cell of the CA has a memory and an additional processing unit attached to it. The CA was implemented over a Cayley tree, with each node denoting a cell. We've demonstrated that the model is capable of handling the searching problem. The model can decide the search's conclusion by sensing just one node (the root node).

A brief survey of cellular automata, elementary cellular automata, artificial life, and computational and societal applications of cellular automata that are relevant to this research work is briefly reviewed in Chapter~\ref{chap2}. Introduction of Temporally Stochastic Elementary Cellular Automata, its classes and dynamics are reported in Chapter~\ref{chap3}. This was the first step in the exploration of the spaces of the temporally stochastic CAs. Firstly, we have identified that some of the stochastic CAs are affected by the temporal noise, but they are not sensitive to temporal noise rate. However, even these CAs have shown a diverse set of results. On the other hand, temporally stochastic CAs that are sensitive to temporal noise have demonstrated phenomena like phase transition and class transition. It is noteworthy that stochastic CAs with (at least one) chaotic rule have often displayed lower resistance during phase transition (i.e. the critical value of the noise rate is low). But during phase transition, the stochastic CAs devoid of any chaotic rule have shown greater resilience (i.e. critical value of noise rate is high). This is another exciting finding from the study.

In Chapter~\ref{chap4}, we have proposed a variant of CAs, termed Temporally Stochastic CAs (TSCAs), in which, instead of one local rule, two rules (default rule f and noise rule g) are utilized. After analyzing their dynamics, we have identified the convergent TSCAs that have been used to design two-class pattern classifiers. In this context, in comparison to existing common algorithms, the proposed design of a TSCA-based two-class pattern classifier offers competitive performance.

Chapter~\ref{chap5} introduces a new type of problem known as the affinity classification problem, where we build a dedicated machine that is integrated into a two-dimensional cellular automaton with periodic boundary conditions and Moore neighborhood dependence. Our model may be described by the four parameters $K, \phi(x), \psi(x),$ and $p$ and has affection capabilities up to a convergence point, all-1 or all-0. We can lead to a self-system using this paradigm. We are aware that any species may live and evolve due to self-healing. We may argue that our model behaves something like a naturally occurring biological system and that it has become intelligent since it has this property and makes decisions democratically.

The last chapter, Chapter~\ref{chap6}, was devoted to exploring how a new model may be used to resolve a well-known search problem. In this model, each CA cell has some additional processing capability, which aids in resolving the given situation.

\section{Future Directions}
Here are a few intriguing future research initiatives that may be conducted as a result of our current work,
\begin{enumerate}
	\item The most comprehensive future possibilities of these temporally stochastic CAs are discussed in Chapter~\ref{chap3}. However, in this research, our primary method of investigating these CAs was experimental. Therefore, there is still a need for investigation into the precise theoretical explanation of the temporally stochastic CA.
	\item The natural extensions of the work, which is discussed in Chapter~\ref{chap4}, on the sensitivity to temporally stochastic CAs and pattern classification, include,
	\begin{itemize}
		\item Here, we have only experimentally explored the convergent TSCAs. What
		can be said for theoretical understanding behind the convergence?
		\item What can be said for classification time (i.e., convergence time) of these
		TSCAs?
		
	\end{itemize}
	\item In Chapter~\ref{chap5}, we must decide whether our model possesses other aspects of one ’s life that an intelligent machine must have. The Moore neighborhood is the only one that has been taken into consideration here, therefore it is still unclear what type of behavior may result from changing the area's dependence on the regulations. By changing the parameters of our model, several additional behaviors could appear. In addition, our concept may be applicable for a number of additional applications outside self-healing systems. These questions still need to be answered in the future.
	\item Chapter~\ref{chap6} shows the model's ability to address the searching problem. The model may decide the search's conclusion by sensing just one node. It proves how effective the model is. Finding answers to further computational challenges, such as the sorting problem, the greatest number problem, etc., is the main goal of our next research.
\end{enumerate}




\cleardoublepage
\phantomsection
\addcontentsline{toc}{chapter}{Author's Publications}
\chapter*{Author's Statements}
\label{chap8}

\hspace{0.1in}\textbf{Accepted Paper}\\
\begin{itemize}
\item Souvik Roy, Subrata Paul and Sukanta Das, ``Temporally Stochastic Elementary Cellular Automata : classes and dynamics'' \textit{International Journal of Bifurcation and Chaos}, 2022.
\end{itemize}

\vspace{0.4in}
\hspace{0.0in}\textbf{Preprint}\\
\begin{itemize}
	\item Kamalika Bhattacharjee, Subrata Paul and Sukanta Das, ``Affinity Classification Problem by Stochastic Cellular Automata'' \textit{arXiv preprint   arxiv.2207.05446}, 2022.
\end{itemize}

\vspace{0.4in}
\hspace{0.0in}\textbf{Submitted Papers}\\
\begin{itemize}
\item Subrata Paul, Souvik Roy and Sukanta Das, ``Pattern Classification with Temporally
Stochastic Cellular Automata'' \textit{AUTOMATA 2022} , 2022.

\item Subrata Paul, Sukanta Das and Biplab K. Sikdar, ``Searching with Cellular Automata on Cayley Tree'' \textit{AUTOMATA 2022} , 2022.
\end{itemize}

\pagestyle{plain}
\bibliographystyle{IEEEtran}

\cleardoublepage
\phantomsection
\addcontentsline{toc}{chapter}{Bibliography}
\bibliography{Thesis}
\end{document}